\documentclass[iop,twocolappendix]{emulateapj}

\usepackage{multirow}
\usepackage{subfigure}
\usepackage{natbib}
\usepackage{graphicx}
\usepackage{amsmath}

\def \hh{H$_2$}
\def \hho{H$_2$O}
\def \hhso{H$_2^{16}$O}
\def \hheo{H$_2^{18}$O}

\slugcomment{Draft: \today}

\shorttitle{Water absorption in the Milky Way}
\shortauthors{Flagey et al.}

\begin{document}

\title{Water absorption in Galactic translucent clouds:  conditions and history of the gas derived from Herschel$^\star$/HIFI {\it PRISMAS} observations}
\thanks{$^\star${\it Herschel} is an ESA space observatory with science instruments provided by European-led Principal Investigator consortia and with important participation from NASA}

\author{N. Flagey\altaffilmark{1}}
\author{P. F. Goldsmith\altaffilmark{1}}
\author{D. C. Lis\altaffilmark{2}}
\author{M. Gerin\altaffilmark{3}}
\author{D. Neufeld\altaffilmark{4}}
\author{P. Sonnentrucker\altaffilmark{5}}
\author{M. De Luca\altaffilmark{3}}
\author{B. Godard\altaffilmark{3}}
\author{J. R. Goicoechea\altaffilmark{6}}
\author{R. Monje\altaffilmark{2}}
\author{T. G. Phillips\altaffilmark{2}}

\email{nflagey@jpl.nasa.gov}

\altaffiltext{1}{Jet Propulstion Laboratory, California Institute of Technology, 4800 Oak Grove Drive, Pasadena, CA 91109, USA}
\altaffiltext{2}{California Institute of Technology, 1200 East California Boulevard, Pasadena, CA 91125, USA}
\altaffiltext{3}{LERMA, UMR 8112 du CNRS, Observatoire de Paris, \'Ecole Normale Sup\'erieure, UPMC \& UCP, France}
\altaffiltext{4}{Dept. of Physics \& Astronomy, Johns Hopkins Univ. 3400 N. Charles St., Baltimore, MD 21218, USA}
\altaffiltext{5}{Space Telescope Science Institute, 3700 San Martin Drive, Baltimore, MD 21218, USA}
\altaffiltext{6}{Centro de Astrobiolog\'ia (CSIC-INTA), 28850, Torrej\'on de Ardoz, Madrid, Spain}

\begin{abstract}
We present Herschel/HIFI observations of the three ground state transitions of \hho\ (556, 1669 and 1113~GHz) and \hheo\ (547, 1655 and 1101~GHz) -- as well as the first few excited transitions of \hho\ (987, 752 and 1661~GHz) -- towards six high-mass star-forming regions, obtained as part of the PRISMAS (PRobing InterStellar Molecules with Absorption line Studies) Guaranteed Time Key Program. Water vapor associated with the translucent clouds in Galactic arms is detected in absorption along every line of sight in all the ground state transitions. The continuum sources all exhibit broad water features in emission in the excited and ground state transitions. Strong absorption features associated with the source are also observed at all frequencies except 752~GHz. We model the background continuum and line emission to infer the optical depth of each translucent cloud along the lines of sight. We derive the column density of \hho\ or \hheo\ for the lower energy level of each transition observed. The total column density of water in translucent clouds is usually about a few $10^{13}~\rm{cm^{-2}}$. We find that the abundance of water relative to hydrogen nuclei is $1\times10^{-8}$ in agreement with models for oxygen chemistry in which high cosmic ray ionization rates are assumed. Relative to molecular hydrogen, the abundance of water is remarkably constant through the Galactic plane with X(\hho)~$= 5\times10^{-8}$, which makes water a good traced of \hh\ in translucent clouds. Observations of the excited transitions of \hho\ enable us to constrain the abundance of water in excited levels to be at most 15\%, implying that the excitation temperature, $T_{ex}$, in the ground state transitions is below 10~K. Further analysis of the column densities derived from the two ortho ground state transitions indicates that $T_{ex}\simeq5$~K and that the density $n($\hh$)$ in the translucent clouds is below $10^4~\rm{cm^{-3}}$. We derive the water ortho-to-para ratio for each absorption feature along the line of sight and find that most of the clouds show ratios consistent with the value of 3 expected in thermodynamic equilibrium in the high temperature limit. However, two clouds with large column densities exhibit a ratio that is significantly below 3. This may argue that the history of water molecules includes a cold phase, either when the molecules were formed on cold grains in the well-shielded, low temperature regions of the clouds, or when they later become at least partially thermalized with the cold gas ($\sim25$~K) in those regions; evidently, they have not yet fully thermalized with the warmer ($\sim50$~K) translucent portions of the clouds.

\end{abstract}

\section{Introduction}

The ratio of the ortho to para spin modifications of a molecule containing identical atoms is a valuable tracer of its chemical history. The original formation of the species in question, as well as its subsequent history, affect the relative abundance of the molecules having spins parallel (ortho) and antiparallel (para). The different spin modifications are treated as almost entirely separate species as radiative transitions are generally highly forbidden, and collisions that convert one spin modification to another are very slow, with their rate reflecting the density and temperature experienced by the molecules throughout their history. The observed ortho-to-para ratio (OPR) is thus sensitive to physical conditions throughout the lives of the observed molecules.  

A variety of effects complicate using the OPR to probe the history of molecules in dense clouds. Excitation and radiative transfer effects make it difficult to compare accurately the column densities of the different spin modifications. A complex chemical network with numerous formation and destruction pathways make it difficult to disentangle to what other molecules' OPR that of a given species is related \citep[e.g.][]{Kahane1984}. The physics of spin exchange processes for binary collisions, but even more so for surface interactions \citep[e.g][]{Buntkowsky2006}, is very complex and not fully understood. Despite these challenges, the OPR has been observed using molecules in a variety of conditions and the observed ratios used to model current and past conditions. For example, the rotational transitions of molecular hydrogen are readily observed in shock--heated gas. \citet{Neufeld1998} studied \hh\ in HH54, and derived shock conditions from the observed OPR. Subsequent modeling by \citet{Wilgenbus2000} indicated that a complex set of shocks was required and that the solution was not unique. Much better data was obtained subsequently \citep{Neufeld2006}, but interpreting the significance and time dependence of the OPR remains a challenge. Studying molecular absorption in interstellar clouds is a potentially powerful probe of the OPR, as well as of the physical and chemical conditions in these regions. As discussed at length elsewhere, the relatively low density in these regions makes the collision rate modest, and thus for many species it is a good approximation to assume that only the molecular ground state has appreciable population. Since the background sources are generally of small angular size, at velocities for which the absorption may be optically thick, the line should have a flat-bottomed appearance, which is a valuable confirmation that the continuum level has been determined accurately, and helps define the velocities for which the emission is optically thin.

Water, like \hh, is a molecule with an ortho and para form. Water vapor has been widely observed in diffuse and translucent interstellar clouds, by means of absorption line spectroscopy towards bright submillimeter continuum sources. Observations carried out using the {\it Infrared Space Observatory} \citep[ISO, ][]{Kessler1996}, {\it Submillimeter Wave Astronomy Satellite} \citep[SWAS][]{Melnick1999} and the {\it Herschel Space Observatory} \citep{Pilbratt2010} have revealed typical water abundances of a few $10^{-8}$ relative to H$_2$ \citep[e.g.][]{Neufeld2002, Neufeld2010, Sonnentrucker2010}. However, significantly higher water abundances, of a few $10^{-7}$, have been reported for diffuse molecular and translucent clouds in the Galactic Center region \citep[][]{Neufeld2000, Cernicharo2006, Sonnentrucker2011}.

In the diffuse interstellar medium, the destruction of water is dominated by photodissociation. Three separate water formation mechanisms have been posited to date \citep[e.g.][and references therein]{Hollenbach2009, Hollenbach2012}. Much details about water in the interstellar medium can be found in \citet{vanDishoeck2011} in their presentation of the early results of the {\it Herschel} WISH key program, mainly dedicated to the study of water in star forming regions and protostars. First, water can be produced by means of a gas-phase ion-neutral chemistry that is driven by cosmic ray ionization. The chemistry is initiated by cosmic ray ionization of H or H$_2$, followed by the formation of OH$^+$ via the sequence $\rm H_2^+(H_2, H)H_3^+(O,H_2)OH^+$ or $\rm H^+(O,H)O^+(H_2,H)OH^+$; the $\rm OH^+$ ions thereby produced undergo a series of hydrogen atom abstraction reactions, leading to $\rm H_3O^+$, which ultimately produces water vapor as a dissociative recombination product $\rm OH^+(H_2,H)H_2O^+(H_2,H)H_3O^+(e,H)H_2O$. The fraction of $\rm H_3O^+$ recombinations that lead to $\rm H_2O$ has been measured in several laboratory experiments, the most recent of which -- performed in ion-storage rings -- suggest a value $\sim 17 - 26 \%$ \citep{Neau2000, Jensen2000}. A second formation mechanism, the importance of which was emphasized by \citet{Hollenbach2009}, is the hydrogenation of oxygen atoms on grain surfaces, followed by the release of water ice by photodesorption. Finally, enhancements in the production of water are predicted to occur in regions heated by shocks or turbulent dissipation to temperatures above $\sim200$~K by a sequence of two neutral-neutral reactions, both possessing activation energy barriers: $\rm O(H_2,H)OH(H_2,H)H_2O$.

\citet{Neufeld2000} observed the ground--state transitions of ortho--\hhso\ and ortho--\hheo\ in absorption towards Sgr B2, while \citet{Neufeld2002} observed the \hhso\ isotopologue towards W51. \citet{Plume2004} carried out a similar study towards W49A. \citet{Gerin2010} have observed CH in absorption towards W31(C), W49(N), and W51, obtaining information on the physical characteristics of the line of sight absorbing clouds. \citet{Lis2010} analyzed \hho\ and \hheo\ data from the HIFI instrument \citep{deGraauw2010} on {\it Herschel} for lines of sight towards Sgr B2(M) and W31(C), to determine the OPR ratio in different velocity components. Most of the results were consistent with the statistical high--temperature limiting ratio of 3. However, for velocities corresponding to the expanding molecular ring, this study found OPR = 2.35 $\pm$\ 0.35, which could imply equilibration at a lower temperature.

In the present paper we extend the work of \citet{Lis2010} to the remaining PRISMAS lines of sight, towards DR21(OH), G34.3+0.1, W28(A), W33(A), W49(N), and W51, which have been utilized as background sources previously, but only now with the capability of the HIFI instrument on Herschel, can the ground state transitions of ortho-- and para--\hho\ be observed with the high signal to noise ratio required to determine the OPR in the interstellar clouds along these lines of sight. In \S \ref{sec:obs} we present the observations. In \S \ref{sec:res} we discuss the difficult problem of establishing the proper background spectrum to use in determining the absorption optical depths. The reader not interested in the details of the modeling can skip this section. In \S \ref{sec:disc} we discuss our results for both the abundance of water and the OPR in these translucent clouds. We summarize our results in \S \ref{sec:ccl}.

\section{Observations}
\label{sec:obs}

\begin{figure}[t]
  \centering
  \includegraphics[angle=90,width=\linewidth]{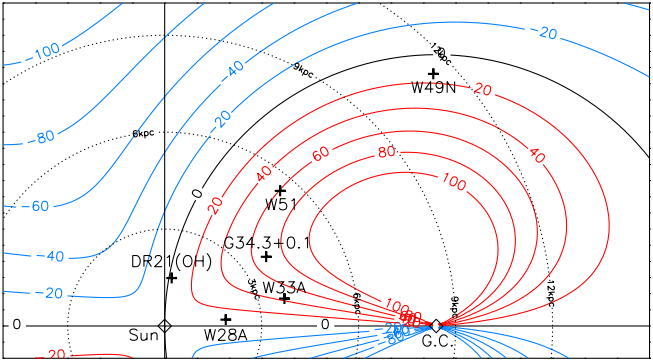}
  \caption{Position of the sources in the Milky Way. The red and blue curves indicate the LSR velocities. The dotted circles show heliocentric radii every 3~kpc. The dashed line indicates a Galactic longitude of $\pm90\degr$. \label{fig:sources}}
\end{figure}

\begin{figure}[t]
  \centering
  \includegraphics[width=\linewidth]{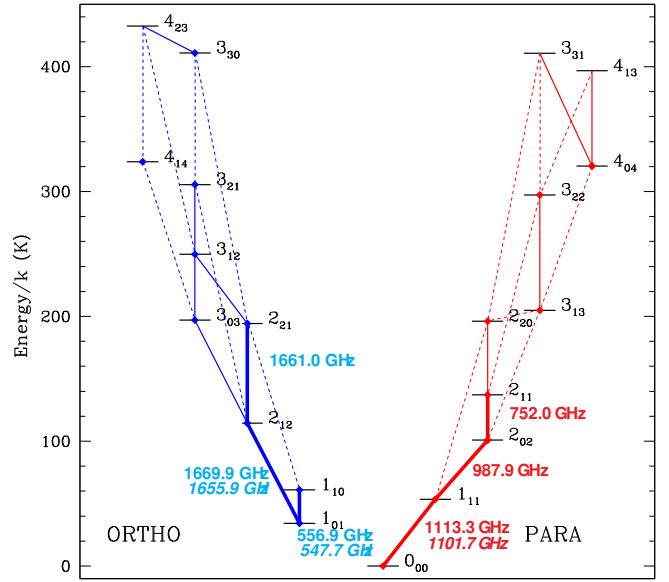}
  \caption{Energy level diagram for ortho and para water. The transitions within the HIFI spectral coverage are indicated by solid lines, while those outside of this range are indicated by dashed lines. The transitions observed in PRISMAS are indicated by heavy lines, labeled with the frequency of \hhso\ and \hheo\ (in italics) when observed.}
  \label{fig:h2odiag}
\end{figure}

We present here all the HIFI \citep{deGraauw2010} observations of the ortho-- and para--\hho\ and \hheo\ towards six of the eight sources in the Guaranteed Time Key Program PRISMAS (P.I. Maryvonne Gerin). The sources are DR21(OH), G34.3+0.1, W28(A), W33(A), W49(N) and W51. Their coordinates, distances and velocities are listed in Table \ref{tab:sources}. Figure \ref{fig:sources} shows their approximate positions in the Milky Way. The observed water lines are the three ground state transitions of \hho\ and \hheo\ ($2_{12}-1_{01}$, $1_{10}-1_{01}$, and $1_{11}-0_{00}$) as well as two higher excitation transitions of p-\hho\ ($2_{02}-1_{11}$ and $2_{11}-2_{02}$) and one higher excitation transition of o-\hho\ ($2_{21}-2_{12}$). Their frequencies, quantum and pseudo-quantum numbers, and energies are listed in Table \ref{tab:h2olines}. The ortho ground state is 34~K above the para ground state that we use as a reference for all the energy levels. Figure \ref{fig:h2odiag} shows the water energy diagram with the transitions observed with HIFI.

The data were obtained between March 2010 and April 2011 using the dual beam switch observing mode. We use the HIFI wide band spectrometer (WBS) which provides a spectral resolution of 1.1 MHz ($0.6\ \rm{km~s^{-1}}$ at 557~GHz and $0.2\ \rm{km~s^{-1}}$ at 1669~GHz) over a 2.4 to 4~GHz IF bandwidth. Each \hho\ and \hheo\ line has been observed with three different local oscillator frequencies (LOF) and two different polarisations (H and V). All the observations have been reduced using HIPE \citep{Ott2010} pipeline version 6. For a given \hho\ or \hheo\ line, three different LOFs are not enough to deconvolve and isolate a single sideband (SSB) spectrum. Therefore, the \hho\ and \hheo\ lines are contaminated by features from the other sideband. In some cases, these contaminating lines fall near the \hho\ and \hheo\ lines. We do not combine the two polarisations immediately, because they target positions a few arcseconds away from each other, and the receivers and detectors are different, so this may introduce systematic differences.

\begin{table*}[!t]
  \caption{Sources observed. \label{tab:sources}}
  \begin{center}
    \begin{tabular}{l l l l l r r r}
      \hline
      \hline
      Source & R.A. & Decl. & $l$ & $b$ & Distance & Emission & Absorption \\
      &  &  &  &  &  & velocity & velocity range \\
      & (J2000) & (J2000) & ($^\circ$) & ($^\circ$) & (kpc) & (km~s$^{-1}$) & (km~s$^{-1}$) \\
      \hline
      DR21(OH) & $20^h39^m01.1^s$ & $+42^\circ19'43''$ & 81.7 & +0.54 & 1.5 & -4 & -5 to 15 \\
      G34.3+0.1 & $18^h53^m18.7^s$ & $+01^\circ14'58''$ & 34.3 & +0.1 & 3.8 & 58 & 10 to 70 \\
      W28A & $18^h00^m30.4^s$ & $-24^\circ04'00''$ & 5.9 & -0.39 & 1.9 & 10 & -10 to 30 \\
      W33A & $18^h14^m39.4^s$ & $-17^\circ52'00''$ & 12.9 & -0.26 & 3.8 & 37 & 20 to 45 \\
      W49N & $19^h10^m13.2^s$ & $+09^\circ06'12''$ & 43.2 & +0.01 & 11.4 & 2 & 0 to 70 \\
      W51 & $19^h23^m43.9^s$ & $+14^\circ30'30''$ & 49.5 & -0.38 & 5.5 & 55 & 0 to 75 \\
      \hline
    \end{tabular}
    \tablecomments{We give the equatorial and Galactic coordinates, and distances. We also give the velocity of the water emission peak and the velocity range over which water absorption is detected. References for distances are \citet{Rygl2012} for DR21(OH), \citet{Fish2003} for G34.3+0.1, \citet{Velazquez2002} for W28A, \citet{Faundez2004} for W33A, \citet{Gwinn1992} for W49N, and \citet{Reid2009} for W51.}
  \end{center}
\end{table*}

Hereafter, we use W51 as an example. Figure \ref{fig:w51} shows the double sideband spectra of the nine water lines towards this source. In the Appendix, Figures \ref{fig:dr21} to \ref{fig:w49} show the double sideband spectra of the nine water lines for the five other PRISMAS sources. All the spectra presented in this paper show similar properties. Differences of about 1~K between the two polarisations are detected in the continuum intensity for a few spectra (e.g. W51 at 752~GHz, W49N at 547~GHz, G34.3+0.1 at 987~GHz). We first summarize the detection of absorption and emission features and characterize the interstellar clouds. We then discuss each transition from the most to the least excited in the water diagram.

\begin{table*}[!t]
  \caption{Water lines observed. \label{tab:h2olines}}
  \begin{center}
    \begin{tabular}{l r c c c c c c c c c}
      \hline
      \hline
      & Frequency & $J_{K_{-1}K_{1}}(u)$ & $J_{K_{-1}K_{1}}(l)$ & $E_{u}/k$ & $E_{l}/k$ & $g_l/g_u$ & $log(A_{ul})$ & $\Delta v$ & \multicolumn{2}{c}{RMS} \\
      & (GHz) & & & (K) & (K) & & & (m/s) & (mK) & (\%) \\
      \hline
      o-\hho & 1669.905 & $2_{12}$ & $1_{01}$ & 114 (80) & 34 (0) & 3/5 & -1.252 & 197 & 69--357 & 2.4--6.5\\
      & 556.936 & $1_{10}$ & $1_{01}$ & 61 (27) & 34 (0) & 1 & -2.461 & 592 & 5--44 & 0.9--3.6 \\
      & 1661.008 & $2_{21}$ & $2_{12}$ & 194 (160) & 114 (80) & 1 & -1.514 & 199 & 74--348 & 2.4--6.8 \\
      \hline
      p-\hho &1113.343 & $1_{11}$ & $0_{00}$ & 53 & 0 & 1/3 & -1.734 & 296 & 25--105 & 1.6--3.2 \\
      &987.927 & $2_{02}$ & $1_{11}$ & 101& 53 & 3/5 & -2.333 & 334 & 33--133 & 2.0--4.2 \\
      &752.033 & $2_{11}$ & $2_{02}$ & 137 & 101 & 1 & -2.152 & 439 & 19--75 & 1.5--4.6 \\
      \hline
      o-\hheo & 1655.868 & $2_{12}$ & $1_{01}$ & 114 (80) & 34 (0) & 3/5 & -1.263 & 199 & 65--248 & 2.4--4.8 \\
      & 547.676 & $1_{10}$ & $1_{01}$ & 61 (27) & 34 (0) & 1 & -2.483 & 602 & 6--27 & 1.3--2.8 \\
      \hline
      p-\hheo & 1101.698 & $1_{11}$ & $0_{00}$ & 53 & 0 & 1/3 & -1.748 & 299 & 32--137 & 2.2--4.1 \\
      \hline
    \end{tabular}
    \tablecomments{For each transition we give its frequency, the upper and lower level designations ($J_{K_{-1}K_{1}}$), and energies expressed as equivalent temperatures relative to the ground state of the para form of that isotopologue ($T$), the degeneracy factor ratio ($g_l/g_u$), and the spontaneous emission rate ($A_{ul}$). We also give the ortho energy level relative to their ground state in parentheses and the equivalent velocity resolution for the HIFI spectral resolution of 1.1~MHz in the last column. The spectroscopic data are from the JPL database \citep{Pickett1998}. The RMS columns give the typical range of the estimated rms in the observed spectra, away from the main absorption and emission features, in mK, and relative to the single sideband continuum intensity, in \%.}
  \end{center}
\end{table*}

\subsection{Absorption and emission features}
The six PRISMAS sources we present show \hho\ and \hheo\ in emission at 752, 987, 556, and 547~GHz near the systemic velocities of the background sources. Some of them also show \hho\ in emission at 1661~GHz (DR21(OH), W28A, W33A), 1669~GHz (DR21(OH), W28A, W33A) and 1113~GHz (DR21(OH), W28A, W33A, W49N). \hheo\ is seen in emission at 1655~GHz (W28A) and 1101~GHz (W28A, W33A). Absorption in the ground state transitions of \hho\ and \hheo\ is detected towards all sources though it is sometimes faint or difficult to discern at 547~GHz (e.g. towards W51). Absorption in the excited transition of \hho\ is detected towards all sources at 1661~GHz and possibly detected at 987~GHz. At 752~GHz, no absorption is apparently detected towards any of the PRISMAS sources.

The velocity of the emission components relates to the velocity of the source in the Galaxy while that of an absorption component depends on the velocity of the line of sight cloud in the Galaxy. The molecular hydrogen column density in the interstellar clouds have been derived from observations of CH and HF towards the PRISMAS lines of sight by \citet{Godard2012}, assuming a column density ratio of 0.4 for HF/CH, and an abundance ratio $n(\rm{HF})/n($\hh$)=3.6\times10^{-8}$. They also give the atomic hydrogen column density obtained by various authors with the VLA interferometer. The total hydrogen column density of a given cloud is usually between a few $10^{20}~\rm{cm^{-2}}$ and a few $10^{21}~\rm{cm^{-2}}$ though some are lower limits. This corresponds to $A_V$ of a fraction of, to a few magnitudes. The fraction of molecular hydrogen ($2N($\hh$)/(N($H$)+2N($\hh$))$) is typically 30\%, but varies significantly (from less than 20\% to more than 50\%). The interstellar clouds we detect in \hho\ absorption therefore are translucent clouds. The typical density and temperature in those clouds are about $100~\rm{cm^{-3}}$ and 50 to 100~K \citep[e.g.][]{Rachford2002}. We note here that the water distribution in velocity space has been shown to follow that of HF \citep[see e.g.][]{Neufeld2010, Sonnentrucker2010}, implying that water, unlike OH$^+$ and \hho$^+$ probably arises in the gas that is mainly molecular.

\begin{figure*}[!t]
  \centering \subfigure[] {\label{}
    \includegraphics[angle=90,width=.31\linewidth]{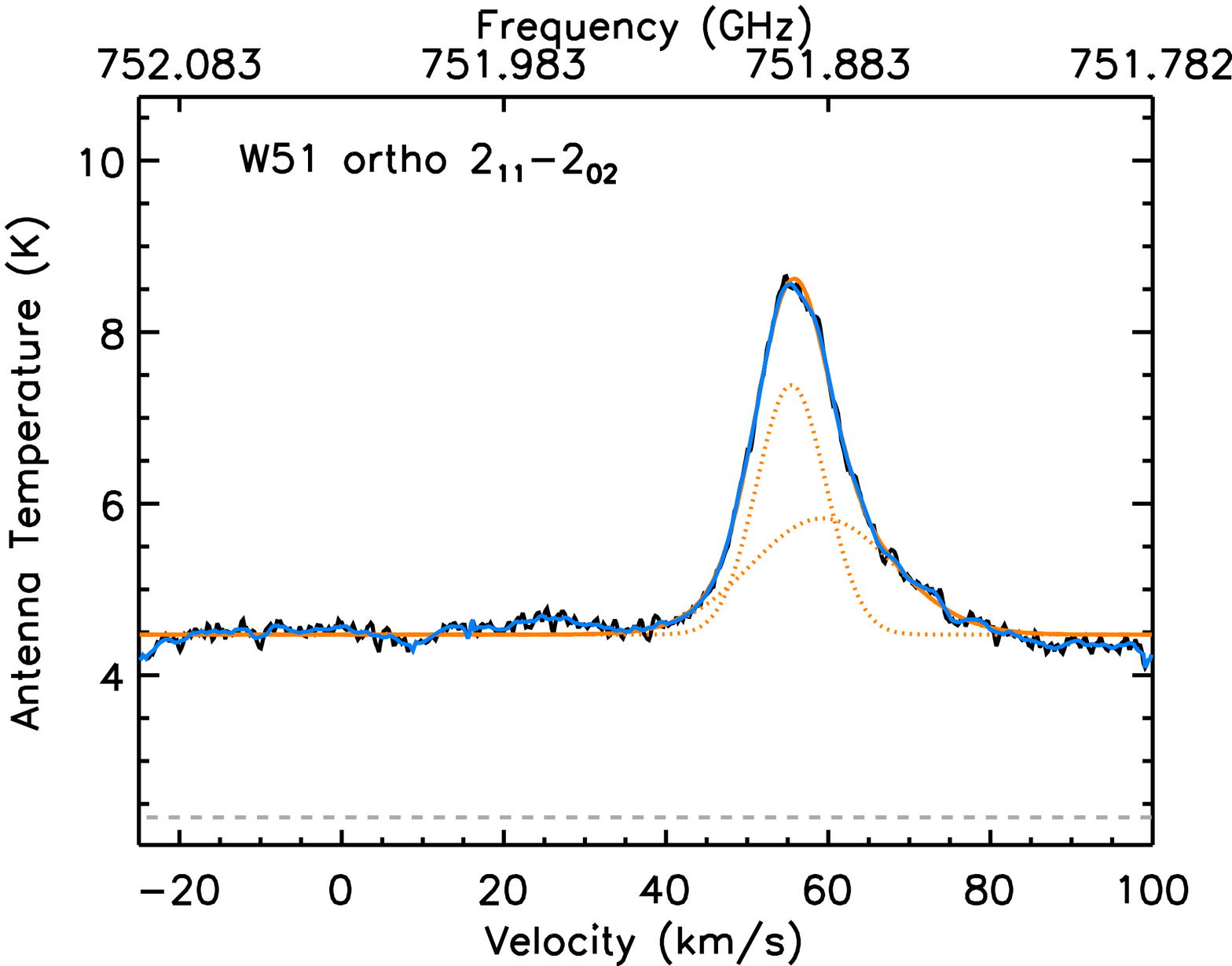}}
  \subfigure[] {\label{}
    \includegraphics[angle=90,width=.31\linewidth]{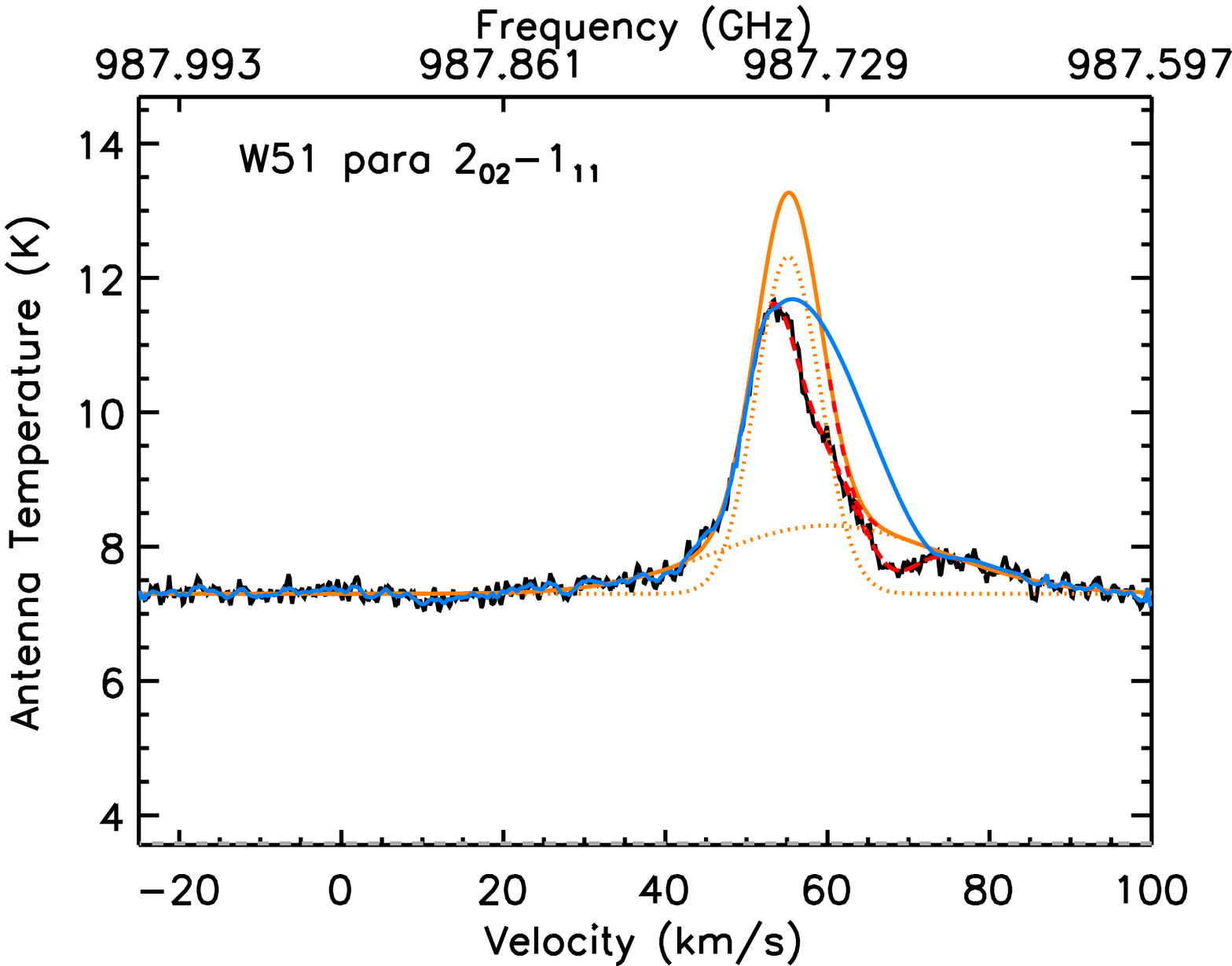}}
  \subfigure[] {\label{}
    \includegraphics[angle=90,width=.31\linewidth]{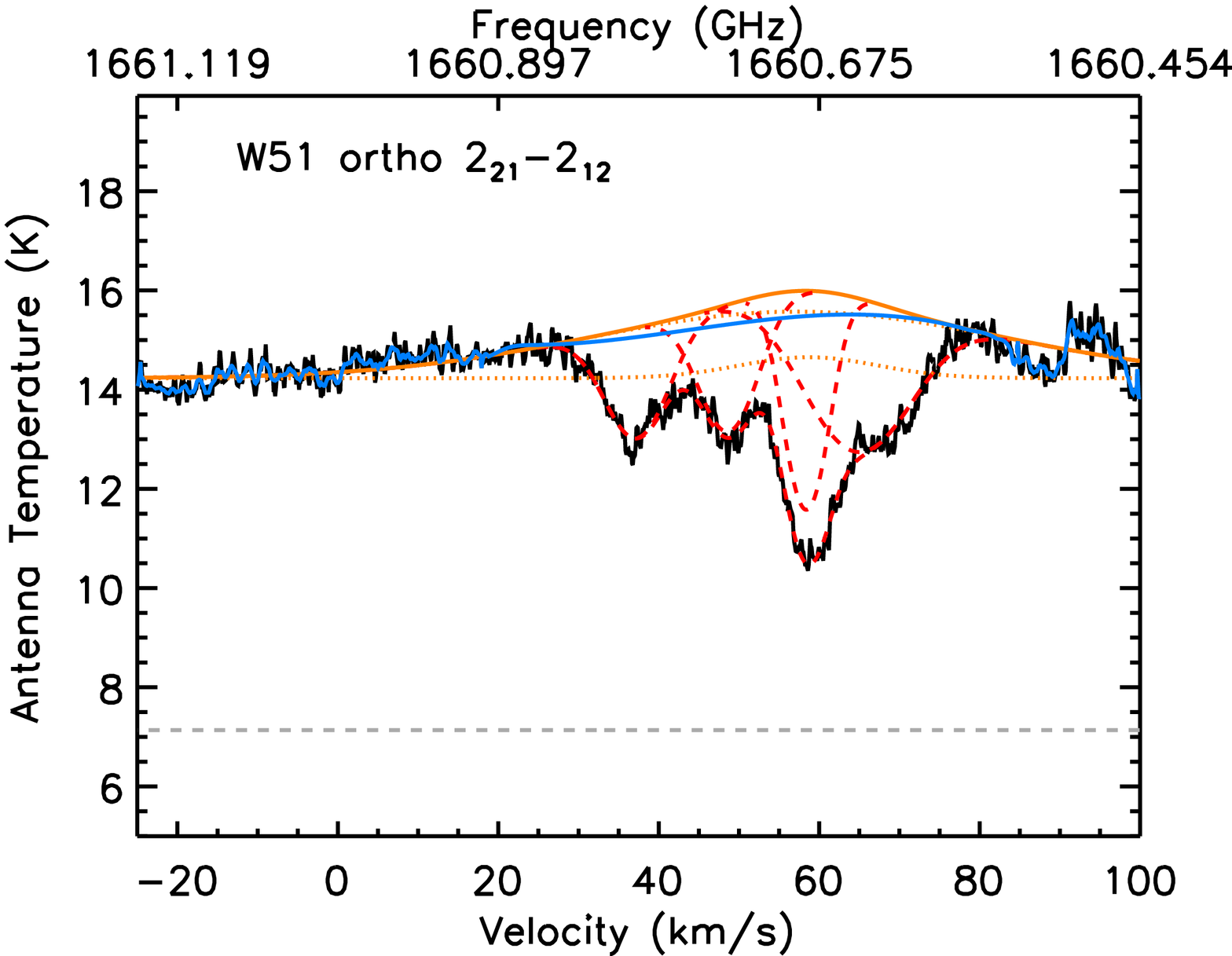}}
  \subfigure[] {\label{}
    \includegraphics[angle=90,width=.31\linewidth]{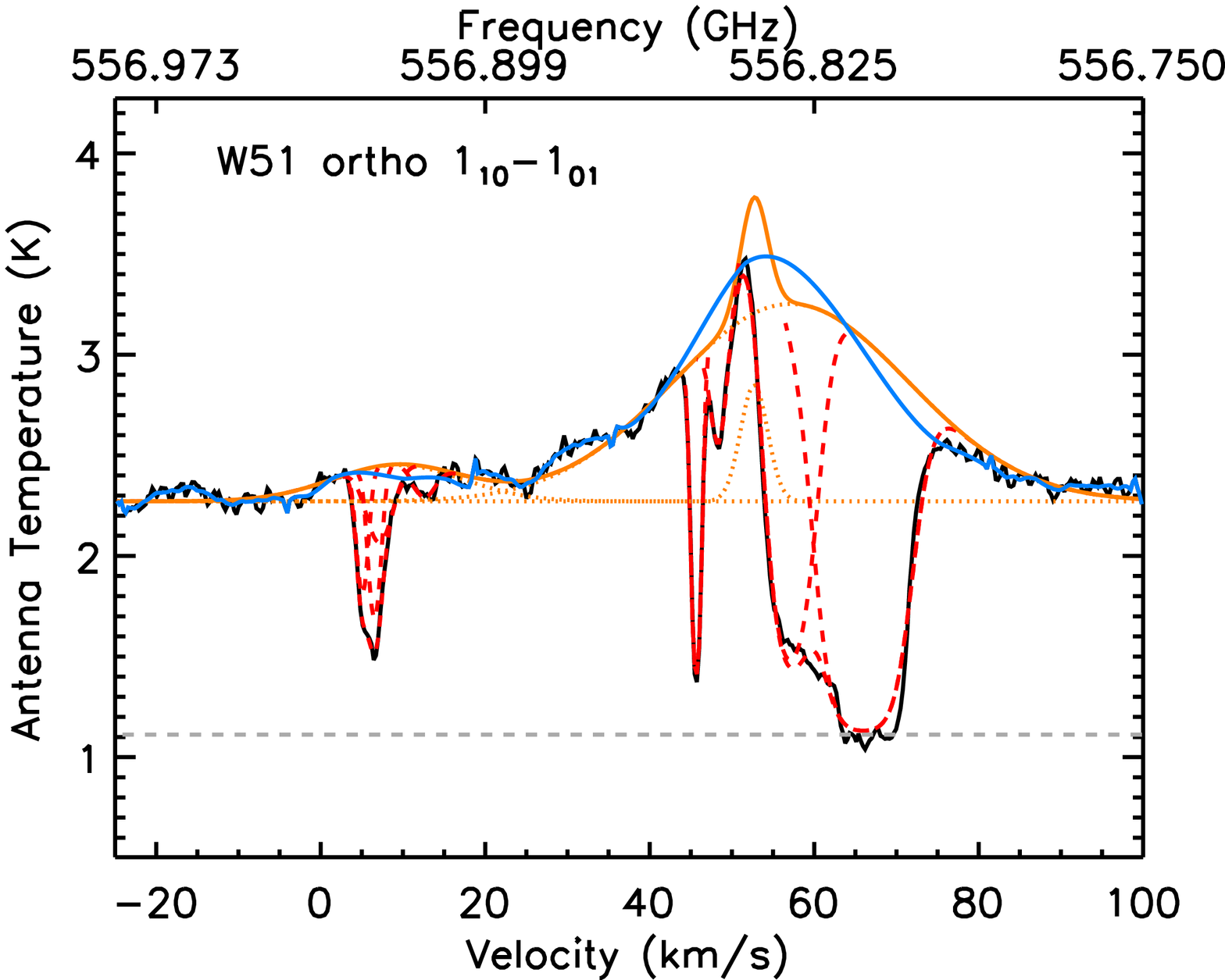}}
  \subfigure[] {\label{}
    \includegraphics[angle=90,width=.31\linewidth]{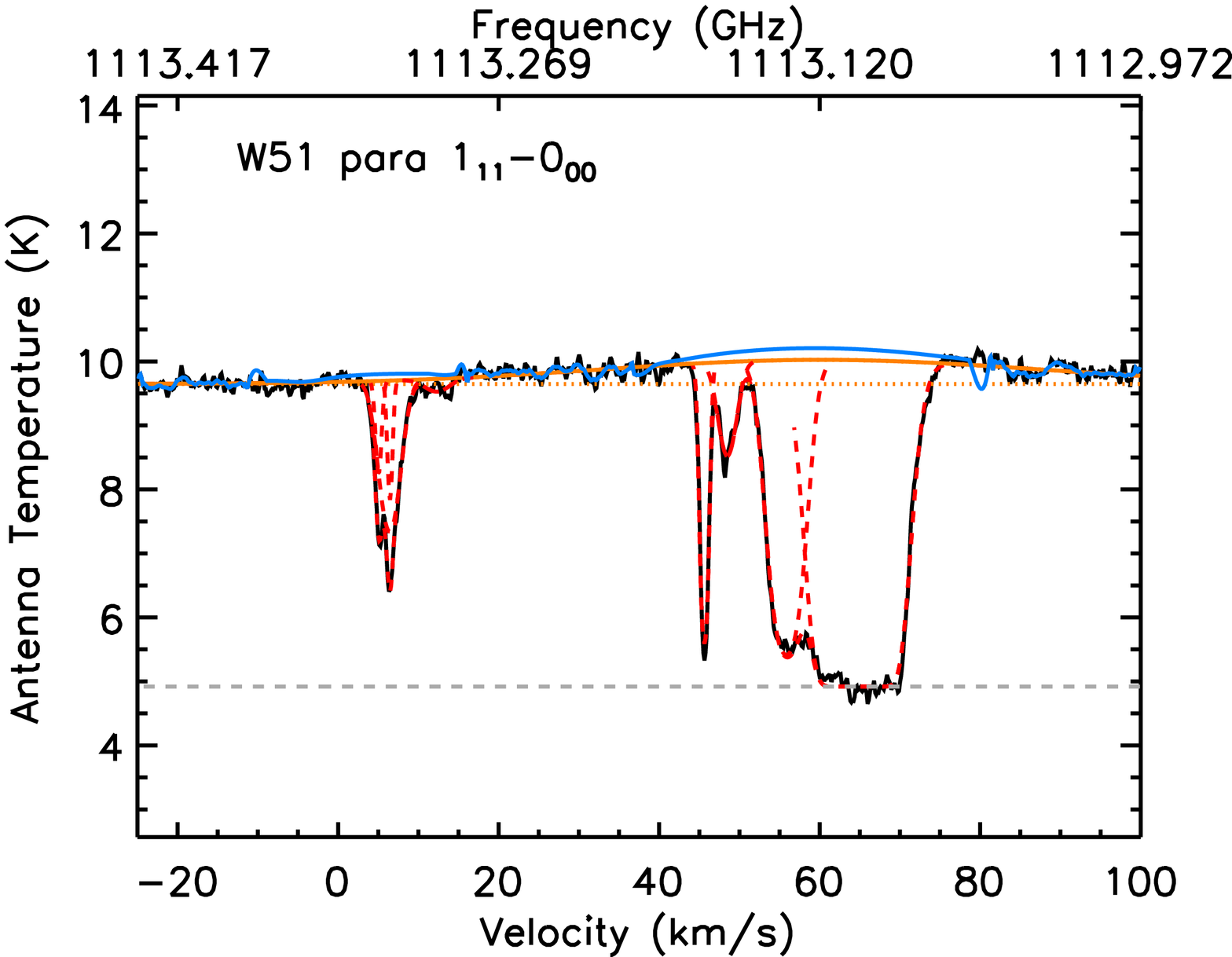}}
  \subfigure[] {\label{}
    \includegraphics[angle=90,width=.31\linewidth]{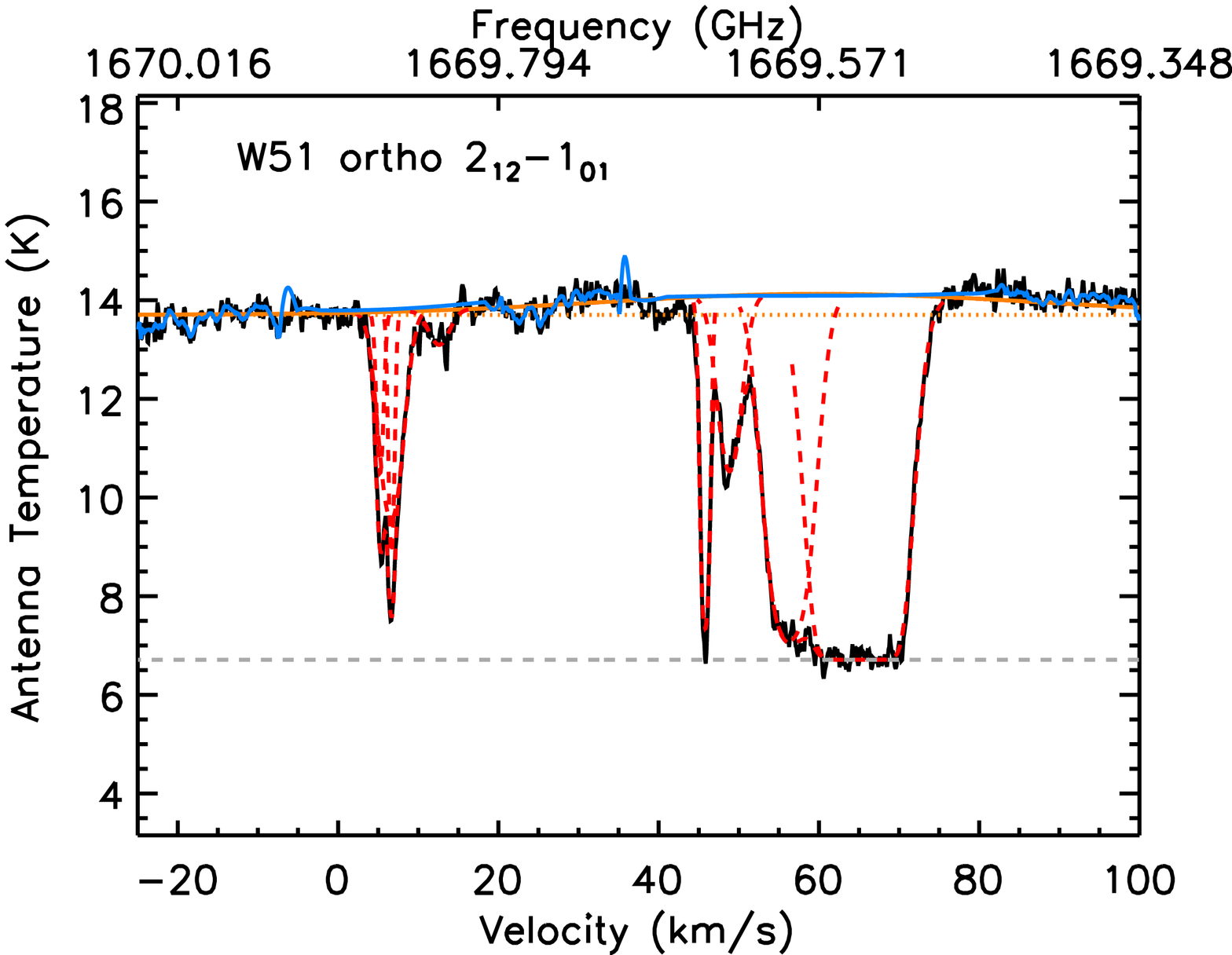}}
  \subfigure[] {\label{}
    \includegraphics[angle=90,width=.31\linewidth]{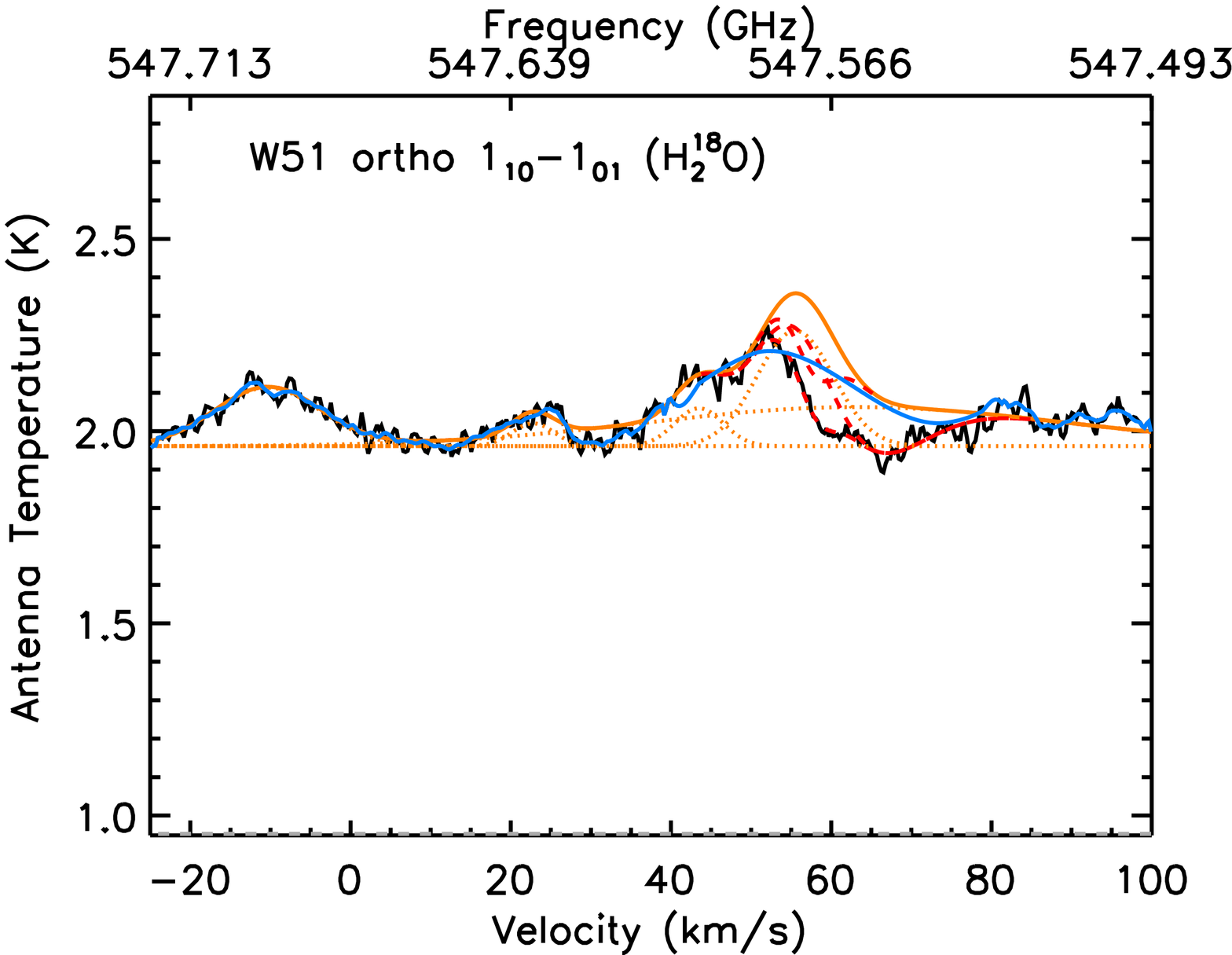}}
  \subfigure[] {\label{}
    \includegraphics[angle=90,width=.31\linewidth]{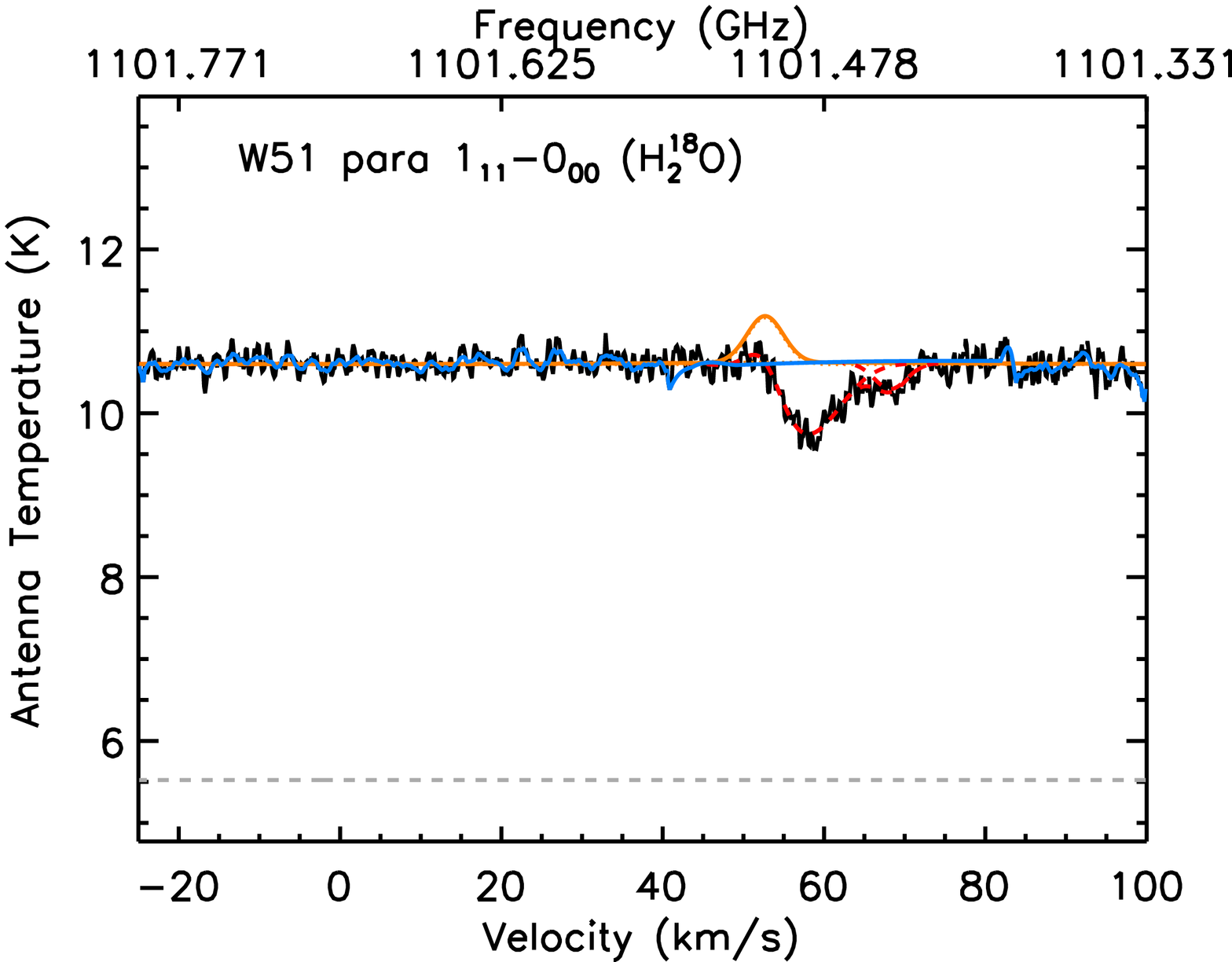}}
  \subfigure[] {\label{}
    \includegraphics[angle=90,width=.31\linewidth]{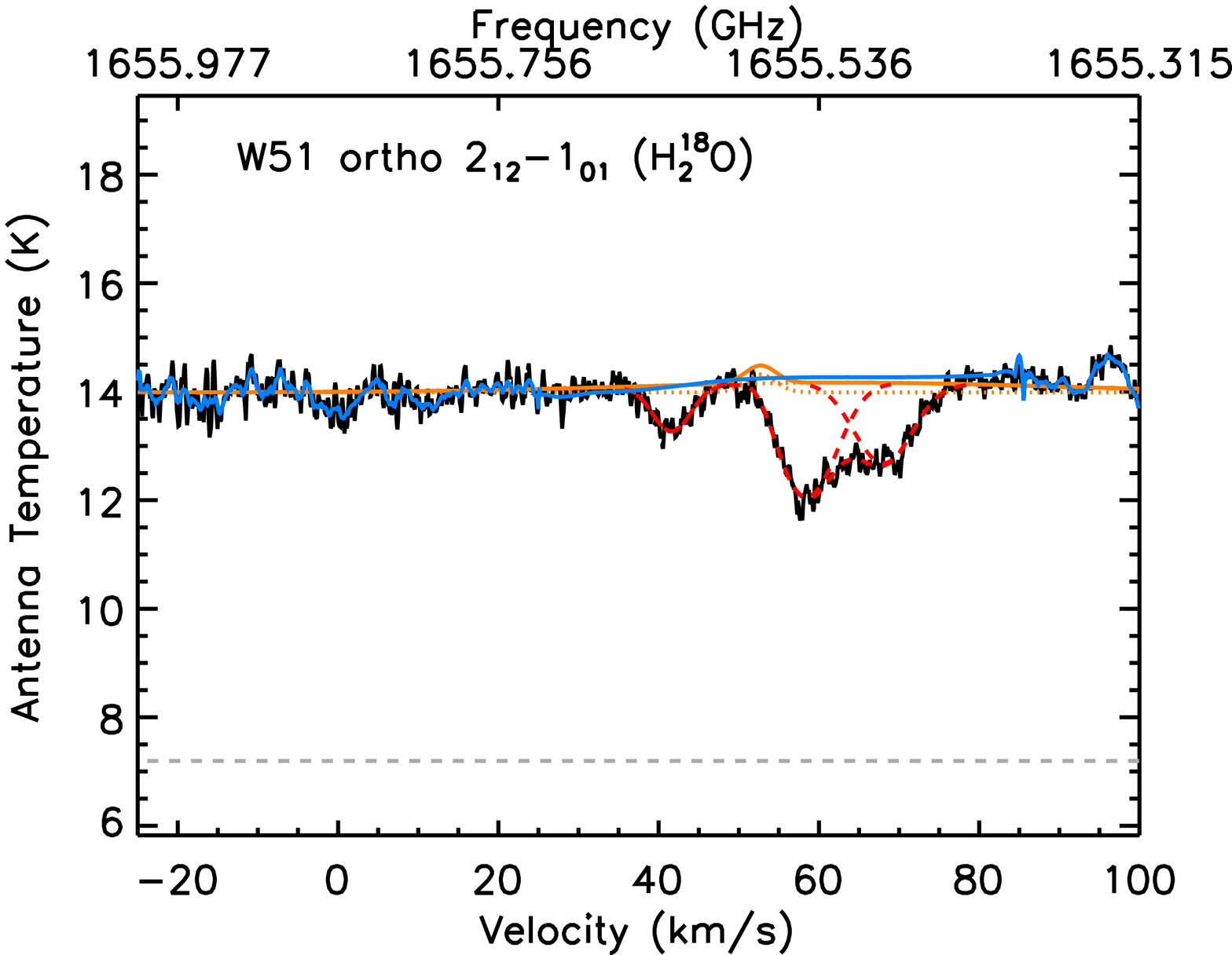}}
  \caption{Double sideband water line spectra towards W51. The top row shows the excited transitions of \hho, the middle row shows the ground transitions of \hho, and the bottom row shows the ground transitions of \hheo. In each panel, we show the H polarisation spectrum (black solid line), the interpolated continuum (blue solid line), the best-fit continuum (orange solid line), the emission components (orange dashed lines), the best fit (red solid line), the absorption components (red dashed lines), and the zero continuum level for the single sideband (grey dashed line). Some of the features seen are not due to water molecules (see text for details). For instance, the 45~km~s$^{-1}$ emission feature at 547~GHz is likely due to $^{34}$SO$_2$ and methanol. The 40~km~s$^{-1}$ absorption feature at 1655~GHz is due to the strongest water absorption at 1661~GHz. The 35 and 50~km~s$^{-1}$ absorption features at 1661~GHz are likely due to CH.}
  \label{fig:w51}
\end{figure*}

\subsection{Excited p-\hho\ transition at 752~GHz}
The only water line that does not show any indication of absorption is that of p-\hho\ at 752~GHz. Indeed, it is the highest excitation line of water observed with HIFI in the PRISMAS program\footnote{Higher energy transitions are observed with PACS, but not velocity resolved.}, with a lower level at 101~K above the para ground state. Therefore, the water molecules in clouds along the line of sight, which are believed to be relatively unexcited, do not absorb photons at this frequency. The asymetry of the emission profile indicates the presence of multiple emission components. The position of the emission peak defines the velocity of the source. We report this velocity in Table \ref{tab:sources} for each source. These velocities are in very good agreement with values from literature.

\subsection{Excited p-\hho\ transition at 987~GHz}
The next highest excitation line of p-\hho\ is at 987~GHz. Its spectrum resembles that of p-\hho\ at 752~GHz except that it shows some sign of absorption. The absorption feature at 987~GHz is more significant towards DR21(OH) than towards G34.3 and W49N, and it is weak towards W33A and W51. However, given the equivalent temperature of the lower level of this transition (53~K), and the apparent agreement in velocity with the emission component, we assume this absorption component is due to warm water vapor in the background source's envelope or outflow.

\subsection{Excited o-\hho\ transition at 1661~GHz}
On the ortho side of the water energy diagram, the first excited transition is that at 1661~GHz. Its lower level is at a lower equivalent temperature than that of the highest excited transition of the p-\hho\ at 752~GHz when measured relative to the ground state of o-\hho\ (80~K vs 101~K). The o-\hho\ transition is always detected in absorption and towards three sources in emission. Transitions of CH at $\sim$1661~GHz, about 100~MHz or 20~km~s$^{-1}$ from that of \hho\, are likely detected towards DR21(OH), G34.3+0.1, W49N and W51. We identify the o-\hho\ line thanks to the matching of its velocity with that of components seen in the spectrum of other \hho\ and \hheo\ transitions. Similar to what we observe at 987~GHz, we always find a good agreement between the velocity of the o-\hho\ absorption feature and that of the emission feature at 752~GHz, which indicates this component arises from the background source's envelope or outflow rather than from the ISM clouds along the line of sight.

\subsection{Ground state \hho\ transitions at 1113, 1669 and 556~GHz}
The three \hho\ ground state lines, the p-\hho\ line at 1113~GHz and the o-\hho\ lines at 1669 and 556~GHz, resemble each other to a significant degree for each source. They all exhibit multiple absorption features with similar positions, shapes and intensities. Optically thick and thin absorption features are found towards each source. Some absorption components are wide (e.g. $\sim 20\ \rm{km~s^{-1}}$ towards W51 at about 60~$\rm{km~s^{-1}}$), some are much narrower (e.g. $\lesssim 1\ \rm{km~s^{-1}}$ towards G34 at about 27~$\rm{km~s^{-1}}$). We indicate the range of absorption velocities for each source in Table \ref{tab:sources}. We note here that for each PRISMAS source, the velocity and width of the emission feature seen at 752 and 987~GHz is such that it covers at least partially the range of velocities of the absorption components.

Emission from the wings of water lines is detected at 556~GHz towards each source, and is visible at 1113~GHz and 1669~GHz only towards DR21(OH), W28A and W33A. However, given the position, width and intensity of the absorption features detected towards G34.3+0.1, W49N and W51, we cannot claim there is no water emission at those two frequencies. Only a few absorption features are significantly displaced from potential emission contamination (e.g. towards G34.3+0.1 at about 10 and 30~$\rm{km~s^{-1}}$). All of the absorption features have velocities that match those of the Galactic arms crossed by the lines of sight, as well as the velocity of the arm in which the source is located. However, we cannot claim that all features are associated with the foreground ISM and are not intrinsic to the source. For instance, the components with velocities lower than 30~$\rm{km~s^{-1}}$ have been suggested to be due to W49 itself \citep[see][]{Plume2004}. We detail the possible association of each absorption feature with a Galactic arm below.

\subsection{Ground state \hheo\ transitions at 1113, 1669 and 556~GHz}
The three \hheo\ ground state lines (the para transition at 1101~GHz and the ortho transitions at 1655 and 547~GHz) closely resemble each other for a given source. The line profiles usually exhibit only one absorption feature at a velocity close to that of the absorption feature seen at 987~GHz. An emission feature that resembles that observed at 752 and 987~GHz is always detected in the spectrum at 547~GHz, though it is significantly weaker. Consequently, the spectrum at 547~GHz closely resembles that at 987~GHz. This is particularly true towards DR21(OH), W28A, and W49N. The emission component is also visible at 1101~GHz towards W28A and W33A, and at 1655~GHz towards W28A. The abundance of \hheo\ relative to \hho\ is generally assumed to be about 1/500 though it has been shown to increase by about a factor 2 towards the inner Galaxy \citep[e.g.][]{Penzias1981, Wilson1994}. Therefore the \hheo\ absorption features associated with translucent clouds are not detected.

\subsection{Features from other sideband and from other molecules}
\label{lab:contam}
As mentioned previously, features from the other sideband or from other molecules may in a few cases appear very close to those of \hho\ and \hheo. The most common case in our analysis is that of the CH (N=2, J=5/2-3/2) transition at 1661.1~GHz, at about $-20~\rm{km~s^{-1}}$ from that of \hho\ at 1661~GHz. Due to the difficulty of deconvolving the two sidebands, a combination of the CH and the \hho\ features also appear in the \hheo\ transition at 1655~GHz. Other contaminants include the SO$_2$ 28$_{6,22}$-28$_{5,23}$, $^{34}$SO$_2$ 21$_{3,19}$-20$_{2,18}$, and the methanol 15$_{1,15}$−15$_{2,14}$, v$_t$ = 1 transitions in the \hheo\ spectrum at 547~GHz at -70~$\rm{km~s^{-1}}$, +35~$\rm{km~s^{-1}}$, and -10~$\rm{km~s^{-1}}$, respectively, relative to the \hheo\ feature (e.g. towards G34.3+0.1, W28A and W49N).

\section{Results}
\label{sec:res}

To estimate the OPR of the water vapor in the translucent clouds, we need to measure the column density of ortho and para water along the line of sight, which in turn requires us to determine the optical depth as a function of velocity. The main difficulty that we face in estimating the optical depth of the water lines is due to the background emission. The sources have been chosen for their strong dust continuum. However their spectra also show emission of water lines. The velocity and width of the water emission and absorption components is such that they usually overlap significantly, over at least a few tens of km~s$^{-1}$. Therefore, a careful estimate of the water emission lines from the source is required to infer the optical depth of the water lines from the clouds along the line of sight. This section provides many details on our analysis and the reader not interested in those details is invited to directly go to the discussion section.

\subsection{Optical depth}
\label{lab:optdep}

The optical depth is $\tau = -\ln(I/I_0)$ where $I$ is the observed single sideband antenna temperature and $I_{0}$ is the signal sideband pseudo-continuum, which comprises a dust continuum component $I_{dust,SSB}$ and the emission features. $I_{dust,ISB}$ denotes the dust continuum component of the image sideband and $I_{dust,DSB}$ the dust continuum component of the double sideband spectrum. We treat $I_{dust,SSB}$, $I_{dust,ISB}$ and $I_{dust,DSB}$ as constants and $I_{dust,DSB} = I_{dust,SSB} + I_{dust,ISB}$. For instance, towards W51, at 556~GHz, $I_{dust,DSB} \simeq 2.3$~K, and $I_{dust,ISB} \simeq 1.1$~K, as given by the antenna temperature in the plateau of optically thick absorption features. Since we do not deconvolve the spectra, we treat all the emission features as part of the signal sideband pseudo-continuum $I_0$. Towards W51, at 556~GHz, $I_0$ is thus represented by all the emission features on top of a continuum $I_{dust,SSB} \simeq 1.2$~K. The actual values of the continua depend on the sideband gains $g_{ISB}$ and $g_{SSB}$, linked by the relation $g_{ISB}+g_{SSB} = 1$.

In this paper, we use two methods to measure the pseudo-continuum $I_0$. The first method uses the information provided by the higher excitation transitions (e.g. at 752, 987 and 1661~GHz) about the water emission components in terms of amplitude, velocity position and width, to better constrain the continuum for the lower transitions. We give details of this method in what follows. From this, we obtain a first estimate of the single sideband pseudo-continuum intensity $I_{0,bf}$, parameters for all absorption and emission components, and the sideband gain ratio. The second method consists of interpolating each spectrum over the velocity ranges where the absorption features are detected with a spline function. We manually select regions and points to be connected for each spectrum. We use the same velocity points for both polarisations.  Since some spectra may show significant noise, we first smooth all the velocity ranges with a 10 pixel wide window. With this method, we obtain an estimate of the double sideband pseudo-continuum intensity from which we later subtract the best estimate of the image sideband dust continuum determined using the sideband gain ratio. We therefore obtain a second estimate of the single sideband pseudo-continuum intensity $I_{0,int}$.

\subsubsection{Best fit method}
\label{lab:fitmeth}

In this section, we detail the method we use to fit the data in order to infer the optical depth of each cloud along the PRISMAS lines of sight. For each water transition we model the observed intensity as follows:

\begin{equation}
I = \frac{1}{g_{SSB}}\left(g_{ISB}~I_{ISB} + g_{SSB}~I_{SSB}\right),
\end{equation}

\noindent where:
\begin{align}
g_{ISB} =& 1-g_{SSB}, \\
I_{ISB} =& I_{dust,ISB}, \\
I_{SSB} =& \left(I_{dust,SSB} + \sum_i a_i e^{-(v-v_i)^2/2\sigma_i^2}\right) \nonumber \\
& \times \exp\left(-\sum_j a_j e^{-(v-v_j)^2/2\sigma_j^2 }\right),
\end{align}

\noindent where the sums over $i$ and $j$ correspond to the emission and absorption features, assumed to be gaussians, respectively. The fitting procedure therefore aims at obtaining the best values for $g_{SSB}$, $I_{dust,ISB}$, $I_{dust,SSB}$, and the gaussian parameters.

We start with a fit to the spectrum at 752~GHz as only water emission, without absorption, is seen at this frequency. This way, we define the properties of the emission components (position, width and intensity). We use as few components as possible -- usually two to three -- to obtain a satisfactory fit of the spectrum. We use the MPFIT package\footnote{http://purl.com/net/mpfit} in IDL \citep{Markwardt2009}. We give an initial guess specific to each source.

We then move downwards in energy in the water level diagram and fit the spectrum at 987~GHz. We assume the properties of the emission components remain roughly similar to those defined at 752~GHz: the position of each emission component can change by up to 3 km~s$^{-1}$ while the width can change by a factor 2. The intensity of each emission component is free to vary, as long as it remains positive. We then add one or two absorption components to fit the spectrum at 987~GHz. For each absorption component, we use a single gaussian profile to model the corresponding optical depth. We set the initial parameters (position, width and intensity) for each component manually. As for the emission component, we allow the position to be within 3 km~s$^{-1}$ of the initial guess and the width to be within a factor 2.

Then, we move to the \hho\ ground state transitions, where the absorption features due to the clouds along the line of sight are detected. We use the spectrum at 556~GHz to define the absorption components, in addition to that already defined at 987~GHz. We again use as few gaussian profiles as possible for each of the broad absorption components. We usually define one component at each local minimum of the observed antenna temperature. For a given source, we use the results from the fit of the spectrum at 556~GHz as the initial guess for both the fits at 1113 and 1669~GHz. Therefore, the number of emission components and absorption components are the same for the three \hho\ ground state lines. Additionally, the initial parameters of the components (position, width) are identical for the 1113 and 1669~GHz spectra. We manually adjust the intensity of the emission components for each spectrum independently. Our method is thus consistent with the observed similarity between those three spectra. In a few cases at 1669~GHz, we have to use additional absorption features to account for contamination due to other species, either from the same or the other sideband (see section \ref{lab:contam}).

We finally fit the \hheo\ lines at 547, 1101 and 1655~GHz as well as the \hho\ line at 1661~GHz using the results from the fit of the spectrum at 987~GHz as the initial guess. As discussed above, the spectra at 547 and 1661~GHz, and to a lesser extent at 1655 and 1101~GHz, are very similar to that at 987~GHz. We manually adapt the emission components intensity to account for the abundance difference between \hho\ and \hheo. In a few cases for the spectrum at 1655 and 1661~GHz, we have to use additional absorption features to account for contamination due to other species, either from the same or the other sideband (see section \ref{lab:contam}).

We perform all these fits separately for each polarisation. However, we use the same initial guess for both polarisations observed in a given source and a given water line. We combine the results from both polarisations only when deriving the water column density (see section \ref{lab:colden}).

\subsubsection{Gain and continuum ratio}
Since we work on the double sideband spectrum, we have to take into account possible variations in the continuum due to different sideband gains ($g_{SSB}\ne0.5$) and to the intrinsic spectral shape of the dust continuum ($I_{SSB}\ne I_{ISB}$). At 752 and 987~GHz, and for the \hheo\ transitions, this is not critical since there is no optically thick water absorption. Therefore, we assume the sideband gains to be both equal to 0.5. We have also no way to constrain the dust continuum intensities $I_{dust,ISB}$ and $I_{dust,SSB}$ at these frequencies and in return, they do not affect the results of the fit. We therefore let them vary within a default range of $\pm$5\% away from $I_{dust,DSB}/2$. On the other hand, for the ground state transitions, the water absorption features that are optically thick allow us to constrain the sideband gain and the dust continuum.
 
We add $g_{SSB}$ as a free parameter in the fit for the ground state transitions and allow it to vary within 5\% of 0.5. We justify here these limits. At a velocity where the foreground absorption is optically thick, $I_{SSB}=0$ which leads to $(1-g_{SSB})~I_{dust,ISB} = g_{SSB}~I_{dust,DSB}$. The limits on the gain value then give $0.9~I_{dust,ISB} < I_{dust,DSB} < 1.1~I_{dust,ISB}$. We then compare $I_{dust,DSB}$, as given by the median of the spectrum away from emission or absorption features, to $I_{dust,ISB}$, as given by the single minimum value of the spectrum where the water absorption is optically thick, taking into account the standard deviation of the spectrum $\sigma$. For all our spectra, we find that the constraints on $I_{dust,DSB}$ and $I_{dust,ISB}$ are respected within 3-$\sigma$ at most. This is very satisfying since we estimate $I_{dust,ISB}$ with the single minimum value point in each spectrum rather than with an average of the intensity in the plateau of optically thick absorption features. The values we find for $g_{SSB}$ when performing the fits are very close to 0.5 and slightly decrease with increasing frequency. When averaging over all the PRISMAS sources and both polarisations, we find $g_{SSB}=0.51\pm0.02$, $0.51\pm0.01$, and $0.50\pm0.02$ at 556, 1113, and 1669~GHz, respectively. We do not find any difference between the two polarisations or any trend when comparing the sources, except that $g_{SSB}$ is systematically the lowest towards W28A with values below 0.5, which may indicate some contamination by the OFF position. \citet{Roelfsema2012} report in-orbit performance of the HIFI instrument. The measured sideband gain ratio is usually between 0.45 and 0.55, throughout the entire frequency range of HIFI. Our findings are therefore in agreement with expectations.

We also add the intensities $I_{dust,ISB}$ and $I_{dust,SSB}$ as free parameters in the fit of the ground state transitions to estimate the variation of the dust continuum between the two sidebands. We estimate the variations of the dust continuum over the entire frequency range covered by our data, from 547 to 1669~GHz. We find that the antenna temperature follows a law in $\nu^{1.7}$, except towards W28A where it is closer to $\nu^{2.4}$. If we limit this analysis to the frequencies below 1600~GHz, the antenna temperature follows a law in $\nu^{2.3}$ except towards W28A where it varies in $\nu^{2.8}$. Therefore, at 556~GHz, where the effect is the strongest, the difference between the dust continuum in the upper sideband and in the lower sideband, which are separated by 12~GHz, should be, at the most, about 6\% ($(556+12)^{2.8}/556^{2.8}\simeq1.06$).

At 556~GHz, the lower sideband contains the water features therefore we expect $I_{dust,ISB} > I_{dust,SSB}$. We thus assume $I_{dust,ISB}$ is about 5\% larger than the mean value of the continuum, as given by $I_{dust,DSB}/2$, and $I_{dust,SSB}$ is about 5\% smaller. We allow both parameters to vary by up to 5\% around those values, which means that in the extreme cases, they can be either equal or about 20\% apart from each other. At 1669~GHz, the upper sideband contains the water features therefore we expect $I_{dust,SSB} > I_{dust,ISB}$. We thus use the same limits as defined above but on opposite parameters. At 1113~GHz, we combine data from lower and upper sidebands therefore we expect $I_{dust,ISB} \simeq I_{dust,SSB}$. We thus let vary the parameters within $\pm$5\% of $I_{dust,DSB}/2$. The results from the fit confirm our expectations with $I_{dust,ISB} / I_{dust,SSB} = 1.05\pm0.05$, $1.01\pm0.05$, and $0.91\pm0.05$ at 556, 1113, and 1669~GHz, respectively. As for $g_{SSB}$ we do not find any systematic between polarisations and we find that apart from W28A, for which $I_{dust,ISB} / I_{dust,SSB}$ is almost systematically the highest, there are no obvious trends when comparing sources.


\begin{table}[t]
  \centering
  \caption{Best-fit gaussian parameters for the absorption features towards W51 and column density in the lower level of each transition.}
  \begin{tabular}{l c c c c c c}
    \hline
    \hline
    Transition & $v_0$ & FWHM & $\tau_0$ & $N_l($\hho$)$\\
    (GHz) & (km~s$^{-1}$) & (km~s$^{-1}$) & & ($\times 10^{12}~\rm{cm^{-2}}$)\\
    \hline
    987~GHz &     58.1 &     7.02 &    0.350 &     21 $\pm$     2 \\
    &     67.8 &     8.87 &    0.264 &     13 $\pm$     8 \\
    \hline
    1661~GHz &     58.6 &     6.15 &    0.810 &     69 $\pm$     14 \\
    &     66.0 &     13.6 &    0.442 &     77 $\pm$     11 \\
    \hline
    556~GHz &     5.03 &     1.50 &    0.545 &     3.9 $\pm$    0.5 \\
    &     6.51 &     1.07 &    0.564 &     2 $\pm$     2 \\
    &     6.73 &     3.18 &    0.713 &     10 $\pm$     4 \\
    &     12.6 &     3.00 &    0.132 &     2 $\pm$    0.5 \\
    &     45.7 &     1.18 &     1.84 &     10.1 $\pm$    0.2 \\
    &     48.5 &     2.51 &    0.373 &     4 $\pm$     2 \\
    &     56.9 &     5.32 &     1.75 &     43 $\pm$     2 \\
    &     65.7 &     7.58 &     4.97 &     185 $\pm$     24 \\
    \hline
    1113~GHz &     5.08 &     1.22 &    0.457 &    1.0 $\pm$    0.5 \\
    &     6.46 &    0.706 &    0.509 &    0.84 $\pm$   0.03 \\
    &     6.49 &     2.62 &    0.616 &     3.4 $\pm$    0.9 \\
    &     12.8 &     2.56 &   0.0497 &    0.3 $\pm$   0.1 \\
    &     45.6 &    0.961 &     2.04 &     4.6 $\pm$   0.1 \\
    &     48.5 &     2.65 &    0.360 &     2.2 $\pm$    0.2 \\
    &     56.1 &     4.19 &     2.44 &     24 $\pm$     3 \\
    &     64.9 &     6.18 &     16.9 &     244 $\pm$     25 \\
    \hline
    1669~GHz &     5.24 &    0.896 &    0.497 &     1.9 $\pm$    0.7 \\
    &     6.58 &     2.95 &     1.05 &     15 $\pm$     1 \\
    &     6.65 &    0.657 &     1.04 &     3 $\pm$     1 \\
    &     12.4 &     3.16 &   0.0991 &     1.5 $\pm$    0.2 \\
    &     45.8 &     1.07 &     2.43 &     12.1 $\pm$    0.3 \\
    &     48.8 &     3.17 &    0.649 &     9.6 $\pm$   0.1 \\
    &     56.4 &     4.97 &     3.04 &     71 $\pm$     4 \\
    &     65.0 &     6.52 &     17.6 &     534 $\pm$     3 \\
    \hline
    547~GHz &     57.9 &     6.04 &    0.157 &     4.7 $\pm$    0.9 \\
    &     64.9 &     16.4 &    0.141 &     10.8 $\pm$   0.1 \\
    \hline
    1101~GHz &     58.1 &     8.47 &    0.194 &     3.7 $\pm$    0.9 \\
    &     67.6 &     4.19 &   0.0659 &    0.7 $\pm$  0.1 \\
    \hline
    1655~GHz &     58.0 &     5.70 &    0.318 &     9 $\pm$     4 \\
    &     66.0 &     9.23 &    0.240 &     10 $\pm$     3 \\
    \hline
  \end{tabular}
  \tablecomments{The uncertainty in the column density in the lower level of the transition takes into account the uncertainty in the continuum and the standard deviation of the antenna temperature. The column densities are derived assuming $T_{ex}\ll h\nu/k$.}
  \label{tab:w51gg}
\end{table}

\subsubsection{Results}

We here present the results for the sightline towards W51. The results for the other lines of sight can be found in the appendix. The continuum estimates are all shown in Figure \ref{fig:w51}. All the observations are very well reproduced with the best-fit method. The reduced $\chi^2$ is usually very close to 1.

The two methods we use to estimate the continuum $I_0$ provide us with two ways to infer the optical depth along each line of sight. Figure \ref{fig:w51_od} compares the optical depth inferred from each method for the three ground state \hho\ transitions. On the one hand, the best-fit method provides us with the gaussian parameters for each absorption features. We give these parameters in Table \ref{tab:w51gg}. Uncertainties on the gaussian parameters are, in 80\% of the cases, lower than 0.2~km~s$^{-1}$ on the position, lower than 0.4~km~s$^{-1}$ on the width, and lower than 0.1 on $\tau_0$. The sum of all the absorption gaussian components gives the optical depth as a function of velocity. By construction, the estimate of the optical depth with the best-fit method is thus always zero at 752~GHz. Also, there is no noise in the inferred optical depth curves. On the other hand, we combine the interpolated continuum with the observed antenna temperature to compute $\tau = -\ln(I/I_{0,int})$. Since $I_{0,int}$ is a smoothed interpolation of the spectrum, the inferred estimate of the optical depth shows the same noise as the data. Consequently, the optical depth at 752~GHz derived from the interpolated continuum is not zero. Also, towards W51, the noise in the data leads to a residual of about a few 0.01 in the optical depth derived with the interpolated continuum for the ground state transitions. We use these residuals to estimate upper limits for the column densities of undetected absorption features (e.g. excited transitions of \hho\ in translucent clouds).


\subsection{Column density}
\label{lab:colden}

We compute the column density $N_l$ of \hho\ and \hheo\ in a given lower energy level using:
\begin{equation}
  N_l = 8\pi \frac{g_l}{g_u} \frac{\nu^3}{c^3} \frac{\tau\Delta v}{A_{ul}} \frac{1}{1-\exp(\frac{-h\nu}{kT_{ex}})},
\end{equation}

\noindent where $g_l/g_u$ is the ratio of the degeneracy factors of the lower and upper levels, $\nu$ is the water line rest frequency, $A_{ul}$ is the spontaneous emission rate of the transition, $T_{ex}$ is the excitation temperature, and $\Delta v$ is the width of the feature. The transition parameters are given in Table \ref{tab:h2olines}. If we assume $T_{ex}\ll h\nu/k$, then the column density is given by:
\begin{equation}
\label{eq:2}
  N_l = 25.1 \frac{g_l}{g_u} \frac{\nu^3}{c^3}  \frac{\tau}{A_{ul}} \Delta v,
\end{equation}

\noindent in SI units. Therefore the corresponding expressions for the column densities of each \hho\ transitions, in units of $10^{12}~\rm{cm^{-2}}$, with $\Delta v$ the FWHM of a gaussian profile in km~s$^{-1}$, and the associated conditions on the excitation temperature are:
\begin{align}
  N_{0_{00}} &= 2.33~\tau({\rm 1113~GHz})  \Delta v &{\rm for}~T_{ex}\ll53~K,\\
  N_{1_{01}} &= 4.66~\tau({\rm 556~GHz})  \Delta v &{\rm for}~T_{ex}\ll27~K,\\
  N_{1_{01}} &= 4.66~\tau({\rm 1669~GHz})  \Delta v &{\rm for}~T_{ex}\ll80~K,\\
  N_{1_{11}} &= 9.23~\tau({\rm 987~GHz})  \Delta v &{\rm for}~T_{ex}\ll47~K,\\
  N_{2_{02}} &= 5.63~\tau({\rm 752~GHz})  \Delta v &{\rm for}~T_{ex}\ll36~K,\\
  N_{2_{12}} &= 13.9~\tau({\rm 1661~GHz})  \Delta v &{\rm for}~T_{ex}\ll79~K,\\
  \intertext{Similarly, for the \hheo\ transitions, we have:}
  N_{0_{00}} &= 2.33~\tau({\rm 1101~GHz})  \Delta v &{\rm for}~T_{ex}\ll53~K,\\
  N_{1_{01}} &= 4.66~\tau({\rm 547~GHz})  \Delta v &{\rm for}~T_{ex}\ll26~K,\\
  N_{1_{01}} &= 4.66~\tau({\rm 1655~GHz})  \Delta v &{\rm for}~T_{ex}\ll79~K,
\end{align}

\begin{figure*}[!t]
\centering
\subfigure[Optical depth]
	{\label{fig:w51_od}
	\includegraphics[angle=90,width=.475\linewidth]{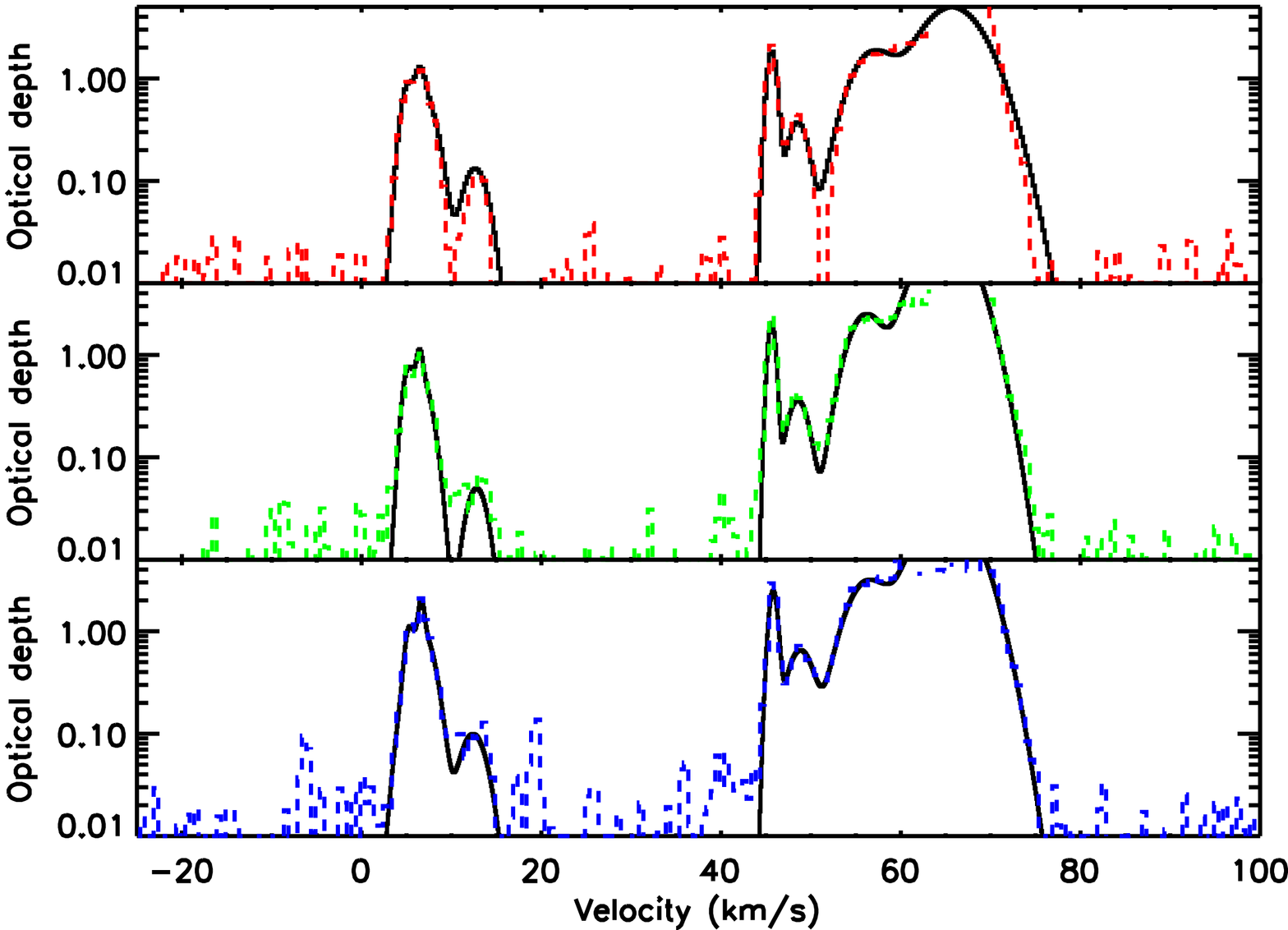}}
\subfigure[Column density in the lower level]
	{\label{fig:w51_cd}
	\includegraphics[angle=90,width=.475\linewidth]{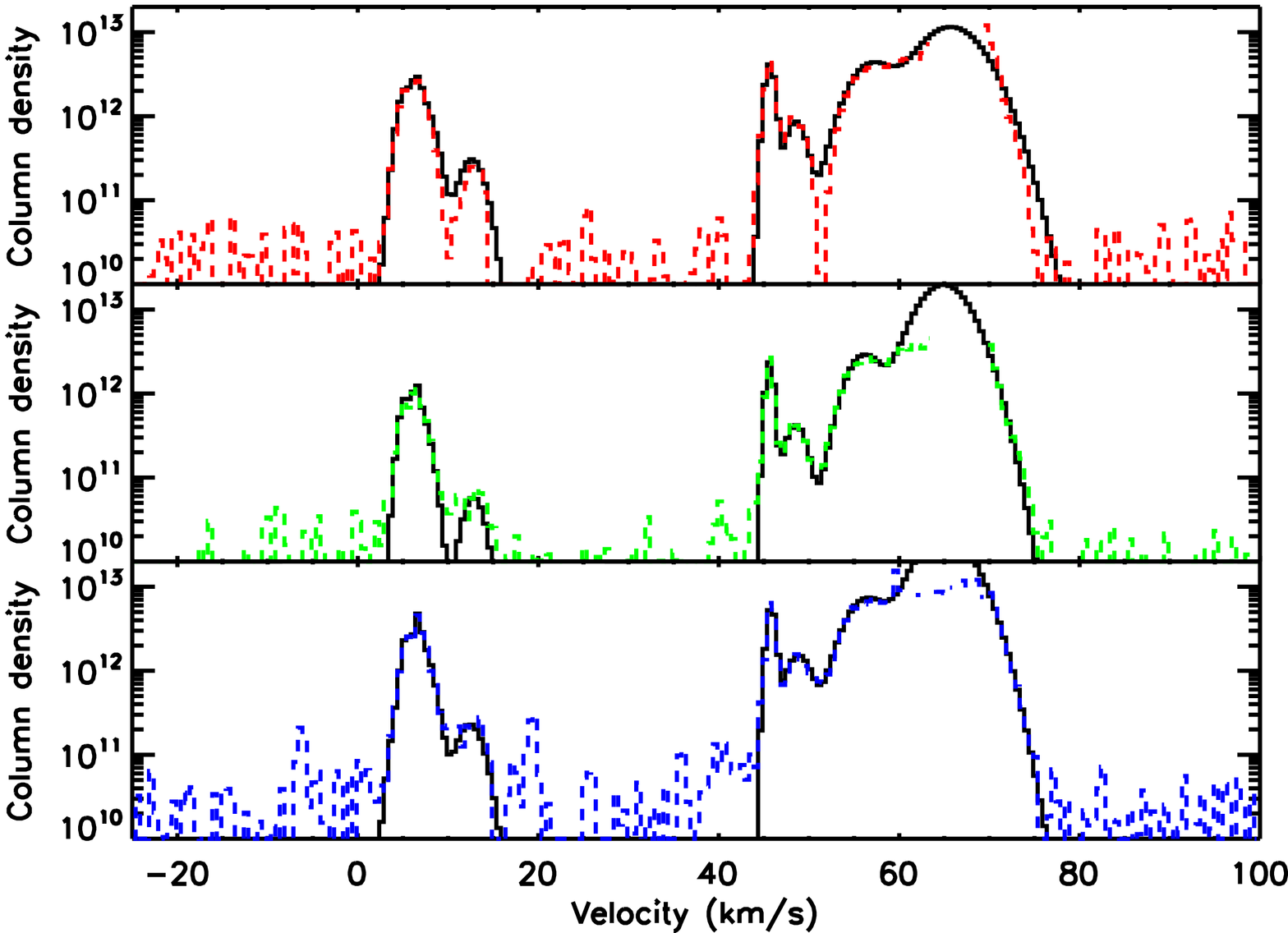}}
\caption{Optical depth and column density in the lower level for the ground state \hho\ transitions towards W51. In each panel, the black solid line is inferred from the gaussian components and the dashed colored line is derived from the interpolated continuum. The optical depth is shown at the native velocity resolution of the spectrum while the column density is shown with a common velocity bin of 0.5~km~s$^{-1}$. The top, middle and bottom panels show the 556, 1113, and 1669~GHz transitions, respectively.}
\label{fig:w51_odcd}
\end{figure*}
 
Since we have two estimates of the optical depth, derived from two estimates of the continuum, we infer two estimates of the column density. In Table \ref{tab:w51gg}, we give the water column density in the lower level of each transition for each gaussian absorption component using the gaussian parameters ($\tau_0$ for $\tau$ and FWHM for $\Delta v$) from the best fit. The column densities that we report are the average of the two polarisations. The absorption features associated with the source (i.e. outflow, envelope) are always optically thick for the ground state \hho\ transitions and are the only ones detected in the excited transitions of \hho\ and the ground state transitions of \hheo. The optical depth and \hho\ column density derived for the ground rotational levels of these absorption features are thus unreliable while those of the excited transitions of \hho\ and the ground state transitions of \hheo\ are reasonable, though usually highly uncertain. On the other hand, the absorption features associated with the translucent clouds are only detected in the ground state \hho\ transitions and not in the \hho\ excited transitions or ground state \hheo\ transitions. Therefore, for these absorption components, by construction, the column density in the \hho\ excited levels and the \hheo\ ground state levels is zero. Figure \ref{fig:w51_cd} shows the column density curves for the ground states transitions towards W51 for a common velocity resolution of $0.5\ \rm{km~s^{-1}}$. In Table \ref{tab:W51gg_nh2o} we then give the total water column density $N($\hho$) = N_{0_{00}} + N_{1_{01}}+N_{1_{11}} + N_{2_{02}} +N_{2_{12}}$ for each absorption component identified by the velocity position of its peak. We only compute the total column density of water for the absorption components associated with translucent clouds. The total column density of water for the absorption component associated with the source is not only highly uncertain but also not representative of the true column density of water, which also comprises the molecules that are responsible for the emission features and that we do not constrain. For the interstellar absorption features, $N_{1_{11}}$, $N_{2_{02}}$, and $N_{2_{12}}$ are zero by construction. Therefore $N($\hho$) = N_{1_{01}}+N_{0_{00}}$ and the fractional abundance of \hho\ in the ground states, $f_{g}($\hho$) = (N_{1_{01}}+N_{0_{00}}) / N($\hho$)$, is thus equal to 1.  The column density in the ortho ground state $N_{1_{01}}$ is given by two \hho\ transitions (556 and 1669~GHz for \hho). We show in the discussion (see section \ref{sec:556vs1669}) that the 1669~GHz transition gives a significantly better estimate of $N_{1_{01}}$ than the 556~GHz transition. The values presented in Table \ref{tab:W51gg_nh2o} and in the Appendix take this into account. In most cases, the total \hho\ column density is about a few times $10^{12}$ to a few times $10^{13}~\rm{cm^{-2}}$.

\begin{table*}[t]
  \centering
  \caption{Column density, OPR, and abundance of water towards W51 for the translucent clouds.}
  \begin{tabular}{c c c | c c c c c c c c}
    \hline
    \hline
    \multicolumn{3}{c|}{Gaussian decomposition} & \multicolumn{8}{c}{Column density curves} \\
    $v$ & $N($\hho$)$ & OPR & $v$ & $N($\hho$)$ & $f_{g}($\hho$)$ & OPR$^+$ & OPR$^g$ & $N($\hh$)$ & X(\hho) & $N($\hheo$)$\\
    (km~s$^{-1}$) & ($10^{12}~\rm{cm^{-2}}$) & & (km~s$^{-1}$) & ($10^{12}~\rm{cm^{-2}}$) & & & & ($10^{20}~\rm{cm^{-2}}$) & ($10^{-8}$) & ($10^{12}~\rm{cm^{-2}}$)\\
    \hline
    5 &    2.8 $\pm$   0.9  &   2 $\pm$ 2 & \multirow{4}{*}{3 to 15} & \multirow{4}{*}{27.3$_{-  0.6}^{+  4}$} & \multirow{4}{*}{$>$0.88} & \multirow{4}{*}{3.0$\pm$0.8} & \multirow{4}{*}{3.5$\pm$0.2} & \multirow{4}{*}{4.3$\pm$1.4} & \multirow{4}{*}{6$_{-2}^{+3}$} & \multirow{2}{*}{$<$0.9} \\
    6 &    3 $\pm$   1  &   3 $\pm$ 1 & \\
    7 &    18 $\pm$   2  &   4 $\pm$ 2 & \\
    13 &    1.9 $\pm$   0.2  &  4.6 $\pm$ 0.7 & \\
    \hline
    46 &    16.7 $\pm$   0.3  &  2.7 $\pm$ 0.1 & \multirow{2}{*}{44 to 51} & \multirow{2}{*}{30$_{-  1}^{+  31}$} & \multirow{2}{*}{$>$0.49} & \multirow{2}{*}{5$\pm$2} & \multirow{2}{*}{3.2$\pm$0.1} & \multirow{2}{*}{2.6$\pm$0.8} & \multirow{2}{*}{12$_{-4}^{+15}$} & \multirow{2}{*}{$<$0.8} \\
    49 &   11.8 $\pm$   0.2  &  4.3 $\pm$ 0.5 & \\
    \hline
  \end{tabular}
  \tablecomments{The left hand side of the table gives values derived from the gaussian decomposition of the optical depth while the right hand side gives values derived from the optical depth curve. The upper hand error bars on $N($\hho$)$ and $X($\hho$)$ are derived from upper limits on the column density in the excited levels and should also be viewed as upper limits. The molecular hydrogen column densities $N($\hh$)$ are derived from CH observations assuming a column density ratio of 0.4 for HF/CH, and an abundance ratio $n(\rm{HF})/n($\hh$)=3.6\times10^{-8}$ \citep{Godard2012}. See section \ref{sec:opr} for details about the OPR values and the meaning of the error bars.}
  \label{tab:W51gg_nh2o}
\end{table*}

In order to estimate more accurately the possible contribution of the excited transition to the total column density of water, we convert the optical depth curves derived from the interpolated continuum into column density curves. Since a velocity bin is not a gaussian profile, the column density inferred from Equation \ref{eq:2} needs to be multiplied by a correction factor $2\sqrt{2\ln(2)} / \sqrt{2\pi} \simeq 0.94$ and $\Delta v$ is given by the velocity bin width. Figure \ref{fig:w51_cd} compares the inferred column density curves for the ground states transitions towards W51 for a common velocity resolution of $0.5\ \rm{km~s^{-1}}$ to that previously derived from the gaussian components. For the ground state \hho\ transitions, the two estimates of the column density are very similar at the position of the absorption features. Away from these absorption features and for the other transitions (excited transitions of \hho, ground state transitions of \hheo), the column density curves only show the residual due to the small difference between $I$ and $I_{0,int}$. With these curves, we compute the water column density of each lower level, for each absorption feature, by integrating the column density curves over the velocity range that best covers the absorption feature. A given absorption feature usually combines several gaussian components, as shown in Table \ref{tab:W51gg_nh2o}. For the excited levels of \hho\ (i.e. $N_{1_{11}}$, $N_{2_{02}}$, and $N_{2_{12}}$) and the ground states of \hheo\ the derived values are upper limits. They usually are about a few $10^{11}~\rm{cm^{-2}}$. For the ortho ground state level of \hho\ (i.e. $1_{01}$), we use the column density inferred from the transition at 1669~GHz rather than that at 556~GHz, for the reasons detailed in the discussion (see section \ref{sec:556vs1669}). In Table \ref{tab:W51gg_nh2o}, we give the total water column density over several velocity ranges. For the translucent clouds, the total column density $N($\hho$)$ is thus given as $N_{0_{00}} + N_{1_{01}}$ with a lower uncertainty $\delta(N_{0_{00}} + N_{1_{01}})$ and an upper uncertainty $\delta(N_{0_{00}} + N_{1_{01}}) + N_{1_{11}}+N_{2_{02}}+N_{2_{12}}$ that shall be viewed as an upper-limit. For each velocity range, we then compute the lower-limit of the fractional abundance of water in the ground states $f_{g}($\hho$) = \min(N_{1_{01}}+N_{0_{00}}) / \max(N($\hho$))$. We use $\min(N_{1_{01}}+N_{0_{00}}) = N_{0_{00}} + N_{1_{01}} - d(N_{0_{00}} + N_{1_{01}})$ and $\max(N($\hho$)) = \min(N_{1_{01}}+N_{0_{00}})+ N_{1_{11}}+N_{2_{02}}+N_{2_{12}}$ as we have to use a consistent value for the column density in the ground states. We also give the upper-limit of the \hheo\ total column density by adding the upper limits of $N_{0_{00}}$ and $N_{1_{01}}$, where $N_{1_{01}}$ is the average from the 547 and 1655~GHz spectra, weighted by their respective uncertainty.  The column density curve also allows us to compute the water abundance relative to that of \hh\ inferred from CH and HF absorption lines by \citet{Godard2012}. The velocity ranges we use to compute the water column density are usually very close to those obtained from the CH and HF observations. We therefore give the column density of \hh\ from \citet{Godard2012} for the closest velocity range from their Table 4. We chose to report the values they derive from the CH observations (their column $d$) as they saturate significantly less often than those derived from the HF observations. We finally compute the water abundance for each velocity range. We discuss the water column density and abundance in section \ref{sec:disc_abun} to \ref{sec:disc_exc}.

\subsection{Ortho-to-para ratio}
\label{sec:opr}

We compute the ortho-to-para ratio $OPR = (N_{1_{01}}+N_{2_{12}})/(N_{0_{00}}+N_{1_{11}}+N_{2_{02}})$. Using the gaussian decomposition of the optical depth, for translucent clouds, the OPR simplifies into $OPR = N_{1_{01}}/N_{0_{00}}$. We report this value as $OPR^g$ in Table \ref{tab:W51gg_nh2o}. Using the column density curves, we compute two extreme values of OPR assuming (1) there are no water molecules in the lower energy levels of the transitions not observed (i.e. for instance $N_{1_{10}}$, $N_{2_{21}}$, and $N_{2_{11}}$ are null) and either (2a) the column density in the ortho excited levels is null while that in the para excited levels is exactly the upper limit (i.e.  $OPR = N_{1_{01}} / (N_{0_{00}} + N_{1_{11}} + N_{2_{02}})$), or (2b) the column density in the para excited levels is null while that in the ortho excited levels is exactly the upper limit (i.e.  $OPR = (N_{1_{01}}  + N_{2_{12}}) / N_{0_{00}}$). We define $OPR^+$ as the mean of these two values. Half their difference is added to the uncertainty on $OPR^g$ to define the uncertainty on $OPR^+$. We report this very conservative value in Table \ref{tab:W51gg_nh2o}. By defintition, $OPR^g$ is thus included in the range defined by $OPR^+\pm dOPR^+$.


\subsection{Uncertainties}
\label{sec:unc}

In the following, we detail our estimation of the uncertainties in the optical depth, column density and OPR. Every time we average two estimates of a value (e.g. column density from two polarisations), we take into account the respective uncertainty into account, using a weighting equal to the inverse of the respective uncertainty squared: \begin{equation}
  <X>=\sum_i^n(x_i/\sigma_i^2)/\sum_i^n(1/\sigma_i^2).
\end{equation}
\noindent The uncertainty in the weighted mean is:
\begin{equation}
  \sigma^2=\frac{\sum_i^n( (x_i-<X>)^2/\sigma_i^2)}{(n-1)~\sum_i^n(1/\sigma_i^2)}.
\end{equation}
\noindent Every time we add values (e.g. to compute the total water column density), we add the uncertainties in quadrature: $\sigma^2=\sum_i^n(\sigma_i^2)$.

One source of uncertainty in $\tau$ is due to the noise $\delta I$ in the observed antenna temperature $I$, while the second source of uncertainty is the difference $\delta I_0$ between the two estimates of the continuum $I_{0,bf}$ and $I_{0,int}$. $\delta I_0$ is not constant for a given spectrum and usually is about a fraction of Kelvin. Therefore, $\delta I_0/I_0$ usually is about a few percent. To estimate $\delta I$, we measure the standard deviation of $I$ away from emission and absorption features. Consequently, $\delta I$ is constant for a given spectrum. At 547 and 556~GHz, $\delta I$ ranges from 0.006~K towards DR21(OH) and W33A, to 0.027~K towards W51. At 1655, 1661 and 1669~GHz, $\delta I$ ranges from 0.07~K towards W33A to 0.350~K towards G34.1+0.1. There is a slight difference, usually less than a few percent but up to $\sim$25\%, between the value of $\delta I$ inferred from each polarisation spectrum, though we do not see any systematic effect. The characteristic value for $\delta I/I$ is about a few percent. $\delta I$ and $\delta I/I$ can later be significantly reduced by working at a common lower velocity resolution. We chose to work with a velocity bin size of 0.5~km~s$^{-1}$ which is slightly smaller than the worst resolution of $\sim0.6$~km~s$^{-1}$ at 556 and 547~GHz.

We perform the best fit to the observations taking the uncertainty $\delta I$ into account. The resulting parameters for the gaussian components ($\tau_0$, FWHM and $v_0$) and their uncertainty are computed by the MPFIT package. 
The uncertainty in the column density then is $\delta N_{0_{00}} = 2.33 \times (\delta \tau_0 \times FWHM + \delta FWHM \times \tau_0)$ and similarly for each transition and each polarisation separately. We then add the contribution of $\delta I_0$. The corresponding uncertainty in the optical depth is $\delta \tau_{cont} = (\tau_{bf}-\tau_{int})=\ln{(I_{0,bf}/I_{0,int})}$ which is not constant as a function of velocity. Therefore, for each gaussian component, we compute the mean value of $\delta \tau_{cont}$ over a velocity range centered on $v_0$ and as wide as the FWHM. We then convert $\delta \tau_{cont}$ into $\delta N_{0_{00}} = 2.33 \times \delta \tau_{cont} \times FWHM$ and similarly for each transition and each polarisation separately. We then quadratically sum both sources of uncertainty in $N$ and average both polarisations. We average the gaussian parameters the same way. We report the resulting values in Table \ref{tab:w51gg}. In Table \ref{tab:W51gg_nh2o}, we give the total column density and average OPR of water for each gaussian component.

In Table \ref{tab:W51gg_nh2o}, we give the results from the integration of the column density curves, then used to compute the OPR and water abundance relative to \hh. The uncertainty associated with the column density curve is $\delta N = N/ \tau \times (\delta I/I + \delta I_0/I_0)$ for each transition and each polarisation separately. It is therefore not constant as a function of velocity. When integrating over a given velocity range, we quadratically sum the uncertainty associated with each velocity bin before averaging the results from the two polarisations. When averaging the column density derived from each polarisation, we note that in the very large majority of cases, the difference between the two values is lower than the uncertainty and almost always less than 3 times the uncertainty.

\subsection{Comparing the two methods}
\label{sec:disc_met}

\begin{figure*}[!t]
  \centering
  \subfigure[]
  {\label{fig:h2ovsh2o}
    \includegraphics[angle=90,width=.475\linewidth]{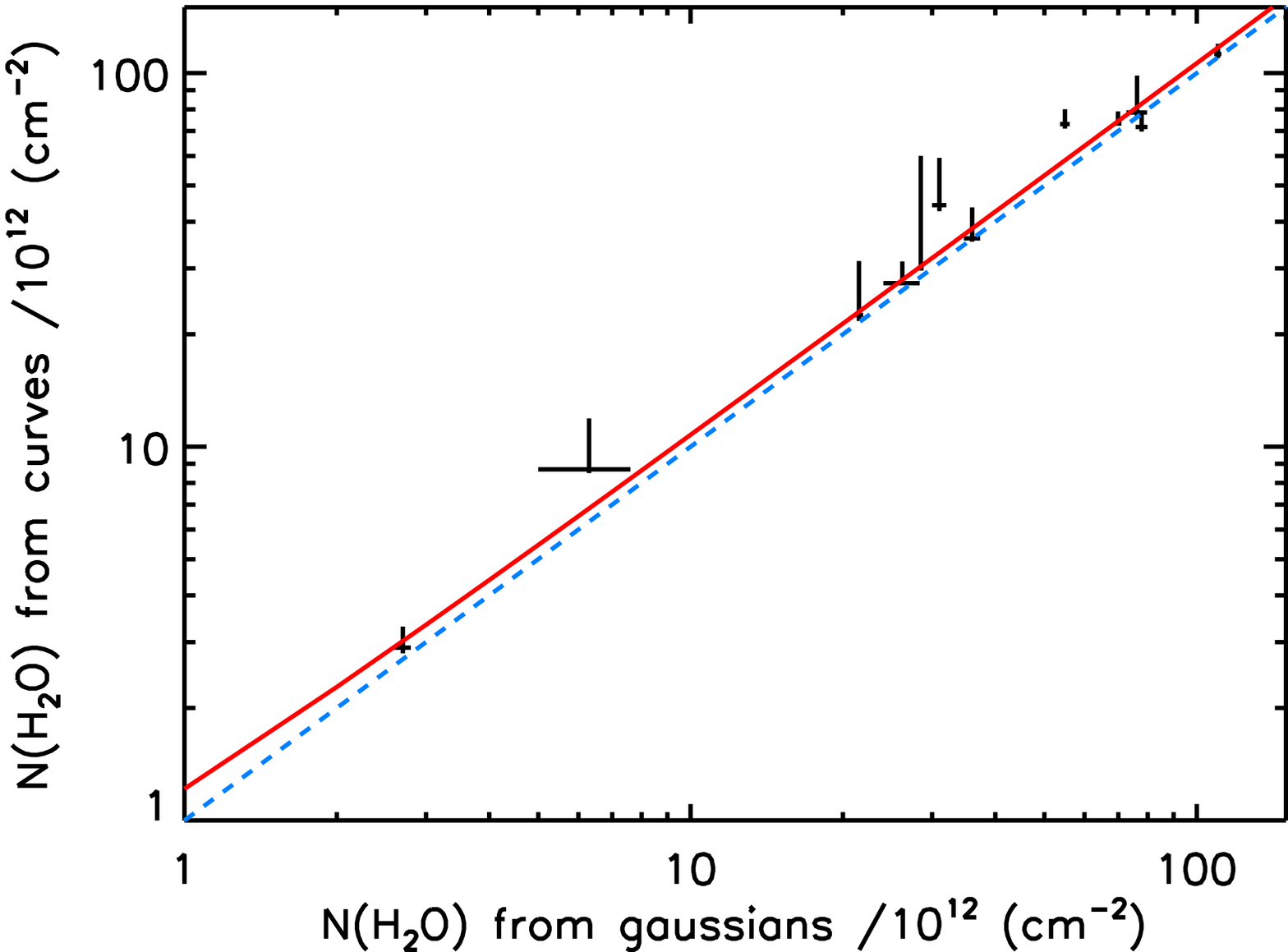}}
  \hfill
  \subfigure[]
  {\label{fig:h2oh2ovsh2o}
    \includegraphics[angle=90,width=.475\linewidth]{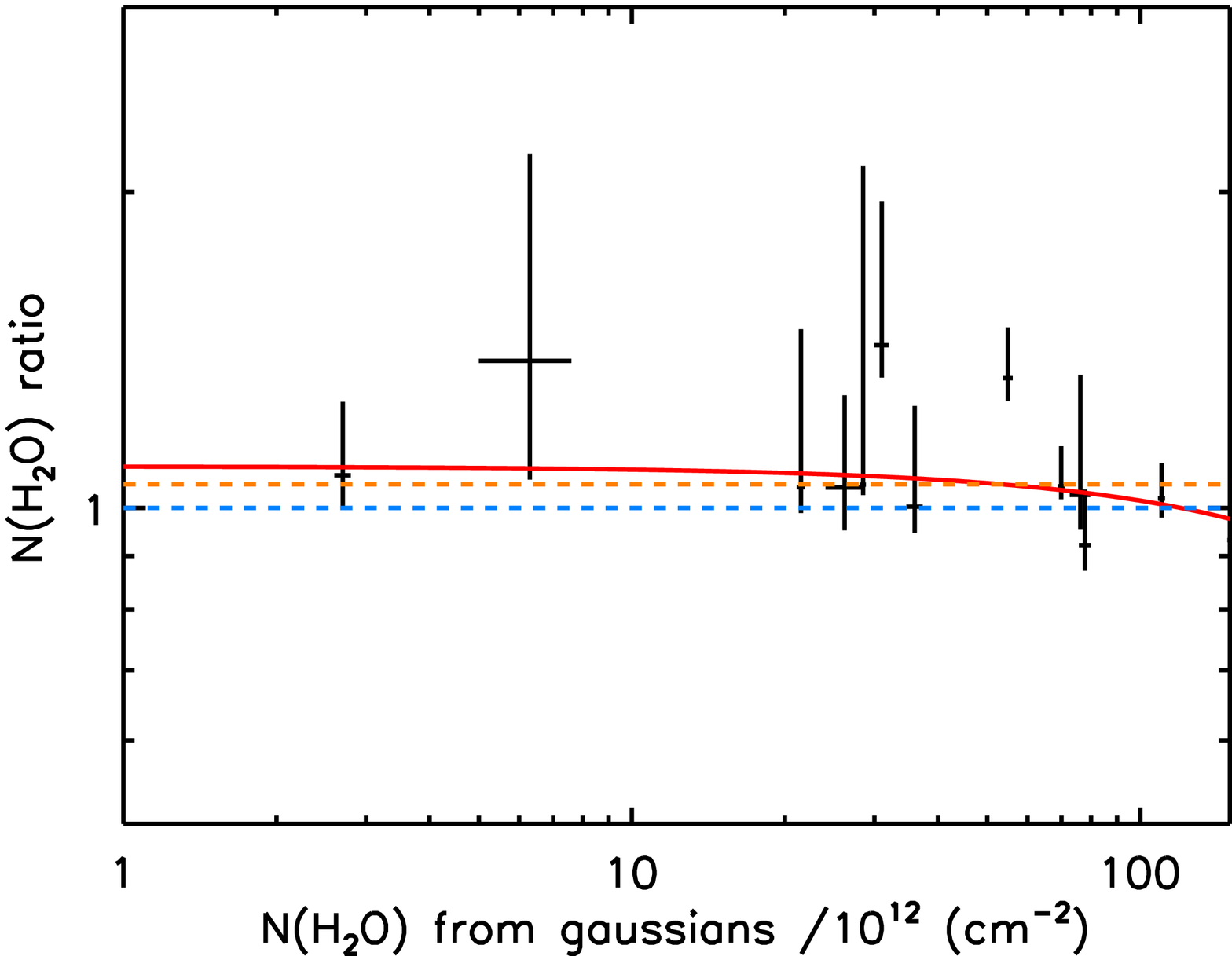}}
  \caption{Comparison between the total column density of \hho\ inferred from the gaussian profiles and from the column density curves. Each data point represents one absorption feature group. In each panel, the red solid line is the best linear fit to the data, the blue dashed line is a column density ratio equal to unity, and the orange dashed line is the average column density ratio}
  \label{fig:h2oh2o}
\end{figure*}
 
We measure the total column density of water using the gaussian profiles and using the column density curves. Differences between the two methods may arise from contamination due to non-water absorption feature, possible water in the excited levels, nearby water absorption feature associated with the clouds, or broad water absorption profiles associated with the source. For instance, when measuring the total column density towards W51 with the curve (see Figure \ref{fig:w51_cd}), the feature at 49~km~s$^{-1}$ is obviously contaminated by the interstellar feature at 46~km~s$^{-1}$ and by the broad feature between 55 and 80~km~s$^{-1}$ associated with the source. Additionally, when integrating the curve, we compute the possible contribution from excited levels, assuming these are upper limits. Finally, the use of finite limits of integration may underestimate the column density inferred from the curve. Consequently, there might be some scattering and/or systematics between the total column density of water derived with each method.

Figure \ref{fig:h2oh2o} compares the values obtained for each group of absorption features with the gaussian profiles and the column density curves. In sum, we have 13 groups of absorption features, each of which could be an isolated cloud or a group of clouds. In figure \ref{fig:h2ovsh2o} we show that over almost two orders of magnitude, the total column densities inferred from each method are very well correlated. The best linear fit to the data leads to a slope of 1.06$\pm$0.01, if we only take into account the uncertainty in the ground state population. The slope is 1.03$\pm$0.04 if we add the upper limits on the excited states population to the total column densities. This means there is a good agreement between the two methods but that a slight systematic effect may lead to lower value of the total column density of water when using the gaussian components alone. Figure \ref{fig:h2oh2ovsh2o} then shows the ratio of the column density derived from the two methods as a function of the total column density of water. We detect no clear trend though larger column densities seem to give a better agreement between the two methods. A linear fit leads to an intercept of $\sim1.1$ and a slope of $\sim10^{-3}$. The average ratio is 1.05$\pm$0.01 or 1.03$\pm$0.04 depending on the uncertainties we use. We conclude that there is a systematic effect, whose origin may be one of the explanations given above. However, we also show here that the differences between the two methods are not significant (less than 5\%). Both methods thus appear to yield equally reliable measurements of the total column density of water.

\section{Discussion}
\label{sec:disc}

In this section, we discuss in detail the water column density, abundance, and OPRs found along the PRISMAS lines of sight. We first discuss the presence of water in the excited levels and quantitatively analyze the differences between the two ortho ground state transitions to derive constraints on the physical conditions within the translucent clouds. We then summarize our findings regarding the total water column density, its abundance relative to molecular hydrogen, and discuss possible variations throughout the Galaxy. We finally analyse the OPR, and how these relate to the physical conditions the \hho\ molecules have interacted with in their history.

\subsection{On the abundance of \hheo\ and the detection of \hho\ in the excited transitions}
\label{sec:disc_exc}

For the first time, Herschel/HIFI enables the observations of all the \hheo\ ground state transitions as well as the first few excited transitions of both ortho- and para-\hho. The only absorption features that are unequivocally detected in those transitions are related to the background source and not discussed further in this paper. The absorption features associated with the translucent clouds only lead to upper limits of the total column density of \hheo\ and to that of the column density of \hho\ in the excited transitions.

The abundance ratio between $^{16}$O and $^{18}$O is commonly accepted to be $\sim$500, though significant variations have been observed through the Galaxy. Therefore, the optical depth of \hheo\ is $\sim$500 times lower than that of \hho\ for any given velocity bin. This implies that either the \hheo\ optical depth is too small to be detected or the \hhso\ optical depth is large and the absorption feature is saturated. Therefore, we can only determine the column density of one of the isotopic form of water at a time and consequently, can not derive the \hho-to-\hheo\ isotopic abundance from our data. For the translucent clouds, we only measure $N($\hho$)$ and provide upper limits on $N($\hheo$)$, usually around $10^{12}~\rm{cm^{-2}}$. The inferred lower limits on the $^{16}$O-to-$^{18}$O abundance ratio are always consistent with, though significantly lower than 500.

The excited transitions of water are all detected in emission and, with the exception of the 752~GHz para-\hho\ transition, in absorption at the velocity of the source. The translucent clouds only exhibit absorption features in the ground transitions and only upper limits on the column density of water in the excited transitions can be computed. Therefore we can estimate the fraction of the population of water molecules that lies in the ground state. We define $f_g = (N_{0_{00}}+N_{1_{01}})/(N_{0_{00}} + N_{1_{11}} + N_{2_{02}} + N_{1_{01}} + N_{2_{12}})$ for each absorption component. We therefore assume there are no water molecules above the excited levels observed with Herschel. For the gaussian components, $f_g=0$ by construction as no gaussian are required to fit the excited transition spectrum for the translucent clouds. For the column density curves, we compute the upper limits of $N_{2_{02}}$, $N_{1_{01}}$, and $N_{2_{12}}$ and use them to derive a lower limit on $f_g$ (see Table \ref{tab:W51gg_nh2o}). Towards the six PRISMAS sources, the lower limit on $f_g$ varies between 0.50 and 0.98. Averaged over all the interstellar features where we measure it, the lower limit on $f_g$ is 0.88, if weighted by the total water column density of each absorption feature. We also compute the fractional abundance of water in the ground state for the ortho and para form of \hho\ independently for each absorption feature and do not find any significant difference between them though we notice a better correlation between $f_g$ and the ortho fractional abundance than the para fractional abundance.

These constraints on the population in the excited levels thus enable us to estimate the excitation temperature. We have:
\begin{align}
  \frac{N_{2_{12}}}{N_{1_{01}}} = \frac{5}{3} \exp(-\frac{80}{T_{ex}})
\end{align}
\noindent for the ortho water, and:
\begin{align}
  \frac{N_{2_{02}}}{N_{1_{11}}} = \frac{5}{3} \exp(-\frac{48}{T_{ex}}) ~ &; ~ \frac{N_{1_{11}}}{N_{0_{00}}} = 3 \exp(-\frac{53}{T_{ex}}),
\end{align}
\noindent for the para water. We assume here that a single $T_{ex}$ can properly describe the water populations we observe in the first few ortho or para levels but treat the ortho and para-\hho\ independently. We then subsitute these expression of the column density ratios into the expression of $f_o$ and $f_p$, the fractional abundances in the ground state levels of the ortho and para water, respectively. We thus infer the following relations between the excitation temperature and the fractional ground abundances:
\begin{align}
  \frac{1}{f_o} &= 1 + \frac{5}{3} \exp \left( -\frac{80}{T_{ex}}\right), \nonumber \\
  \frac{1}{f_p} &= 1 + 3 \exp \left( -\frac{53}{T_{ex}}\right) + 5 \exp \left( -\frac{101}{T_{ex}}\right),
  \label{eq:frac}
\end{align}

Since the fractional abundances are lower limits, the excitation temperatures we compute are upper limits. The para-\hho\ transitions are more sensitive to $T_{ex}$ as the para 1113~GHz transition corresponds to a smaller temperature difference (53~K) than the ortho 1669~GHz transition (80~K). As a consequence, the upper-limit on the excitation temperature derived from the para water is almost always more stringent. The upper limits on the excitation temperature we find for the 13 translucent clouds range from about 13 to 26~K. The best situation corresponds to the cloud at $16-28$~km~s$^{-1}$ towards W28A. If instead of assuming a single $T_{ex}$ for the first three para levels, we compute it using $N_{1_{11}}$ and $N_{0_{00}}$, we find almost exactly the same results, since the correction due to the population in the $2_{02}$ level is not significant.



\subsection{On the discrepancy between the two ortho ground state transitions}
\label{sec:556vs1669}

While SWAS observed the ortho-water ground state transitions at 547 and 556~GHz and ISO observed the ortho-water ground state transitions at 1655 and 1669~GHz, the HIFI instrument on board Herschel allows us for the first time to observe all these transitions with the same instrument. If the assumption of low excitation temperature ($h\nu\gg kT_{ex}$) is verified, then our estimates of $N_{1_{01}}$ inferred from either the 556 or the 1669~GHz transition for \hho\ should be the same. 

\begin{figure*}[!t]
  \centering
  \subfigure[]
  {\label{fig:o556vso1669}
    \includegraphics[angle=90,width=.475\linewidth]{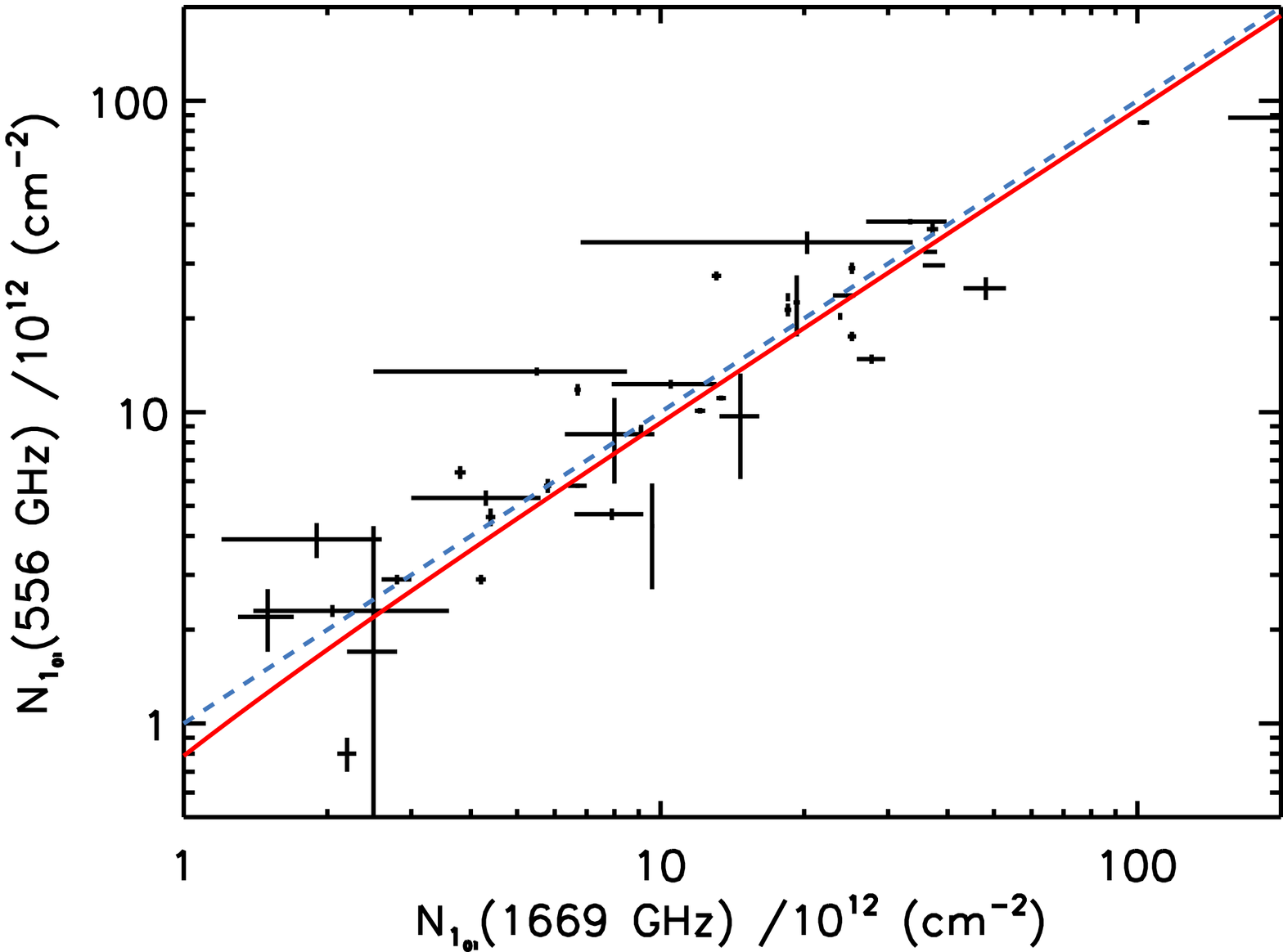}}
  \hfill
  \subfigure[]
  {\label{fig:o556o1669vso1669}
    \includegraphics[angle=90,width=.475\linewidth]{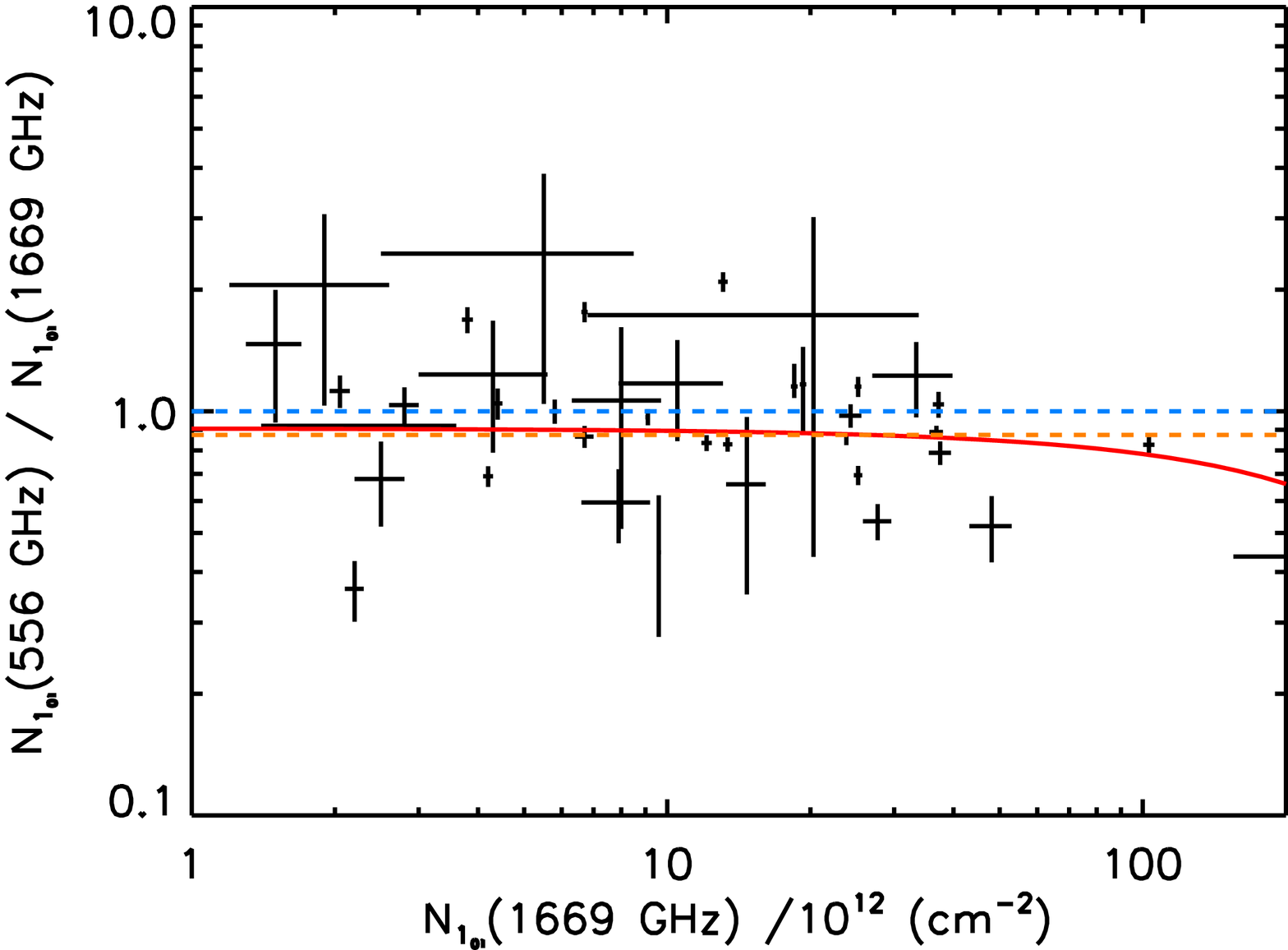}}
  \caption{Comparison between the column density of the $1_{01}$ level inferred from the 556~GHz transition ($N_{1_{01}}(556~GHz)$) and the 1669~GHz transition ($N_{1_{01}}(1669~GHz)$). Each data point represents one absorption gaussian profile. In each panel, the red solid line is the best linear fit to the data, the blue dashed line is a column density ratio equal to unity, and the orange dashed line is the average column density ratio.}
  \label{fig:ortho}
\end{figure*}

Figure \ref{fig:o556vso1669} shows the column density of the $1_{01}$ level inferred from the 556~GHz transition ($N_{1_{01}}(556~GHz)$) as a function of that inferred from the 1669~GHz transition ($N_{1_{01}}(1669~GHz)$) for each gaussian component, not associated with the source, in the best fit of the spectra. We also show a 1-to-1 correlation and the best linear fit to the observations. While the two estimates of the column density are very well correlated, there is a clear systematic effect. The 556~GHz transition statistically leads to a lower column density than the 1669~GHz transition. The best linear fit has a slope of $0.94\pm0.01$ and an intercept of $-0.2\pm0.3$. We do the same comparison with the values derived from the column density curves and find a slightly better correlation (not shown here) and $N_{1_{01}}(556~GHz)$ remains statistically lower than $N_{1_{01}}(1669~GHz)$. Figure \ref{fig:o556o1669vso1669} then shows the ratio between the column density inferred from each transition as a function of $N_{1_{01}}(1669~GHz)$ to look for second order variations. The discrepancy is even clearer on this plot and seems to increase as the column density increases, though the scatter is significant at all column densities. A linear fit to the data gives a slightly decreasing slope of $\sim10^{-3}$ and an intercept of $\sim0.9$. The average value of $N_{1_{01}}(556~GHz)$-to-$N_{1_{01}}(1669~GHz)$ is between 0.87$\pm$0.03 and 0.89$\pm$0.04 whether we use the gaussian components or the column density curves. We draw the reader's attention to the fact that the plots only show the gaussian absorption component associated with the translucent clouds but that some of these might be more affected than others by the emission from the source. We removed the few most extreme data points before running the fit. It is however clear that a systematic discrepancy of about 10\% exists between the two estimates of the ortho ground state column density. To interprete this discrepancy, we explore several possibilities.

\begin{figure*}[!t]
  \centering
  \subfigure[Towards DR21(OH)]
  {\label{fig:dr21tex}
    \includegraphics[angle=90,width=.315\linewidth]{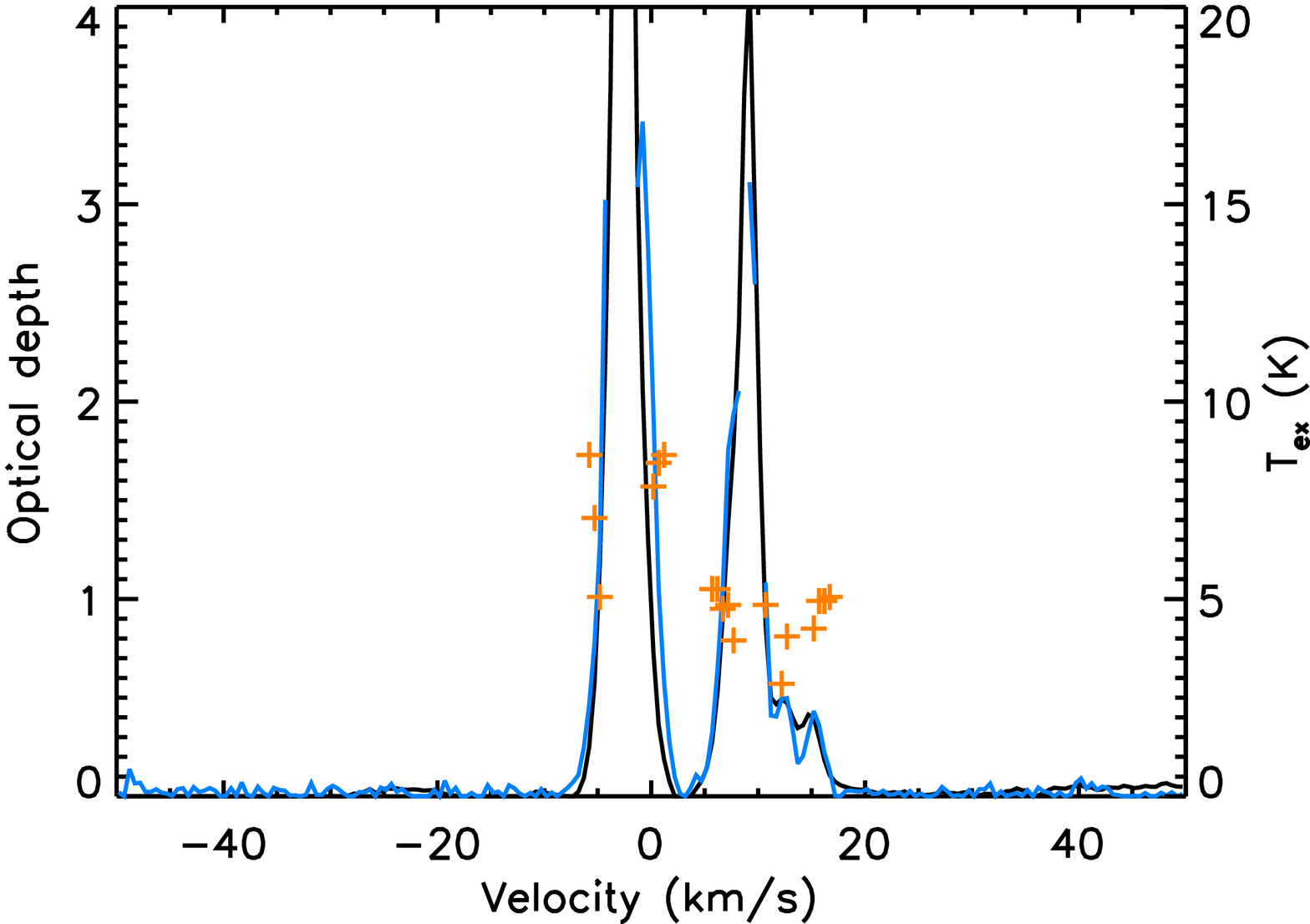}}
  \subfigure[Towards G34.3+0.1]
  {\label{fig:g34tex}
    \includegraphics[angle=90,width=.315\linewidth]{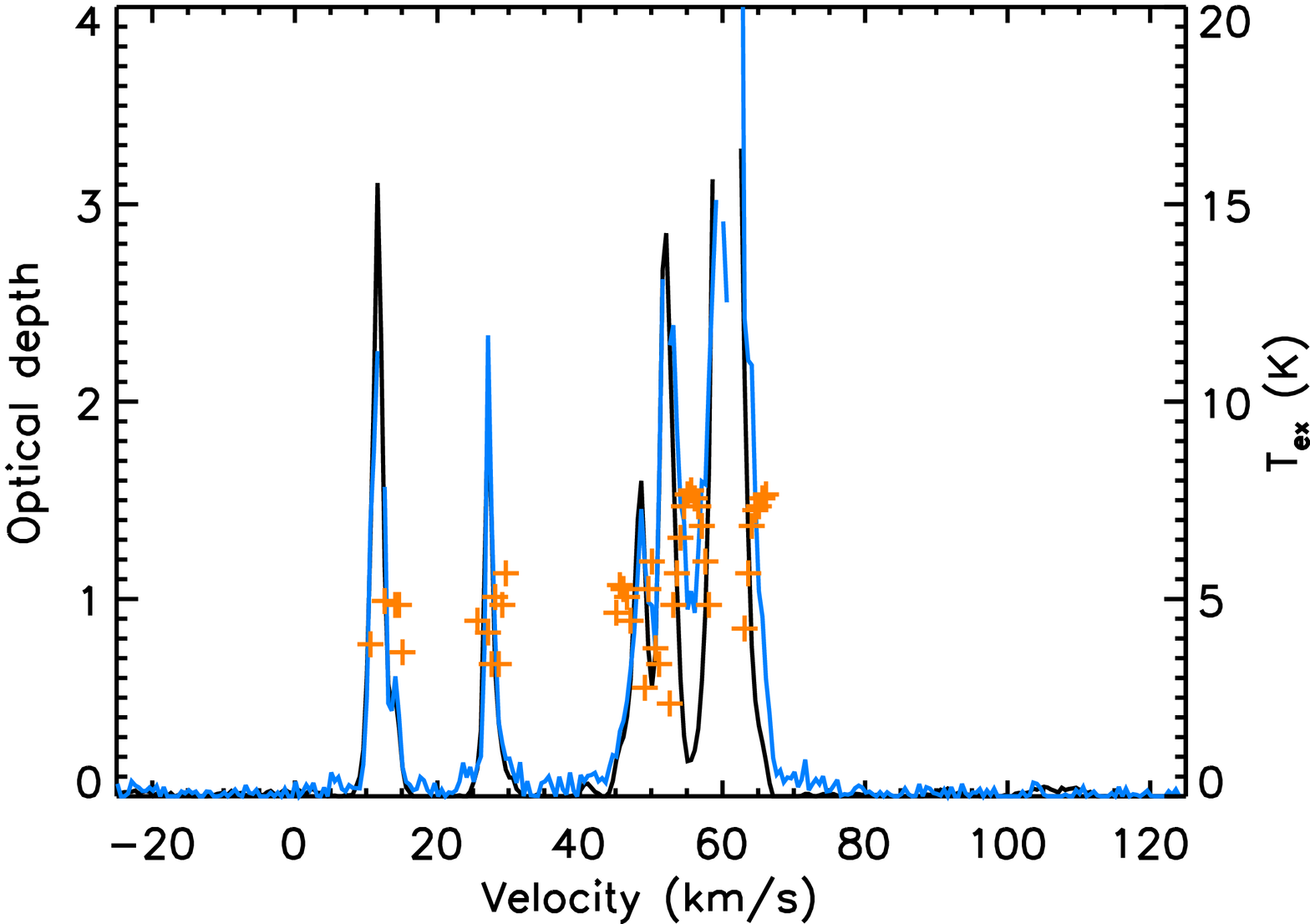}}
  \subfigure[Towards W28A]
  {\label{fig:w28tex}
    \includegraphics[angle=90,width=.315\linewidth]{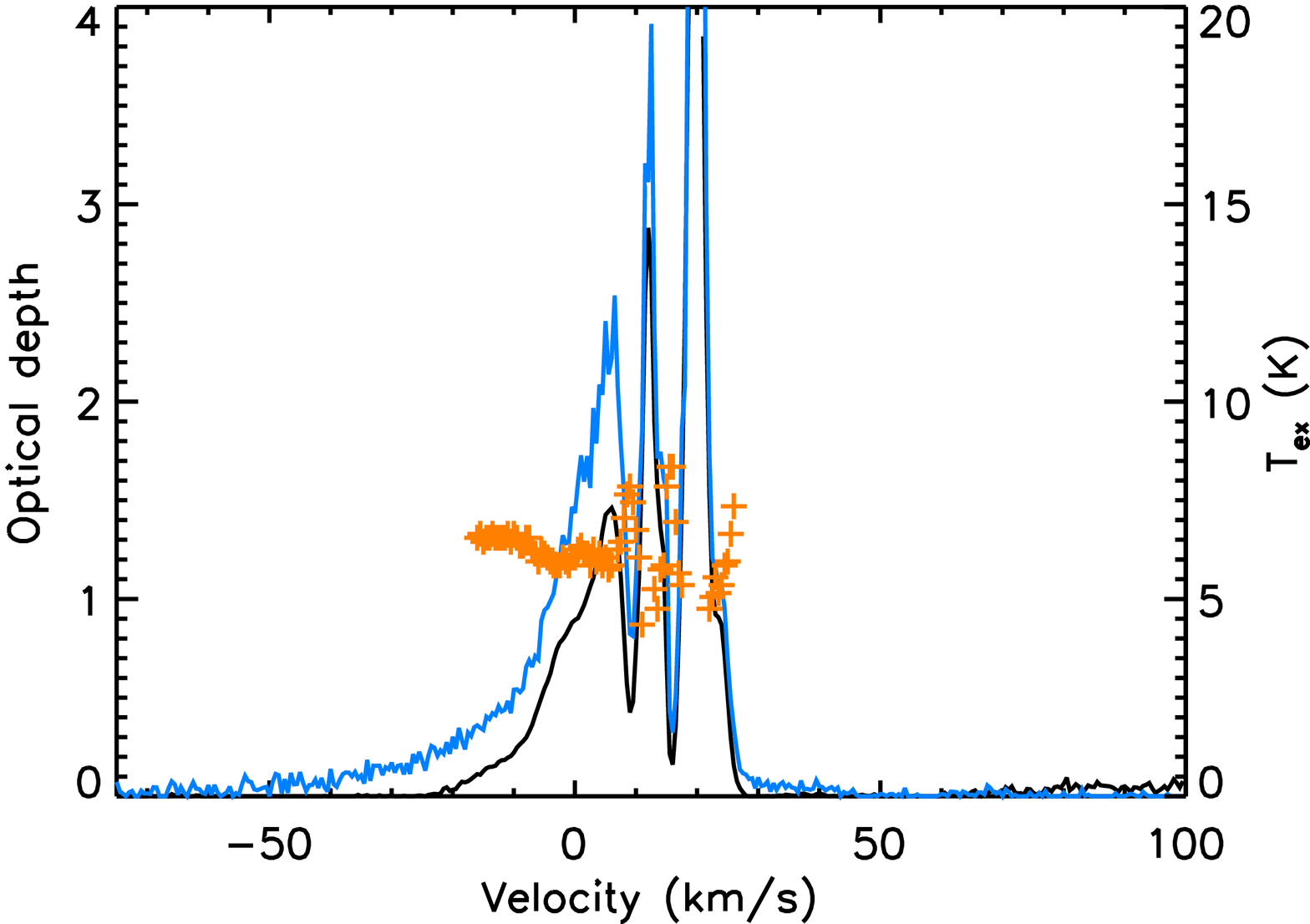}}
  \subfigure[Towards W33A]
  {\label{fig:w33tex}
    \includegraphics[angle=90,width=.315\linewidth]{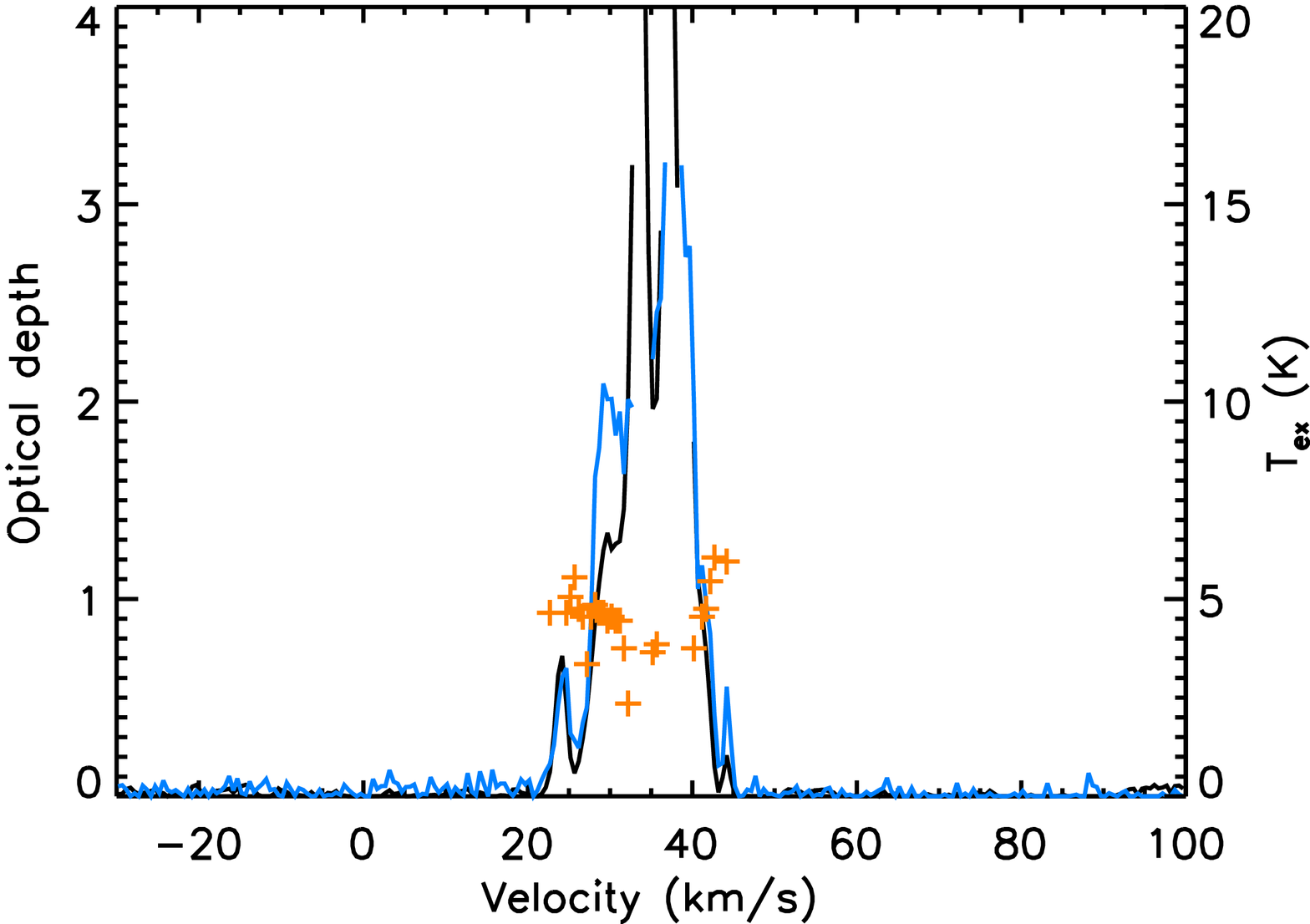}}
  \subfigure[Towards W49N]
  {\label{fig:w49tex}
    \includegraphics[angle=90,width=.315\linewidth]{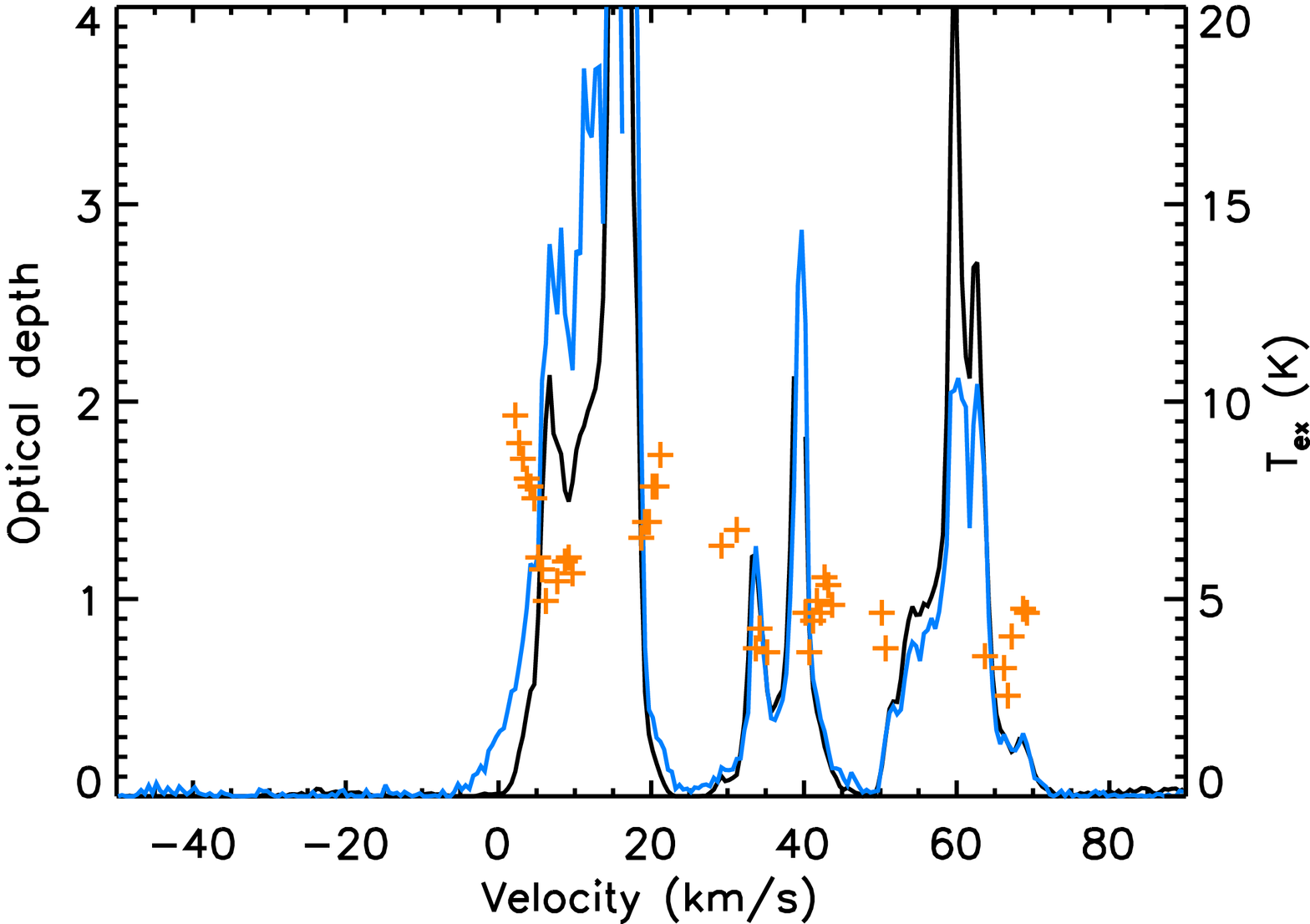}}
  \subfigure[Towards W51]
  {\label{fig:w51tex}
    \includegraphics[angle=90,width=.315\linewidth]{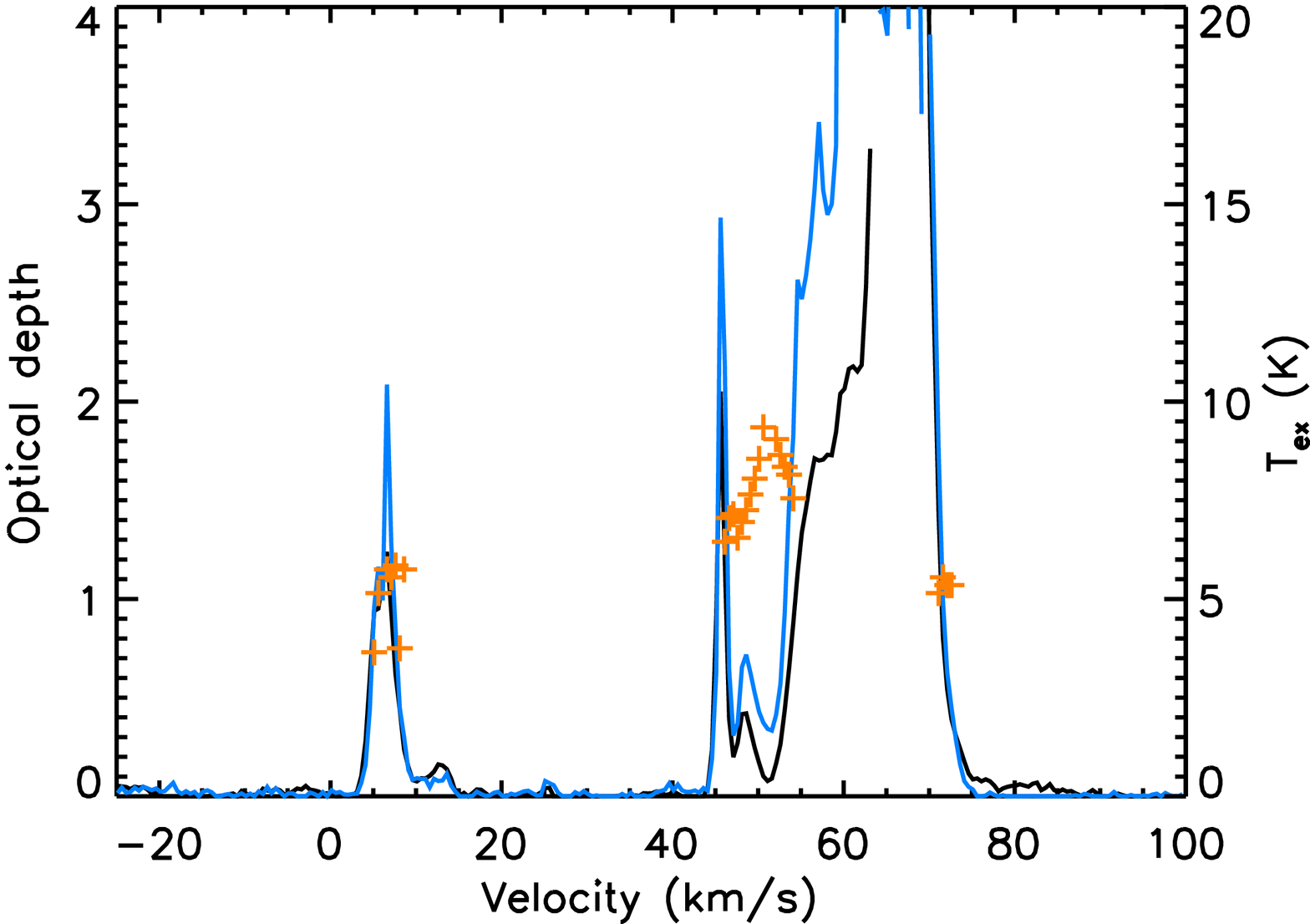}}
  \caption{Comparison between the optical depth derived from the 556~GHz (in black) and the 1669~GHz (in blue) transitions under the low $T_{ex}$ assumption. The values of $T_{ex}$ (right y-axis) required to make these optical depths consistent are indicated by the orange crosses over the regions where the optical depth is between 0.1 and 2.5.}
  \label{fig:tex}
\end{figure*}
 
First, the assumption on the low excitation temperature (see equation \eqref{eq:2}) may not be always true. Therefore, in the general case, we do not find $\tau_{556} = \tau_{1669}$ because $\tau \ne \ln{ (I/I_0)}$ but:
\begin{equation}
  \tau = -\ln{ \left( \frac{I - B_\nu(T_{ex})}{I_0-B_\nu(T_{ex})} \right) },
\end{equation}
\noindent where $B_\nu$ is the Planck function. The condition $T_{ex}\ll h\nu/k$ is more easily verified at 1669 ($h\nu/k\simeq80$~K) than 556~GHz ($h\nu/k\simeq27$~K). Additionally, the dust continuum is usually significantly higher at 1669 than 556~GHz. In our data, we observe that the continuum intensity is between 6 and 16 times higher at higher frequency. As a consequence, the optical depth inferred from the 1669~GHz spectrum assuming a low $T_{ex}$ should be closer to the optical depth derived without any assumption on $T_{ex}$, than that derived from the 556~GHz transition. To confirm this and to estimate $T_{ex}$ in the translucent clouds, we look for the value of $T_{ex}$ that gives

\begin{align}
  &-\ln{ \left( \frac{I_{0,556}\exp{(-\tau_{556})} - B_{556} (T_{ex})}{I_{0,556}-B_{556} (T_{ex})} \right) } \nonumber \\
= &-\ln{ \left( \frac{I_{0,1669}\exp{(-\tau_{1669})} - B_{1669} (T_{ex})}{I_{0,1669}-B_{1669} (T_{ex})} \right) }
\end{align}

\noindent where $I_{0,556}$ and $I_{0,1669}$ are the continuum intensities as inferred from the best-fit at 556 and 1669~GHz, respectively, and $\tau_{556}$ and $\tau_{1669}$ are the optical depth we derived under the assumption $T_{ex}\ll h\nu/k$. In this equation we can neglect the contribution of the CMB whose antenna temperature at 556~GHz is less than 2~mK. We also assume here that no other effect can produce any discrepancy between $\tau_{556}$ and $\tau_{1669}$ (see below for other explanations). For all six sources, over regions where the optical depth is larger than 0.1 and smaller than 2.5, we then find that the averaged value of $T_{ex}$ ranges from 4.5~K towards W33A to 6.7~K towards W51. The required $T_{ex}$ varies within each cloud (see Figure \ref{fig:tex}). These expected values on the excitation temperatures are in agreement with the constraints derived from the upper limits on the column density of water in the excited levels (see section \ref{sec:disc_exc}) though they are more stringent. Assuming that collisional excitation is responsible for populating the excited levels, and that the excitation temperature is 5~K, the fractional abundance of water in the excited levels (1-$f_o$) and (1-$f_p$), as derived from Equation \ref{eq:frac}, would be about $2~10^{-7}$ for ortho-\hho\ and about $8~10^{-5}$ for para-\hho. These constraint are significantly lower than the upper limits derived from the observations of the excited transitions.

The constraints on the excitation temperatures provide us with a constraint on the density in the translucent clouds along the line of sight. We use the computer program Radex \citep{vanderTak2007} and the associated Leiden Atomic and Molecular Database \citep[LAMDA][]{Schoier2005} to compute the excitation temperature of ortho- and para-\hho. The collision rates between \hho\ and \hh\ are from \citet{Dubernet2009} and \citet{Daniel2011}. We use a kinetic temperature of the gas below 100~K and densities of \hh\ lower than $10^6~\rm{cm^{-3}}$. We assume the only partner for collisions is molecular hydrogen. The results, not shown here, indicate that for a kinetic temperature of 50 to 100~K, typical in translucent clouds \citep{Rachford2002}, an excitation temperature of 5~K is found for density smaller than $\sim10^4~\rm{cm^{-3}}$ for ortho-\hho. These upper limits on the density are consistent with, though significantly larger than the densities that characterize translucent clouds ($\sim100~\rm{cm^{-3}}$), and do not rule out the presence of high density clumps in the clouds. The total hydrogen column densities are about a few $10^{21}~\rm{cm^{-2}}$ while the sizes of the clouds are about a few 100~pc at most (see section \ref{sec:556vs1669}), which corresponds to average densities of a few cm$^{-3}$ at least. However, the cloud sizes are overestimated due to likely velocity dispersion within them and the inferred density is averaged along the line of sight while significant variations may occur within each cloud.

\begin{figure}[!t]
  \centering
  \includegraphics[angle=90,width=.9\linewidth]{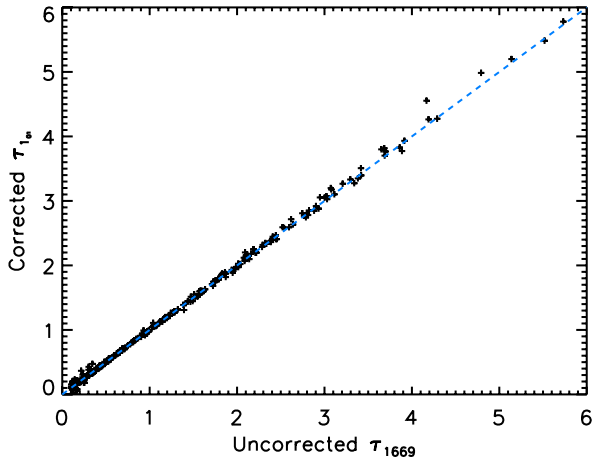}
  \caption{Corrected optical depth for the ortho ground state of water as a function of the uncorrected optical depth derived from the 1669~GHz transition. Each data point represents one velocity channel where the corrected optical depth is above 0.1. All the sources are combined. The dashed blue line is the 1-to-1 correlation.}
  \label{fig:corrtau}
\end{figure}

We also compare the corrected $\tau_{1_{01}}$ to that derived under the assumption of low $T_{ex}$ for both ortho ground transitions. The correlation is almost perfect with $\tau_{1669}$ (see Figure \ref{fig:corrtau}). This plot confirms that the 1669~GHz transition is better than the 556~GHz transition for estimating the column density of water in the ortho ground state. We discuss how this constraint on the excitation temperature translates into a constraint on the density in section \ref{sec:disc_exc}.

Another interpretation of the discrepancy between $\tau_{556}$ and $\tau_{1669}$ relies on the significant amount of water emission at 556~GHz which makes it more difficult to properly constrain the continuum emission and consequently the optical depth and column density for this transition. For instance, towards W51, the peak of the water emission is at least 1~K above the $\sim$1~K continuum at 556~GHz and almost undetectable above the $\sim$7~K continuum at 1669~GHz. The differences in the emission features consequently affect the spectral shape of the absorption components at 556 and 1669~GHz (e.g. towards W51 between 45 and 65~km~s$^{-1}$). The spectral shape is also affected by the velocity resolution ($\sim 200$~m/s at 1669~GHz, $\sim600$~m/s at 556~GHz) and the sensitivity at each frequency. However, these differences should not introduce systematics. The only explanation for the observed systematics would be an underestimation of the water emission from the source at 556~GHz, or an overestimation of the water emission from the source at 1669~GHz. The latter seems difficult to obtain as the two estimates of the continuum, from the best fit and the interpolation, tend to minimize the contribution from water emission. The former seems more plausible as the only constraints on the water emission profile are usually limited to their wings and sometimes a few velocity channels significantly above the dust continuum. A perfect example of this situation is seen towards W51. The true contribution of the water emission to the continuum could obviously be much stronger than that derived from the best fit and the interpolation at 556~GHz but it could hardly be fainter at 1669~GHz (see Figure \ref{fig:w51}). The $N_{1_{01}}(556~GHz)$-to-$N_{1_{01}}(1669~GHz)$ ratio for the two absorption components towards W51 is 0.88 from 3 to 15~km~s$^{-1}$ and 0.70 from 44 to 51~km~s$^{-1}$. Therefore, to explain these discrepancies with an increase of the water emission at 556~GHz, the optical depth should be 1.1 and 1.4 times larger than that derived from the best fit for each absorption component at this frequency. We convert these scaling factors for the optical depth into correction factors for the continuum intensity. From 44 to 51~km~s$^{-1}$ for instance, the corrected continuum $I_{0,corr}$ would be given by $I_{0,corr}/I = (I_{0}/I)^{1.4}$. Averaged over the width of the feature, this implies a continuum about 0.3~K brighter. This correction seems reasonable though a bit large compared to the amplitude of the dust continuum ($\sim$1~K) and that of the water emission inferred from the best-fit ($\sim$1~K at the peak). It thus seems unlikely that an underestimation of the continuum at 556~GHz alone explains the discrepancy observed between the two ortho ground state transitions.

One last interpretation we explore is related to the difference in the beam size of the observations. The half power beam width of HIFI, in band 1 (480-636~GHz) is 41\arcsec, while it is 14.5\arcsec\ in band 6 (1426-1703~GHz). At least two effects could be at play. Either the emission from the background source or the absorption from translucent clouds along the line of sight could vary significantly over angular scales comparable to the beam size. Therefore, the continuum averaged over the beam could probe different environments in the source at 556 and 1669~GHz. This would in turn affect the flux-to-continuum ratios and explain the discrepancy between the two estimates of $N_{1_{01}}$. To explain the discrepant values between the two ortho ground transitions, one would need a stronger continuum at 1669~GHz than at 556~GHz, which can be explained by a stronger dilution at the shorter frequency. To better understand how the beam size affects the observations, we compare our data to those obtained with SWAS at 556~GHz \citep[see e.g.][towards W51 and W49 respectively]{Neufeld2002,Plume2004}. The absorption components associated with the background source are narrower and shallower in the SWAS data than in the HIFI data. In particular, towards W51, the SWAS observations at 556~GHz show only one strong, $\sim$10~km~s$^{-1}$ wide, absorption feature at 65~km~s$^{-1}$ while our HIFI observations show a very broad ($\sim$20~km~s$^{-1}$) feature with two clearly distinct components. Towards W49, the absorption features between 0 and 20~km~s$^{-1}$ are well separated in the SWAS data while they are blended in the HIFI observations. The emission features also seem to change in width and intensity, though it is more difficult to assess based on their wings only. This means that within the SWAS beam of 3.3' by 4.5', at least some of the absorption features due to the outflow or envelope of the source are significantly diluted. Therefore, the angular size of these components is notably smaller than $\sim$4'. We can therefore assume the characteristic scales for the emission components in the background are also significantly smaller than $\sim$4' which could lead to sensible variations in the continuum between the observations at 556 and 1669~GHz. This is also suggested by the difference in the continuum's intensity between the observations made in the two polarisations, which are usually only a few arcseconds away from each other. On the other hand, the absorption features due to the translucent clouds are very similar in the SWAS and HIFI data, in terms of both optical depth and width. The angular size of the line of sight clouds thus has to be at least comparable to the SWAS beam.

We estimate the size of the clouds using an independent approach based on the kinematic distance method. For a given Galactic longitude and latitude, a given radial velocity corresponds to, at most, two heliocentric distances. We thus use the Galactic coordinates of the corresponding source and the velocity range of each absorption component to estimate its size. We convert the lower and higher velocities into distances using:
\begin{equation}
  V_{LSR} = R_0  (\theta/R-\theta_0/R_0)  sin(l),
\end{equation}

\noindent where $R$ and $R_0$ are the Galactocentric distances to the Sun and the cloud, $l$ is the Galactic longitude of the cloud, and $\theta$ and $\theta_0$ are the Galactic rotation velocities at the position of the Sun and the cloud. We use a flat rotation curve with $\theta=\theta_0=254$~km~s$^{-1}$ and a Galactocentric distance to the Sun of 8.4~kpc \citep{Reid2009}. Inherent to the kinematic distance is an ambiguity between a far and a near distances that both correspond to the same $l$ and $V_{LSR}$. For most PRISMAS lines of sight, there is no ambiguity as the distance to the background sources is known. Apart from W49N, all the sources are less than 7~kpc from the Sun and the translucent clouds are all on the near side. We infer the size of each cloud along the line of sight from the difference between the kinematic distances that correspond to its extreme velocities. For instance, the feature between 3 and 10~km~s$^{-1}$ towards W51 corresponds to kinematic distances between 0.20 and 0.66~kpc. The derived size of all the translucent clouds is about a few hundred pc. Assuming the projected size on the sky is similar, this corresponds to angular scales of a few degrees, much larger than the beams of HIFI and SWAS. Naturally, this method does not take into account the local velocity dispersion of each cloud, therefore the sizes we infer should be viewed as upper limits. Additionally, we cannot really discuss the clumpiness of the clouds with the data in hand. Indeed, a velocity resolution of 200~m/s (at 1669~GHz) corresponds to a distance of at least 10 pc along the PRISMAS lines of sight or to an angular size of at least 10 arcminutes, still larger than the HIFI and SWAS beams. Nonetheless, if present, those clumps or similar short angular scale density variations could contribute to the discrepancy between the optical depth infered from the 556 and 1669~GHz transitions.

In this paper, to determine the total column density of water, from which we later derive the water abundance and the OPR, we therefore decide to use the 1669~GHz transition as the best tracer of $N_{1_{01}}$.

\begin{figure*}[!t]
  \centering
  \subfigure[The red solid line is the best linear fit to the data.]
  {\label{fig:h2ovsh2}
    \includegraphics[angle=90,width=.45\linewidth]{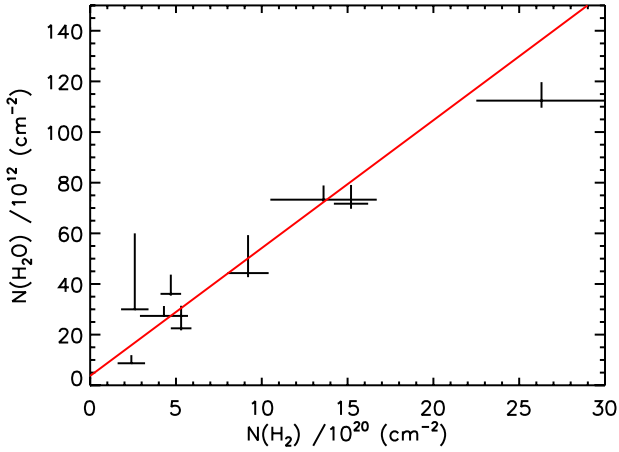}}
  \hfill
  \subfigure[The red solid line is the best linear fit to the data. Arrows show the features for which we only have a lower limit for $N_H$]
  {\label{fig:h2ovsh}
    \includegraphics[angle=90,width=.45\linewidth]{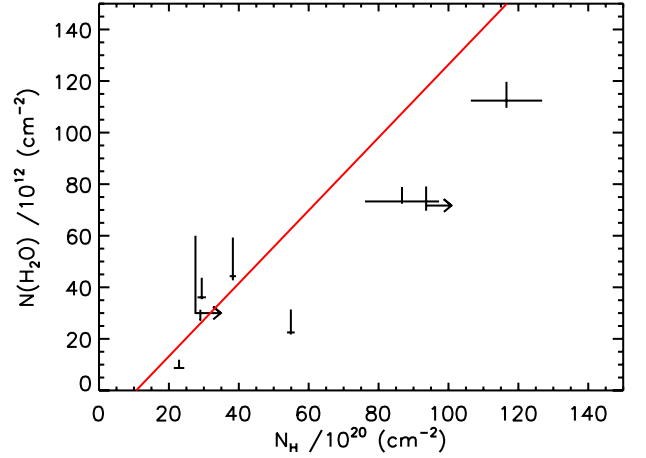}}
  \caption{Total column density of water as a function of the column density of molecular hydrogen and hydrogen nuclei.}
\end{figure*}

\subsection{On the abundance of water}
\label{sec:disc_abun}

 The total column density of water for a given gaussian component associated with a translucent cloud varies between $\sim2\times10^{12}~\rm{cm^{-2}}$ and $\sim130\times10^{12}~\rm{cm^{-2}}$. When we integrate over velocity ranges, we usually combine several gaussian components. Therefore the total column density is usually slightly larger, between $\sim3\times10^{12}~\rm{cm^{-2}}$ and $\sim150\times10^{12}~\rm{cm^{-2}}$.

\citet{Plume2004} analysed the water transitions at 556 and 547~GHz as observed with the {\it Submillimeter Wave Astronomy Satellite} (SWAS) towards W49A, which includes W49N. The \hho\ spectrum at 556~GHz is very similar to the one we present towards W49N except that the feature between about 0 and 20~$\rm{km~s^{-1}}$, associated with the background source, is separated into two narrower components in the SWAS observations. We compare in Table \ref{tab:w49_comp} the column densities obtained by \citet{Plume2004}, \citet{Godard2012} and in the present paper, for water and molecular hydrogen, for interstellar absorption components. We give the values we infer from the gaussian decomposition. \citet{Plume2004} used CO measurements and assumed a CO abundance of $10^{-4}$ to infer \hh\ column density and water abundance. We find that there is a rough agreement, within a factor of a few, between our work and \citet{Plume2004} regarding the column density of water though the SWAS observations seem to indicate the interstellar features are optically thicker than in the HIFI observations. The comparison between the \hh\ column density derived from CO observations and that derived from CH observations leads to a larger scattering than for the water measurements. There seems to be no systematic trend and the ratio between the two estimates of $N($\hh$)$ varies between 0.16 and 2.3. CH is expected to be a more reliable tracer of the \hh\ column than CO in transluscent clouds. The
use of CO as a tracer of \hh\ has been extensively discussed \citep[e.g.][]{Liszt2010}.

As a result, the o-\hho\ abundance inferred from the 556~GHz data in \citet{Plume2004} are always larger, by a factor of at least a few, than in our work. They find abundances between $\sim1\times 10^{-8}$ and $40\times 10^{-8}$ while we obtain abundances between $3.2\times 10^{-8}$ and $7.0\times 10^{-8}$. Besides the method to derive \hh\ abundance, the difference in the water vapor column density and abundance could arise from the significant difference of the beam size between SWAS ($\sim4\arcmin$) and HIFI ($\sim12\arcsec$).

\begin{table}[b]
  \centering
  \caption{Comparison with the results from \citet{Plume2004} towards W49}
  \begin{tabular}{c c c c | c | c c c}
    \hline
    \hline
    & & \multicolumn{3}{c}{Absorption features} \\
    & & \multicolumn{3}{c}{(km~s$^{-1}$)} \\
    & & 33.5 & 39.5 & 53.5 & 59.6 & 63.3 & 68.0 \\
    \hline
    $N($o-\hho$)$ & (a) & 32 & $>$55 & 30 & $>$100 & $>$25 & 13 \\
    $N($o-\hho$)$ & (b)  &\multicolumn{2}{c |}{53} & 26 & 39 & \multicolumn{2}{c}{34} \\
    Ratio & & \multicolumn{2}{c |}{$>$1.6} & 1.2 & $>$2.6 & \multicolumn{2}{c}{$>$1.1} \\
    \hline
    $N($\hh$)$ & (a) & 0.8 & 15 & 1.3 & 4.6 & 28 & 1.6 \\
    $N($\hh$)$ & (c) & 5.9 & 6.6 & 3.7 & \multicolumn{3}{c}{22.6} \\
    Ratio & & 0.16 & 2.3 & 0.35 & \multicolumn{3}{c}{1.5} \\
    \hline
    $X($o-\hho$)$ & (a) & 40 & $>$3.6 & 23 & $>$22 & $>$0.92 & 8.1 \\
    $X($o-\hho$)$ & (b) & \multicolumn{2}{c |}{4.2} & 7.0 & \multicolumn{3}{c}{3.2} \\
    \hline
  \end{tabular}
  \tablecomments{To remain consistent with \citet{Plume2004}, we assume here that $N(o-$\hho$)=N_{1_{01}}$ and that $N_{1_{01}}$ is given by the transitions at 556~GHz. The water column density is given in units of $10^{12}~\rm{cm^{-2}}$, that of \hh\ in units of $10^{20}~\rm{cm^{-2}}$, and the relative abundance of water in units of $10^{-8}$. References: (a) \citet{Plume2004}, (b) this work, (c) \citet{Godard2012}}
  \label{tab:w49_comp}
\end{table}

\citet{Neufeld2002} presented SWAS observations of the water transition at 556~GHz towards W51. For the feature at about 6~km~s$^{-1}$, which is not significantly affected by water emission, they infer a water column density of $1.9\times10^{13}\ \rm{cm^{-2}}$ in good agreement with $1.6\times10^{13}\ \rm{cm^{-2}}$ we obtain here. Assuming an OPR of 3, \citet{Neufeld2002} claimed the total water column density was $2.5\times10^{13}\ \rm{cm^{-2}}$ in excellent agreement with $2.7\times10^{13}\ \rm{cm^{-2}}$ we infer from the ortho and para ground state transitions (from the addition of the 5, 6, and 7~km~s$^{-1}$ gaussian components in Table \ref{tab:W51gg_nh2o}). \citet{Neufeld2002} then derived a water abundance of about $10^{-8}$ relative to H{\sc i} while we derive an abundance of $6\times10^{-8}$ relative to molecular hydrogen. Those water abundances are in agreement with each other within less than a factor 2 since \citet{Godard2012} report for this feature column densities of atomic hydrogen ($14.8\times10^{20}\ \rm{cm^{-2}}$) and molecular hydrogen ($3.5-4.3\times10^{20}\ \rm{cm^{-2}}$) that correspond to a H{\sc i}-to-\hh\ ratio of $3.4-4.2$.

\begin{figure*}[!t]
  \centering
  \subfigure[The dashed orange line is the average of the data.]
  {\label{fig:h2oh2vsh2}
    \includegraphics[angle=90,width=.45\linewidth]{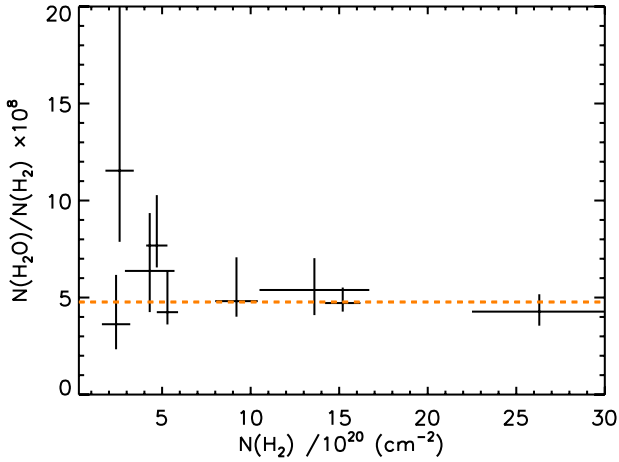}}
  \hfill
  \subfigure[The dashed orange line is the average of the data.]
  {\label{fig:h2ohvsh}
    \includegraphics[angle=90,width=.45\linewidth]{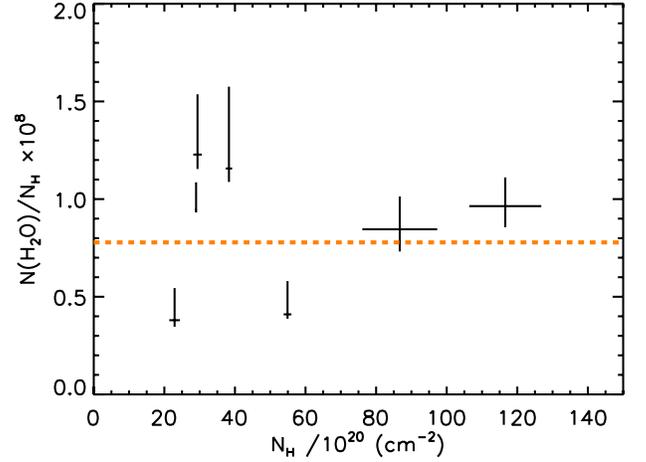}}
  \caption{Abundance of water relative to molecular hydrogen and hydrogen nuclei as a function of $N($\hh$)$ or $N_H$.}
  \label{fig:h2oh2abund}
\end{figure*}

\begin{figure*}[!t]
  \centering
  \subfigure[]
  {\label{fig:h2oh2vsfh2}
    \includegraphics[angle=90,width=.45\linewidth]{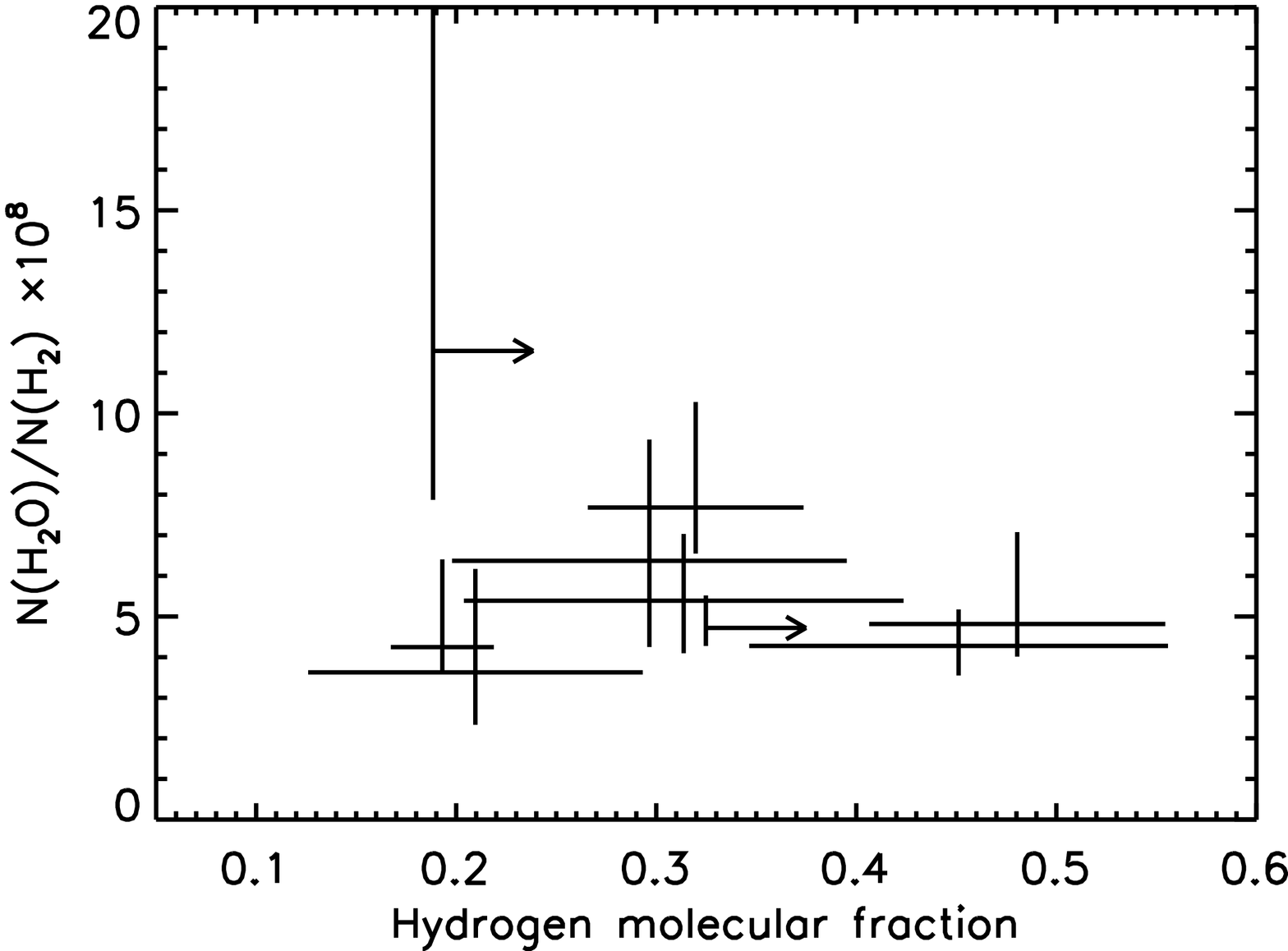}}
  \hfill
  \subfigure[]
  {\label{fig:h2ohvsfh2}
    \includegraphics[angle=90,width=.45\linewidth]{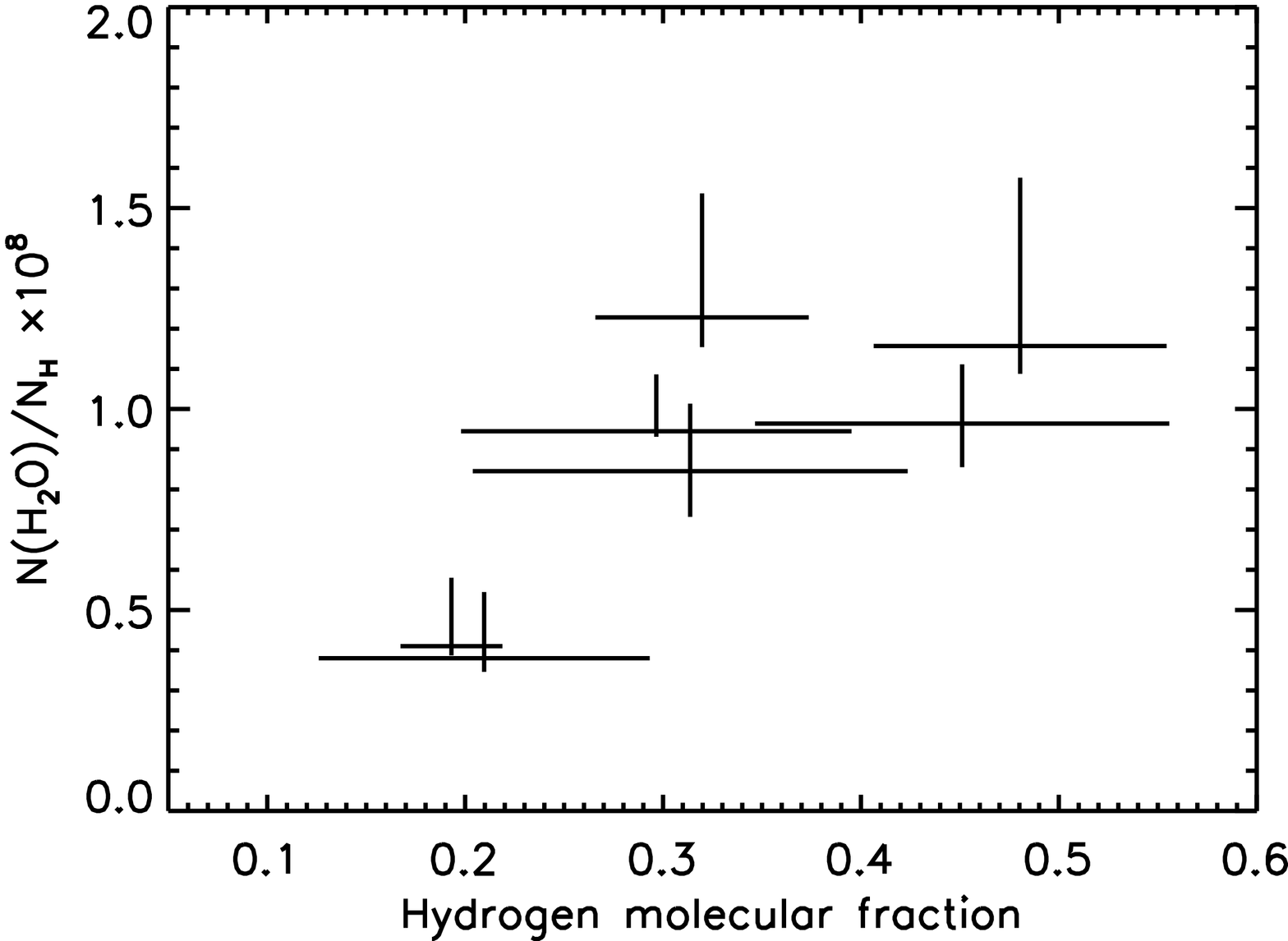}}
  \caption{Abundance of water relative to molecular hydrogen and hydrogen nuclei as a function of the molecular fraction of hydrogen.}
  \label{fig:h2ohabund}
\end{figure*}

We use the \hh\ column densities that \citet{Godard2012} obtained from CH absorption line, assuming HF/CH = 0.4 and HF/\hh=$3.6\times10^{-8}$, since the HF absorption line are more often saturated. We report these \hh\ column densities values in Table \ref{tab:W51gg_nh2o}. Figure \ref{fig:h2ovsh2} shows the total \hho\ column density as a function of the \hh\ column density for nine translucent clouds along the PRISMAS lines of sight. We select those as (1) the velocity position and width of the water emission from the source is not an obvious contamination and (2) \citet{Godard2012} provide a column density for \hh\ (all the sources except W28A). The nine clouds span a range of column density that is similar to our whole sample, from about $2\times10^{20}~\rm{cm^{-2}}$ to $30\times10^{20}~\rm{cm^{-2}}$ for \hh, and from about $10\times10^{12}~\rm{cm^{-2}}$ to $120\times10^{12}~\rm{cm^{-2}}$ for \hho. There is a clear linear correlation between the column densities which indicates a remarkably constant abundance of \hho\ from low to high \hh\ column density clouds. A linear fit to the data gives a slope of $(5.0 \pm 0.4) \times 10^{-8}$ if we assume there are no water molecules in the excited levels and the uncertainty is thus only given by the uncertainty on the ground state column densities. We also show the total column density of water as a function of the total hydrogen column density $N_H$ from \citet{Godard2012} for the same nine clouds, though we only have lower-limits of $N_H$ for two of them (see Figure \ref{fig:h2ovsh}). Again, we find a clear correlation, though not as good as with $N($\hh$)$ suggesting that water is a better tracer of \hh\ than hydrogen. A linear fit gives a slope of $(1.4\pm0.2)\times10^{-8}$ if we use the uncertainty on the water ground levels only. \citet{Hollenbach2012} modeled the production of oxygen bearing molecules (OH$^+$, \hho$^+$, and H$_3$O$^+$) in interstellar clouds and found that, for a density of $100~\rm{cm^{-3}}$, at a depth between $A_V\sim0.1$ and $A_V\sim1$~mag, the abundance of water relative to hydrogen nuclei was almost constant at $\sim1\times10^{-8}$. To obtain this abundance, \citet{Hollenbach2012} used an average incident diffuse interstellar radiation field ($\chi=1$) and a high cosmic ray ionization rate of $2\times10^{-16}~\rm{s^{-1}}$. Our measurements of the water abundance relative to hydrogen nuclei are in very good agreement with their results.

\begin{figure*}[!t]
  \centering
  \subfigure[]
  {\label{fig:h2oh2vsrgal}
    \includegraphics[angle=90,width=.45\linewidth]{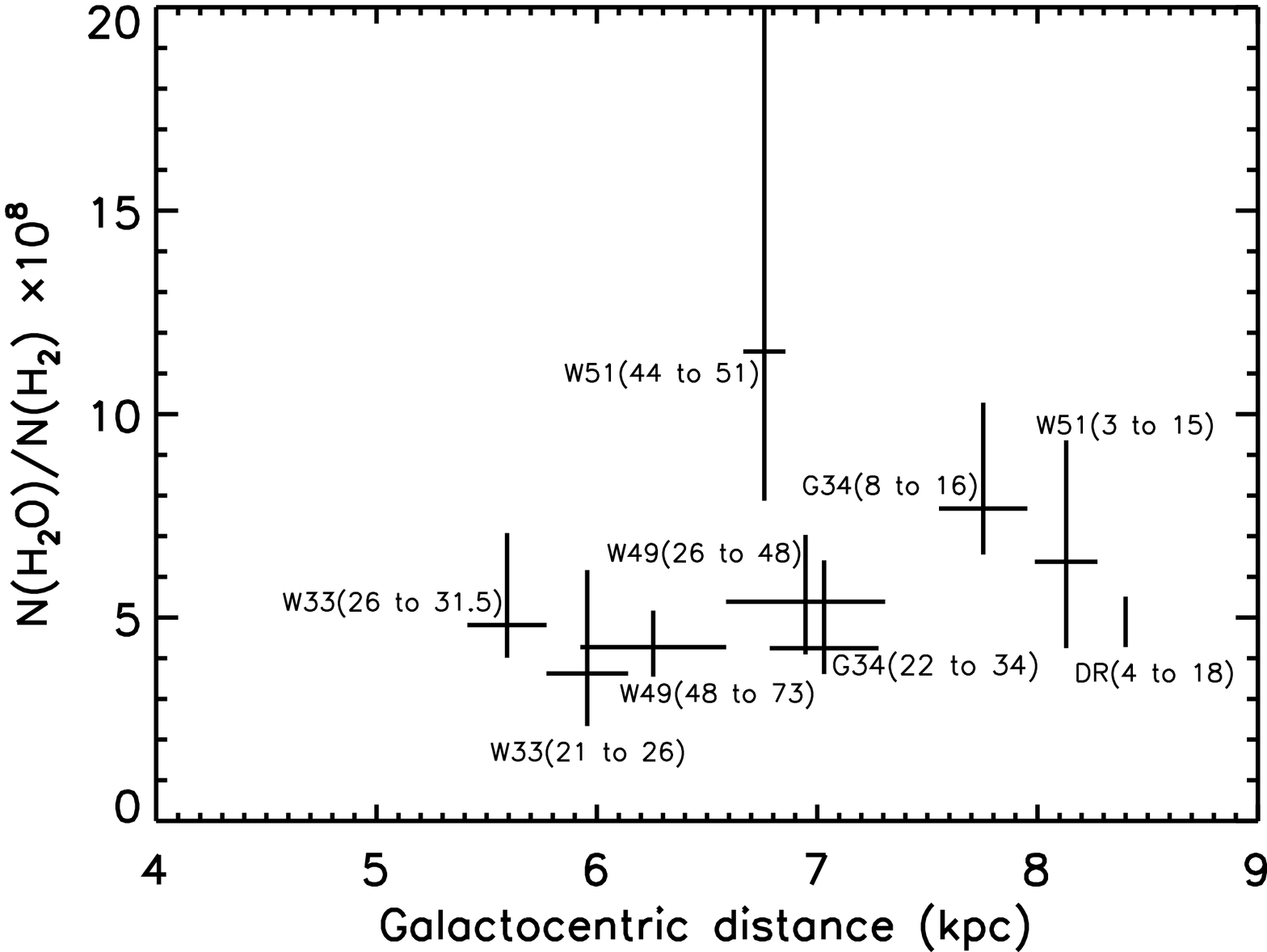}}
  \hfill
  \subfigure[]
  {\label{fig:h2ohvsrgal}
    \includegraphics[angle=90,width=.45\linewidth]{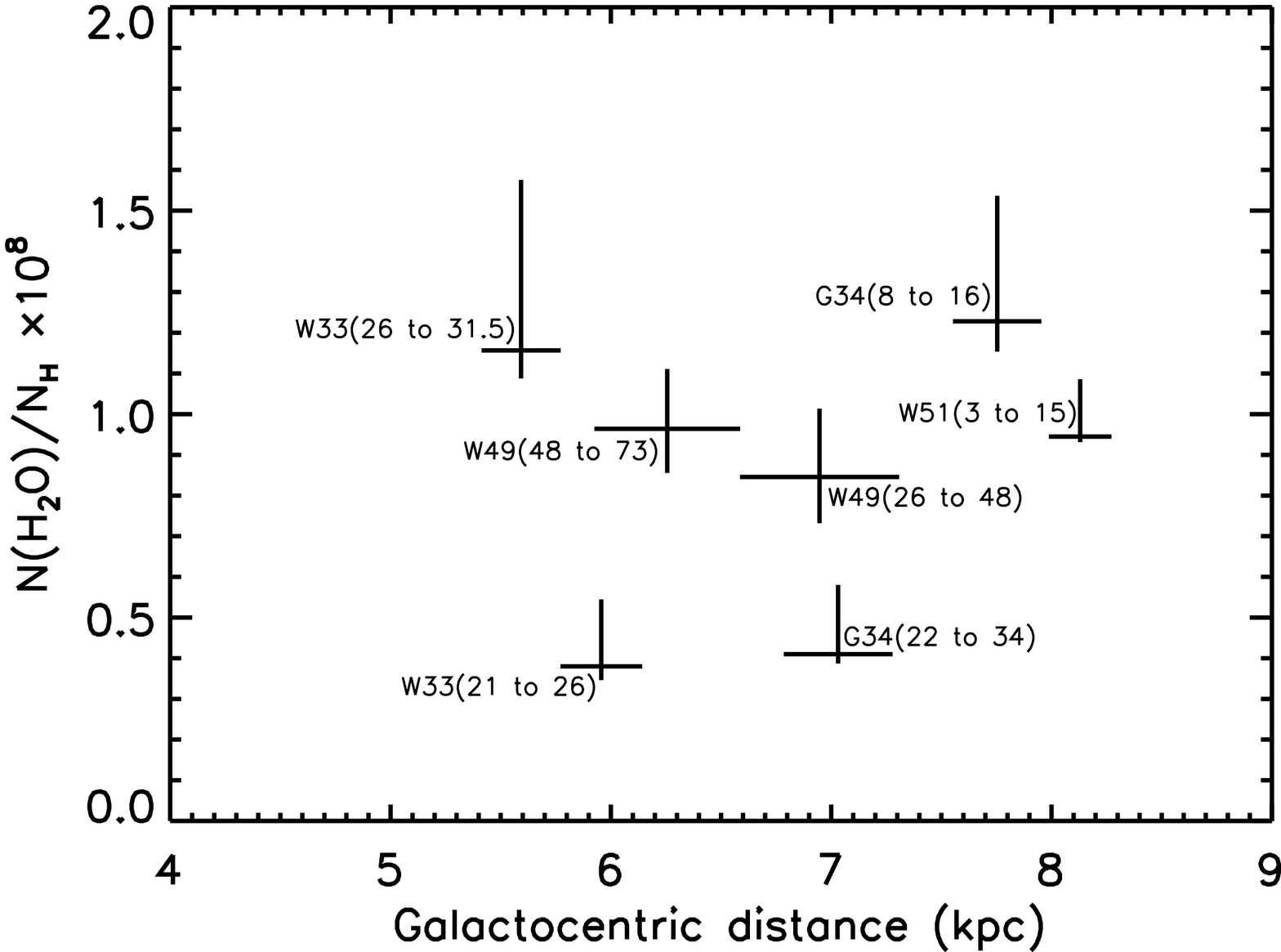}}
  \caption{Abundance of water relative to molecular hydrogen and hydrogen nuclei as a function of the Galactocentric distance. Each data point represents one absorption feature. The label indicate the background source and the velocity for each absorption feature.}
  \label{fig:h2orgal}
\end{figure*}

We then explore possible variations of the water abundance, whose values are indicated in Table \ref{tab:abun} for each cloud. First, we show in Figures \ref{fig:h2oh2vsh2} and \ref{fig:h2ohvsh} the variations of the water abundance, relative to molecular hydogen, or hydrogen nuclei, as a function of $N($\hh$)$, or $N_H$. Neither plot show any sign of a correlation. The average value for $N($\hho$)/N($\hh$)$, taking into account the only uncertainty of each data point due to the ground state column densities, is $(4.8\pm0.3)\times10^{-8}$. The average value for $N($\hho$)/N_H$ is $(0.8\pm0.1)\times10^{-8}$ whatever uncertainties we use. Then, we show in Figures \ref{fig:h2oh2vsfh2} and \ref{fig:h2ohvsfh2} the abundance of water relative to molecular hydrogen, or hydrogen nuclei, as a function of the molecular fraction of hydrogen ($f($\hh$)=2N($\hh$)/N_H$). If $N($\hho$)$ was direclty proportional to $N_H$, then we should see a correlation between $N($\hho$)/N($\hh$)$ and $f($\hh$) = (N_H/2N($\hh$))^{-1}$. Instead, we see that $N($\hho$)/N($\hh$)$ does not exhiti any trend as a function of $f($\hh$)$. On the other hand, $N($\hho$)/N_H$ increases as $f($\hh$)=2N($\hh$)/N_H$ increases, which confirms that water is a better tracer of \hh\ than it is of hydrogen. The water abundance relative to hydrogen nuclei increases from less than $0.5\times10^{-8}$ to about $1.5\times10^{-8}$ as $f($\hh$)$, which is the range of values predicted by \citet{Hollenbach2012} as the visual extinction, therefore the molecular fraction of hydrogen, increases from less than 0.1~mag to about 1~mag. This confirms that $N($\hho$)$ is a good tracer of $N($\hh$)$ in the translucent section of clouds.

\begin{table}[!t]
  \centering
  \caption{Abundance of water relative to molecular hydrogen and to hydrogen nuclei. \label{tab:abun}}
  \begin{tabular}{l c c}
    \hline
    \hline
    Cloud & $N($\hho$)/N($\hh$)$ & $N($\hho$)/N_H$ \\
    \hline
    DR21(OH) (4 to 18 km~s$^{-1}$) & 4.7$^{+0.8}_{0.4}$ & $<$0.77 \\[2pt]
    G34.3+0.1 (8 to 16 km~s$^{-1}$) & 7.7$^{+2.7}_{1.1}$ & 1.2$^{+0.3}_{0.1}$ \\[2pt]
    G34.3+0.1 (22 to 34 km~s$^{-1}$) & 4.2$^{+2.2}_{0.6}$ & 0.41$^{+0.17}_{0.02}$ \\[2pt]
    G34.3+0.1 (42 to 55 km~s$^{-1}$) & $<$10.4  & - \\[2pt]
    W33(A) (21 to 26 km~s$^{-1}$) & 4$^{+2}_{1}$ & 0.38$^{+0.16}_{0.03}$ \\[2pt]
    W33(A) (26 to 31.5 km~s$^{-1}$) & 4.8$^{+2.3}_{0.8}$ & 1.2$^{+0.4}_{0.1}$ \\[2pt]
    W49(N) (26 to 48 km~s$^{-1}$) & 5$^{+2}_{1}$ & 0.8$^{+0.2}_{0.1}$ \\[2pt]
    W49(N) (48 to 73 km~s$^{-1}$) & 4.3$^{+0.9}_{0.7}$ & 1.0$^{+0.1}_{0.1}$ \\[2pt]
    W51 (3 to 15 km~s$^{-1}$) & 6$^{+3}_{2}$ & 0.94$^{+0.1}_{0.01}$ \\[2pt]
    W51 (44 to 51 km~s$^{-1}$) & 12$^{+15}_{4}$ &$<$1.1 \\[2pt]
    \hline
  \end{tabular}
  \tablecomments{$N($\hho$)$ are in units of $10^{12}~\rm{cm^{-2}}$. $N($\hh$)$ and $N_H$ are in units of $10^{20}~\rm{cm^{-2}}$. Abundances are in units of $10^{-8}$.}
\end{table}

We finally explore the possiblity of a variation of the water abundance with Galactocentric distance as the oxygen abundance in the Galaxy has been shown to decrease a few 0.01~dex/kpc with Galactocentric distance over significantly larger range \citep[e.g.][]{Balser2011,Henry2011}. Therefore, we associate each cloud with its Galactocentric distance, inferred from the kinematic distance method, to determine whether the abundance of water also follows a gradient through the Galaxy. We use the whole range of velocity covered by each absorption feature to determine its Galactocentric distance and estimate the uncertainty on this distance. We find that the nine translucent clouds for which we measure a water abundance are located between 5.5 and 8.5~kpc. Figure \ref{fig:h2oh2vsrgal} shows that neither the \hho\ abundance relative to molecular hydroge or that relative to hydrogen nuclei varies significantly with the distance to the Galactic center. Only additional measurements over a wider range of Galactocentric distances might help better constrain a trend, if any.

\subsection{The OPR as a tracer of the gas history}
\label{sec:disc_opr}

\begin{figure}[!t]
  \centering
  \includegraphics[angle=90,width=.9\linewidth]{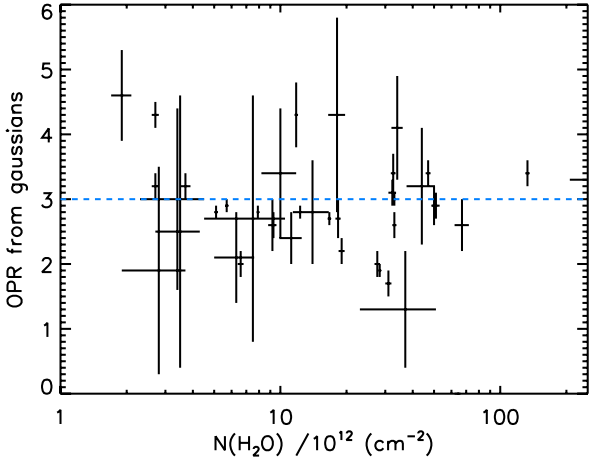}
  \caption{Water ortho-to-para ratio as a function of the total water column density both derived from the gaussian absorption components. The dashed blue line shows the high temperature limiting value of 3.}
  \label{fig:oprvsh2ogauss}
\end{figure}

Figure \ref{fig:oprvsh2ogauss} shows the OPR for each interstellar gaussian absorption feature in the PRISMAS spectra as a function of the \hho\ column density. No obvious trend can be detected and a linear fit to the data leads to a slope of less than $10^{-2}$. The weighted average of the OPR is about 2.8$\pm$0.1. Most features are in close or very close agreement with the high temperature limiting value of 3. Indeed, for 34 out of 41 interstellar absorption components, the observed OPR is within 3-$\sigma$ of a value of 3.0. Only a few gaussian components are either significantly below or above 3. Among them, the most extreme values correspond to the features at 55~km~s$^{-1}$ towards W49N, and at 29~km~s$^{-1}$ towards W33A. These two features have in common that they are blended with other stronger absorption features, including features associated with the background source. As a result, the column density and OPR of water derived for those absorption components might only be correct when summed over a larger velocity range that includes all the blended features.

We then analyse the OPR derived from the column density curves. Hereafter, we use $OPR^g$ as the estimator of the OPR, rather than $OPR^+$, since we expect the water column density in the excited levels to be negligible, assuming $T_{ex}\simeq5$~K. In total, we measure the water OPR for 13 translucent clouds. For these 13 clouds, the average OPR is 2.9$\pm$0.1, in very close agreement with the high temperature limiting value of 3. Out of the 13 clouds, 10 have an OPR less than 3-$\sigma$ away from a value of 3. One of the three other clouds has an OPR value above 3 and correspond to the narrow absorption component at $\sim44$~km~s$^{-1}$ towards W33(A). The OPR inferred for the remaining two clouds is 2.3$\pm$0.1, 2.4$\pm$0.2. These are the clouds towards W49N at $\sim40$~km~s$^{-1}$ and $\sim60$~km~s$^{-1}$. 

\begin{figure}[!t]
  \centering
  \includegraphics[angle=90,width=.9\linewidth]{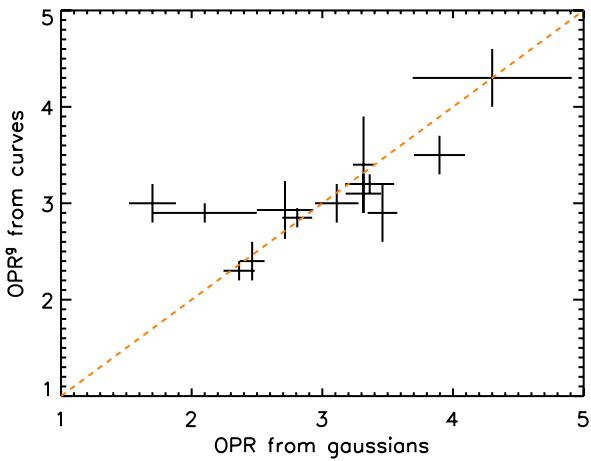}
  \caption{Comparison between the water OPR derived from the gaussian components and that derived from the column density curves, assuming there are no water molecules in the excited levels. The dashed blue line shows where the two OPR ratios are equal.}
  \label{fig:oprvsopr}
\end{figure}

We compare the OPR derived from the gaussian components to that obtained for each cloud, combining the gaussian components accordingly. We find there is almost a perfect agreement between the result from the two methods (see Figure \ref{fig:oprvsopr}). Some discrepancy is notable for a couple of clouds where the gaussians give a lower than 3 OPR while the column density curves leads to an OPR in agreement with a value of 3. The two clouds towards W49N are the only ones that have an OPR significantly lower than 3 for both methods. This statistical analysis on 13 clouds confirms and extends the results from \citet{Lis2010} who analyzed two PRISMAS sightlines and found some values of the water OPR in agreement with the high temperature limiting value of 3 and some values that were significantly below this limit.

\begin{figure*}[!t]
  \centering
  \subfigure[]
  {\label{fig:oprcc0vsh2o}
    \includegraphics[angle=90,width=.45\linewidth]{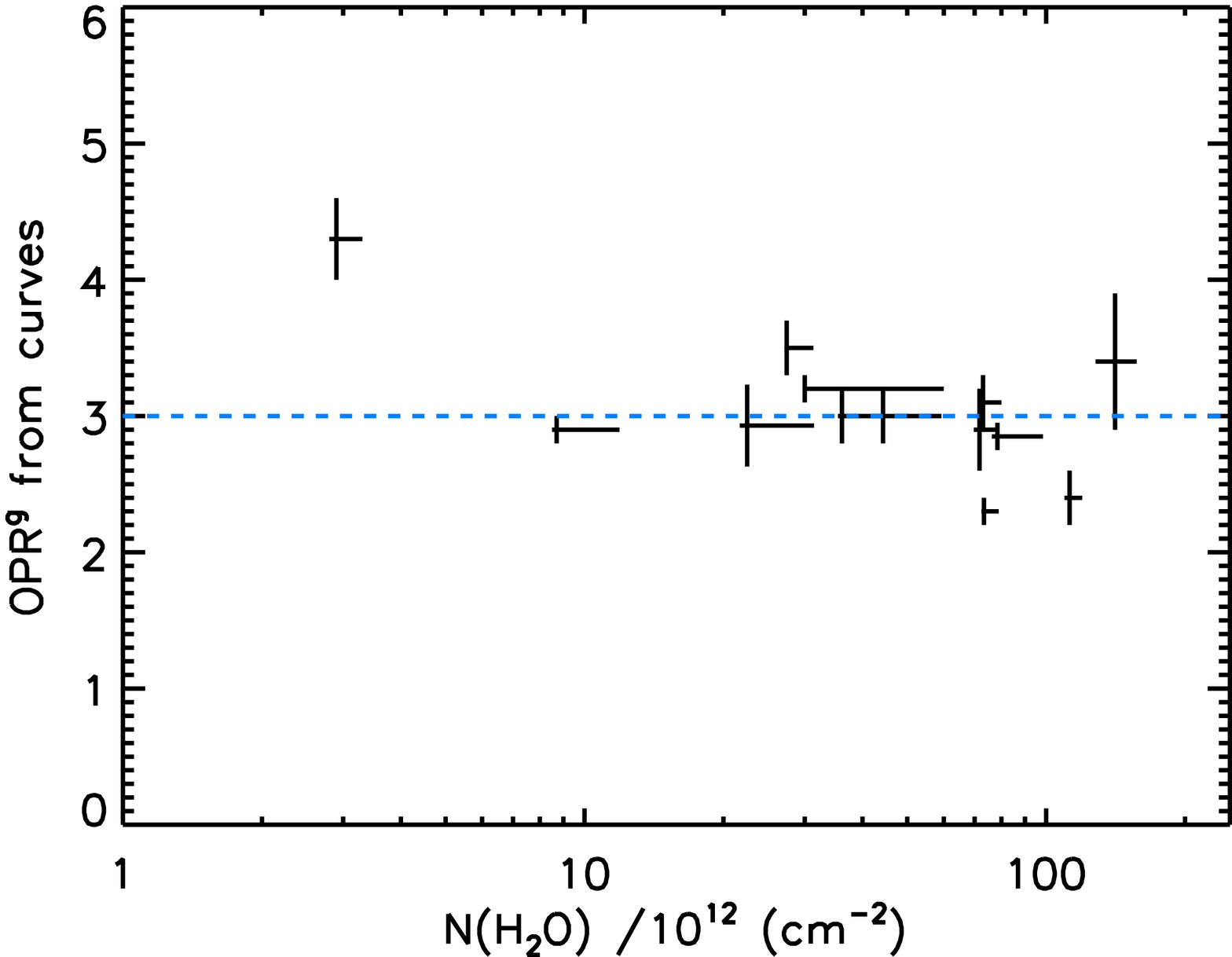}}
  \hfill
  \subfigure[]
  {\label{fig:oprcc0vsh2}
   \includegraphics[angle=90,width=.45\linewidth]{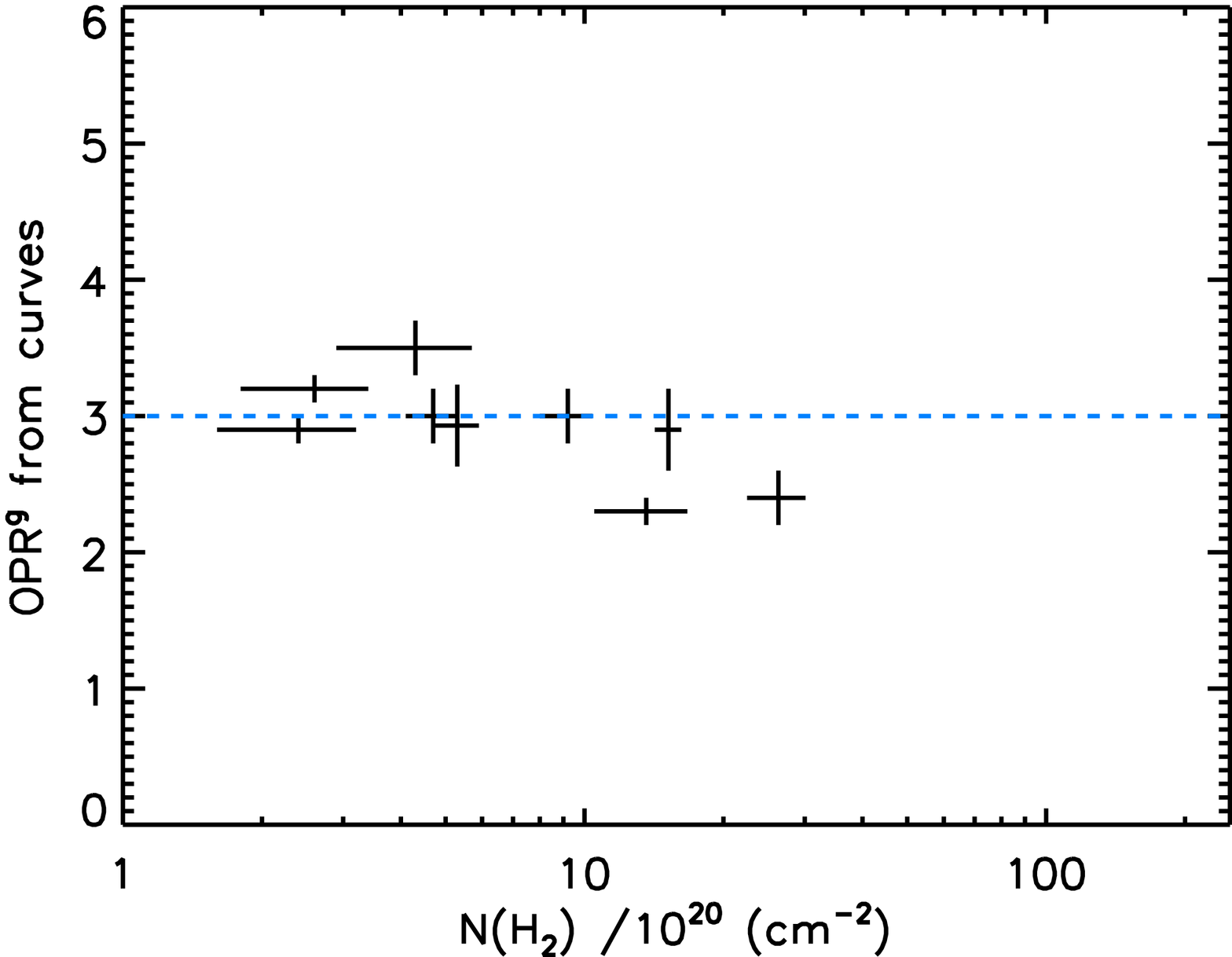}}
  \subfigure[]
  {\label{fig:oprcc0vsh}
   \includegraphics[angle=90,width=.45\linewidth]{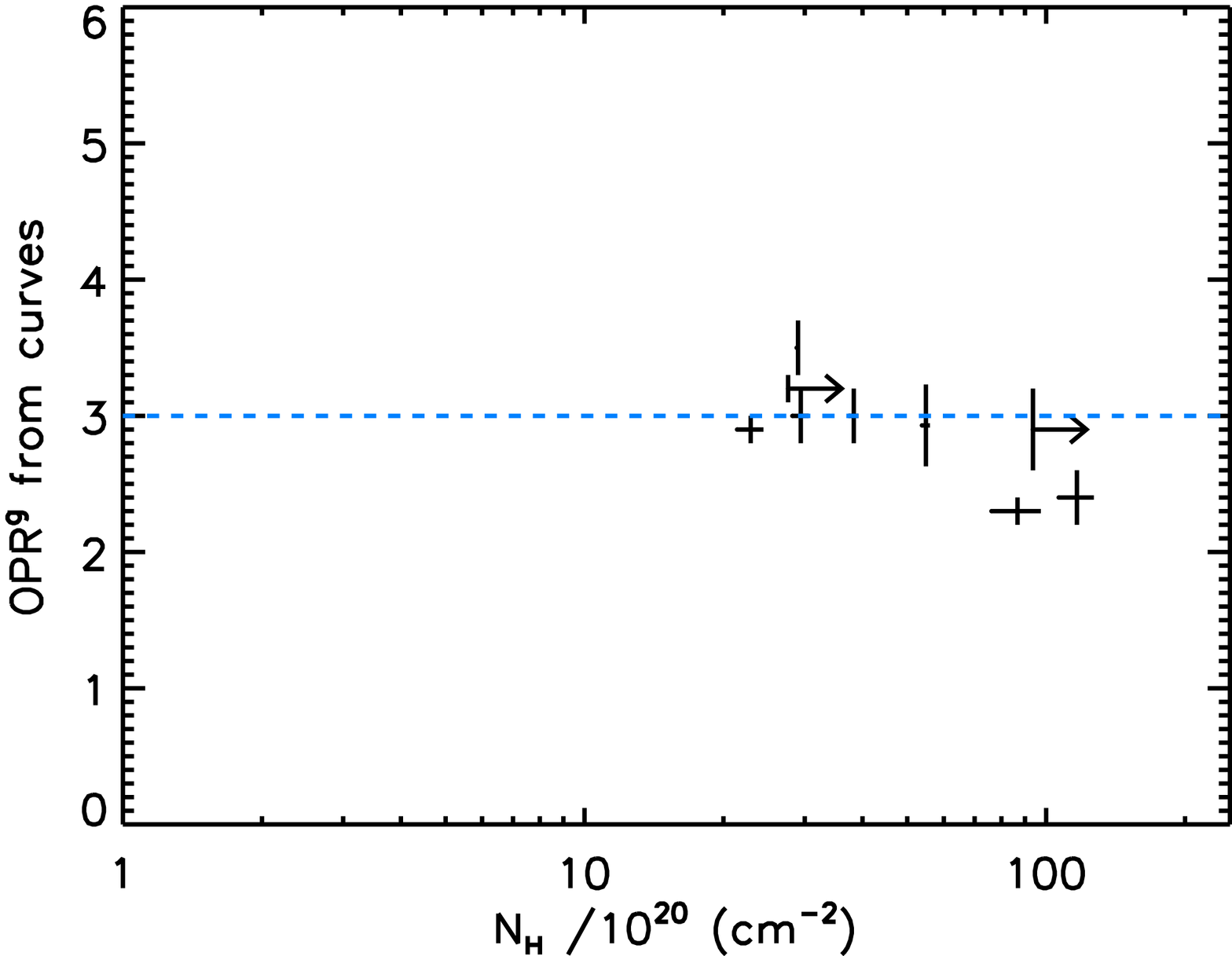}}
  \hfill
  \subfigure[]
  {\label{fig:oprcc0vsfh2}
   \includegraphics[angle=90,width=.45\linewidth]{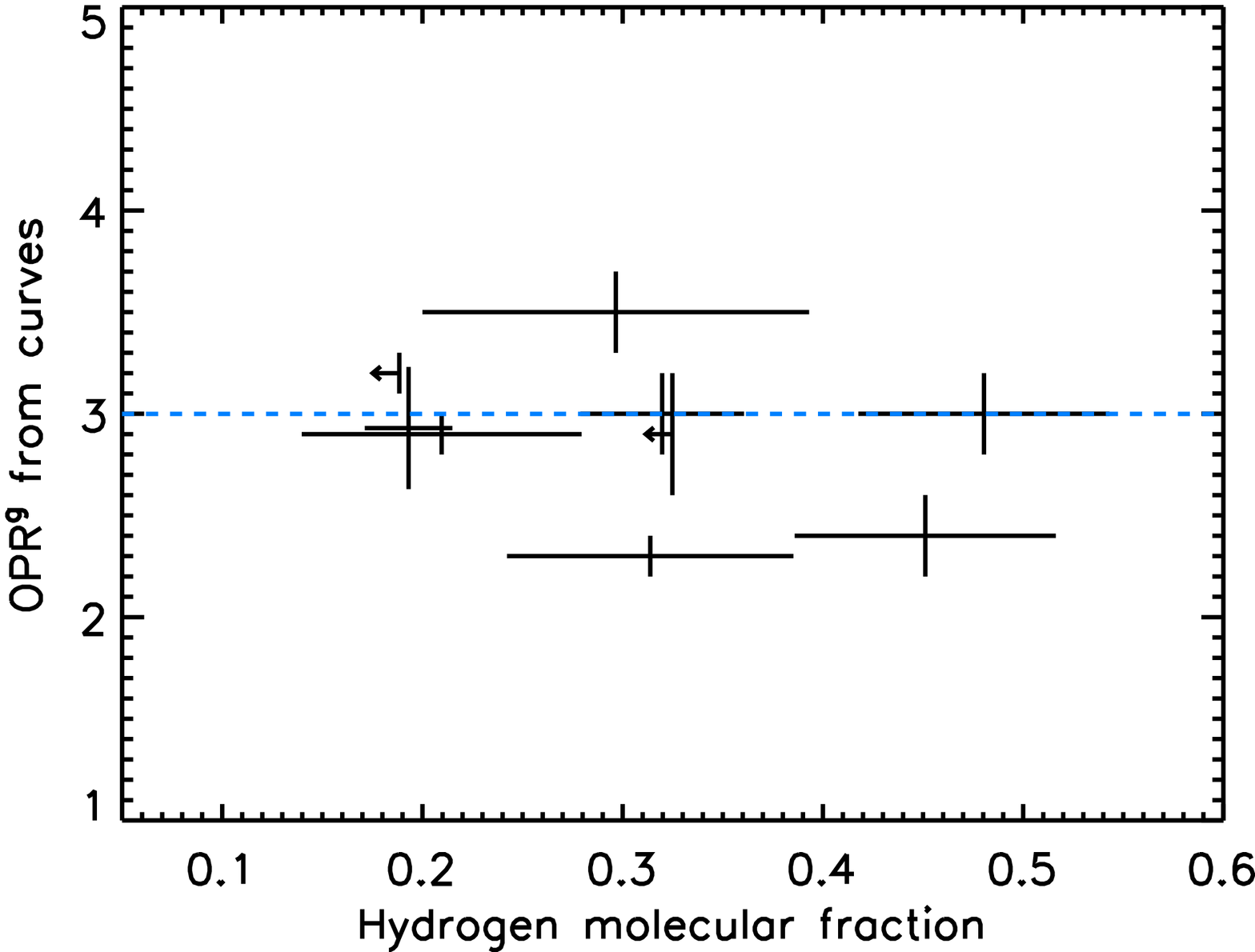}}
  \caption{Water ortho-to-para ratio as a function of the water column density, the molecular hydrogen column density, the hydrogen nuclei column density, and the hydrogen molecular fraction. The dashed blue line shows the high temperature limiting value of 3.}
  \label{fig:oprcc0}
\end{figure*}

Figure \ref{fig:oprcc0} shows the OPR derived for each translucent cloud using the column density curves as a function of the total water column density, the molecular hydrogen column density, the hydrogen nuclei column density, and the hydrogen molecular fraction. We assume the water column density inferred for the excited levels is only due to the noise. There seems to be a slight decrease of the OPR as the column density increases, whether it is that of \hho\, \hh\, or hydrogen nuclei. If true, this trend should imply a decrease of the OPR as $f($\hh$)$ increases. However, Figure \ref{fig:oprcc0vsfh2} hardly shows any sign of correlation. In Figure \ref{fig:oprcc0vsh2o}, \ref{fig:oprcc0vsh2} and \ref{fig:oprcc0vsh}, the two clouds that have an OPR value significantly lower than 3 always appear on the high end of the column density axis. However, the limited number of clouds and amplitude of variations though prevent us from drawing a definitive conclusion regarding a correlation between a low OPR and a high column density.


The two translucent clouds for which we measure an OPR of $\sim$2.3 are observed towards W49N, at the velocities of $\sim40$ and $\sim60$~km~s$^{-1}$ (see Figure \ref{fig:w49}). These clouds span a large velocity range of about 20~km~s$^{-1}$ and combine many velocity sub-components, which explains why we derive a large total column density of water. Neither of them show evidence of saturation though the peak of the sub-component at 40~km~s$^{-1}$ seems to just reach the zero level of intensity at 556, 1113, and 1669~GHz. However, none of the components exhibit a plateau at the zero level of intensity. Additionally, the background emission seems well constrained over the velocity range of these two clouds as the emission features from the background source peak at $\sim$5~km~s$^{-1}$ and appear to be limited to velocity smaller than $\sim30$~km~s$^{-1}$. One peculiarity of these clouds is due to the position of the background source, W49N, which is the only PRISMAS source for which there might be an ambiguity between the near and the far distances derived from the kinematic distance method. Figure \ref{fig:sources} shows that the velocities of 40 and 60~km~s$^{-1}$ can either be related to structures at about 3~kpc or 9~kpc from the Sun. The Perseus and Sagittarius arms are located at those approximate distances towards W49N. Consequently, the absorption features we see towards W49N at those velocities might be a combination of near and far distance clouds. Nonetheless, the inferred OPR values is significantly below 3 at least in some parts of these clouds. The cloud at $\sim40$~km~s$^{-1}$ is clearly detected in the observations of nitrogen hydrides, especially in NH$_3$, for which the OPR is peculiar as well \citep{Persson2012}. At millimeter wavelengths, the cloud is also detected in H$^{13}$CO$^+$, which shows that it has a high column density and properties in the high end of the transluscent cloud regime approaching dark clouds/dense cores. The situation is less clear for the feature at $\sim60$~km~s$^{-1}$ which is probably a mixture of several environments.

We relate the low OPR values to equilibrium temperatures below 50~K. An OPR of $\sim$2.3 corresponds to water vapor in thermal equilibrium at $\sim$25~K. The low equilibrium temperature could be a result of the formation of the water molecules in cold gas, on cold dust grains, or the thermalisation with cold gas of water molecules formed at higher temperature. When water forms in the cold gas phase through H$_3$O$^+$ recombination, the large amount of energy released in this exothermic reaction should lead to spin equilibration of the molecules. Similarly, when water molecules are desorbed from dust grain ice mantles, the energy required to do so is sufficient to populate many ortho and para levels and lead to an OPR of 3. Nonetheless, \citet{Limbach2006} showed that the OPR of water vapor released from a cold surface (e.g. ice mantle of dust grain) could reflect the temperature of the surface and of the gas phase molecules close to the surface. Indeed, the energy required to desorb water molecules from dust particles can be dissipated into the grain as well as the molecule. \citet{Hollenbach2009} also showed than in diffuse regions, molecules of water could be released in a frozen state with a spin temperature equal to that of the grain, which is smaller than that of the gas in diffuse and translucent clouds. Therefore, the released molecules might carry the information about the temperature where they have been formed. However, photodesorption may be the only efficient way to remove \hho\ molecules from cold dust grains and, depending of the exact mechanism (either \hho\ dissociation followed by recombination and escape, or \hho\ dissociation followed by ejection of neighboring \hho\ molecule), the OPR may or may not be preserved \citep{Andersson2008}. Additionally, at high temperature (e.g. hot cores, outflows), the gas phase reactions $\rm O(H_2,H)OH(H_2,H)H_2O$ can also from water molecules that are expected to have an OPR of 3. Whether it was formed in cold gas, warm gas or desorbed from a dust grain ice mantle, a water molecule in the gas phase will subsequently thermalize through collisions with atoms and molecules. During these collisions, proton exchanges may occur and the OPR of the water molecules will then carry information about the temperature of the medium with which they last interacted. Whether the observed OPR of water molecules relate to their formation or their more recent history depends on the efficiency of the thermalisation and spin conversion processes.

In order to allow such spin conversion, a sufficient number of collisions with ions are required. This may relate to a large density or a long timescale. For instance, within a $10^4~\rm{cm^{-3}}$ gas, protonated ions with a fractional abundance of $10^{-8}$, and a rate coefficient of $10^{-9}~\rm{cm^{-3}s^{-1}}$, the OPR equilibriation time scale is about $3\times10^5$~years. In a translucent cloud with a characteristic density of $100~\rm{cm^{-3}}$, and other parameters as above, the OPR equilibriation time scale is about $3\times10^7$~years. The low OPR values we detect in the two clouds towards W49N implies that the temperature of the cold gas ($\sim25$~K) is below the expected temperature in translucent clouds \citep[$\sim50$~K, see e.g.][]{Rachford2002}. We characterize the clouds along the lines of sight as translucent because of the extinction (1-5~mag) derived from the hydrogen column densities that \citet{Godard2012} have reported. In the two clouds towards W49N, the total hydrogen column densities are (86.7$\pm$10.6)$\times10^{12}~\rm{cm^{-2}}$ and (116.6$\pm$10.2)$\times10^{12}~\rm{cm^{-2}}$. The corresponding extinctions are of 4.6$\pm$0.6 and 6.2$\pm$0.5 mag using $R_V=3.1$ and $N_H/E(B-V)=5.8\times10^{21}~\rm{H~cm^{-2}mag^{-1}}$ \citep{Bohlin1978, Rachford2009}. These are sufficient for water molecules to be stuck on dust grains as the extinction threshold to form water ice mantles has been shown to be about 3~mag \citep{Whittet2001}. Therefore the water molecules released from the grains can thermalize with the cold gas in a few $10^5$~years and have an OPR lower than 3. If those water molecules later move to the translucent regions of the clouds, they will not thermalize in less than a few $10^7$~years. This indicates that the water molecules for which we measure an OPR significantly below 3 have (1) either thermalized with the cold gas or been released cold from grains within denser, colder regions of the clouds and (2) have not yet thermalized at the temperature of the warmer, translucent regions of the clouds. 

In the present study, the low value of the OPR found in several translucent clouds may be related to any of the aforementioned interpretations, either based on the formation mechanisms or the thermalization with gas colder than $\sim$50~K. The conclusion is nevertheless that the water molecules have been in a cold environment, be it in the gas phase or on the dust grains.

\section{Conclusion}
\label{sec:ccl}

For the first time, Herschel/HIFI allows the observation of the three ground state transitions of \hho\ and \hheo\ as well as the first three excited transitions of \hho\ in absorption towards six continuum sources in the Galactic plane. The continuum sources are detected at all frequencies in self-absorption associated to an envelope or outflow. They are also detected in broad water line emission at almost all frequencies. We detect the translucent clouds associated with all the Galactic arms along those lines of sight only in the ground state transitions of both ortho and para-\hho.

For each transition and sightline, we model the continuum emission, taking into account the contribution of the water lines and decompose the optical depth into a sum of gaussian profiles. From there, we infer the column density of water in the ground state levels of \hho, and upper limits for the excited levels of \hho\ and the ground state levels of \hheo, for each interstellar absorption feature. The total column density of water ranges from a few $10^{12}$ to about $10^{14}~\rm{cm^{-2}}$. We compare the column density of water to that of molecular hydrogen and find that the water abundance is remarkably constant at $\sim 5\times10^{-8}$. Relative to hydrogen nuclei, the water abundance is $\sim10^{-8}$, a value predicted by models with high cosmic ray ionization rate. No correlation with the hydrogen molecular fraction or Galactocentric distance are found. Water therefore seems to be a good tracer of \hh\ in translucent clouds.

For the first time, the excited transitions of \hho\ are observed. Though no translucent clouds are detected in absorption in those transitions, we infer upper limits on the column density of \hho\ in the corresponding lower levels. We then derive the lower limits on the fractional abundance of water molecules in the ground states and find it is usually above 85\%. Using this parameter, we set constraints on the excitation temperature ($T_{ex}\le12$~K).

For the first time too, both ground state transitions of ortho water have been observed. Some discrepancies are revealed and we suggest several interpretations. First, assuming that the excitation temperature is low ($T_{ex}\ll h\nu/k$) may be wrong. We find that with $T_{ex}\simeq5$~K, both ortho ground state transitions are consistent, and that the water fractional abundance in the excited levels is significantly lower than 1\%. These values are both in agreement with the upper limits derived from the observations of the excited transitions. The excitation temperature then enables us to provide an upper limit for the density in the clouds ($\le10^4~\rm{cm^{-3}}$, which is in agreement with the typical density in translucent clouds ($\sim100~\rm{cm^{-3}}$) but does not rule out the existence of more extinguished regions in the clouds. Other interpretation of the discrepancy between the two ortho ground state transitions may arise from the significant difference in the continuum strength at both frequencies, and/or the presence of clumps in the translucent clouds whose characteristic size would imply different filling factor within beams of 14.5 and 41\arcsec\ though a comparison with SWAS observations makes this last possibility unlikely.

We finally derive the water OPR for each interstellar absorption component. Most clouds have a ratio in agreement with the high temperature limiting value of 3. However, we also found that a few features correspond to an OPR significantly below 3. We discuss several interpretations, in view of the properties of these clouds which suggest the presence of high extinction regions ($\simeq5$~mag) where the water could be stuck on cold grains. Whether the water molecules were released with a ``cold'' OPR or quickly thermalized with the cold gas in the dark sections of the clouds remains unclear though we claim they have not yet thermalized at the temperature of the translucent section of the clouds.

An analysis of more subtle variations within each absorption features, allowed by the sensitivity and resolution of HIFI on Herschel, correlated with the many other molecular tracers of the ISM properties that have been observed within the PRISMAS Guaranteed Time Key Program may eventually allow us to discriminate among the proposed explanations of a low OPR in relatively warm regions.

\begin{acknowledgements}
This work was carried out in part at the Jet Propulsion Laboratory, California Institute of Technology, supported by NASA.\\
M.G. and M.D.L. acknowledge the support from the Centre National de Recherche Spatiale (CNRS), and from ANR through the SCHISM project (ANR- 09-BLAN-231).\\
J.R.G. is supported by a Ram\'on y Cajal research contract and thanks the Spanish MICINN for funding support through grants AYA2009-07304 and CSD2009-00038.\\
The Herschel spacecraft was designed, built, tested, and launched under a contract to ESA managed by the Herschel/Planck Project team by an industrial consortium under the overall responsibility of the prime contractor Thales Alenia Space (Cannes), and including Astrium (Friedrichshafen) responsible for the payload module and for system testing at spacecraft level, Thales Alenia Space (Turin) responsible for the service module, and Astrium (Toulouse) responsible for the telescope, with in excess of a hundred subcontractors.\\
HIFI has been designed and built by a consortium of institutes and university departments from across Europe, Canada and the United States under the leadership of SRON Netherlands Institute for Space Research, Groningen, The Netherlands and with major contributions from Germany, France and the US. Consortium members are: Canada: CSA, U.Waterloo; France: CESR, LAB, LERMA, IRAM; Germany: KOSMA, MPIfR, MPS; Ireland, NUI Maynooth; Italy: ASI, IFSI-INAF, Osservatorio Astrofisico di Arcetri-INAF; Netherlands: SRON, TUD; Poland: CAMK, CBK; Spain: Observatorio Astron\'omico Nacional (IGN), Centro de Astrobiolog\'ia (CSIC-INTA); Sweden: Chalmers University of Technology - MC2, RSS \& GARD; Onsala Space Observatory; Swedish National Space Board, Stockholm University - Stockholm Observatory; Switzerland: ETH Zurich, FHNW; USA: Caltech, JPL, NHSC.\\
HIPE is a joint development by the Herschel Science Ground Segment Consortium, consisting of ESA, the NASA Herschel Science Center, and the HIFI, PACS and SPIRE consortia.
 \end{acknowledgements}

\bibliographystyle{aa} 
\bibliography{../../nicolasflagey} 

\appendix

\section{Observations}
In this section of the appendix, we gather all the spectra (nine transitions) for all the PRISMAS sources except W51, which is presented in the main part of the paper.

\begin{figure*}[!h]
  \centering \subfigure[] {\label{}
    \includegraphics[angle=90,width=.31\linewidth]{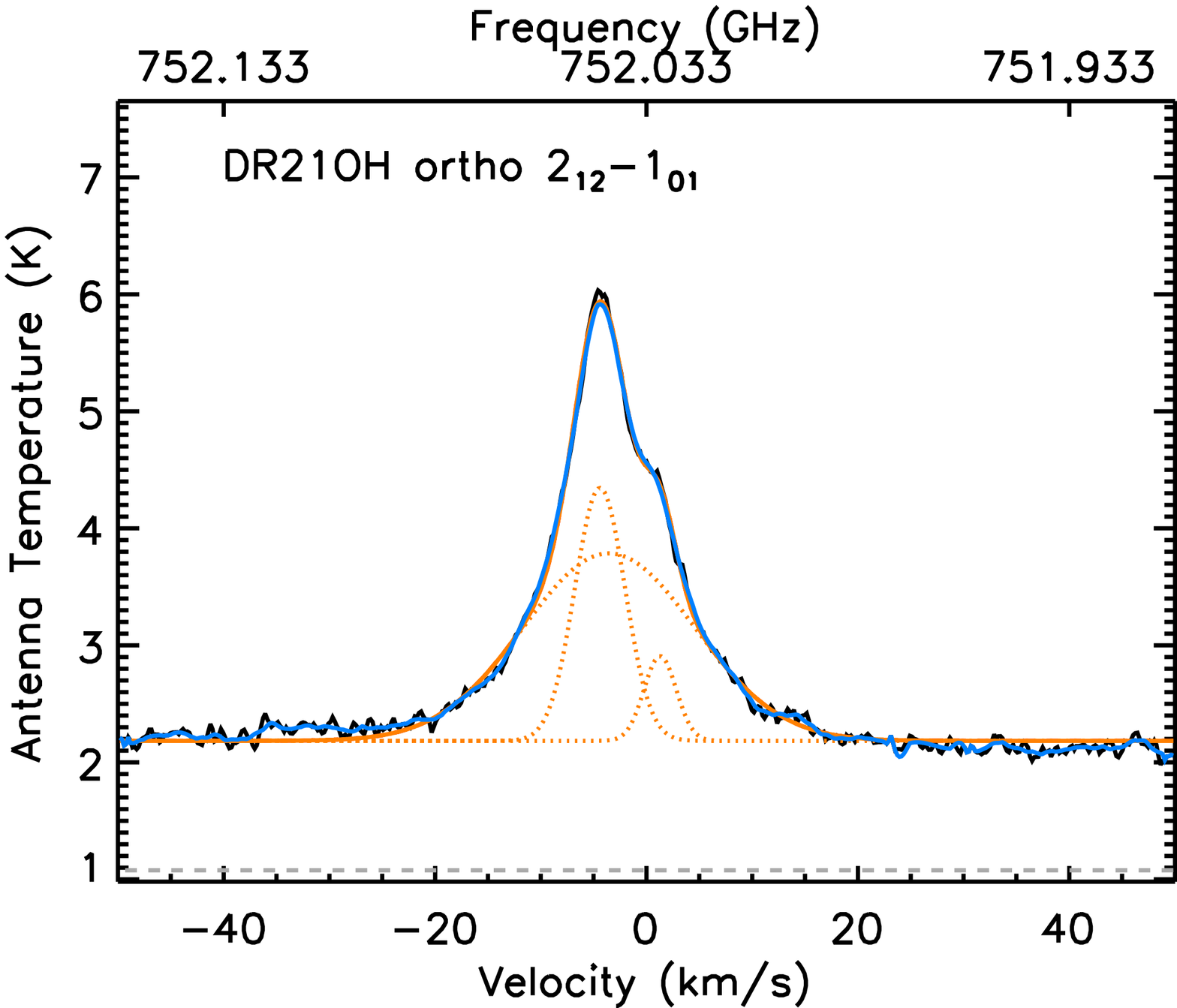}}
  \subfigure[] {\label{}
    \includegraphics[angle=90,width=.31\linewidth]{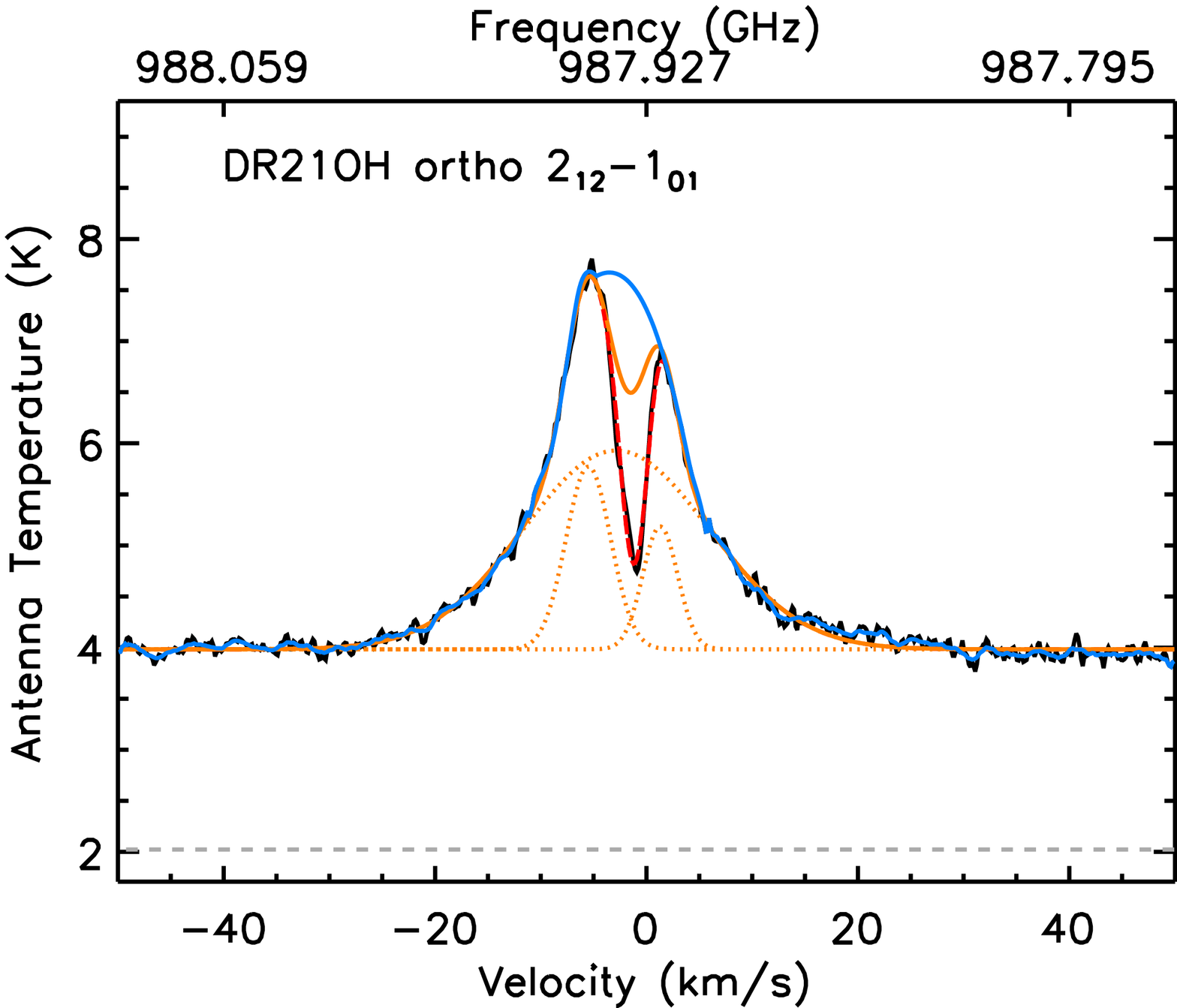}}
  \subfigure[] {\label{}
    \includegraphics[angle=90,width=.31\linewidth]{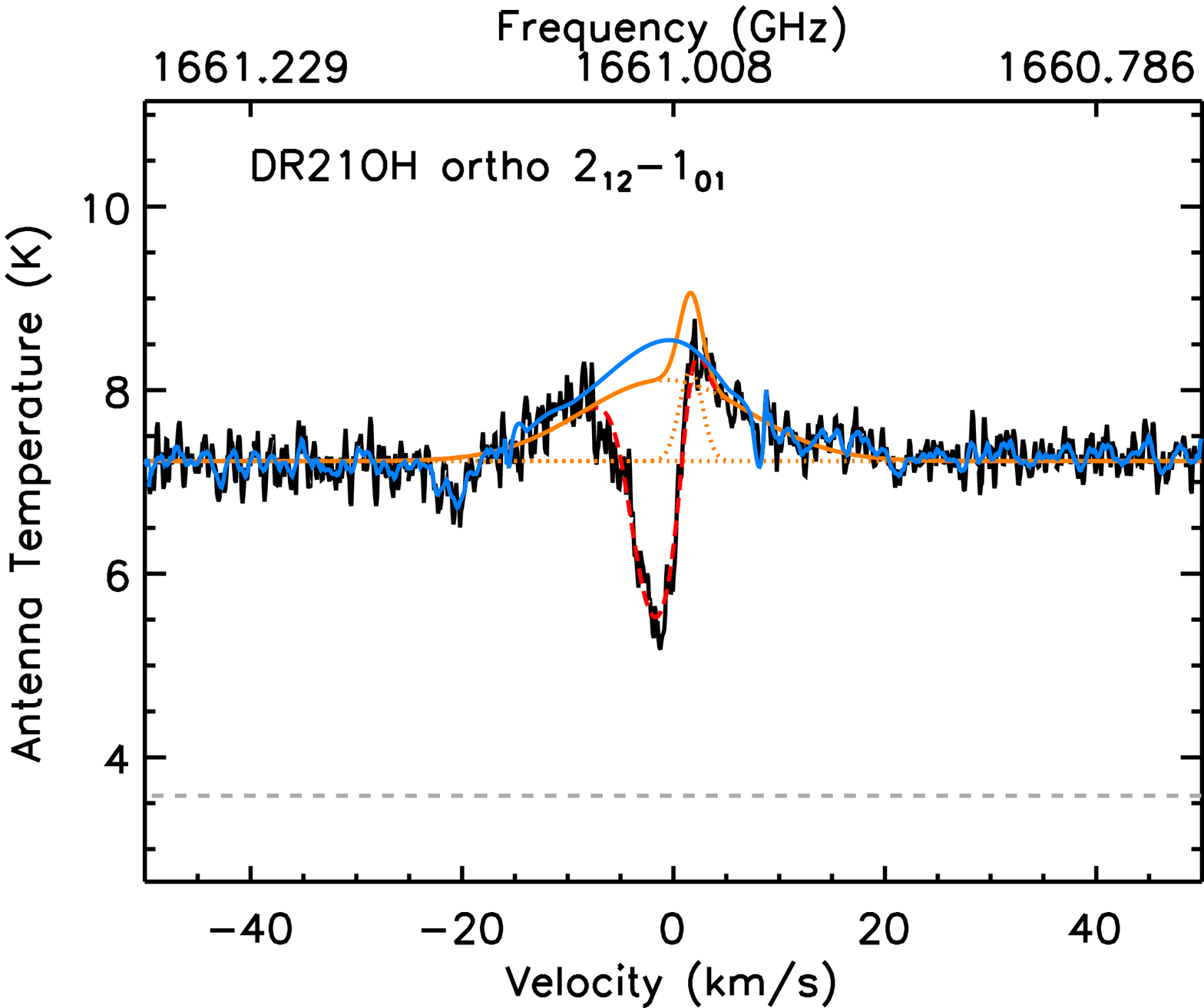}}
  \subfigure[] {\label{}
    \includegraphics[angle=90,width=.31\linewidth]{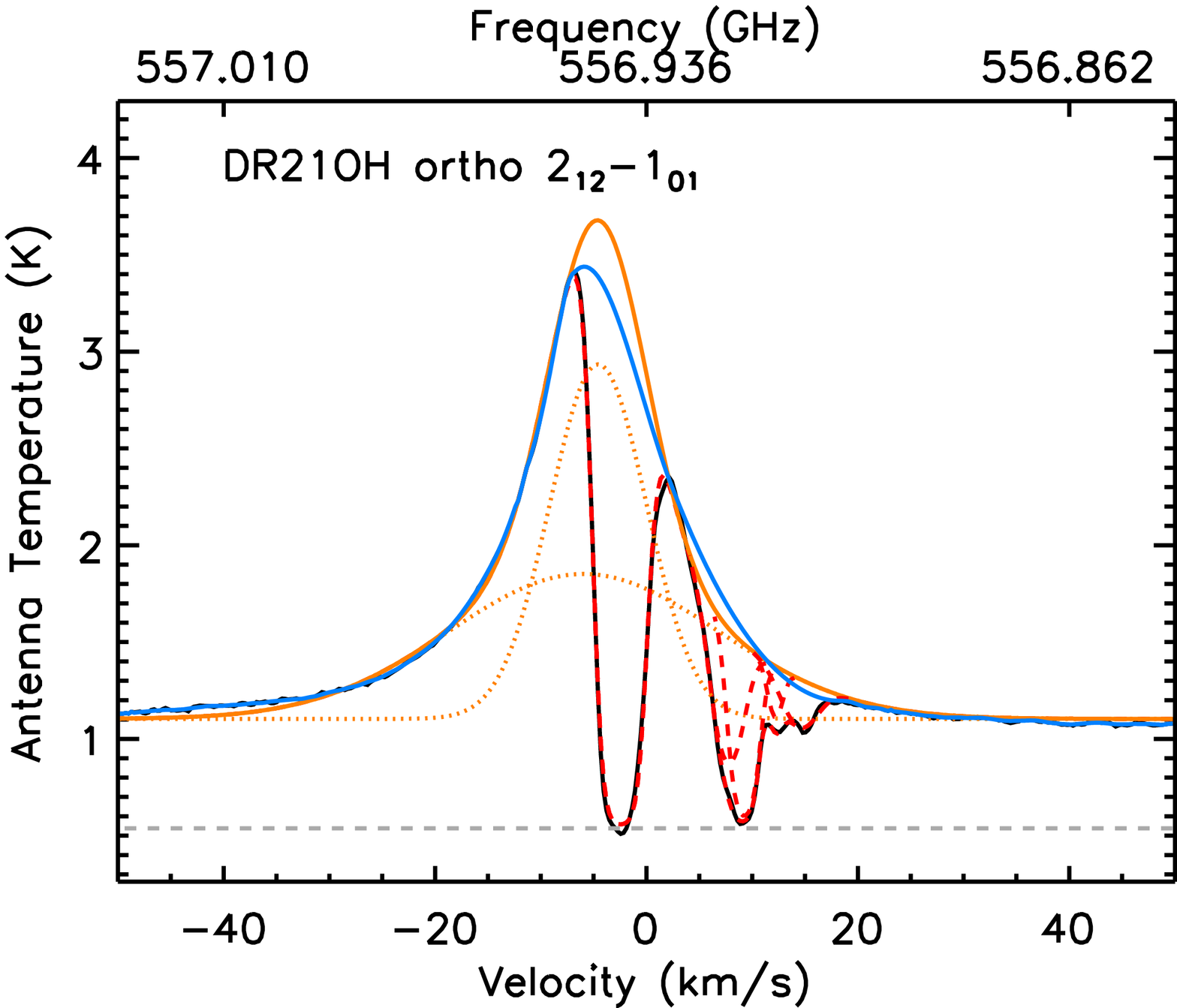}}
  \subfigure[] {\label{}
    \includegraphics[angle=90,width=.31\linewidth]{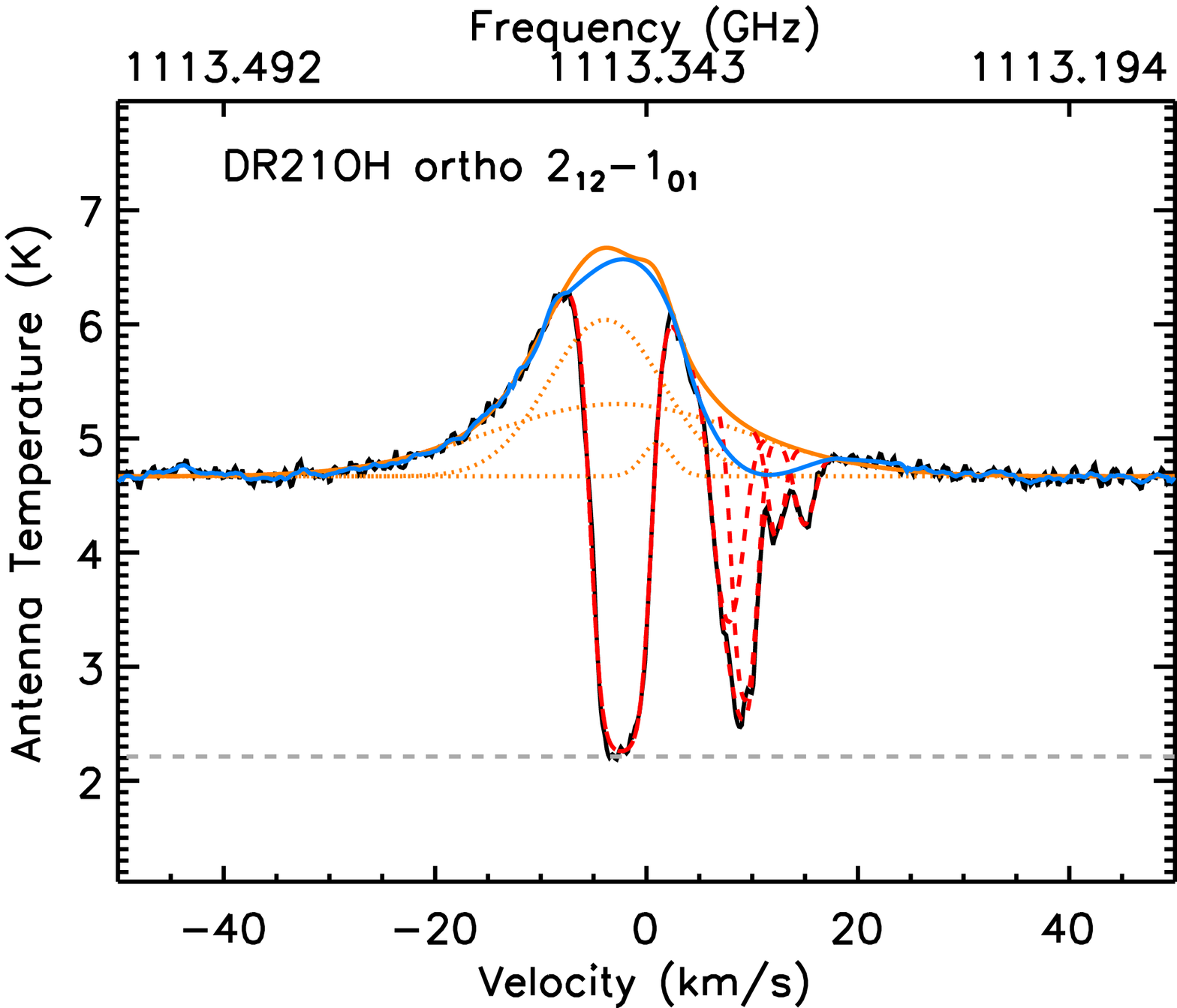}}
  \subfigure[] {\label{}
    \includegraphics[angle=90,width=.31\linewidth]{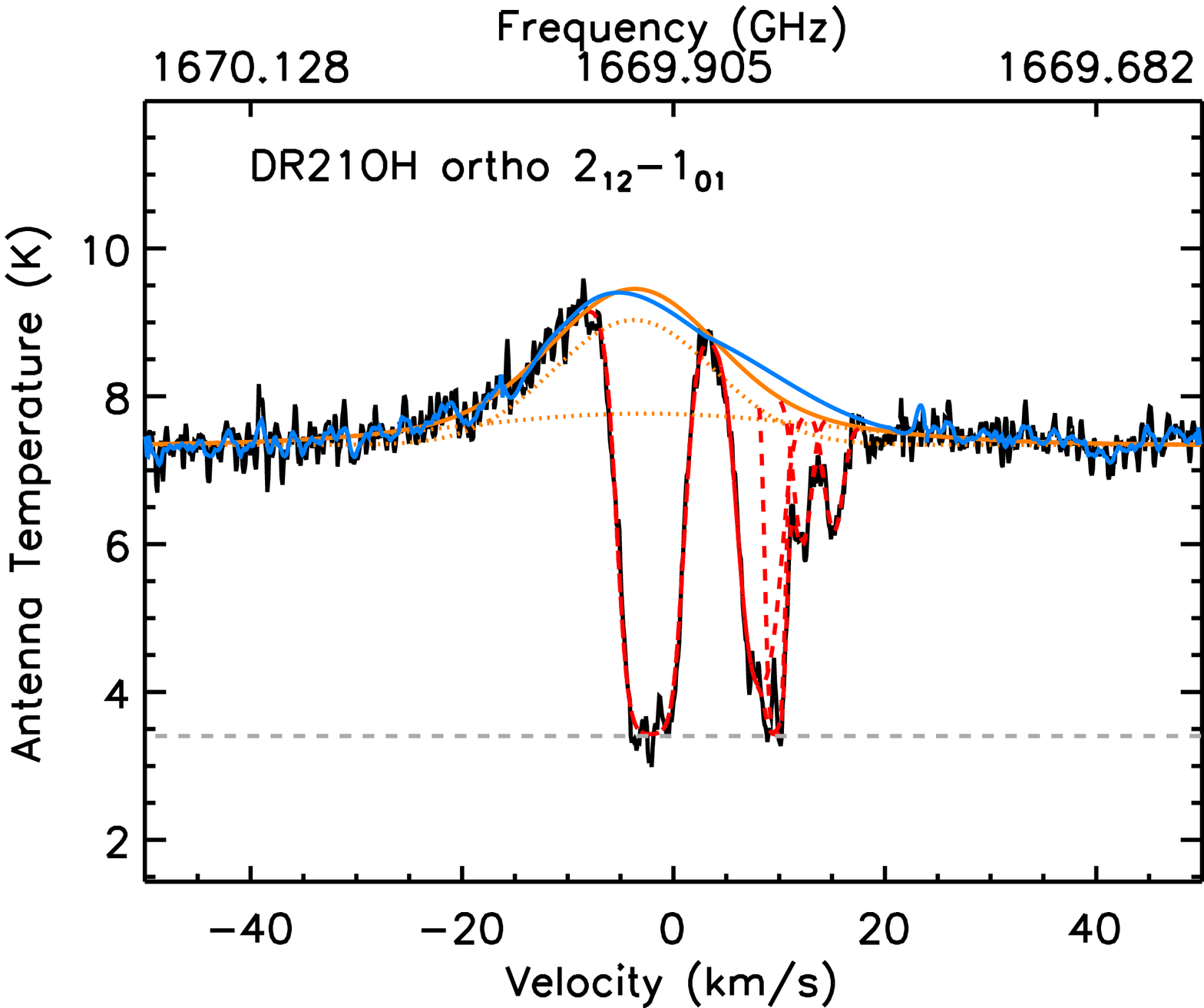}}
  \subfigure[] {\label{}
    \includegraphics[angle=90,width=.31\linewidth]{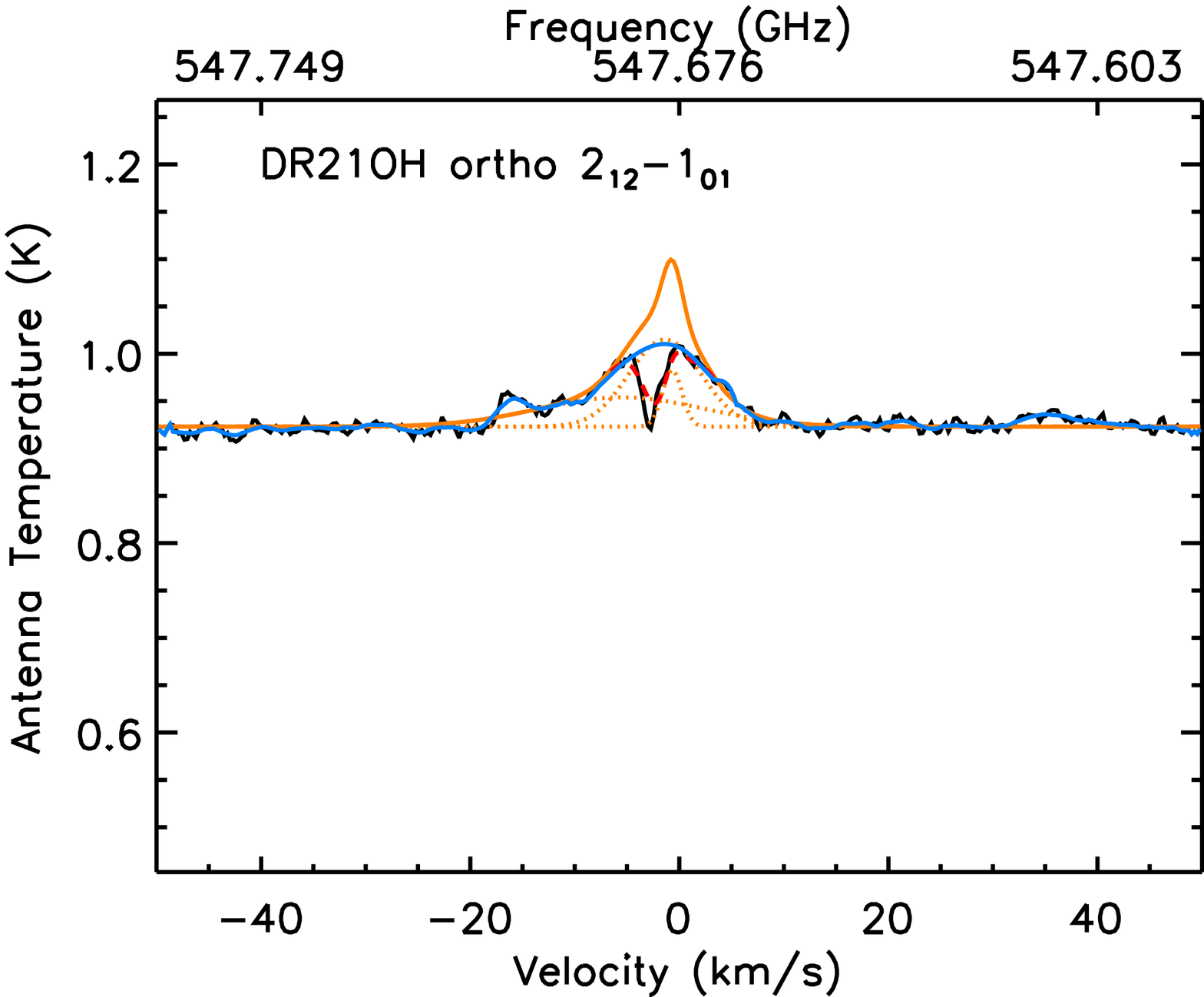}}
  \subfigure[] {\label{}
    \includegraphics[angle=90,width=.31\linewidth]{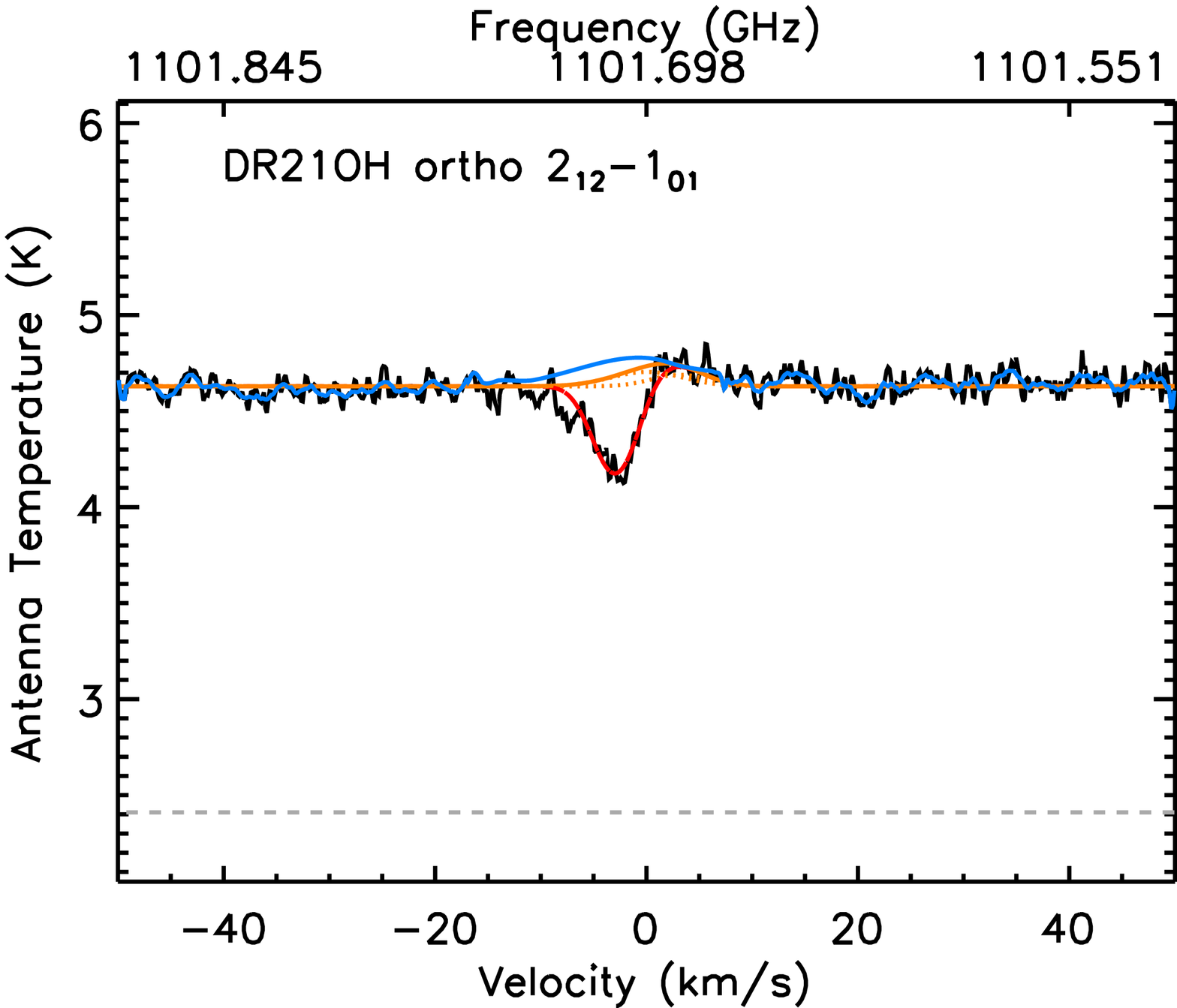}}
  \subfigure[] {\label{}
    \includegraphics[angle=90,width=.31\linewidth]{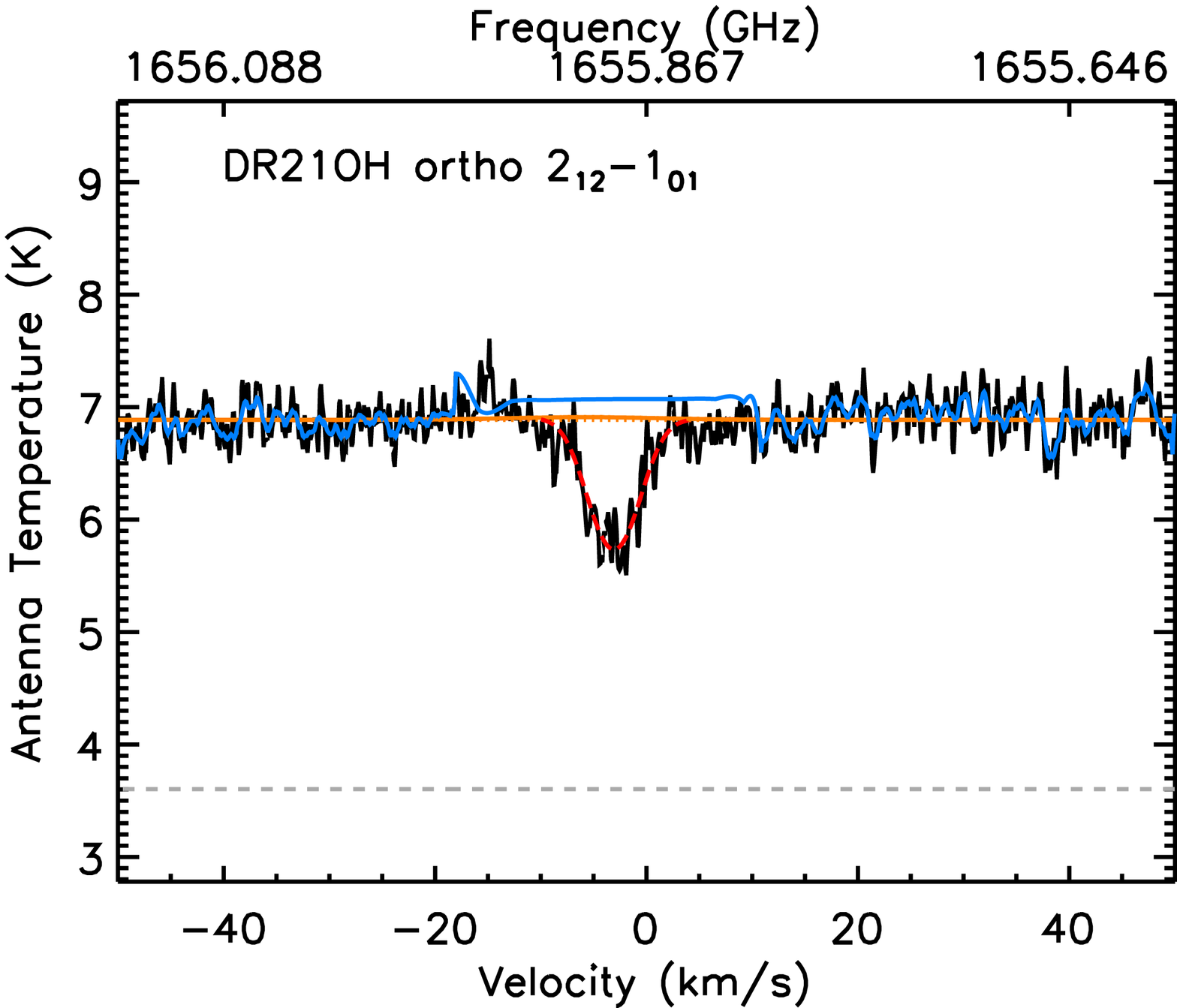}}
  \caption{Same as Figure \ref{fig:w51} but towards DR21(OH).}
  \label{fig:dr21}
\end{figure*}

\begin{figure*}[!t]
  \centering \subfigure[] {\label{}
    \includegraphics[angle=90,width=.31\linewidth]{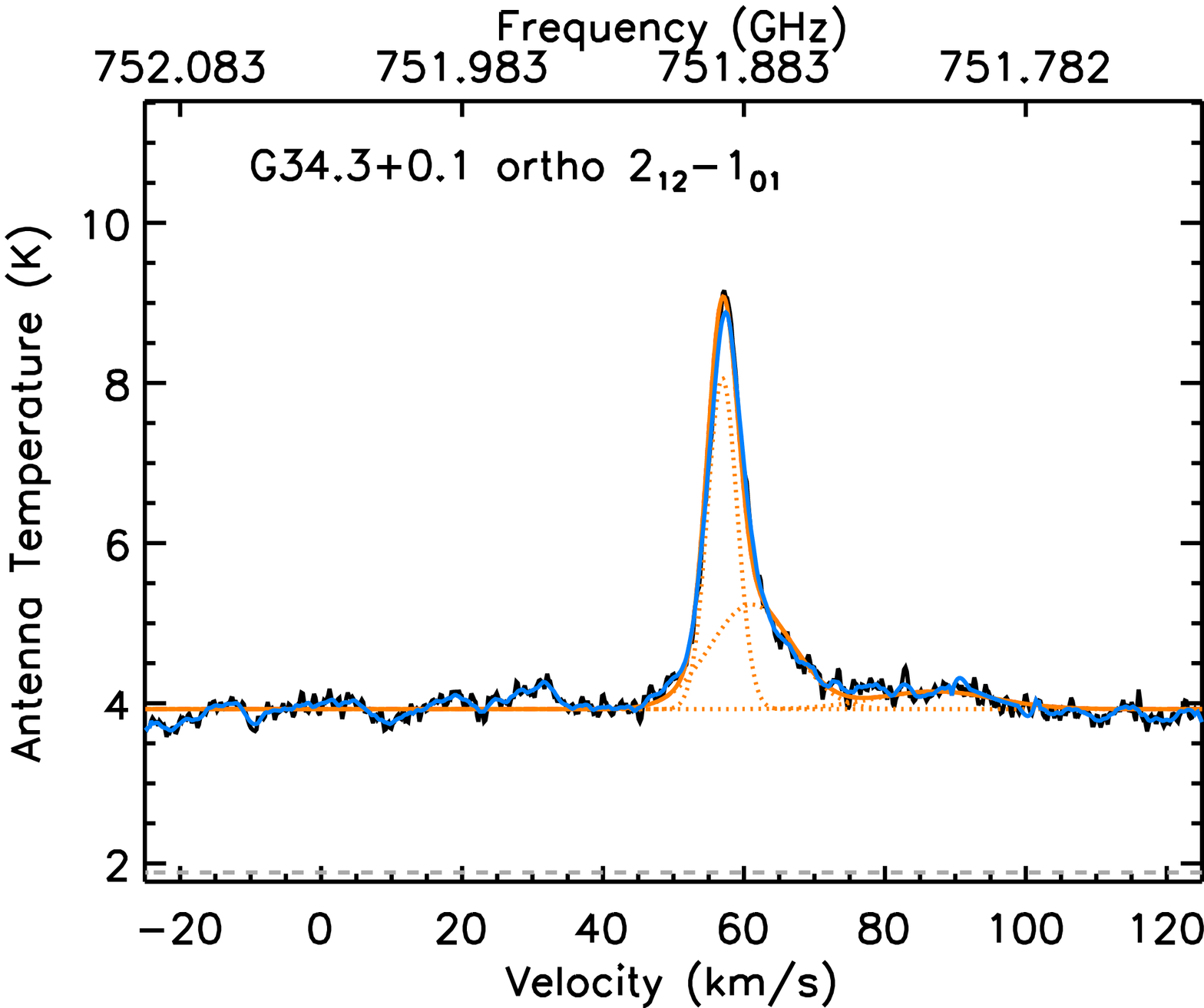}}
  \subfigure[] {\label{}
    \includegraphics[angle=90,width=.31\linewidth]{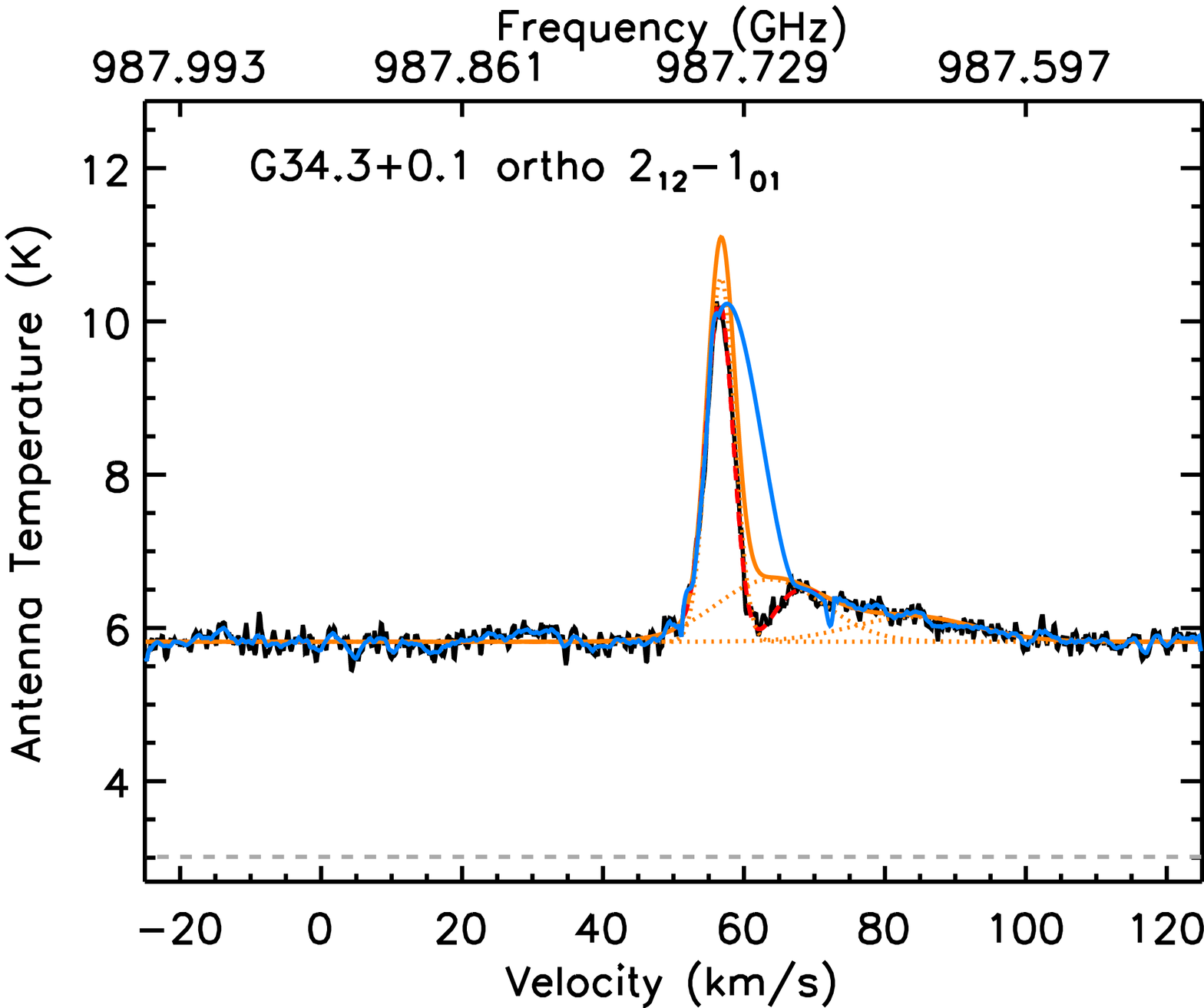}}
  \subfigure[] {\label{}
    \includegraphics[angle=90,width=.31\linewidth]{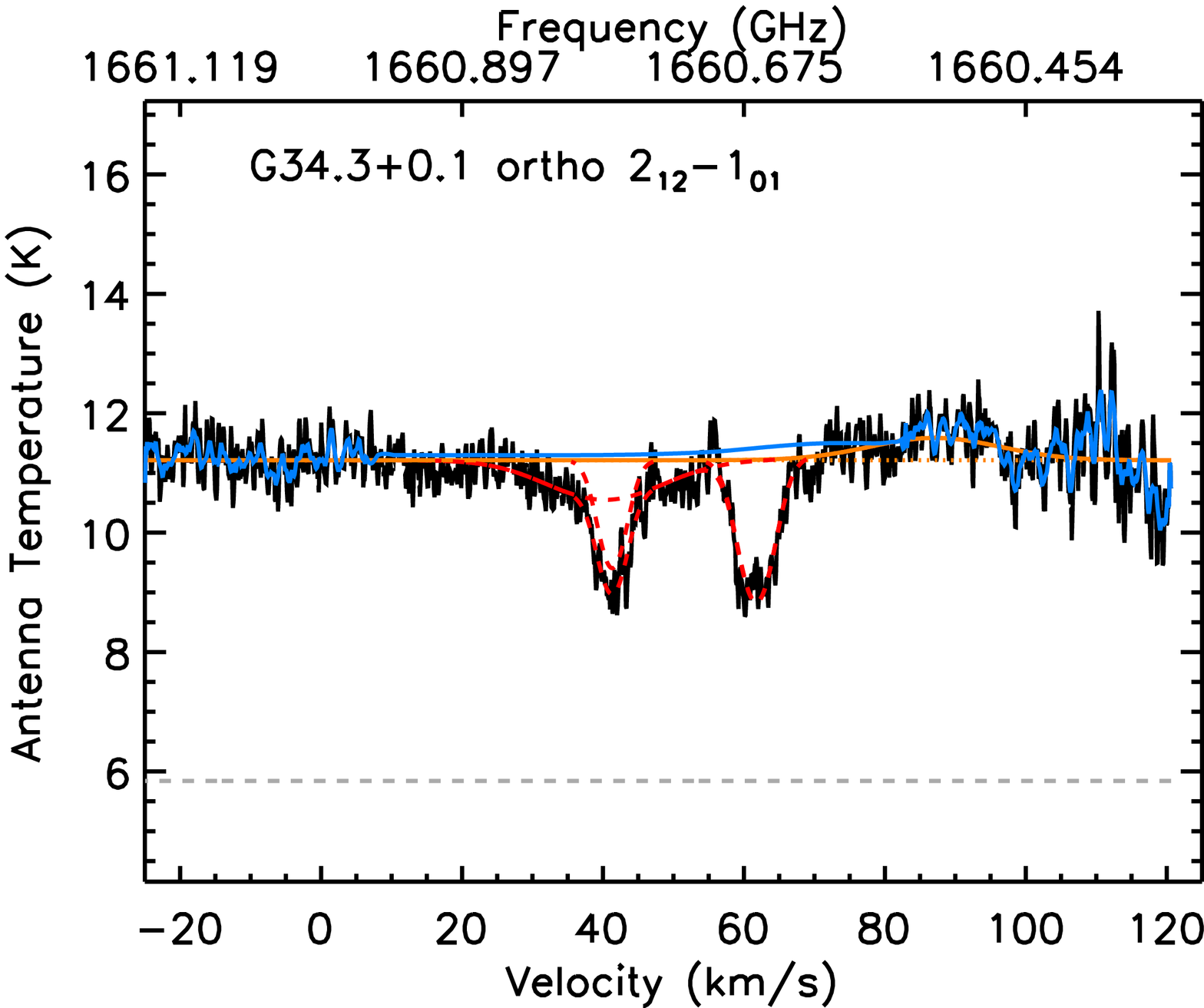}}
  \subfigure[] {\label{}
    \includegraphics[angle=90,width=.31\linewidth]{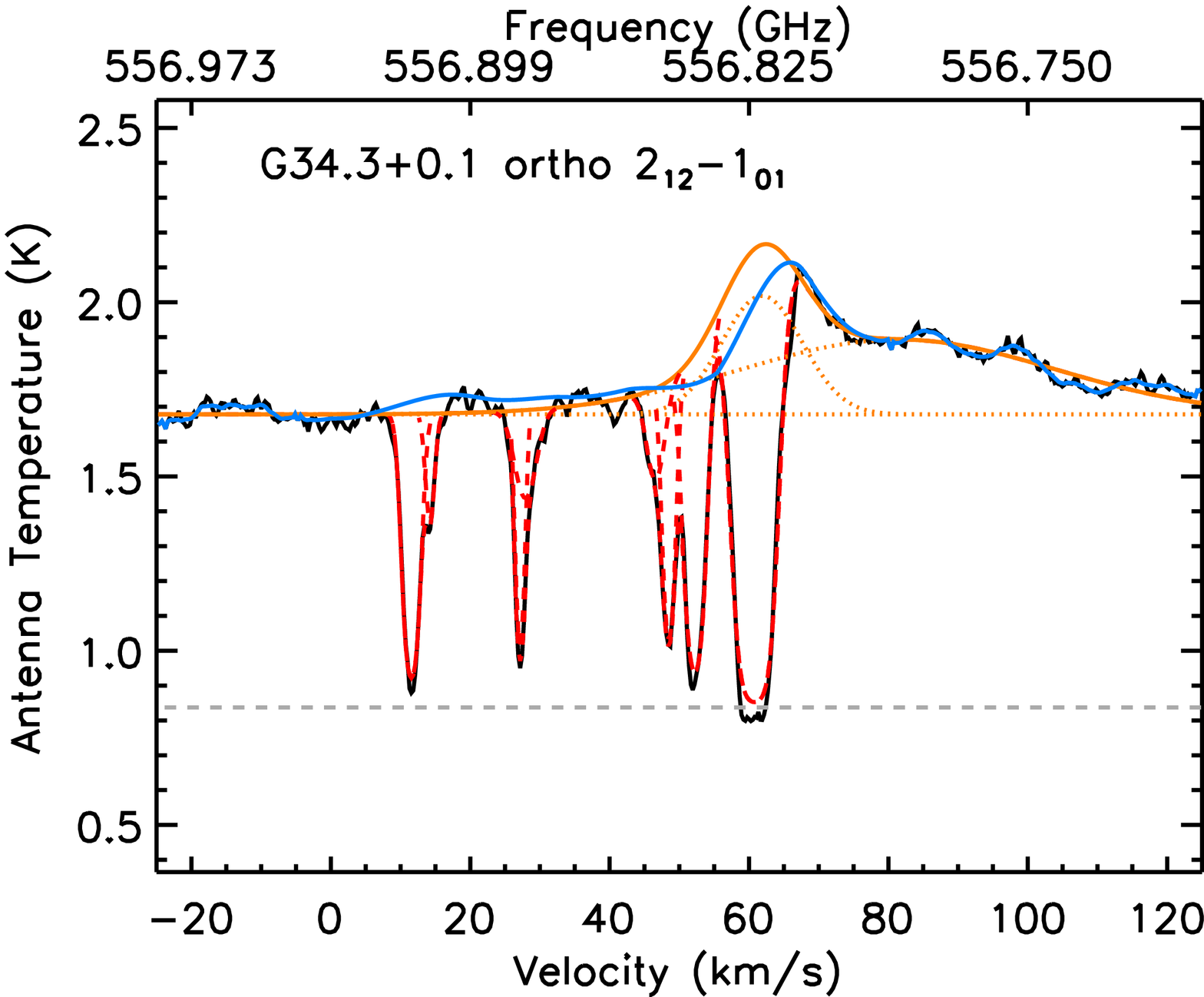}}
  \subfigure[] {\label{}
    \includegraphics[angle=90,width=.31\linewidth]{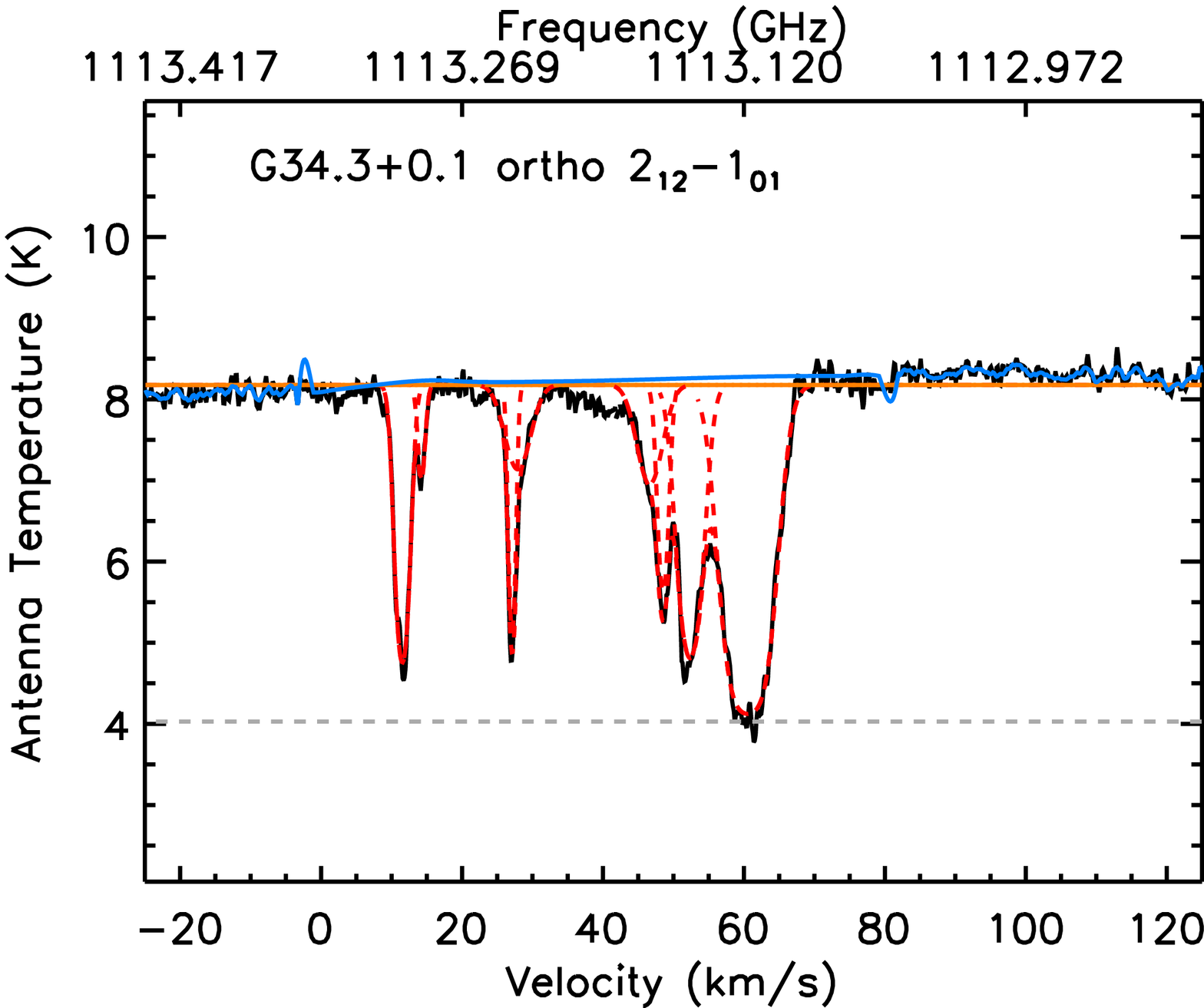}}
  \subfigure[] {\label{}
    \includegraphics[angle=90,width=.31\linewidth]{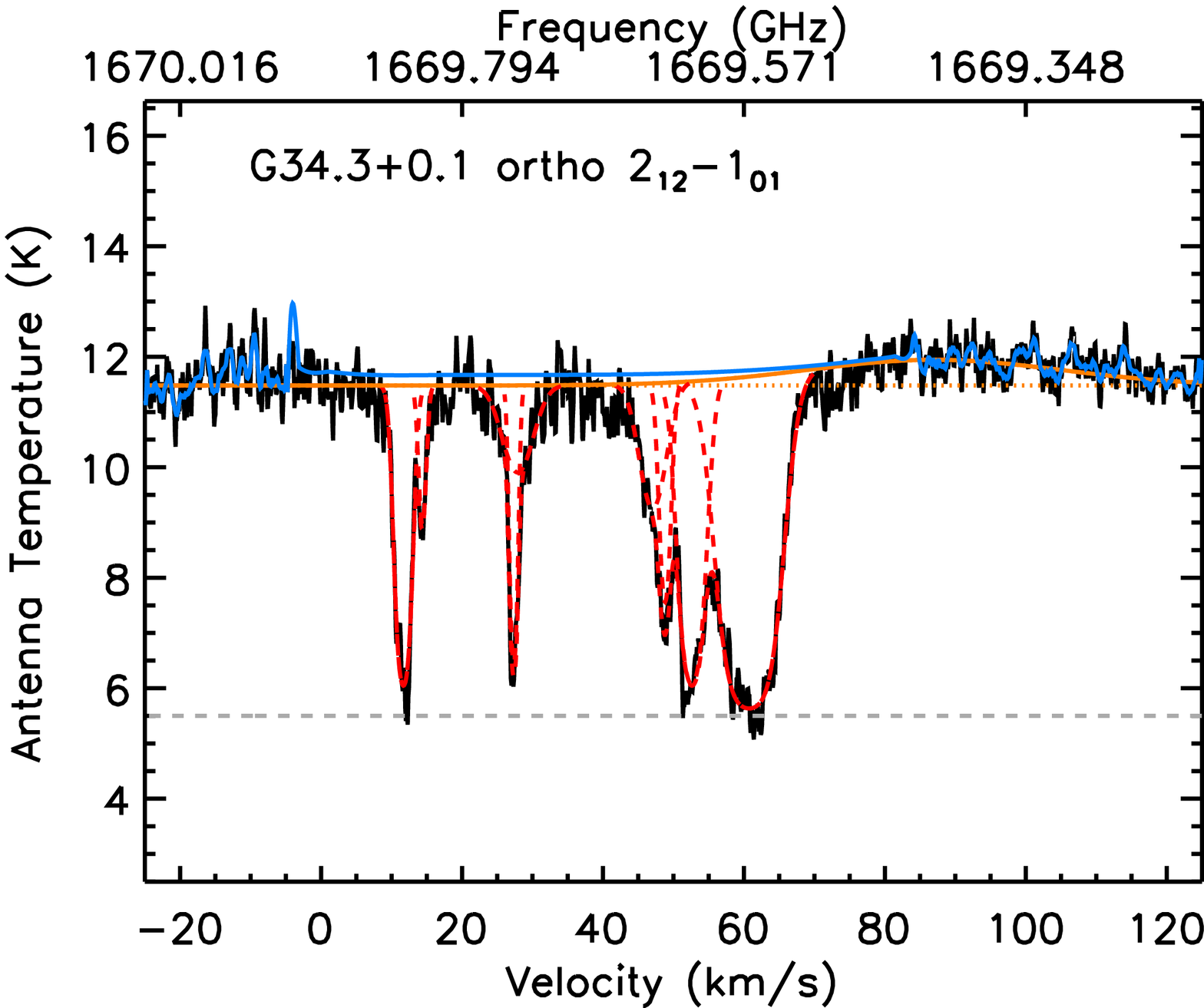}}
  \subfigure[] {\label{}
    \includegraphics[angle=90,width=.31\linewidth]{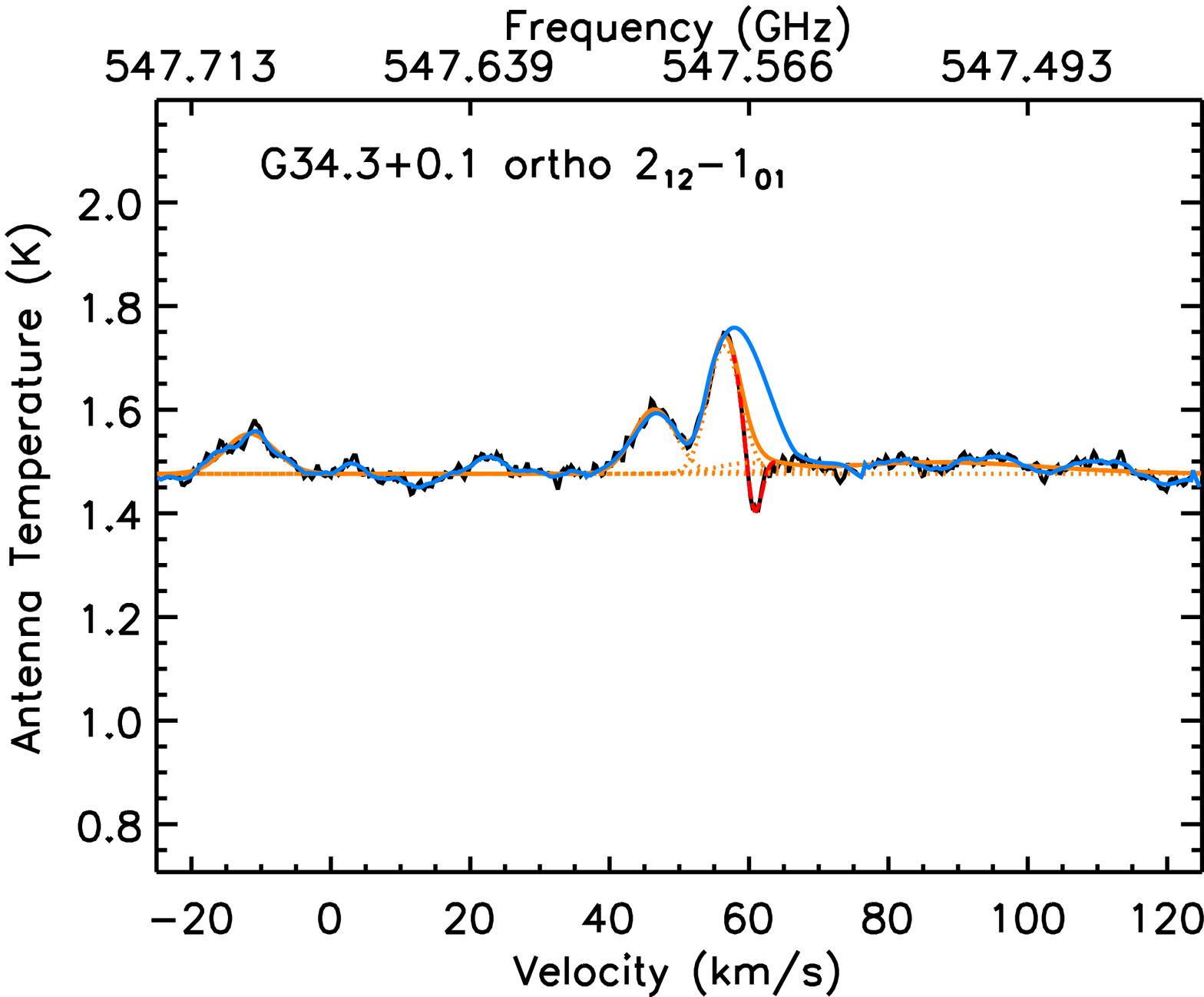}}
  \subfigure[] {\label{}
    \includegraphics[angle=90,width=.31\linewidth]{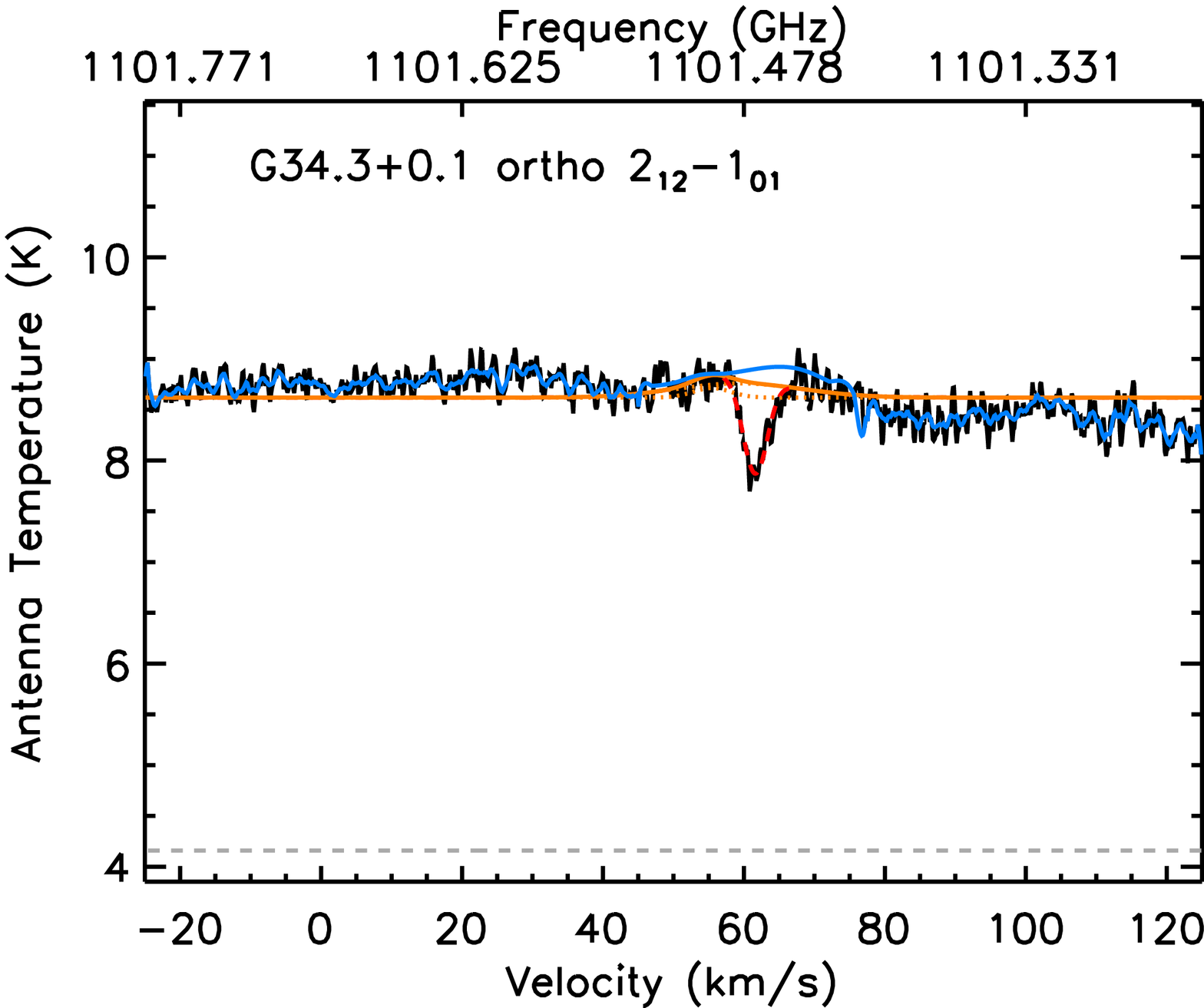}}
  \subfigure[] {\label{}
    \includegraphics[angle=90,width=.31\linewidth]{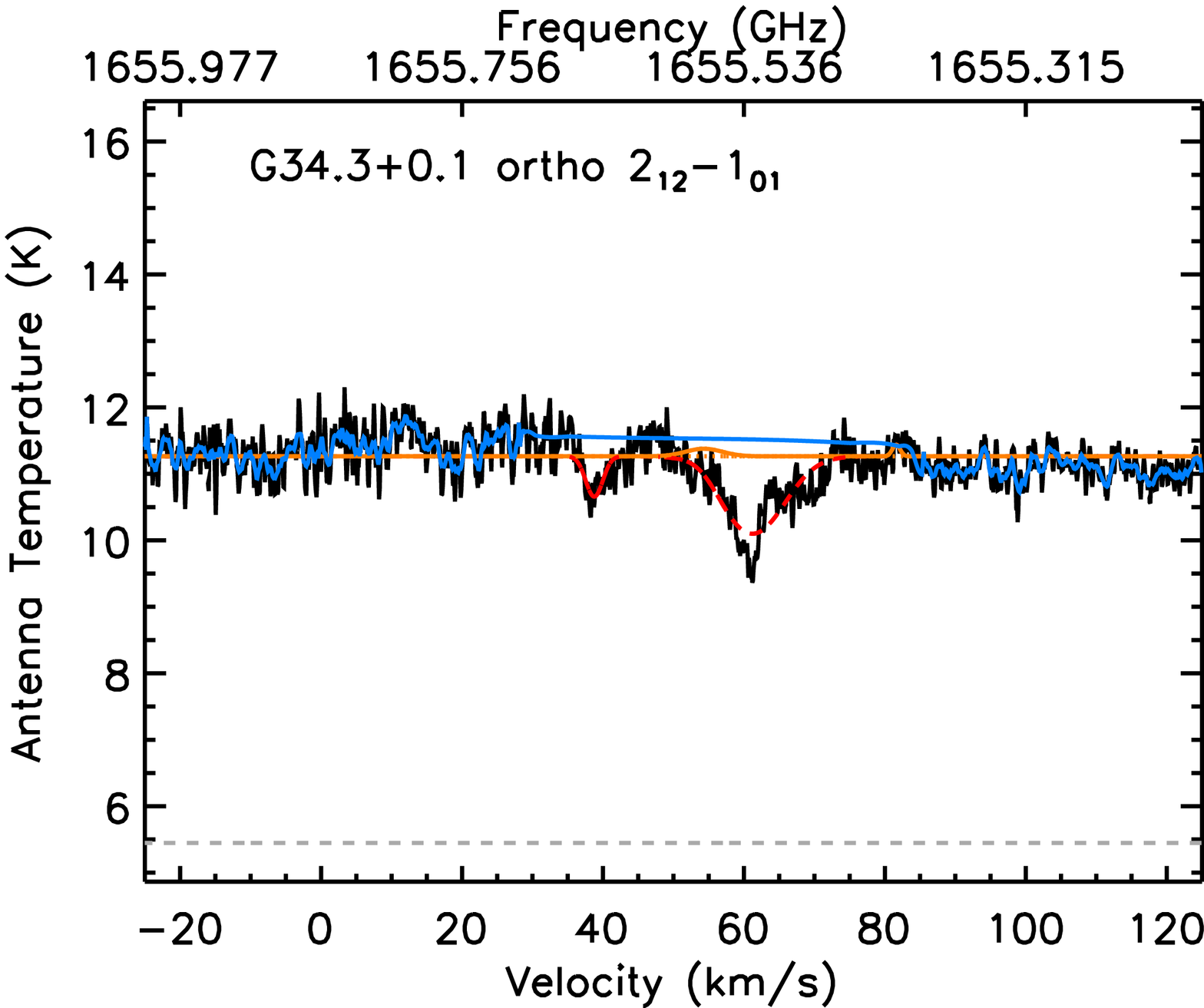}}
  \caption{Same as Figure \ref{fig:w51} but towards G34.3+0.1.}
  \label{fig:g34}
\end{figure*}

\begin{figure*}[!t]
  \centering \subfigure[] {\label{}
    \includegraphics[angle=90,width=.31\linewidth]{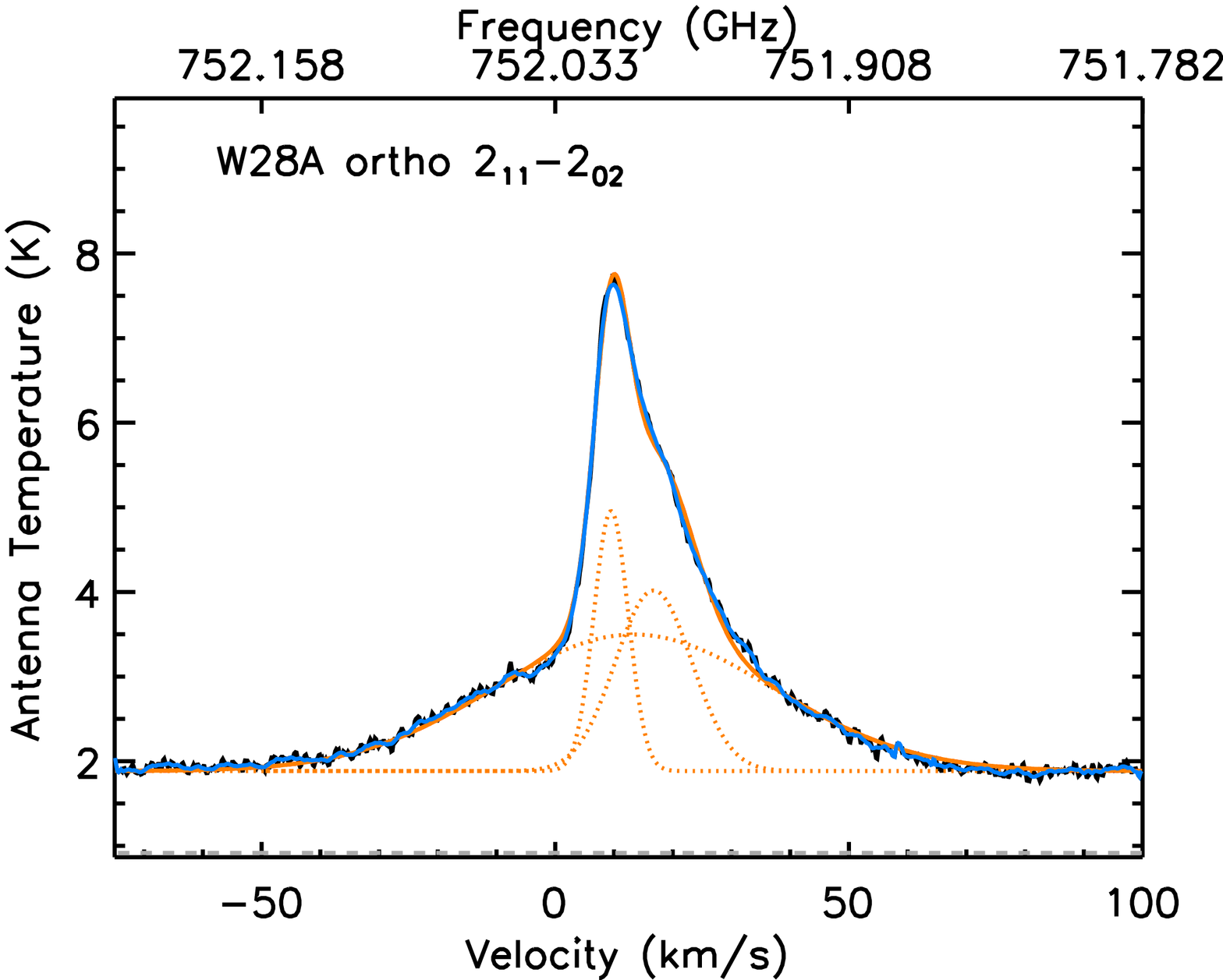}}
  \subfigure[] {\label{}
    \includegraphics[angle=90,width=.31\linewidth]{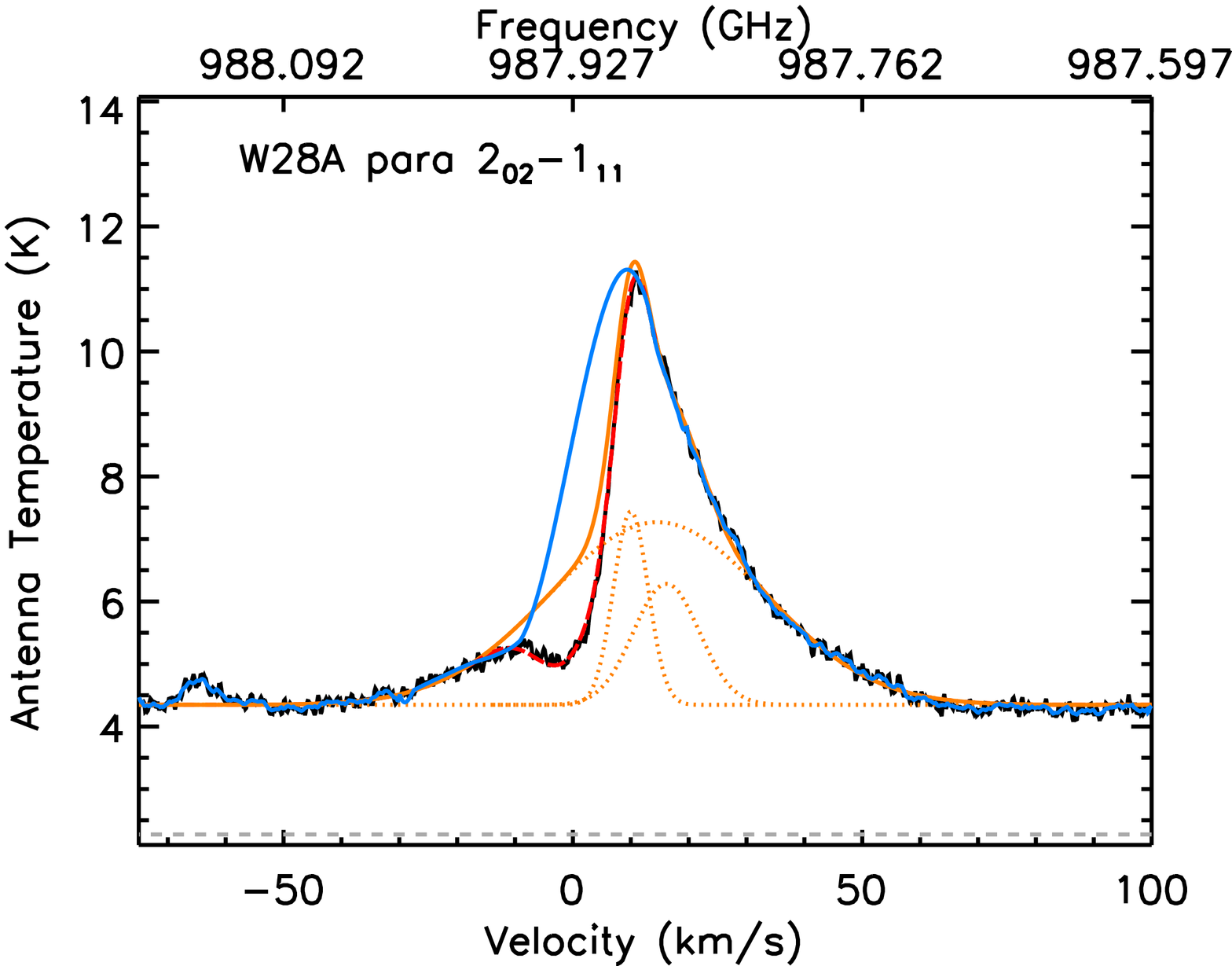}}
  \subfigure[] {\label{}
    \includegraphics[angle=90,width=.31\linewidth]{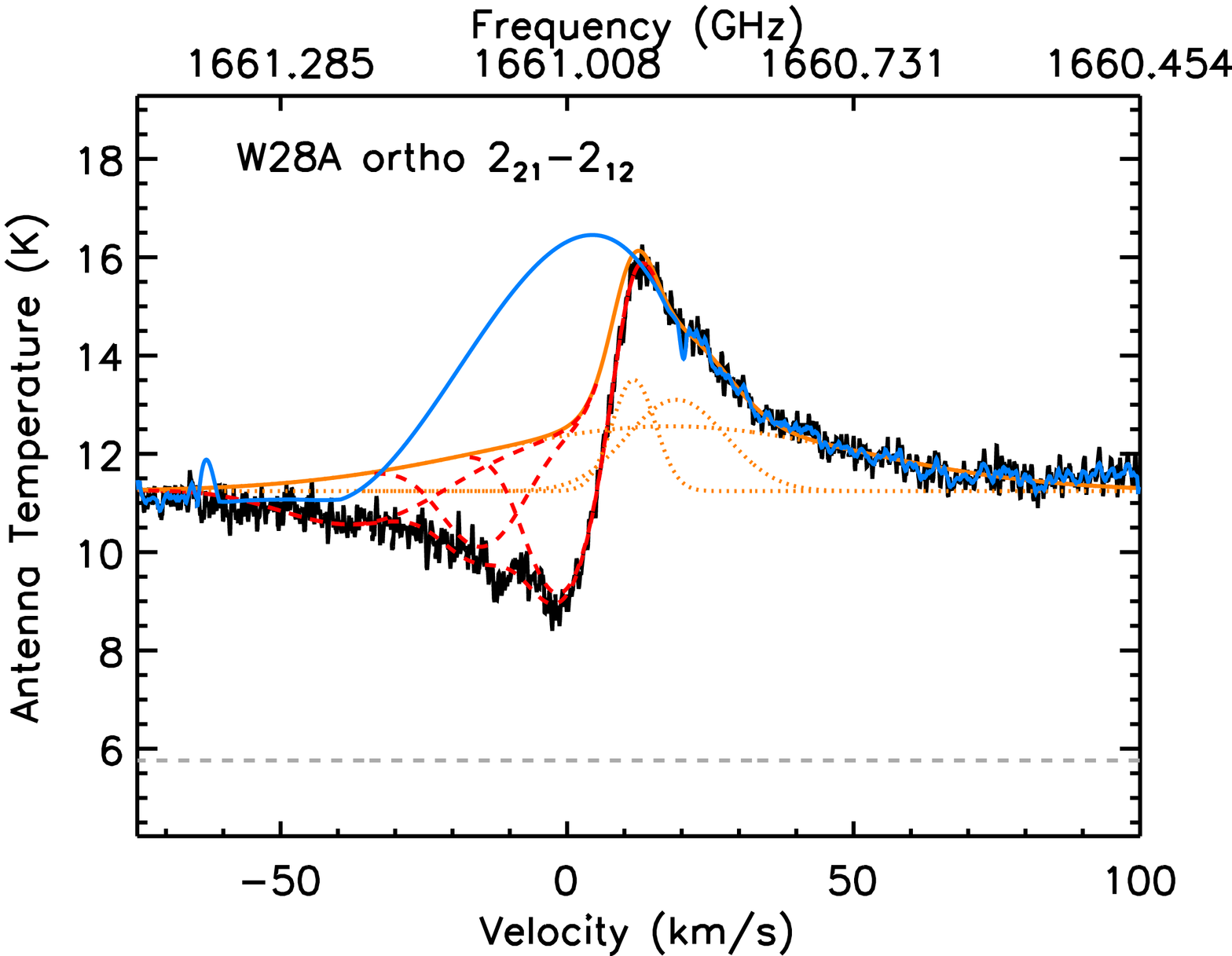}}
  \subfigure[] {\label{}
    \includegraphics[angle=90,width=.31\linewidth]{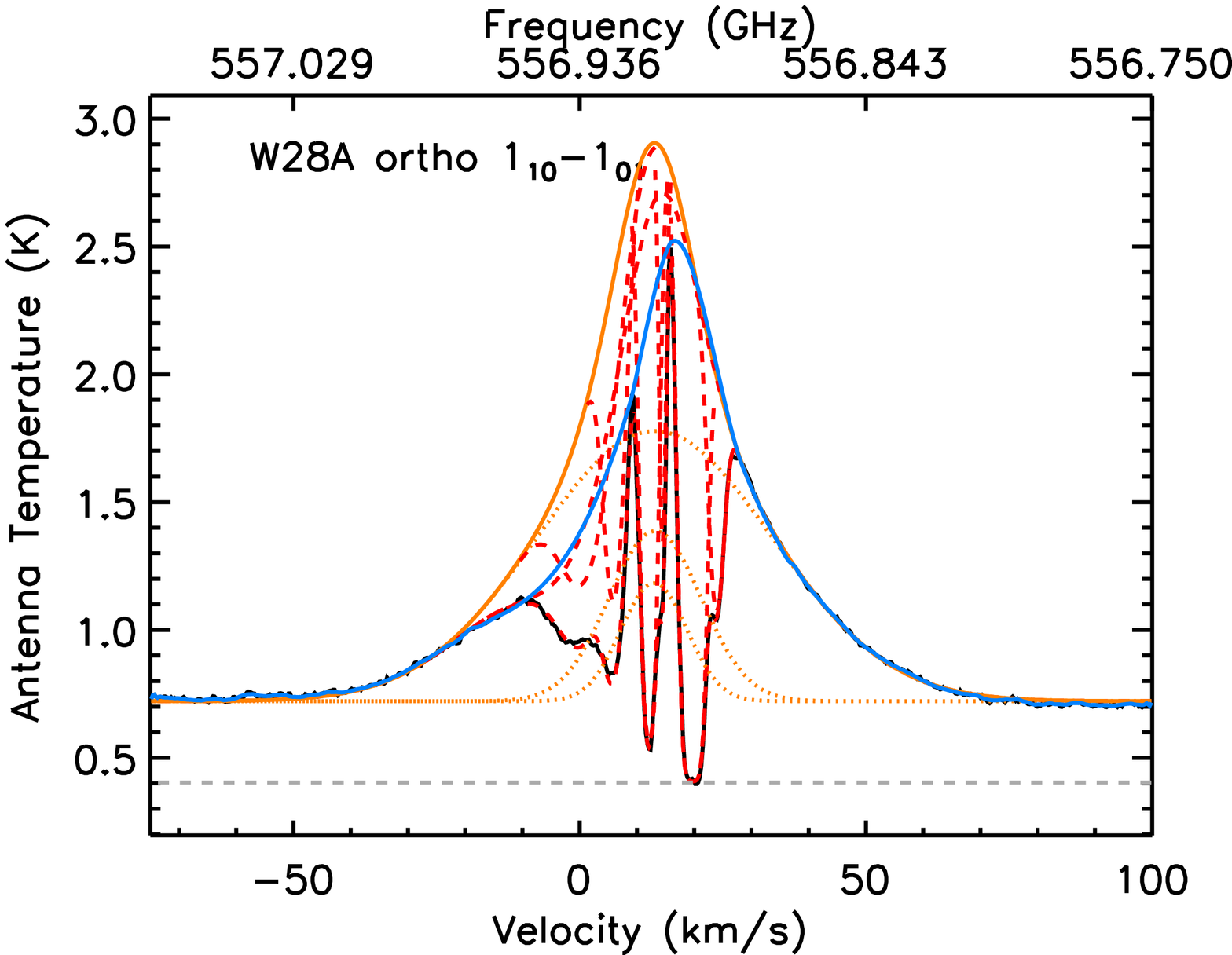}}
  \subfigure[] {\label{}
    \includegraphics[angle=90,width=.31\linewidth]{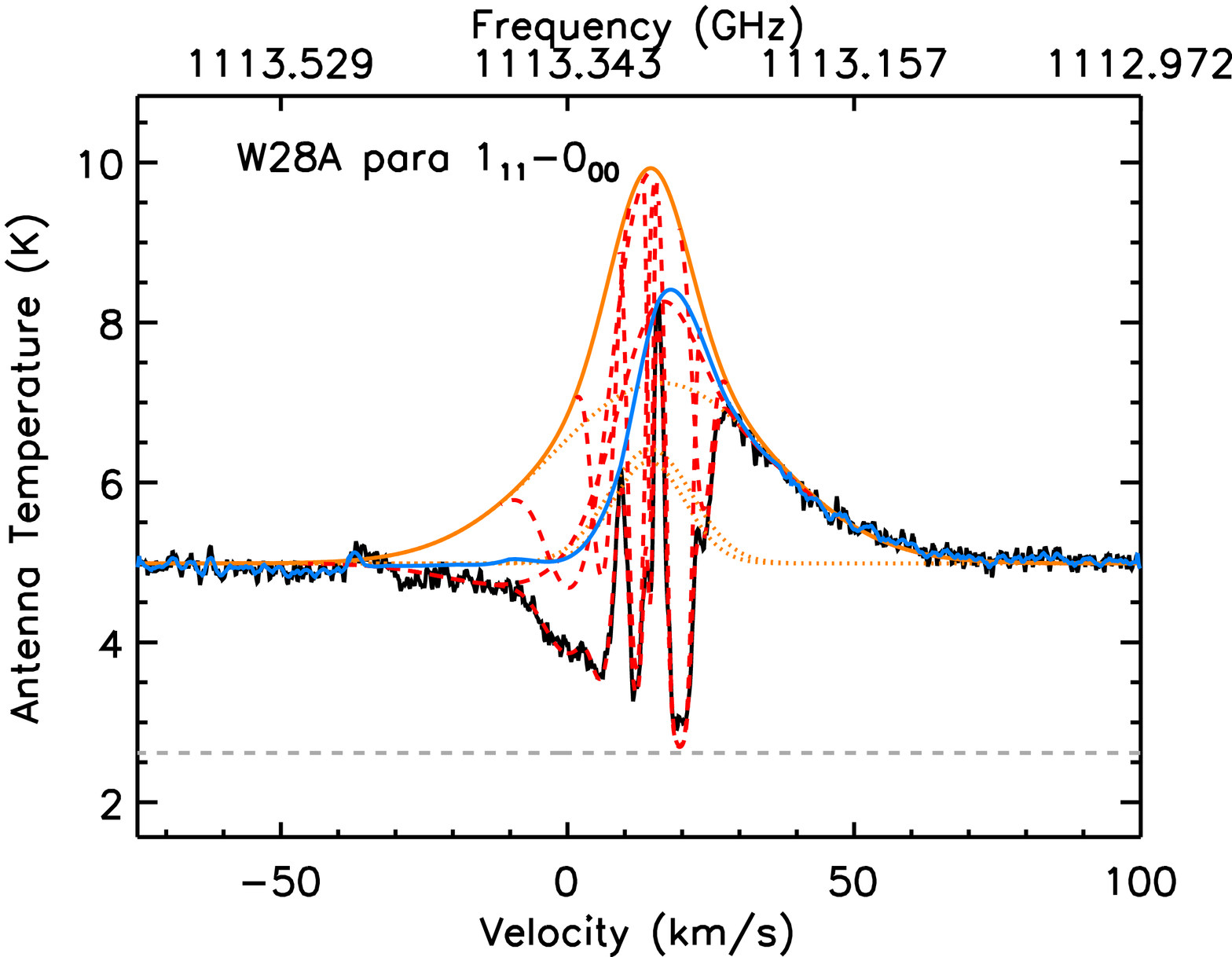}}
  \subfigure[] {\label{}
    \includegraphics[angle=90,width=.31\linewidth]{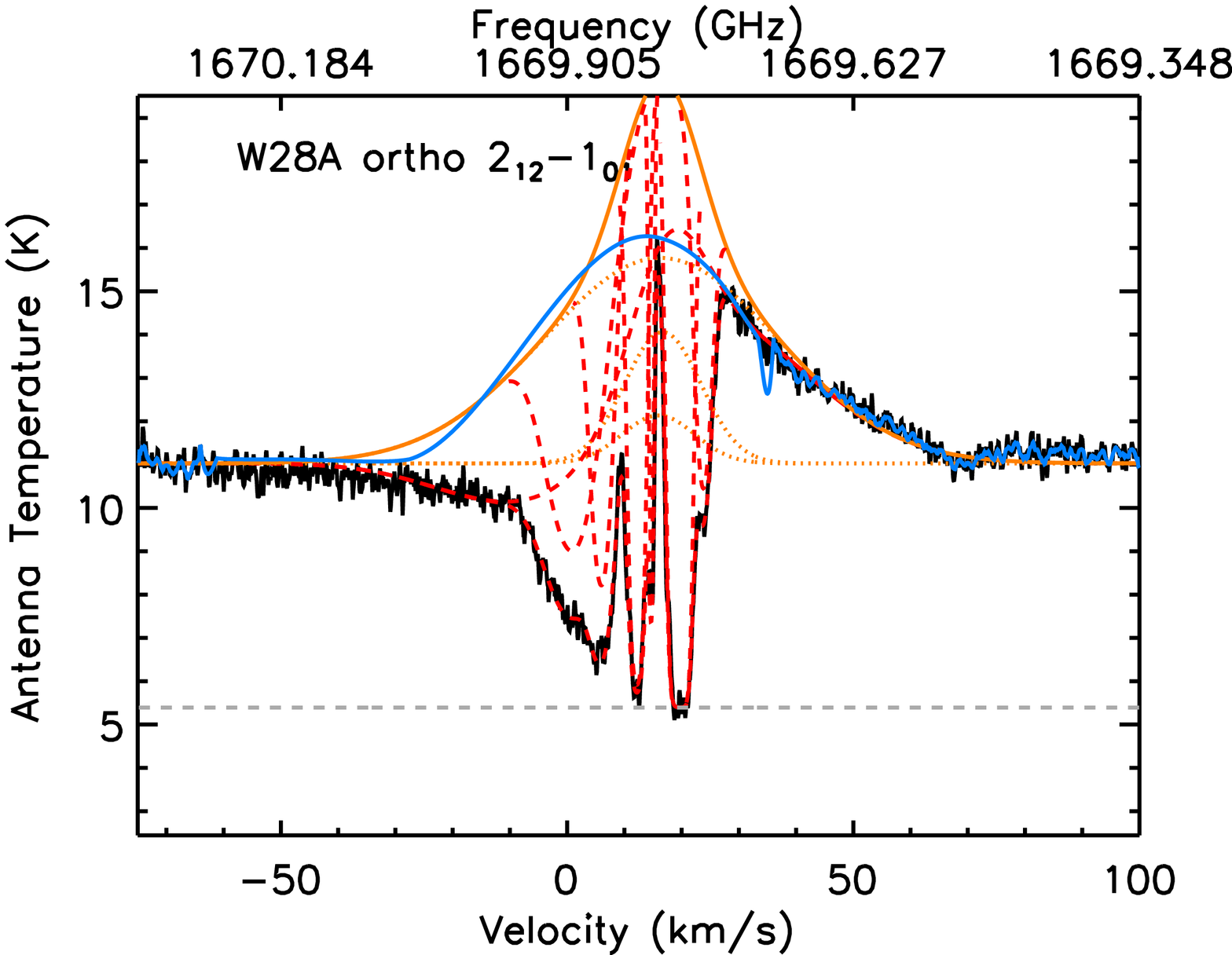}}
  \subfigure[] {\label{}
    \includegraphics[angle=90,width=.31\linewidth]{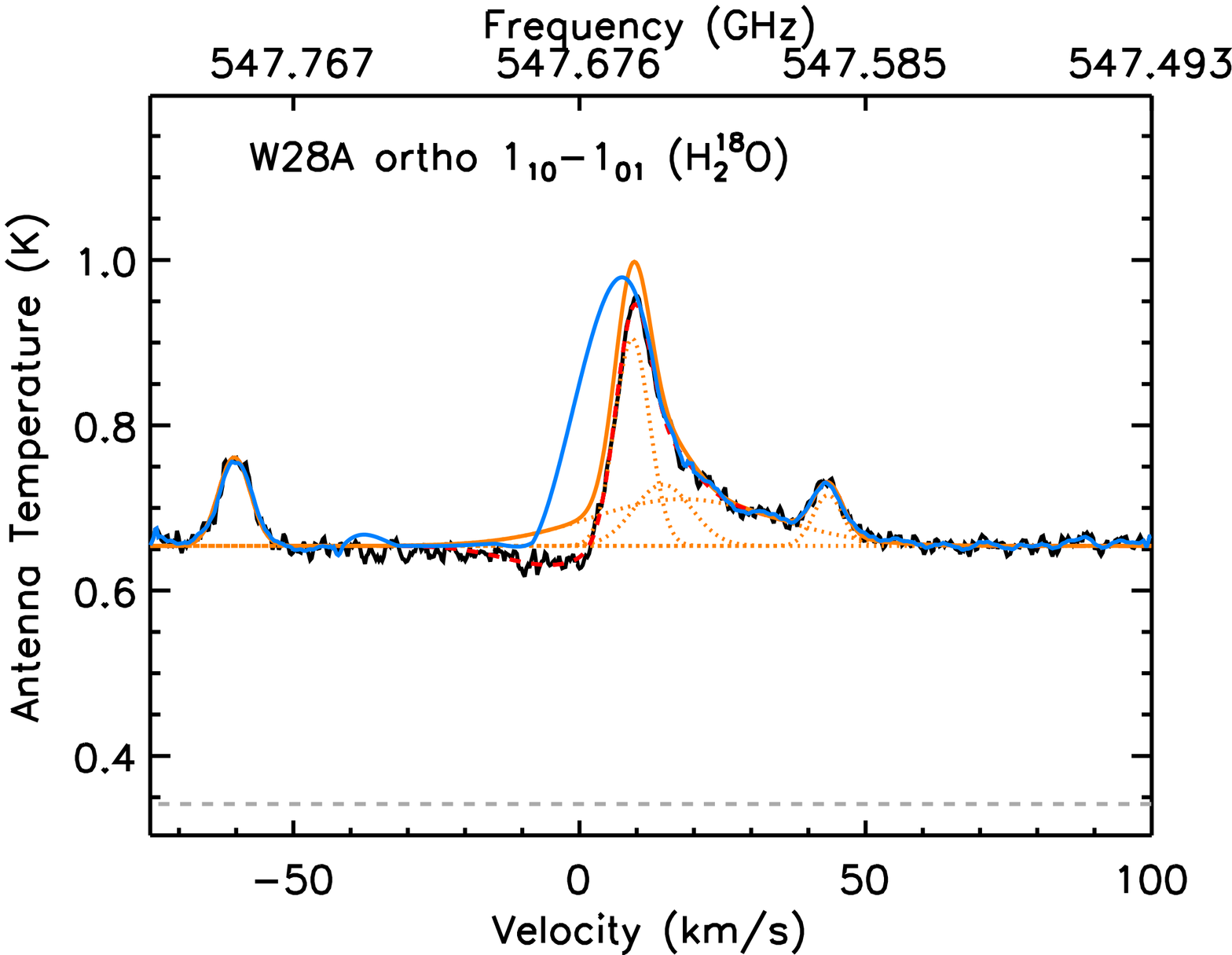}}
  \subfigure[] {\label{}
    \includegraphics[angle=90,width=.31\linewidth]{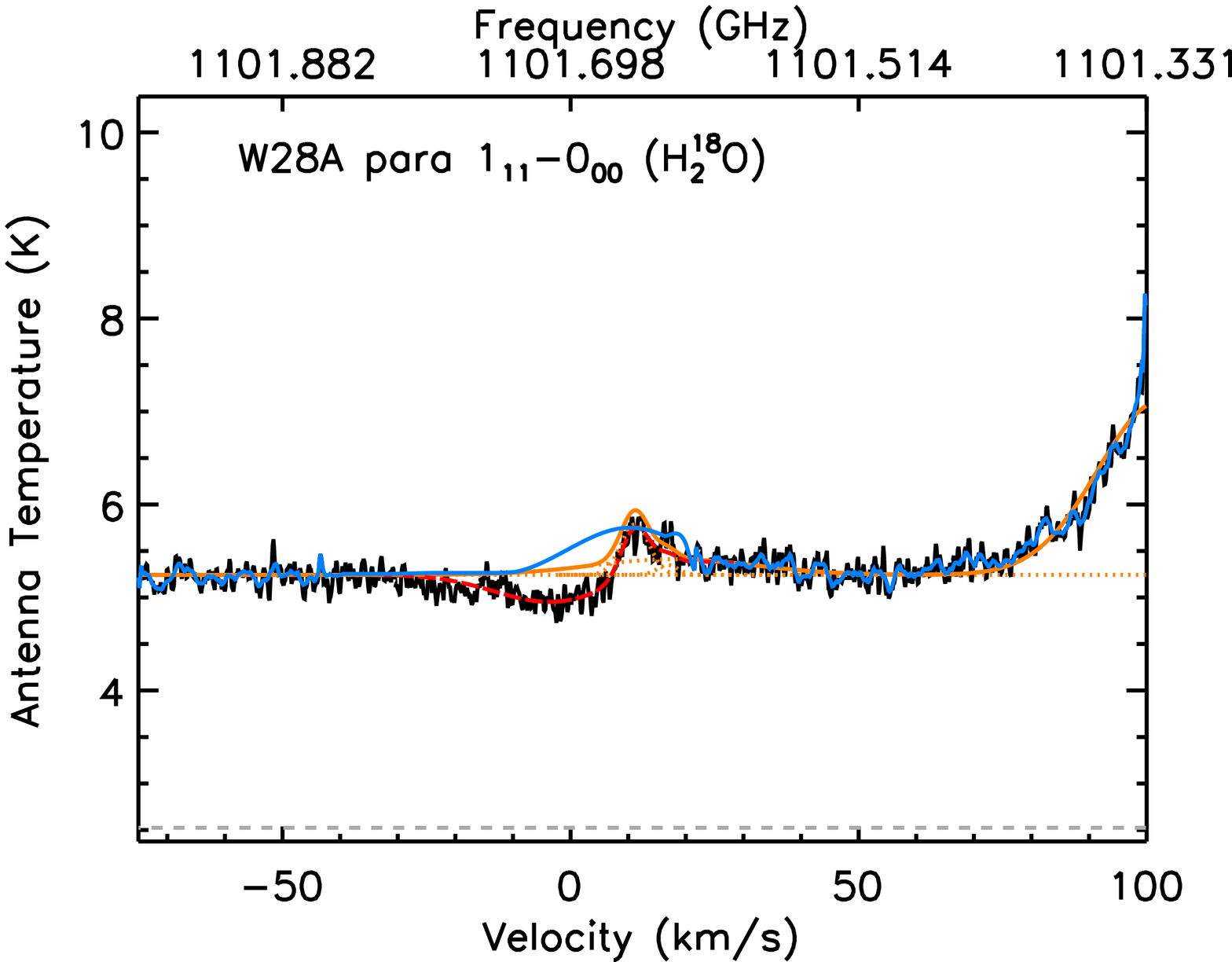}}
  \subfigure[] {\label{}
    \includegraphics[angle=90,width=.31\linewidth]{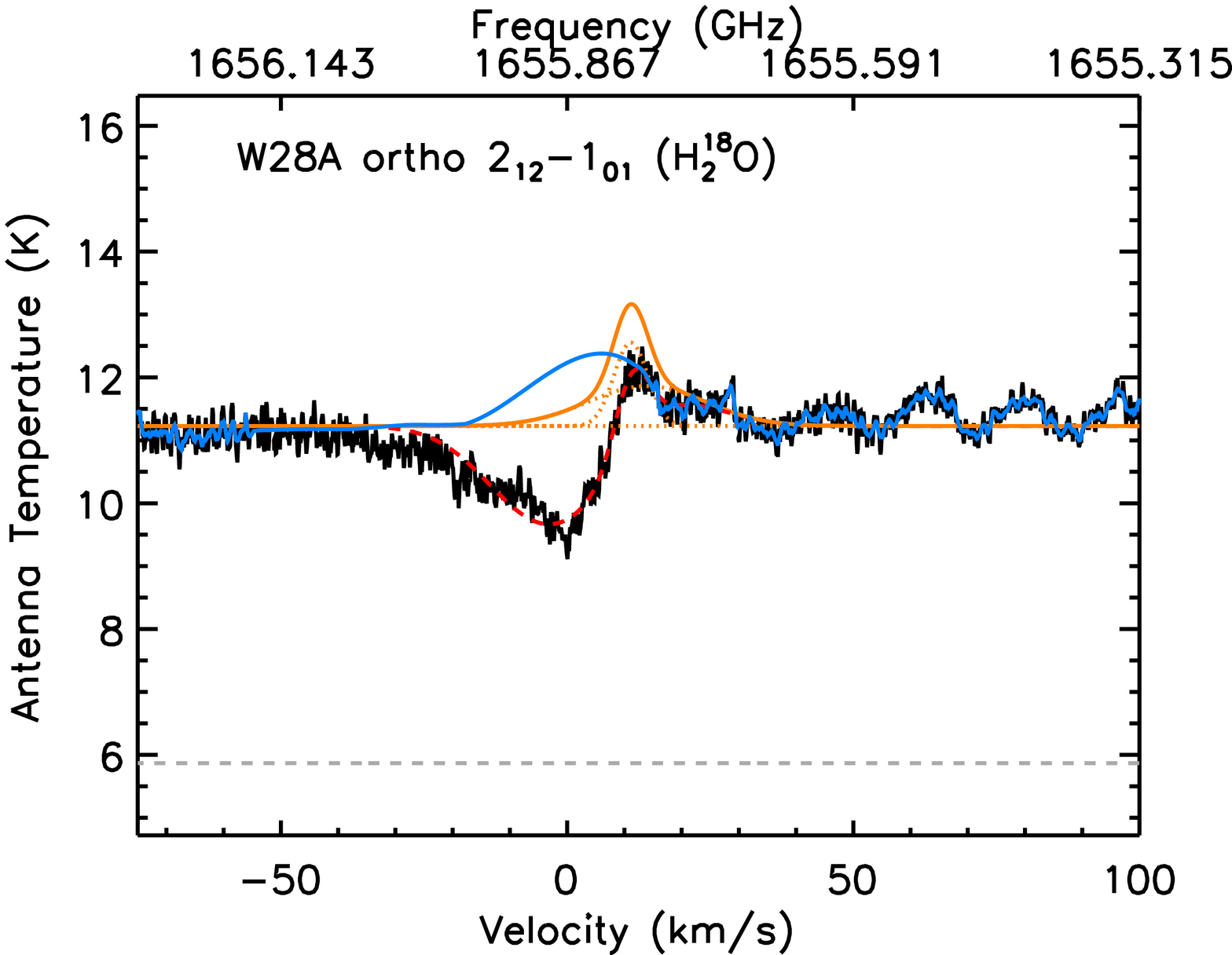}}
  \caption{Same as Figure \ref{fig:w51} but towards W28(A).}
  \label{fig:w28}
\end{figure*}

\begin{figure*}[!t]
  \centering \subfigure[] {\label{}
    \includegraphics[angle=90,width=.31\linewidth]{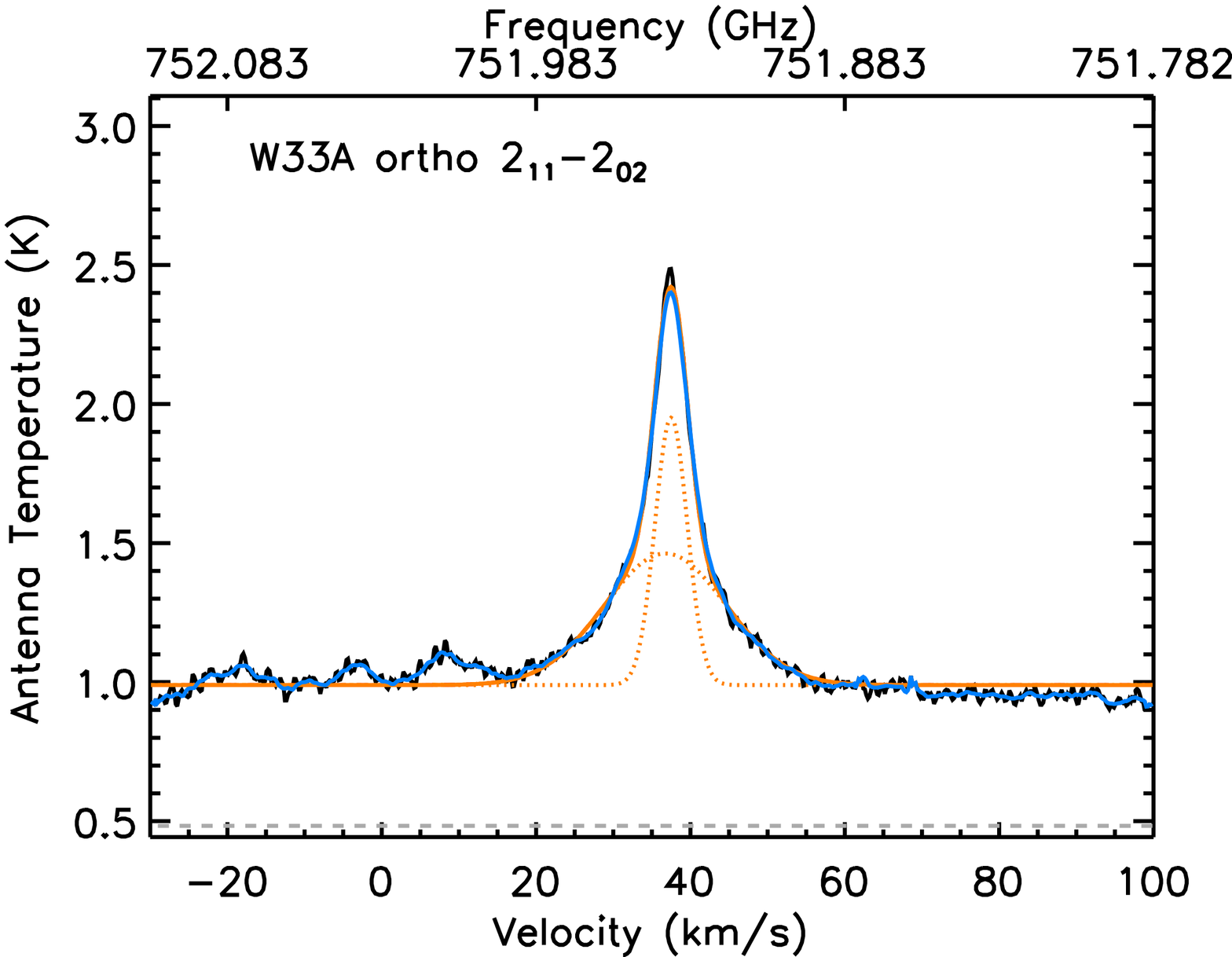}}
  \subfigure[] {\label{}
    \includegraphics[angle=90,width=.31\linewidth]{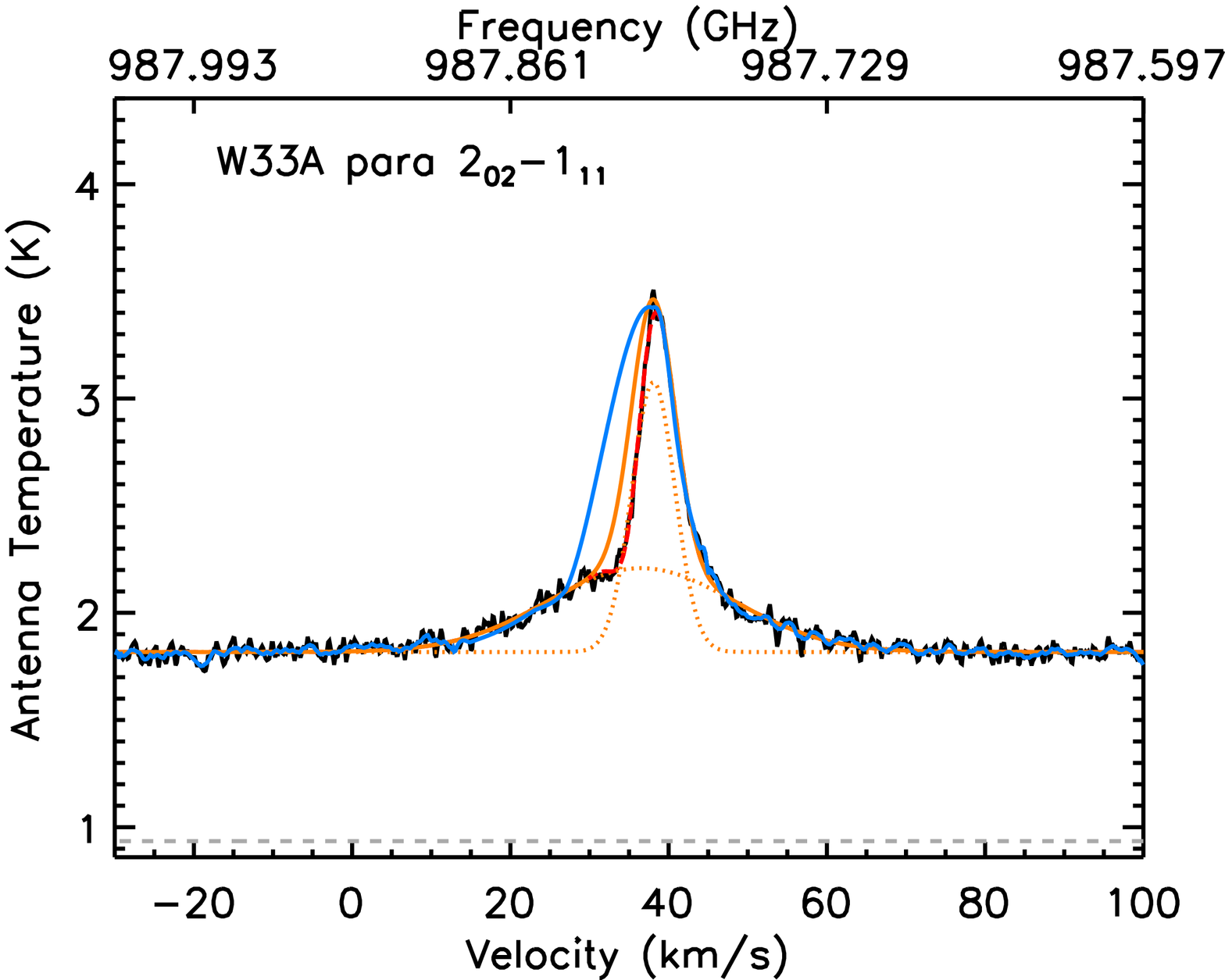}}
  \subfigure[] {\label{}
    \includegraphics[angle=90,width=.31\linewidth]{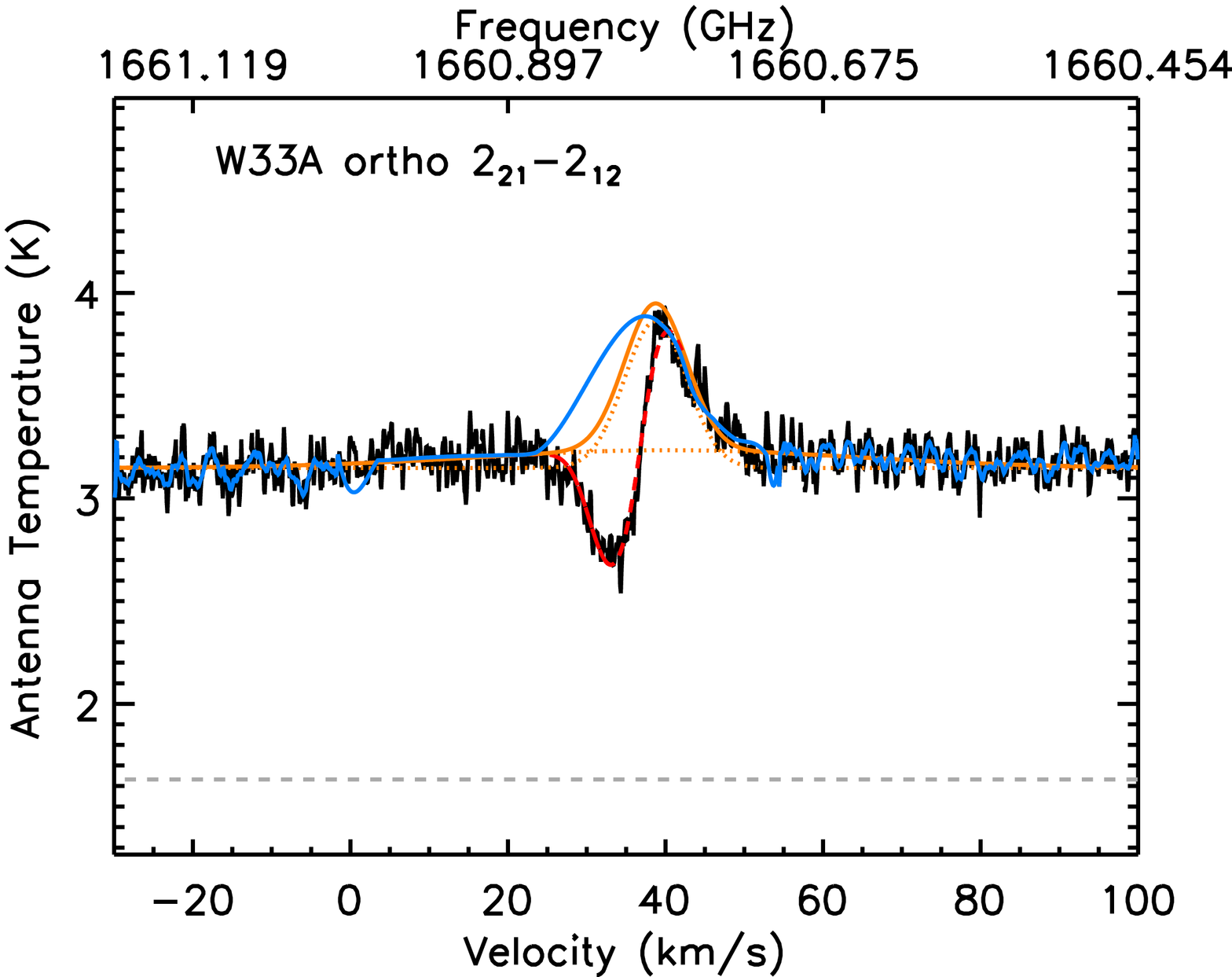}}
  \subfigure[] {\label{}
    \includegraphics[angle=90,width=.31\linewidth]{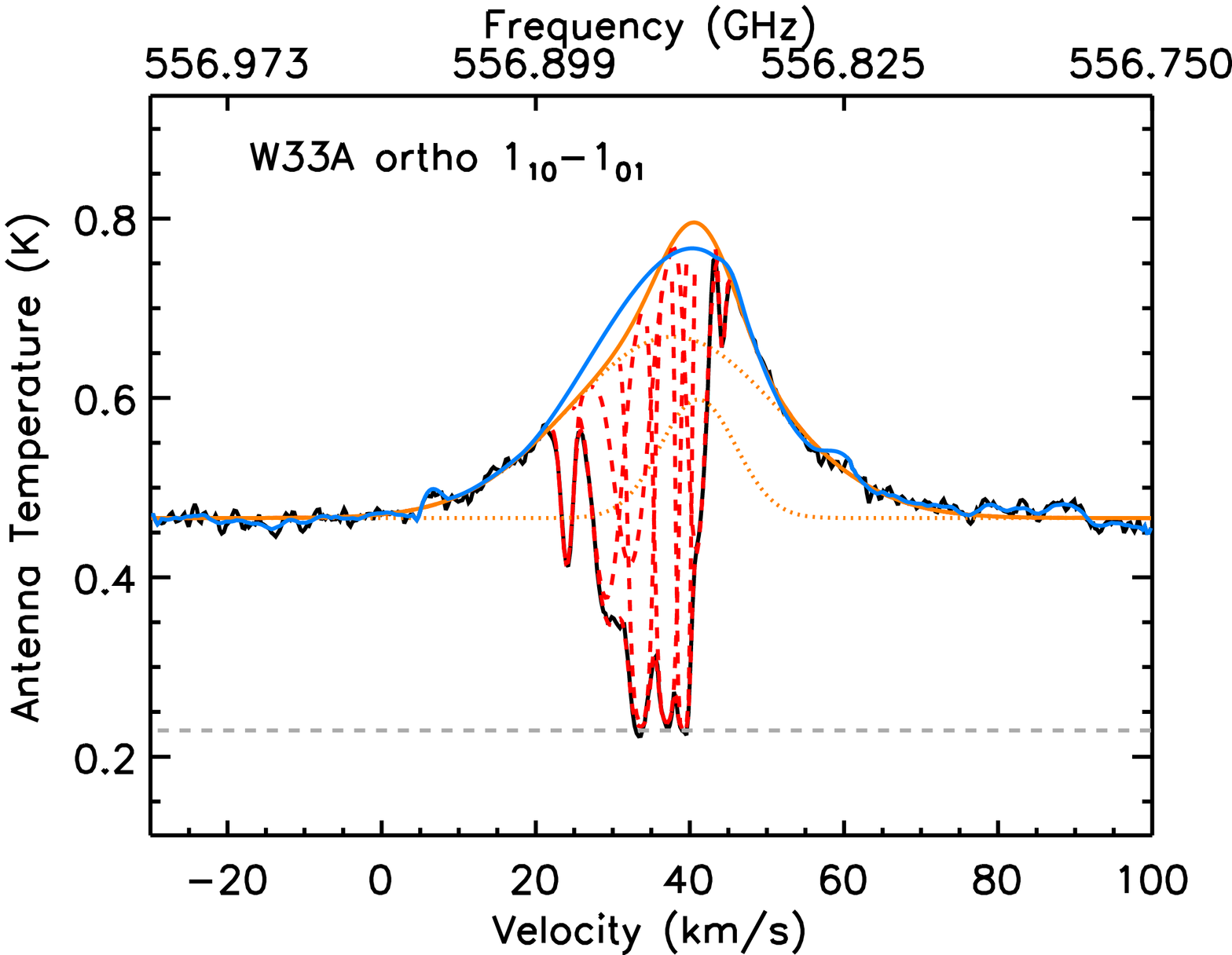}}
  \subfigure[] {\label{}
    \includegraphics[angle=90,width=.31\linewidth]{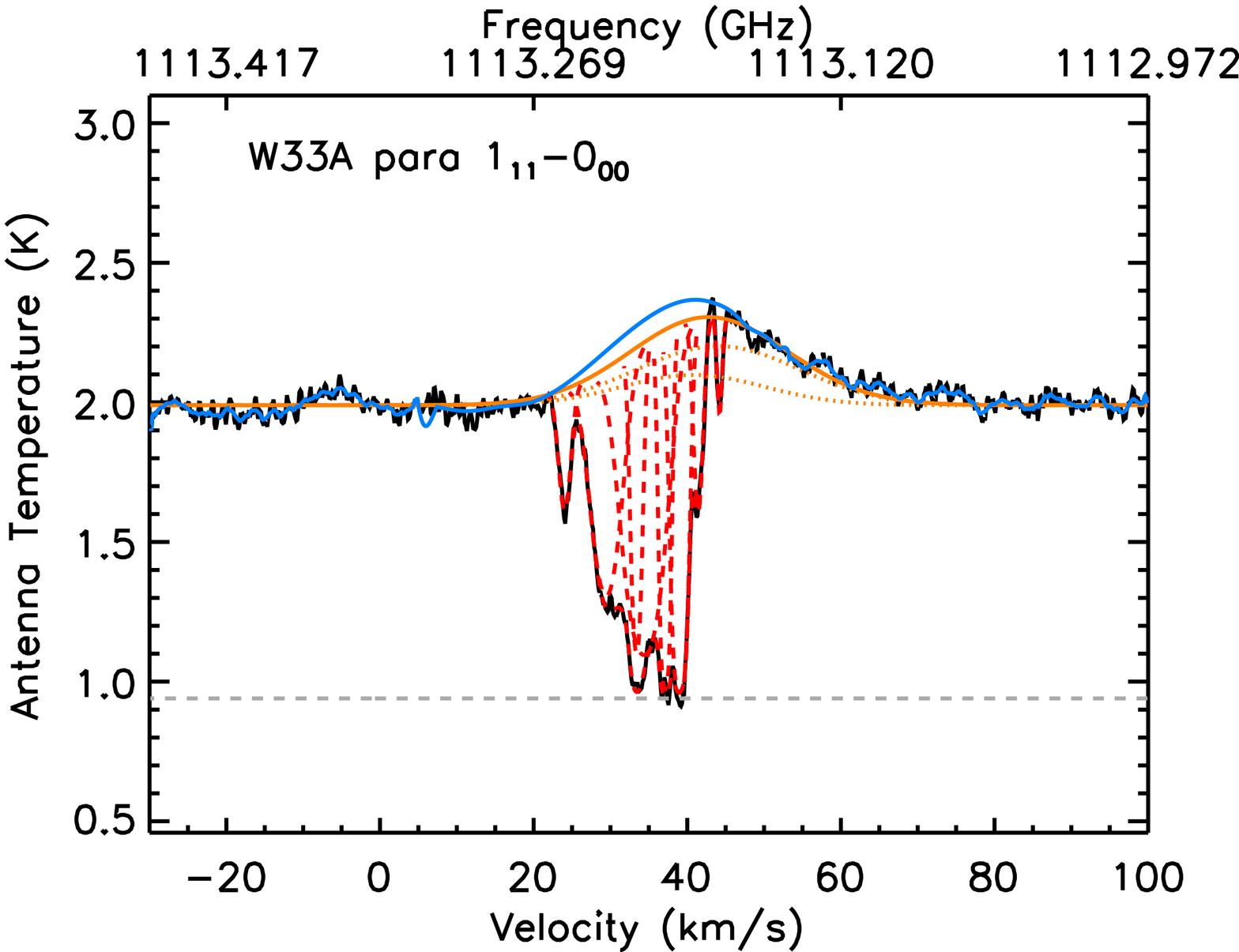}}
  \subfigure[] {\label{}
    \includegraphics[angle=90,width=.31\linewidth]{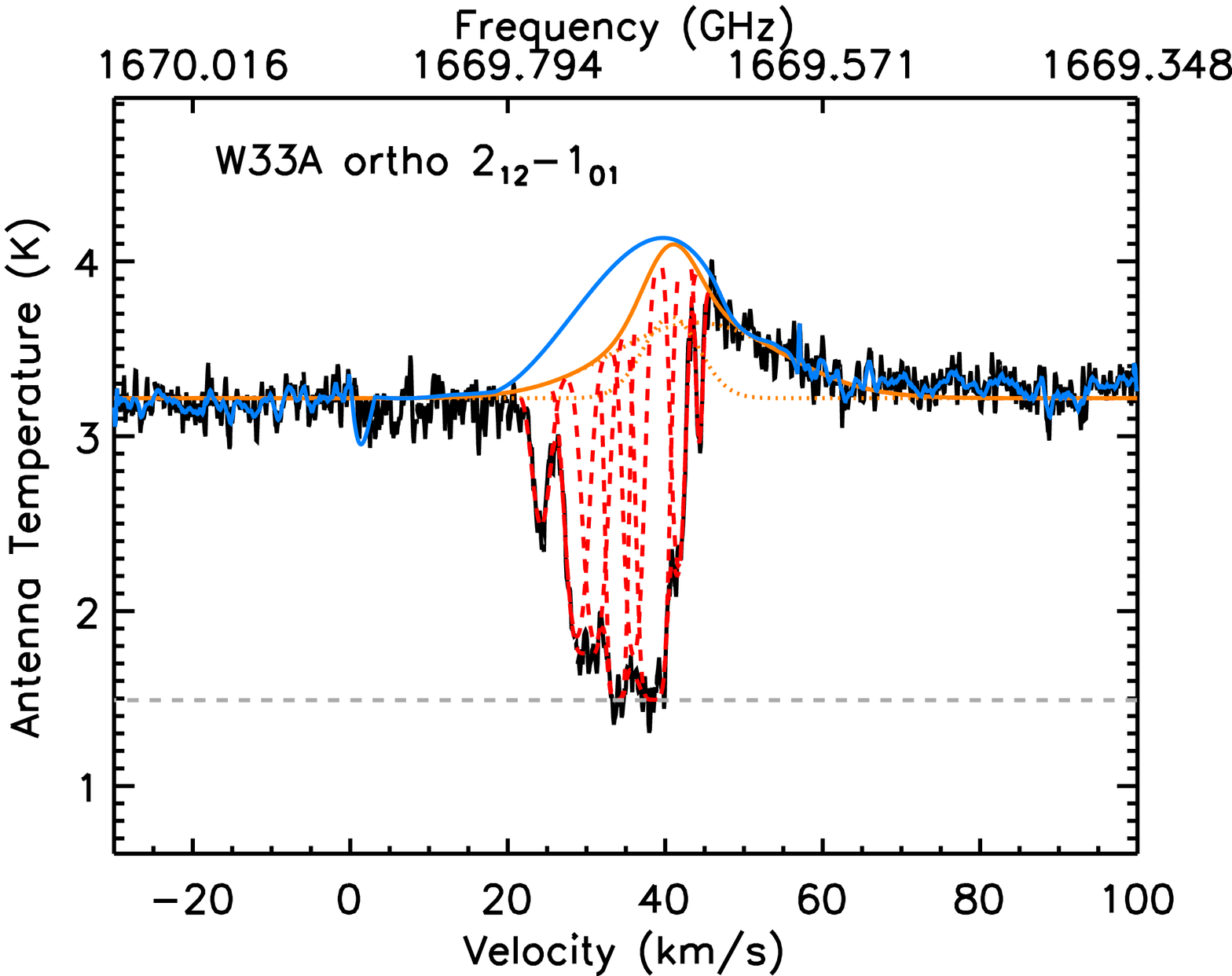}}
  \subfigure[] {\label{}
    \includegraphics[angle=90,width=.31\linewidth]{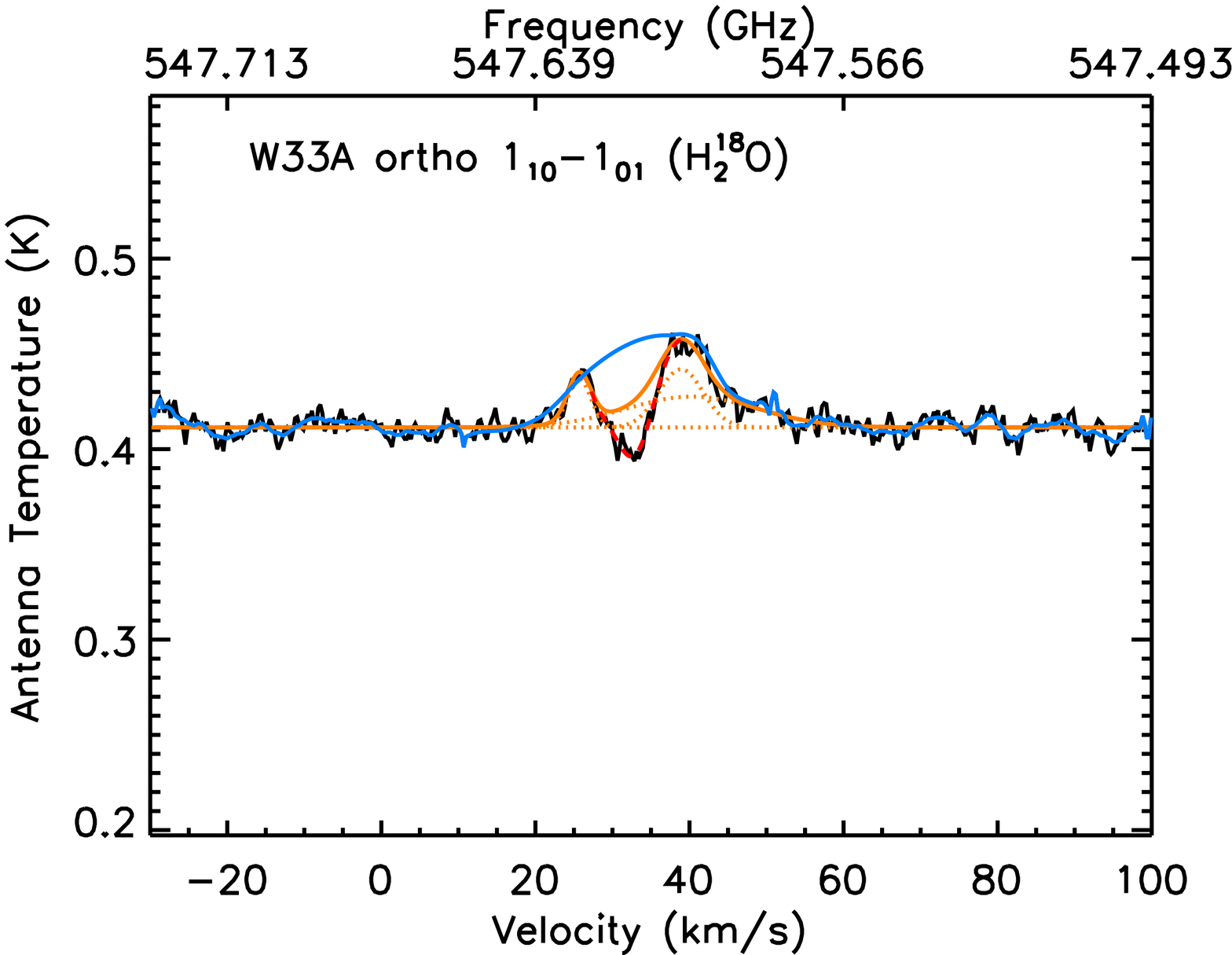}}
  \subfigure[] {\label{}
    \includegraphics[angle=90,width=.31\linewidth]{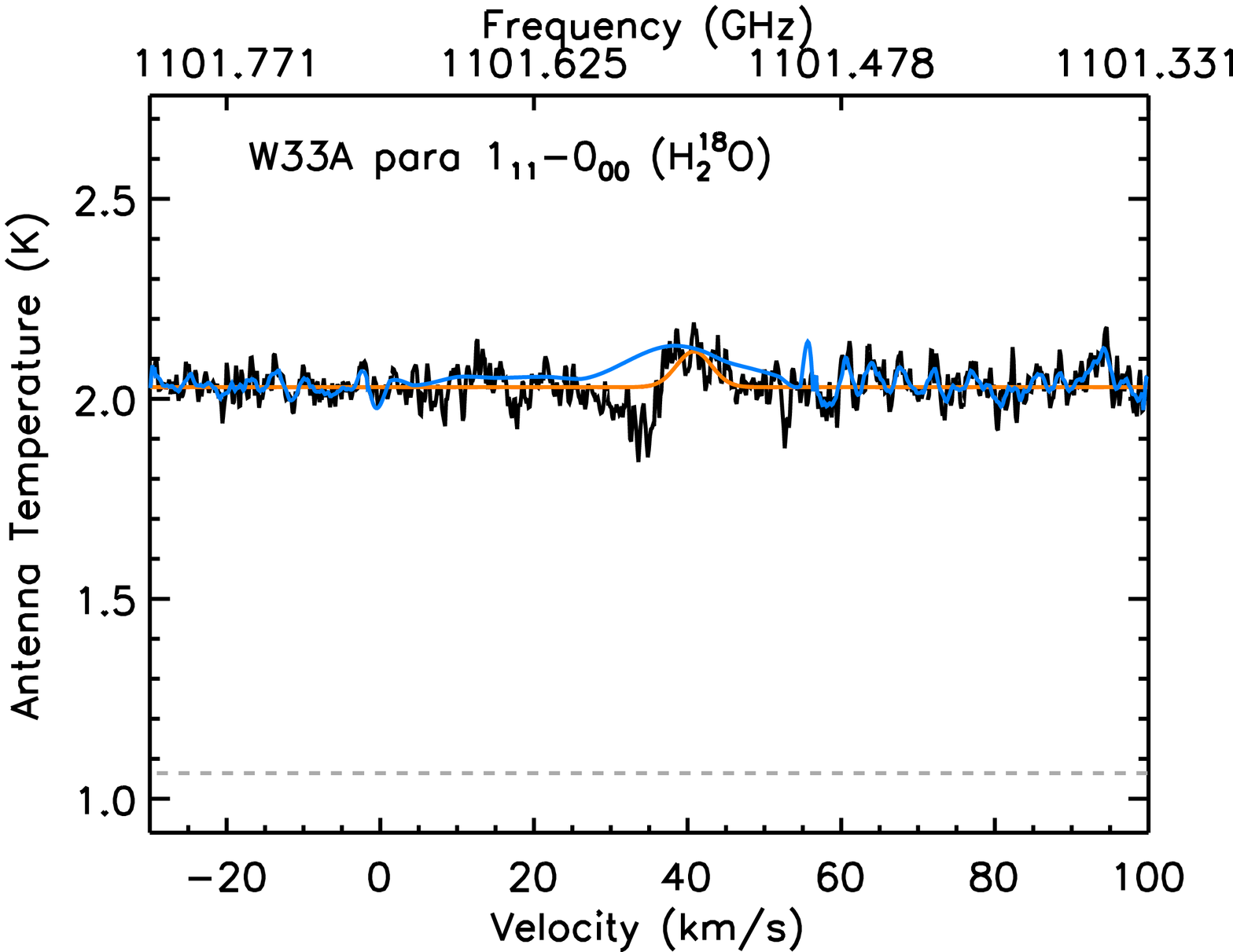}}
  \subfigure[] {\label{}
    \includegraphics[angle=90,width=.31\linewidth]{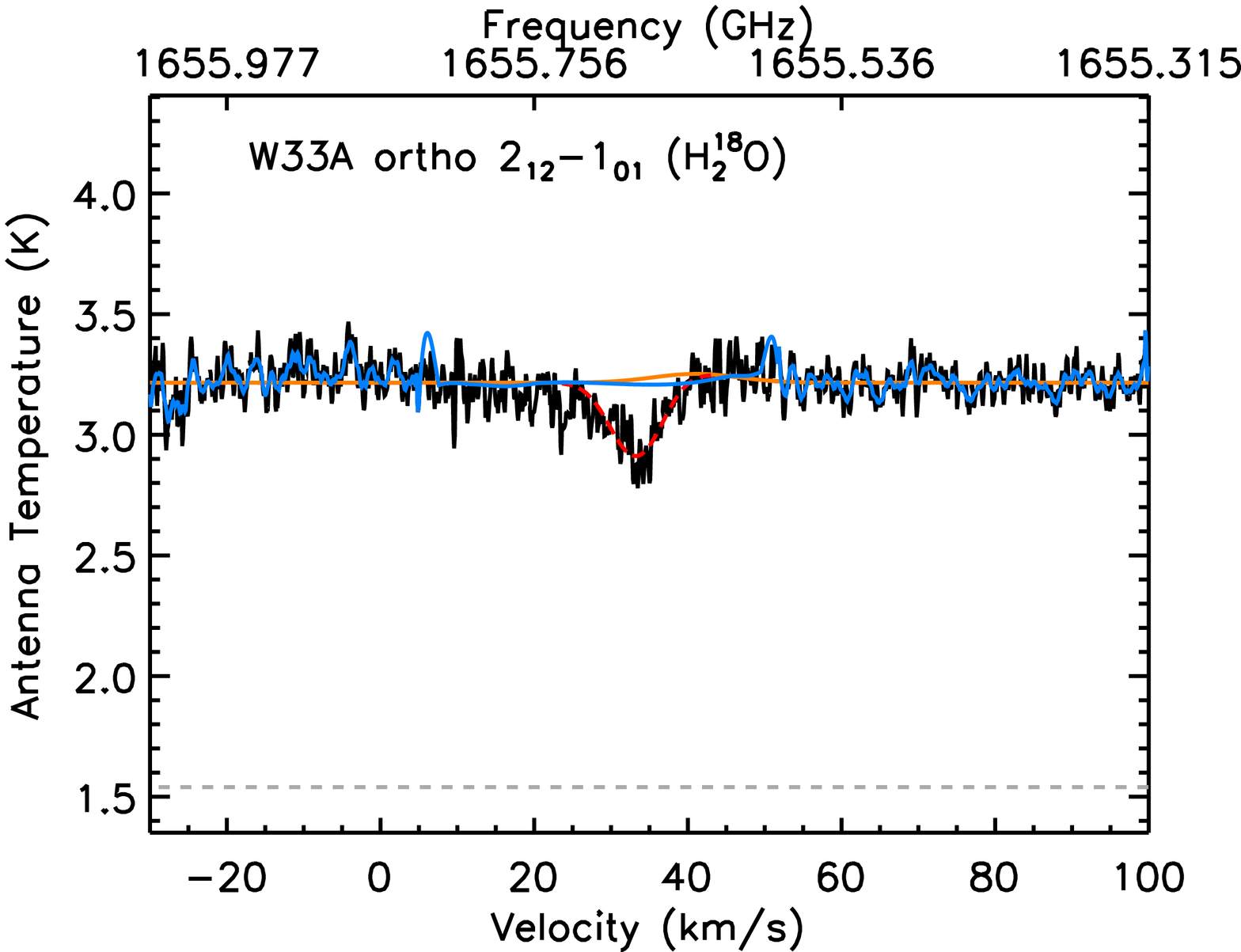}}
  \caption{Same as Figure \ref{fig:w51} but towards W33(A).}
  \label{fig:w33}
\end{figure*}

\begin{figure*}[!t]
  \centering \subfigure[] {\label{}
    \includegraphics[angle=90,width=.31\linewidth]{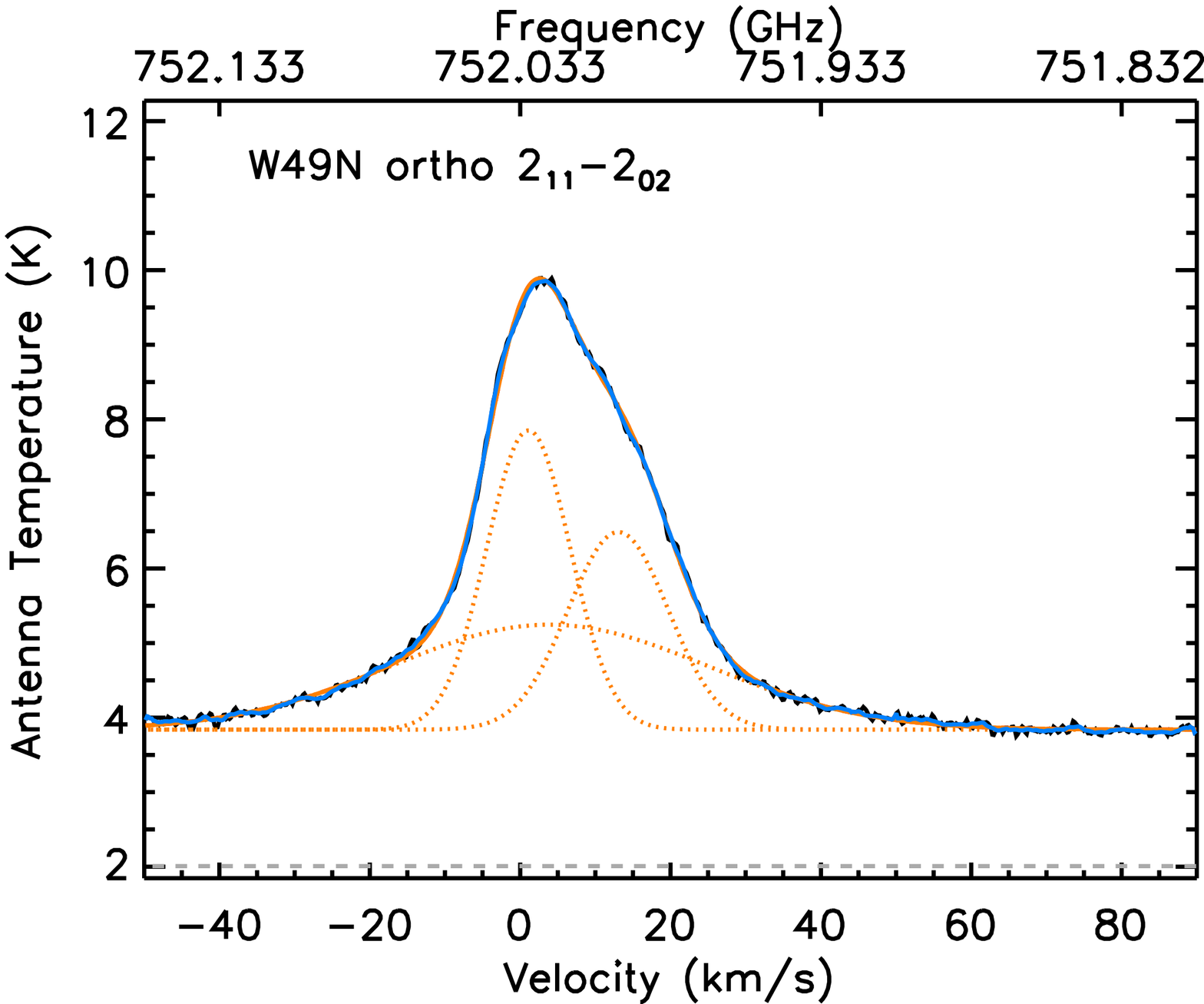}}
  \subfigure[] {\label{}
    \includegraphics[angle=90,width=.31\linewidth]{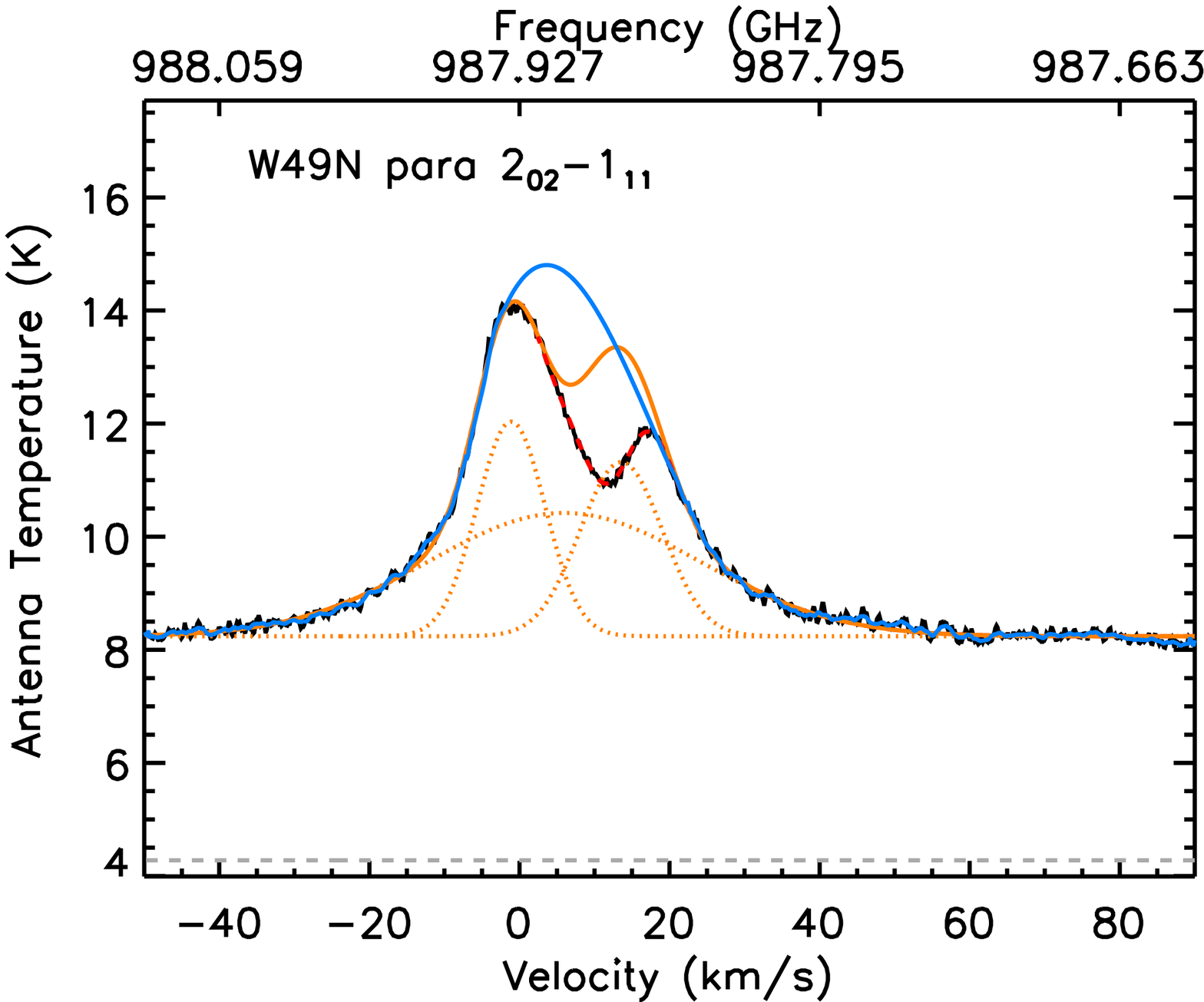}}
  \subfigure[] {\label{}
    \includegraphics[angle=90,width=.31\linewidth]{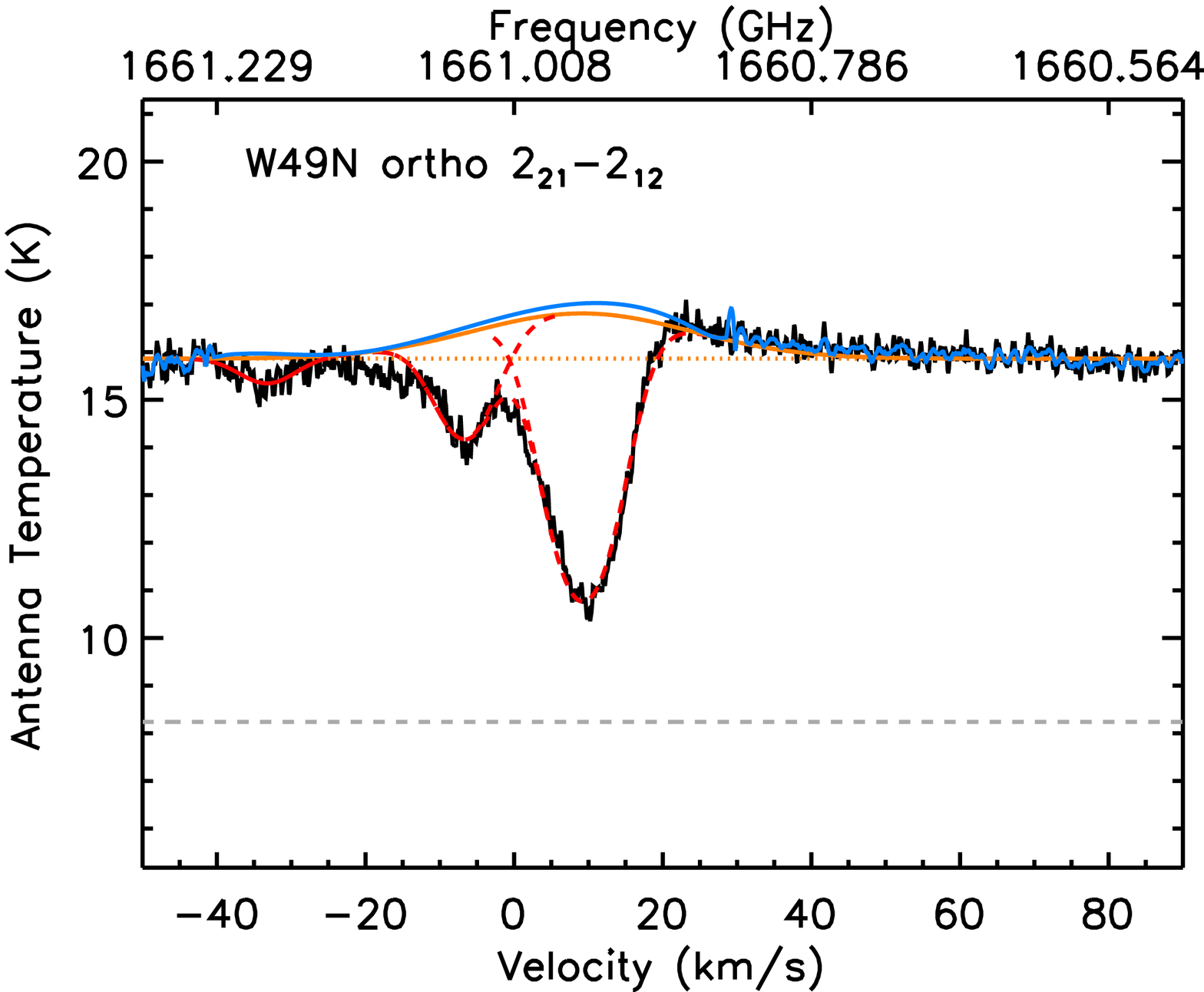}}
  \subfigure[] {\label{}
    \includegraphics[angle=90,width=.31\linewidth]{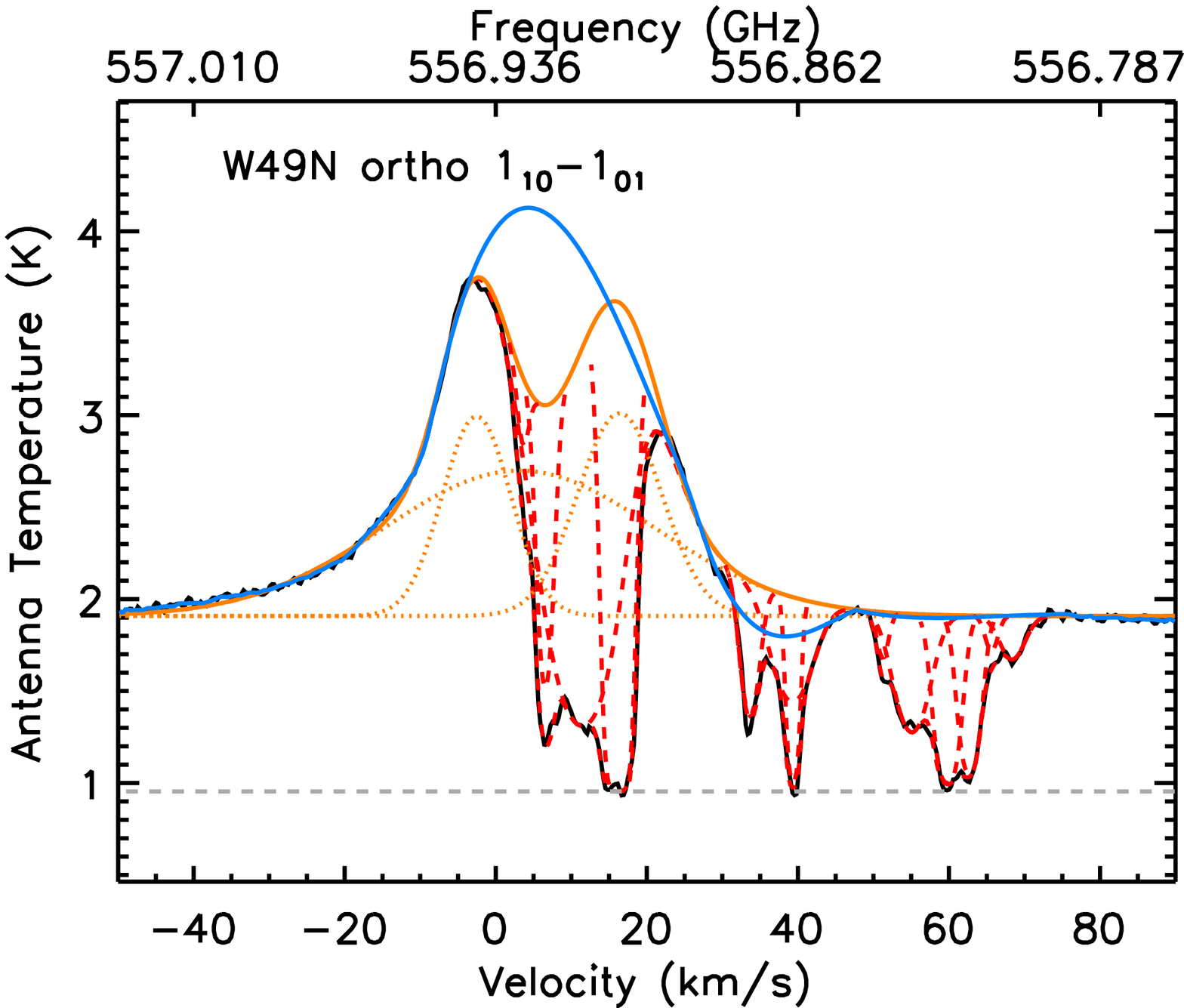}}
  \subfigure[] {\label{}
    \includegraphics[angle=90,width=.31\linewidth]{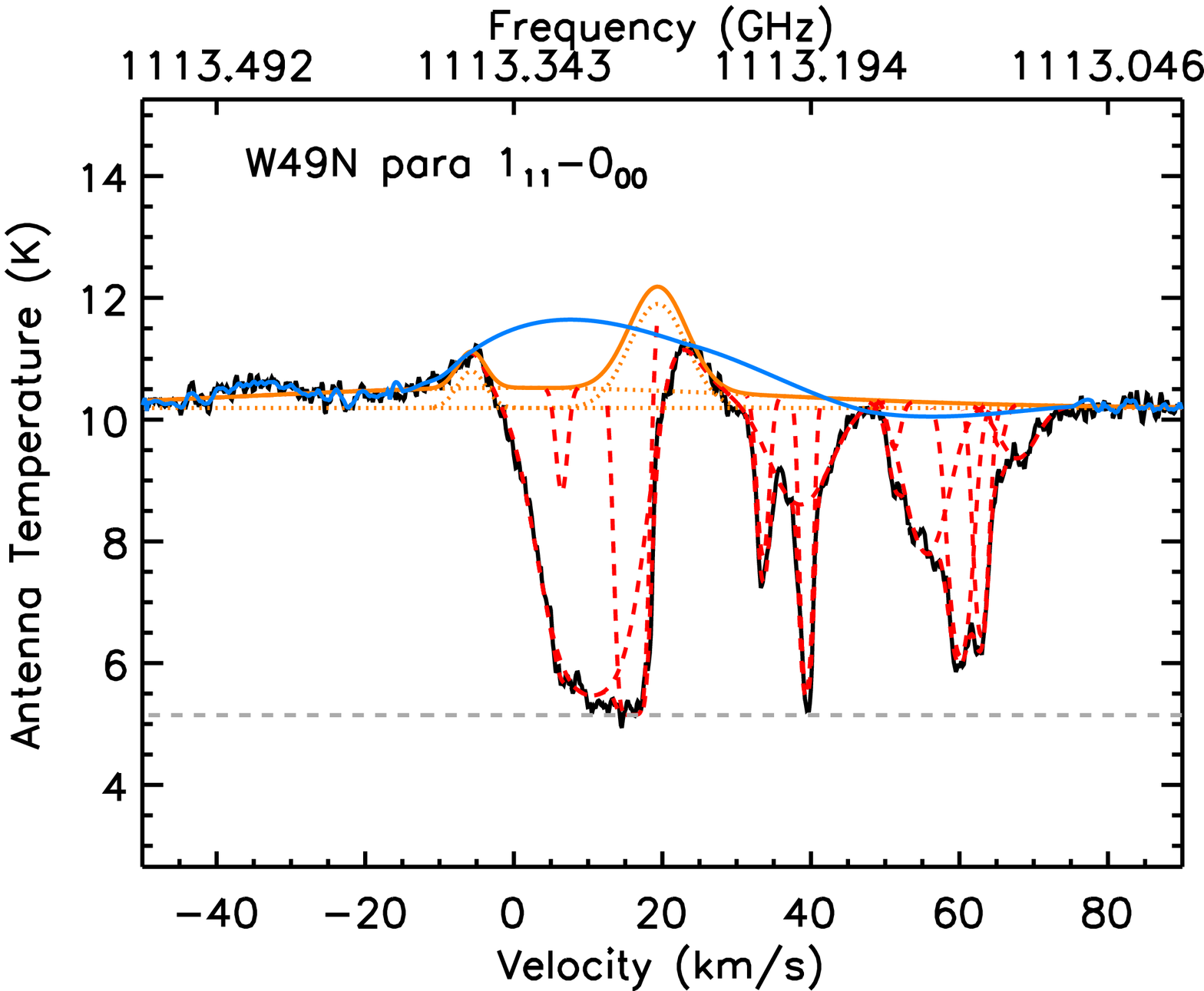}}
  \subfigure[] {\label{}
    \includegraphics[angle=90,width=.31\linewidth]{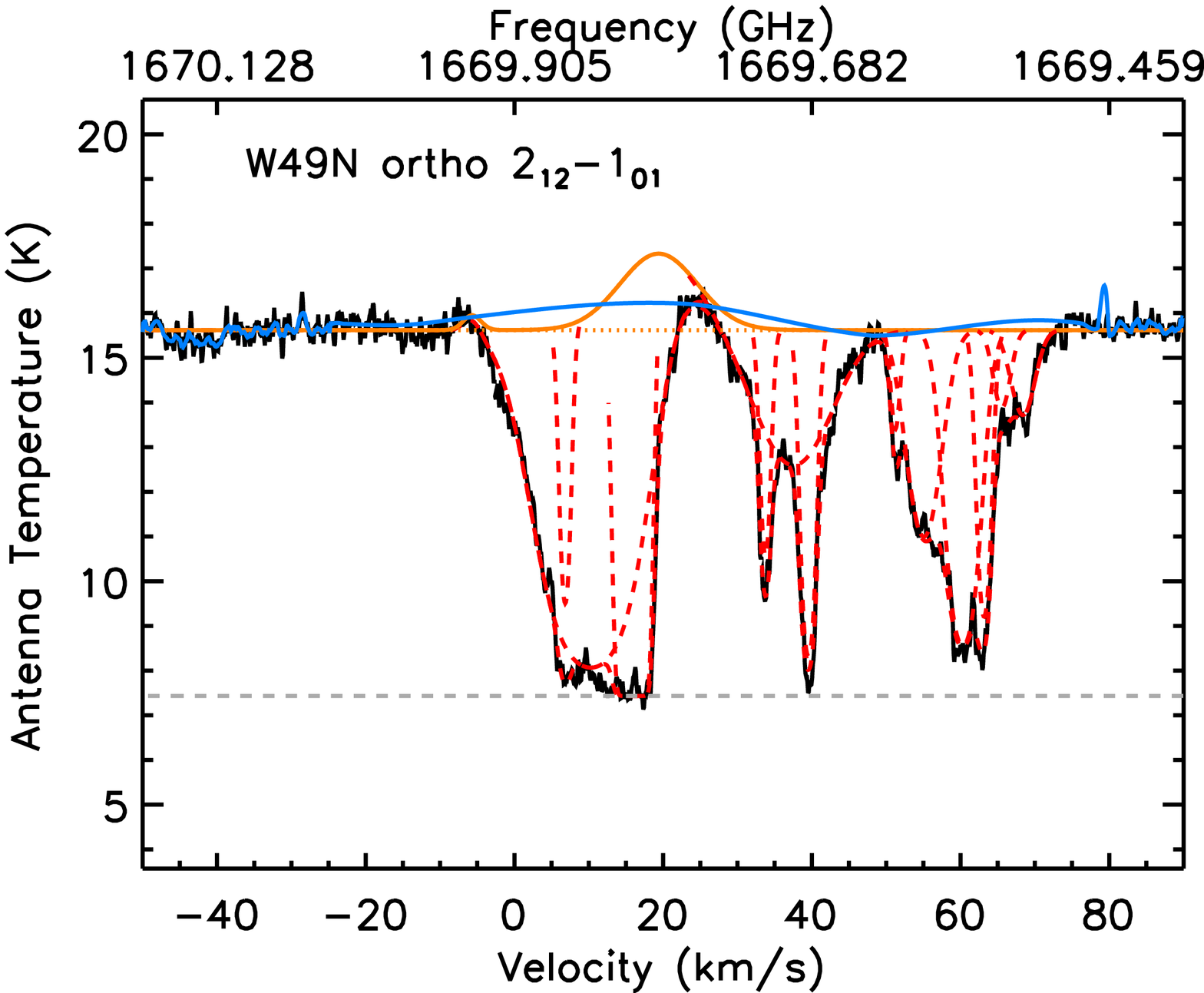}}
  \subfigure[] {\label{}
    \includegraphics[angle=90,width=.31\linewidth]{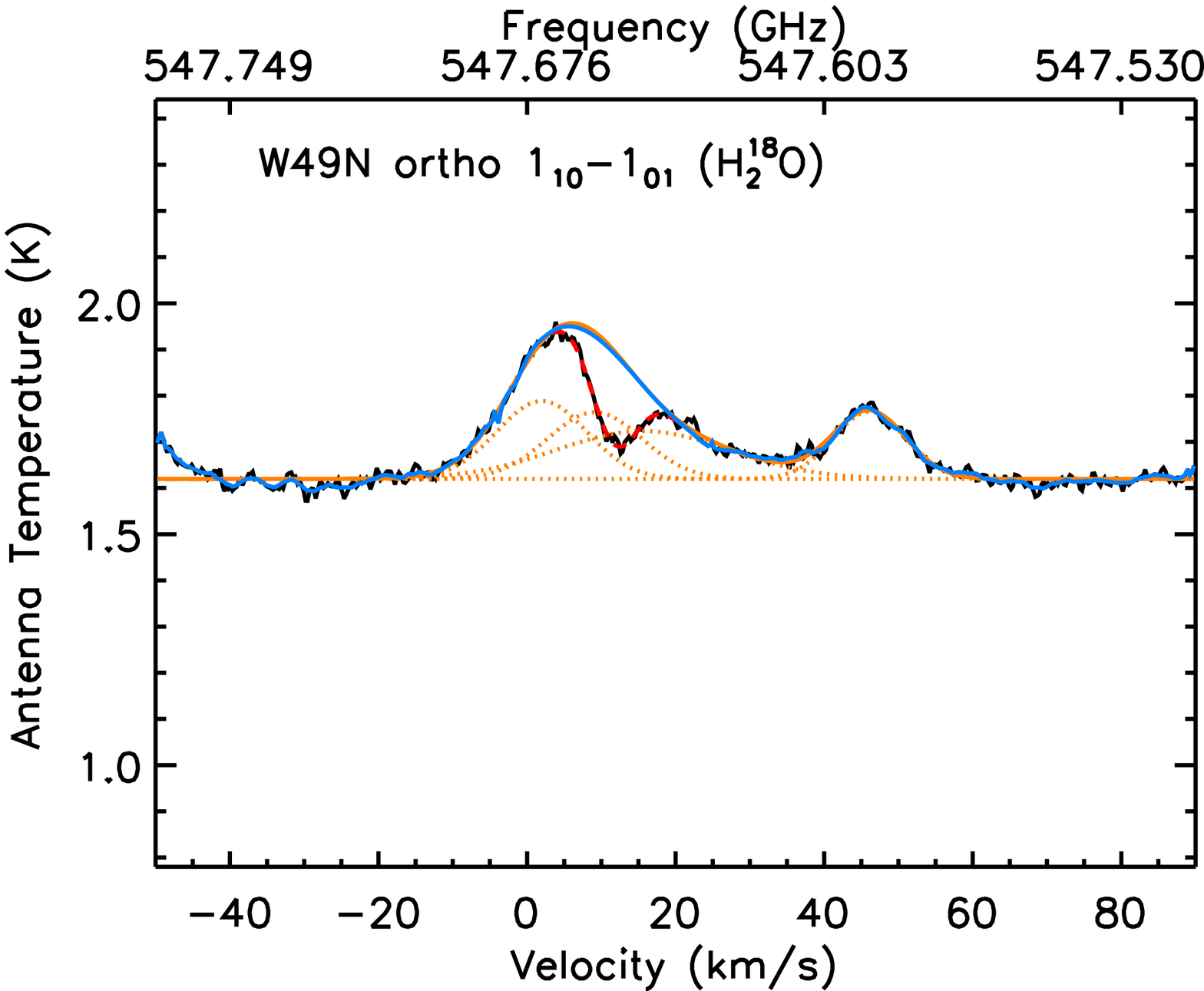}}
  \subfigure[] {\label{}
    \includegraphics[angle=90,width=.31\linewidth]{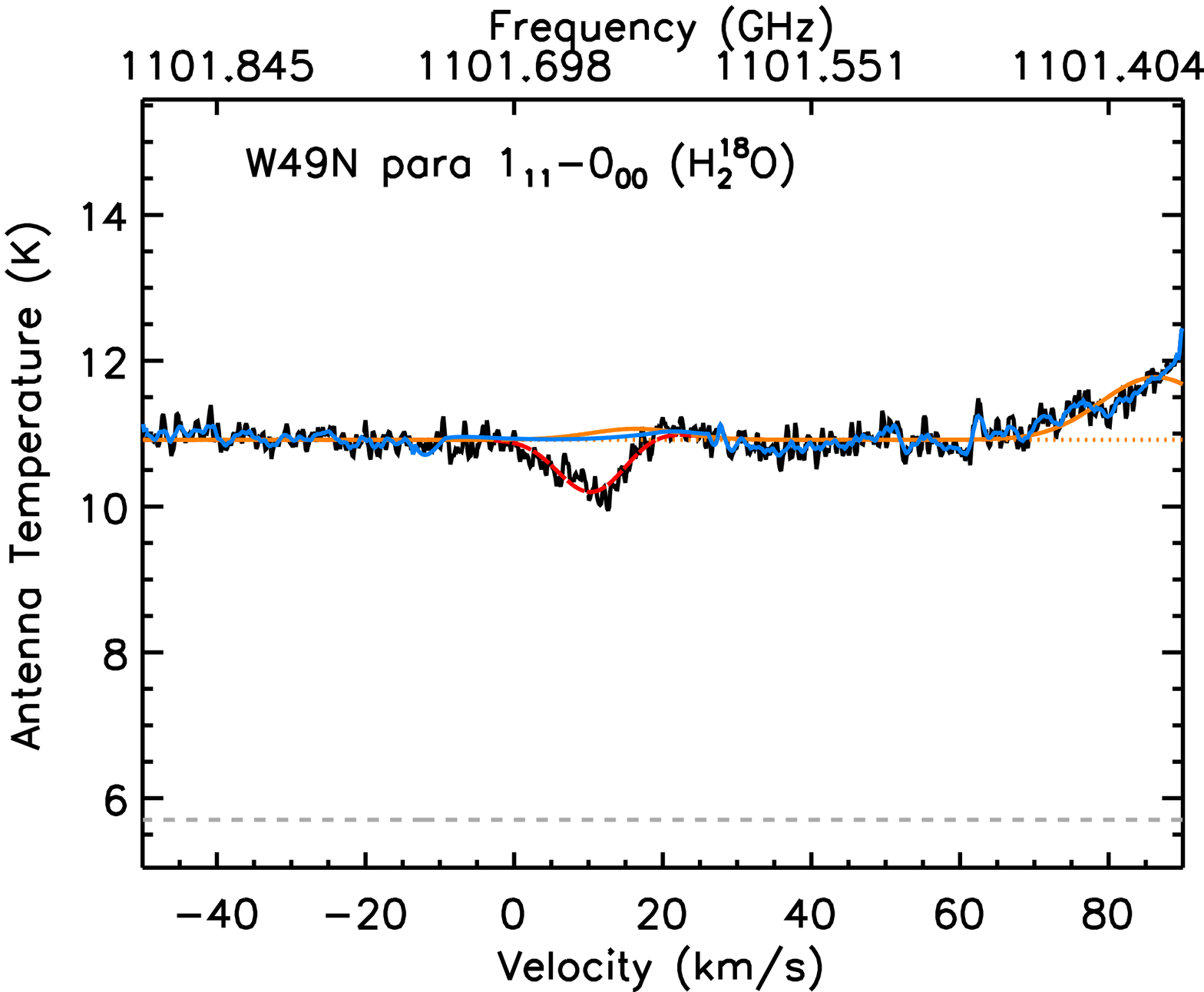}}
  \subfigure[] {\label{}
    \includegraphics[angle=90,width=.31\linewidth]{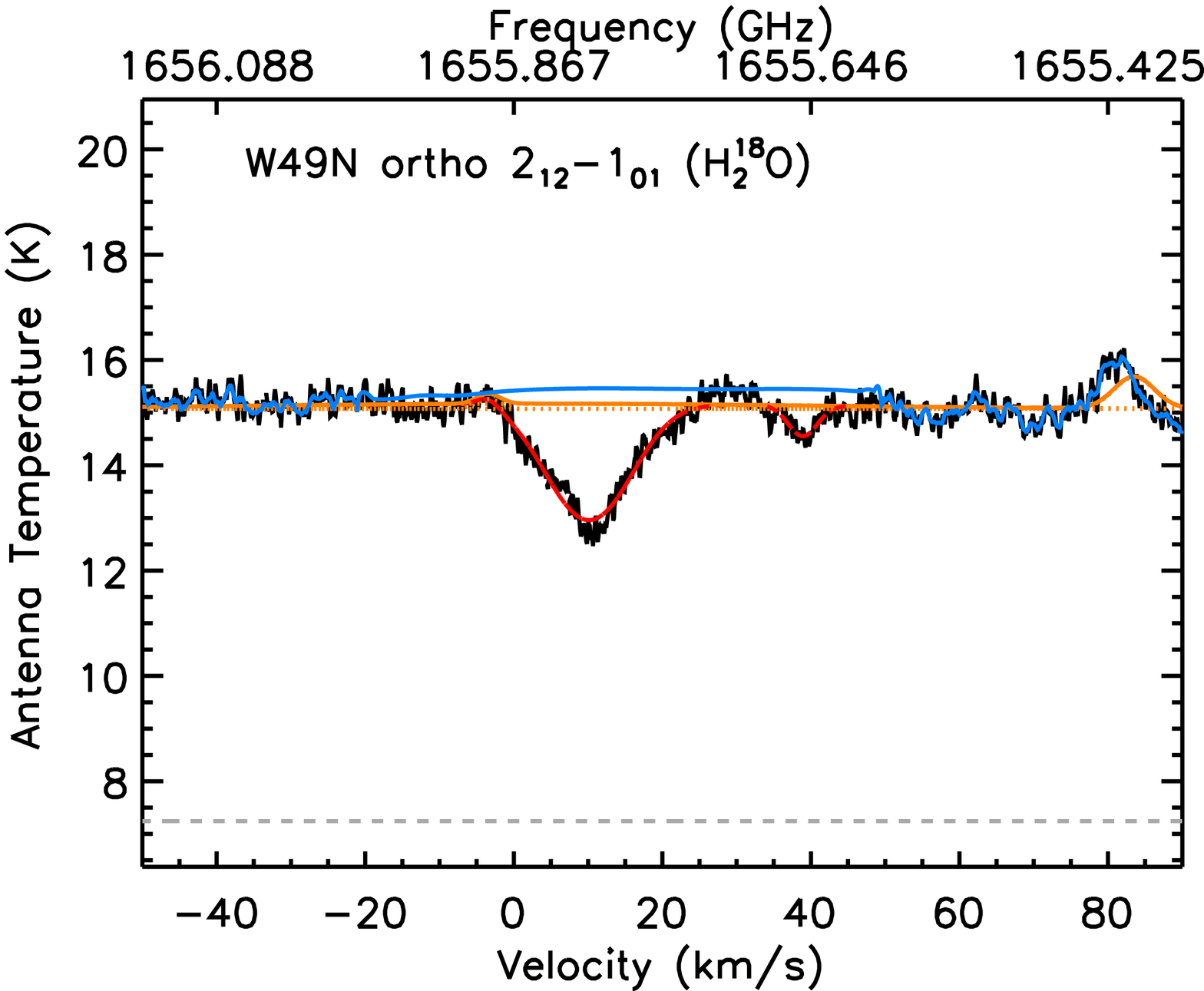}}
  \caption{Same as Figure \ref{fig:w51} but towards W49(N).}
  \label{fig:w49}
\end{figure*}

\clearpage

\section{Results}
In this section of the appendix, we give the column densities of \hho, \hheo, and \hh\ for each absorption feature, the inferred abundance, and OPR, for all the PRISMAS sources except W51, which is presented in the main part of the paper.

\begin{table}[!h]
  \centering
  \caption{Same as Table \ref{tab:w51gg} but towards DR21(OH).}
  \begin{tabular}{l c c c c c c}
    \hline
    \hline
    Transition & $v_0$ & FWHM & $\tau_0$ & $N_l($\hho$)$\\
    (GHz) & (km~s$^{-1}$) & (km~s$^{-1}$) & & ($\times 10^{12}~\rm{cm^{-2}}$)\\
    \hline
    987~GHz &    -1.10 &     3.44 &    0.714 &     15 $\pm$     6 \\
    \hline
    1661~GHz &    -1.67 &     4.28 &    0.846 &     51 $\pm$     1 \\
    \hline
    556~GHz &    -2.48 &     3.21 &     4.99 &     75 $\pm$    1 \\
    &     7.72 &     2.90 &     1.12 &     14.8 $\pm$    0.5 \\
    &     9.21 &     2.21 &     2.75 &     29 $\pm$     1 \\
    &     12.2 &     1.61 &    0.380 &     2.9 $\pm$   0.1 \\
    &     14.5 &     3.58 &    0.393 &     6.4 $\pm$    0.3 \\
    \hline
    1113~GHz &    -2.41 &     3.58 &     4.72 &     40 $\pm$     1 \\
    &     7.87 &     3.14 &    0.946 &     6.7 $\pm$    0.8 \\
    &     9.23 &     2.02 &     1.78 &     8.4 $\pm$   0.1 \\
    &     12.3 &     1.71 &    0.363 &     1.5 $\pm$   0.1 \\
    &     14.9 &     2.07 &    0.281 &     1.4 $\pm$   0.1 \\
    \hline
    1669~GHz &    -2.18 &     3.75 &     5.51 &     96.5 $\pm$    0.4 \\
    &     8.17 &     3.20 &     1.88 &     8 $\pm$     2 \\
    &     9.61 &     1.39 &     3.60 &     25.1 $\pm$    0.5 \\
    &     12.4 &     1.77 &    0.510 &     4.2 $\pm$   0.1 \\
    &     15.3 &     1.84 &    0.436 &     3.8 $\pm$  0.1 \\
    \hline
    547~GHz &    -1.74 &     4.49 &    0.211 &     5 $\pm$     1 \\
    \hline
    1101~GHz &    -2.90 &     4.32 &    0.220 &     2.3 $\pm$    0.4 \\
    \hline
    1655~GHz &    -3.24 &     5.22 &    0.435 &     10.6 $\pm$    0.5 \\
    \hline
  \end{tabular}
  \label{tab:dr21gg}
\end{table}

\begin{table}[t]
  \centering
  \caption{Same as Table \ref{tab:w51gg} but towards G34.3+0.1.}
  \begin{tabular}{l c c c c c c}
    \hline
    \hline
    Transition & $v_0$ & FWHM & $\tau_0$ & $N_l($\hho$)$\\
    (GHz) & (km~s$^{-1}$) & (km~s$^{-1}$) & & ($\times 10^{12}~\rm{cm^{-2}}$)\\
    \hline
    987~GHz &     60.7 &     4.62 &    0.254 &     25 $\pm$     6 \\
    \hline
    1661~GHz &     61.7 &     5.65 &    0.588 &     49 $\pm$     4 \\
    \hline
    556~GHz &     11.6 &     2.10 &     2.39 &     23.7 $\pm$    0.3 \\
    &     14.2 &     1.36 &    0.461 &     2.9 $\pm$   0.1 \\
    &     27.2 &     1.19 &     1.58 &     8.8 $\pm$    0.3 \\
    &     27.8 &     3.33 &    0.367 &     5.8 $\pm$    0.1 \\
    &     46.7 &     2.83 &    0.354 &     4.7 $\pm$    0.2 \\
    &     48.5 &     1.74 &     1.53 &     12.3 $\pm$    0.4 \\
    &     52.2 &     2.64 &     2.38 &     29.6 $\pm$    0.4 \\
    &     60.8 &     4.53 &     4.12 &     88 $\pm$     5 \\
    \hline
    1113~GHz &     11.5 &     2.02 &     1.67 &     7.9 $\pm$    0.2 \\
    &     14.3 &     1.19 &    0.315 &    0.9 $\pm$  0.1 \\
    &     27.1 &     1.04 &     1.33 &     3.2 $\pm$   0.1 \\
    &     28.0 &     4.28 &    0.231 &     2.5 $\pm$    0.3 \\
    &     46.1 &     4.51 &    0.306 &     3.3 $\pm$   0.1 \\
    &     48.5 &     1.80 &    0.918 &     3.8 $\pm$    0.2 \\
    &     52.4 &     3.37 &     1.66 &     13.0 $\pm$    0.4 \\
    &     60.6 &     5.62 &     3.49 &     47 $\pm$     2 \\
    \hline
    1669~GHz &     11.7 &     1.98 &     2.53 &     24 $\pm$     1 \\
    &     14.3 &     1.19 &    0.504 &     2.8 $\pm$    0.2 \\
    &     27.4 &     1.15 &     1.66 &     9.1 $\pm$   0.1 \\
    &     27.9 &     5.46 &    0.251 &     6.7 $\pm$    0.3 \\
    &     45.5 &     4.56 &    0.291 &     8 $\pm$     1 \\
    &     48.7 &     2.16 &     1.22 &     11 $\pm$     3 \\
    &     52.6 &     3.20 &     2.46 &     38 $\pm$     2 \\
    &     60.7 &     6.58 &     3.57 &     110 $\pm$     4 \\
    \hline
    547~GHz &     60.6 &     2.68 &    0.124 &     3 $\pm$    1 \\
    \hline
    1101~GHz &     61.4 &     2.97 &    0.201 &     1.4 $\pm$    0.2 \\
    \hline
    1655~GHz &     61.6 &     10.9 &    0.255 &     12 $\pm$     2 \\
    \hline
  \end{tabular}
  \label{tab:g34gg}
\end{table}

\begin{table}[t]
  \centering
  \caption{Same as Table \ref{tab:w51gg} but towards W28(A).}
  \begin{tabular}{l c c c c c c}
    \hline
    \hline
    Transition & $v_0$ & FWHM & $\tau_0$ & $N_l($\hho$)$\\
    (GHz) & (km~s$^{-1}$) & (km~s$^{-1}$) & & ($\times 10^{12}~\rm{cm^{-2}}$)\\
    \hline
    987~GHz &    -1.25 &     12.4 &    0.377 &     43 $\pm$     2 \\
    \hline
    1661~GHz &    -1.05 &     12.8 &    0.695 &     123 $\pm$     2 \\
    \hline
    556~GHz &    -1.35 &     19.8 &    0.394 &     35.6 $\pm$    0.4 \\
    &     1.29 &     9.40 &    0.575 &     25 $\pm$     2 \\
    &     5.97 &     3.82 &    0.990 &     17.5 $\pm$    0.6 \\
    &     12.1 &     2.54 &     2.77 &     32.7 $\pm$    0.1 \\
    &     14.4 &     1.23 &     1.01 &     5.8 $\pm$    0.3 \\
    &     19.7 &     3.08 &     5.92 &     85 $\pm$     1 \\
    &     23.8 &     2.83 &    0.840 &     11.1 $\pm$    0.2 \\
    \hline
    1113~GHz &    -1.36 &     36.3 &    0.469 &     40.9 $\pm$    0.1 \\
    &     1.20 &     10.0 &    0.809 &     19 $\pm$     1 \\
    &     6.12 &     3.49 &    0.896 &     7.3 $\pm$    0.6 \\
    &     12.1 &     2.46 &     1.87 &     10.7 $\pm$    0.2 \\
    &     14.5 &    0.983 &    0.915 &     2.1 $\pm$   0.1 \\
    &     19.6 &     2.86 &     4.42 &     30 $\pm$     1 \\
    &     23.6 &     3.41 &    0.627 &     4.9 $\pm$    0.4 \\
    \hline
    1669~GHz &    -1.43 &     38.1 &    0.556 &     94 $\pm$     5 \\
    &     1.22 &     9.84 &    0.971 &     48 $\pm$     5 \\
    &     6.05 &     3.92 &     1.37 &     25.2 $\pm$    0.5 \\
    &     12.2 &     2.52 &     3.06 &     37 $\pm$     1 \\
    &     14.7 &    0.885 &     1.39 &     5.8 $\pm$    0.1 \\
    &     19.7 &     2.77 &     7.94 &     103 $\pm$     3 \\
    &     23.7 &     3.17 &    0.897 &     13.4 $\pm$    0.3 \\
    \hline
    547~GHz &    -1.27 &     23.9 &    0.147 &     17 $\pm$     1 \\
    \hline
    1101~GHz &    -2.11 &     23.9 &    0.149 &     8.2 $\pm$    0.9 \\
    \hline
    1655~GHz &    -1.79 &     23.9 &    0.379 &     42.0 $\pm$    0.1 \\
    \hline
  \end{tabular}
  \label{tab:w28gg}
\end{table}

\begin{table}[t]
  \centering
  \caption{Same as Table \ref{tab:w51gg} but towards W33(A).}
  \begin{tabular}{l c c c c c c}
    \hline
    \hline
    Transition & $v_0$ & FWHM & $\tau_0$ & $N_l($\hho$)$\\
    (GHz) & (km~s$^{-1}$) & (km~s$^{-1}$) & & ($\times 10^{12}~\rm{cm^{-2}}$)\\
    \hline
    987~GHz &     34.6 &     3.90 &    0.276 &     10.0 $\pm$    0.3 \\
    \hline
    1661~GHz &     33.6 &     6.96 &    0.392 &     46 $\pm$     7 \\
    \hline
    556~GHz &     24.0 &     1.61 &    0.698 &     5.3 $\pm$    0.3 \\
    &     29.5 &     3.92 &     1.28 &     22 $\pm$     5 \\
    &     32.5 &     4.31 &    0.882 &     17.2 $\pm$    0.5 \\
    &     33.7 &     2.17 &     3.97 &     40 $\pm$    1 \\
    &     37.0 &     2.14 &     4.10 &     40.9 $\pm$    0.8 \\
    &     39.2 &     1.21 &     6.26 &     35 $\pm$     3 \\
    &     40.9 &     1.97 &     1.03 &     8 $\pm$     3 \\
    &     44.2 &    0.821 &    0.218 &    0.8 $\pm$   0.1 \\
    \hline
    1113~GHz &     24.1 &     1.69 &    0.512 &     2.0 $\pm$   0.1 \\
    &     29.6 &     3.93 &     1.24 &     11.5 $\pm$    0.9 \\
    &     33.5 &     1.34 &     2.08 &     6 $\pm$     1 \\
    &     34.4 &     4.78 &     2.10 &     23 $\pm$     2 \\
    &     37.1 &     1.20 &     3.68 &     10.3 $\pm$    0.9 \\
    &     39.0 &     1.74 &     4.04 &     16.3 $\pm$    0.3 \\
    &     41.4 &     1.41 &    0.703 &     2.4 $\pm$    0.2 \\
    &     44.2 &    0.731 &    0.295 &    0.5 $\pm$   0.1 \\
    \hline
    1669~GHz &     24.3 &     2.01 &    0.576 &     4 $\pm$     1 \\
    &     29.3 &     2.48 &     1.73 &     19.3 $\pm$    0.3 \\
    &     31.0 &     6.45 &    0.960 &     21.8 $\pm$    0.3 \\
    &     33.9 &     3.54 &     3.26 &     58 $\pm$     2 \\
    &     37.5 &     2.14 &     3.39 &     33 $\pm$     6 \\
    &     39.4 &     1.87 &     2.88 &     20 $\pm$     13 \\
    &     41.6 &     1.75 &    0.996 &     8 $\pm$     2 \\
    &     44.3 &    0.827 &    0.537 &     2.2 $\pm$   0.1 \\
    \hline
    547~GHz &     33.1 &     4.40 &    0.135 &     3.1 $\pm$   0.1 \\
    \hline
    1101~GHz &     34.3 &     2.62 &   0.0552 &    0.7 $\pm$    0.7 \\
    \hline
    1655~GHz &     33.3 &     7.00 &    0.208 &     7.4 $\pm$    0.2 \\
    \hline
  \end{tabular}
  \label{tab:w33gg}
\end{table}

\begin{table}[t]
  \centering
  \caption{Same as Table \ref{tab:w51gg} but towards W49(N).}
  \begin{tabular}{l c c c c c c}
    \hline
    \hline
    Transition & $v_0$ & FWHM & $\tau_0$ & $N_l($\hho$)$\\
    (GHz) & (km~s$^{-1}$) & (km~s$^{-1}$) & & ($\times 10^{12}~\rm{cm^{-2}}$)\\
    \hline
    987~GHz &     11.9 &     7.78 &    0.398 &     26 $\pm$     4 \\
    \hline
    1661~GHz &    9.39 &     10.4 &     1.22 &     175 $\pm$     2 \\
    \hline
    556~GHz &     3.71 &     1.87 &    0.195 &     1.9 $\pm$    0.4 \\
    &     6.59 &     2.15 &     1.34 &     13.5 $\pm$    0.4 \\
    &     11.5 &     9.13 &     1.84 &     71 $\pm$     9 \\
    &     16.1 &     3.08 &     5.06 &     88 $\pm$     5 \\
    &     33.6 &     2.54 &     1.02 &     11.8 $\pm$    0.5 \\
    &     39.2 &     5.93 &    0.692 &     20.3 $\pm$    0.5 \\
    &     39.4 &     1.49 &     3.08 &     21 $\pm$    1 \\
    &     51.3 &     1.76 &    0.279 &     2.3 $\pm$   0.1 \\
    &     55.1 &     4.71 &     1.07 &     23.4 $\pm$    0.7 \\
    &     59.8 &     2.88 &     2.86 &     39 $\pm$     2 \\
    &     62.8 &     2.66 &     2.21 &     27.4 $\pm$    0.9 \\
    &     65.8 &     1.66 &    0.221 &     1.7 $\pm$    0.2 \\
    &     68.3 &     3.62 &    0.275 &     4.6 $\pm$    0.3 \\
    \hline
    1113~GHz &     6.22 &     1.07 &    0.538 &     2.0 $\pm$    0.3 \\
    &     10.6 &     10.8 &     2.66 &     68 $\pm$     3 \\
    &     15.6 &     2.75 &     9.51 &     61 $\pm$     30 \\
    &     33.5 &     1.67 &    0.675 &     2.6 $\pm$    0.2 \\
    &     38.3 &     9.81 &    0.405 &     9.3 $\pm$    0.7 \\
    &     39.4 &     1.47 &     2.62 &     9.1 $\pm$    0.8 \\
    &     51.2 &     1.29 &    0.194 &    0.6 $\pm$   0.1 \\
    &     55.4 &     6.30 &    0.675 &     9.9 $\pm$    0.3 \\
    &     60.1 &     3.19 &     1.73 &     12 $\pm$    1 \\
    &     62.9 &     1.89 &     1.36 &     5.9 $\pm$    0.5 \\
    &     65.2 &     2.42 &    0.182 &     1.0 $\pm$    0.7 \\
    &     68.3 &     4.52 &    0.202 &     2.2 $\pm$    0.2 \\
    \hline
    1669~GHz &     6.70 &     1.39 &    0.764 &     6 $\pm$     3 \\
    &     10.5 &     11.8 &     2.53 &     138 $\pm$     5 \\
    &     16.1 &     2.53 &     17.5 &     202 $\pm$     47 \\
    &     33.8 &     1.55 &    0.916 &     6.7 $\pm$   0.1 \\
    &     37.6 &     11.2 &    0.459 &     23.8 $\pm$    0.2 \\
    &     39.5 &     1.84 &     2.15 &     18.5 $\pm$    0.3 \\
    &     51.2 &     1.42 &    0.308 &     2.1 $\pm$   0.1 \\
    &     55.0 &     4.75 &    0.836 &     18.5 $\pm$  0.1 \\
    &     60.3 &     4.00 &     1.99 &     37 $\pm$    1 \\
    &     63.2 &     1.90 &     1.47 &     13.1 $\pm$    0.3 \\
    &     65.7 &     2.02 &    0.255 &     2.5 $\pm$    0.3 \\
    &     68.6 &     3.49 &    0.273 &     4.4 $\pm$ 0.1 \\
    \hline
    547~GHz &     12.0 &     7.26 &    0.222 &     7.1 $\pm$    0.5 \\
    \hline
    1101~GHz &     10.5 &     9.68 &    0.126 &     3.1 $\pm$    0.6 \\
    \hline
    1655~GHz &     10.3 &     12.0 &    0.357 &     20.0 $\pm$    0.5 \\
    \hline
  \end{tabular}
  \label{tab:w49gg}
\end{table}

\begin{table*}[t]
  \centering
  \caption{Same as Table \ref{tab:W51gg_nh2o} but towards DR21(OH).}
  \begin{tabular}{l | c c | c c c c c c c c}
    \hline
    \hline
    $v$ & $N($\hho$)$ & OPR & $v$ & $N($\hho$)$ & $f_{g}($\hho$)$ & OPR$^+$ & OPR$^g$ & $N($\hh$)$ & X(\hho) & $N($\hheo$)$\\
    (km~s$^{-1}$) & & & & &\\
    \hline
    8 &    34 $\pm$   2  &   4.1 $\pm$ 0.8 & \multirow{4}{*}{4 to 18} & \multirow{4}{*}{72$_{-  2}^{+  7}$} & \multirow{4}{*}{$>$0.93} & \multirow{4}{*}{2.8$\pm$0.5} & \multirow{4}{*}{2.9$\pm$0.3} & \multirow{4}{*}{15.2$\pm$1.0} & \multirow{4}{*}{4.7$_{-0.4}^{+0.8}$} & \multirow{4}{*}{$<$2.5} \\
    9 &    33.5 $\pm$   0.5  &   3.0 $\pm$ 0.1 & \\
    12 &    5.7 $\pm$   0.1  &   2.9 $\pm$ 0.1 & \\
    15 &    5.1 $\pm$   0.1  &   2.8 $\pm$ 0.1 & \\
    \hline
  \end{tabular}
  \tablecomments{The water column densities $N($\hho$)$ are in $10^{12}~\rm{cm^{-2}}$.}
  \label{tab:DR21gg_nh2o}
\end{table*}

\begin{table*}[t]
  \centering
  \caption{Same as Table \ref{tab:W51gg_nh2o} but towards G34.3+0.1.}
  \begin{tabular}{l | c c | c c c c c c c c}
    \hline
    \hline
    $v$ & $N($\hho$)$ & OPR & $v$ & $N($\hho$)$ & $f_{g}($\hho$)$ & OPR$^+$ & OPR$^g$ & $N($\hh$)$ & X(\hho) & $N($\hheo$)$\\
    (km~s$^{-1}$) & & & & &\\
    \hline
    12 &    32 $\pm$   1  &   3.1 $\pm$ 0.2 & \multirow{2}{*}{8 to 16} & \multirow{2}{*}{36.1$_{-  0.7}^{+  8}$} & \multirow{2}{*}{$>$0.84} & \multirow{2}{*}{2.8$\pm$0.8} & \multirow{2}{*}{3.0$\pm$0.2} & \multirow{2}{*}{4.7$\pm$0.6} & \multirow{2}{*}{7.7$_{-1.1}^{+2.7}$} & \multirow{2}{*}{$<$0.3} \\
    14 &    3.7 $\pm$   0.2  &   3.2 $\pm$ 0.2 & \\
    \hline
    27 &    12.3 $\pm$   0.1  &   2.8 $\pm$ 0.1 & \multirow{2}{*}{22 to 34} & \multirow{2}{*}{22.5$_{-  0.8}^{+ 9}$} & \multirow{2}{*}{$>$0.73} & \multirow{2}{*}{3$\pm$1} & \multirow{2}{*}{2.9$\pm$0.3} & \multirow{2}{*}{5.3$\pm$0.6} & \multirow{2}{*}{4.2$_{-0.6}^{+2.2}$} & \multirow{2}{*}{$<$0.7} \\
    28 &    9.2 $\pm$   0.4  &   2.6 $\pm$ 0.4 & \\
    \hline
    46 &    11.2 $\pm$   1.3  &   2.4 $\pm$ 0.4 & \multirow{2}{*}{42 to 55} & \multirow{2}{*}{78$_{-  2}^{+ 30}$} & \multirow{2}{*}{$>$0.73} & \multirow{2}{*}{3.2$\pm$0.9} & \multirow{2}{*}{2.9$\pm$0.1} & \multirow{2}{*}{$>$10.4} & \multirow{2}{*}{$<$10.4} & \multirow{2}{*}{$<$3.4} \\
    49 &    14 $\pm$   3  &   2.8 $\pm$ 0.8 & \\
    52 &    51 $\pm$   2  &   2.9 $\pm$ 0.2 & \multicolumn{8}{c}{Group of absorption features partly contaminated by the source.} \\
    \hline
  \end{tabular}
  \tablecomments{The water column densities $N($\hho$)$ are in $10^{12}~\rm{cm^{-2}}$.}
  \label{tab:G34gg_nh2o}
\end{table*}

\begin{table*}[t]
  \centering
  \caption{Same as Table \ref{tab:W51gg_nh2o} but towards W28(A).}
  \begin{tabular}{l | c c | c c c c c c c c}
    \hline
    \hline
    $v$ & $N($\hho$)$ & OPR & $v$ & $N($\hho$)$ & $f_{g}($\hho$)$ & OPR$^+$ & OPR$^g$ & $N($\hh$)$ & X(\hho) & $N($\hheo$)$\\
    (km~s$^{-1}$) & & & & &\\
    \hline
    1.0 &    67 $\pm$   5  &   2.6 $\pm$ 0.4 & \multicolumn{8}{c}{Group of absorption features contaminated by the source.} \\
    6.1 &    32.6 $\pm$   0.8  &   3.4 $\pm$ 0.3 & \\
    \hline
    12 &    47 $\pm$   1  &   3.4 $\pm$ 0.2 & \multirow{2}{*}{9.5 to 16} & \multirow{2}{*}{73$_{-  2}^{+ 7}$} & \multirow{2}{*}{$>$0.93} & \multirow{2}{*}{3.0$\pm$0.4} & \multirow{2}{*}{3.0$\pm$0.2} & \multirow{2}{*}{-} & \multirow{2}{*}{-} & \multirow{2}{*}{$<$4.7} \\
    15 &    7.9 $\pm$   0.1  &   2.8 $\pm$ 0.1 & \\
    \hline
    20 &    133 $\pm$   3  &   3.4 $\pm$ 0.2 & \multirow{3}{*}{16 to 28} & \multirow{3}{*}{141$_{-  13}^{+  16}$} & \multirow{3}{*}{$>$0.98} & \multirow{3}{*}{3.3$\pm$0.6} & \multirow{3}{*}{3.4$\pm$0.5} & \multirow{3}{*}{-} & \multirow{3}{*}{-} & \multirow{3}{*}{$<$2.4} \\
    24 &   18.3 $\pm$   0.5  &   2.7 $\pm$ 0.3 & \\
    \hline
  \end{tabular}
  \tablecomments{The water column densities $N($\hho$)$ are in $10^{12}~\rm{cm^{-2}}$.}
  \label{tab:W28gg_nh2o}
\end{table*}

\begin{table*}[t]
  \centering
  \caption{Same as Table \ref{tab:W51gg_nh2o} but towards W33(A).}
  \begin{tabular}{l | c c | c c c c c c c c}
    \hline
    \hline
    $v$ & $N($\hho$)$ & OPR & $v$ & $N($\hho$)$ & $f_{g}($\hho$)$ & OPR$^+$ & OPR$^g$ & $N($\hh$)$ & X(\hho) & $N($\hheo$)$\\
    (km~s$^{-1}$) & & & & &\\
    \hline
    24 &    6.3 $\pm$   1.3  &   2.1 $\pm$ 0.7 & 21 to 26 & 8.7$_{-0.2}^{+3.2}$ & $>$0.74 & 3$\pm$1 & 2.9$\pm$0.1 & 2.4$\pm$0.8 & 3.6$_{-1.3}^{+2.5}$ & $<$1.2 \\
    \hline
    29 &    31 $\pm$  1  &   1.7 $\pm$ 0.2 & 26 to 31.5 & 44.3$_{-  1.6}^{+ 15}$ & $>$0.75 & 3.3$\pm$0.9 & 3.0$\pm$0.2 & 9.2$\pm$1.2 & 4.8$_{-0.8}^{+2.3}$ & $<$2.5 \\
    \hline
    37 &   44 $\pm$  6  &  3.2 $\pm$ 0.9 & \\
    39 &   37 $\pm$  14  & 1.3 $\pm$ 0.9 & \multicolumn{8}{c}{Group of absorption features contaminated by the source.} \\
    41 &   10 $\pm$  2  &  3.4 $\pm$ 1.0 & \\
    \hline
    44 &   2.7 $\pm$  0.1  &   4.3 $\pm$ 0.2 & 43.2 to 45.5 & 2.9$_{-  0.1}^{+  0.4}$ & $>$0.90 & 4.2$\pm$0.8 & 4.3$\pm$0.3 & - & - & $<$0.2 \\
    \hline
  \end{tabular}
  \tablecomments{The water column densities $N($\hho$)$ are in $10^{12}~\rm{cm^{-2}}$.}
  \label{tab:W33gg_nh2o}
\end{table*}

\begin{table*}[t]
  \centering
  \caption{Same as Table \ref{tab:W51gg_nh2o} but towards W49(N).}
  \begin{tabular}{l | c c | c c c c c c c c}
    \hline
    \hline
    $v$ & $N($\hho$)$ & OPR & $v$ & $N($\hho$)$ & $f_{g}($\hho$)$ & OPR$^+$ & OPR$^g$ & $N($\hh$)$ & X(\hho) & $N($\hheo$)$\\
    (km~s$^{-1}$) & & & & &\\
    \hline
    7 &    8 $\pm$   3  &    3 $\pm$ 2 & \\
    16 &    263 $\pm$   56  &    3 $\pm$ 2 & \multicolumn{8}{c}{Group of absorption features contaminated by the source.} \\
    \hline
    34 &    9.3 $\pm$   0.2  &    2.6 $\pm$ 0.2 & \multirow{3}{*}{26 to 48} & \multirow{3}{*}{73.3$_{-  0.9}^{+  6}$} & \multirow{3}{*}{$>$ 0.94} & \multirow{3}{*}{2.2$\pm$0.3} & \multirow{3}{*}{2.3$\pm$0.1} & \multirow{3}{*}{13.6$\pm$3.1} & \multirow{3}{*}{5$_{-  1}^{+  2}$} & \multirow{3}{*}{$<$1.9} \\
    38 &    33.0 $\pm$   0.7  &    2.6 $\pm$ 0.2 & \\
    39 &    27.6 $\pm$   0.8  &    2.0 $\pm$ 0.2 & \\
    \hline
    51 &    2.7 $\pm$  0.1  &    3.2 $\pm$ 0.2 &  \multirow{6}{*}{48 to 73} & \multirow{6}{*}{112$_{-  3}^{+  7}$} & \multirow{6}{*}{$>$ 0.96} & \multirow{6}{*}{2.3$\pm$0.3} & \multirow{6}{*}{2.4$\pm$0.2} & \multirow{6}{*}{26.3$\pm$3.8} & \multirow{6}{*}{4.3$_{-  0.7}^{+ 0.9}$} & \multirow{6}{*}{$<$1.8} \\
    55 &    28.3 $\pm$   0.3  &    1.9 $\pm$ 0.1 & \\
    60 &    50 $\pm$   1  &    2.9 $\pm$ 0.3 & \\
    63 &    19.0 $\pm$   0.6  &    2.2 $\pm$ 0.2 & \\
    65 &    3.5 $\pm$   0.8  &    3 $\pm$ 2 & \\
    68 &    6.6 $\pm$   0.2  &    2.0 $\pm$ 2.2 & \\
    \hline
  \end{tabular}
  \tablecomments{The water column densities $N($\hho$)$ are in $10^{12}~\rm{cm^{-2}}$.}
  \label{tab:W49gg_nh2o}
\end{table*}

\end{document}